\documentclass[aps,rmp,reprint,amsmath,amssymb,graphicx,longbibliography]{revtex4-1}

\usepackage{graphicx}	
\usepackage{mathtools}
\usepackage{braket}
\usepackage{physics}
\usepackage[table,xcdraw]{xcolor}
\usepackage{amssymb}

\renewcommand{\vec}[1]{\mathbf{#1}}

\begin{document}

\title{Electronic structure of quantum materials studied by angle-resolved photoemission spectroscopy}

\begin{abstract}
The physics of quantum materials is dictated by many-body interactions and mathematical concepts such as symmetry and topology that have transformed our understanding of matter. Angle-resolved photoemission spectroscopy (ARPES), which directly probes the electronic structure in momentum space, has played a central role in the discovery, characterization, and understanding of quantum materials ranging from strongly-correlated states of matter to those exhibiting non-trivial topology. Over the past two decades, ARPES as a technique has matured dramatically with ever-improving resolution and continued expansion into the space-, time-, and spin- domains. Simultaneously, the capability to synthesize new materials and apply non-thermal tuning parameters \emph{in-situ} has unlocked new dimensions in the study of all quantum materials. We review these developments, and survey the scientific contributions they have enabled in contemporary quantum materials research.
\end{abstract}

\author{Jonathan A. Sobota}
\affiliation{Stanford Institute for Materials and Energy Sciences, SLAC National Accelerator Laboratory, 2575 Sand Hill Road, Menlo Park, California 94025, USA}
\author{Yu He}
\affiliation{Stanford Institute for Materials and Energy Sciences, SLAC National Accelerator Laboratory, 2575 Sand Hill Road, Menlo Park, California 94025, USA}
\affiliation{Department of Physics, University of California, Berkeley, California 94720, USA}
\author{Zhi-Xun Shen}
\affiliation{Stanford Institute for Materials and Energy Sciences, SLAC National Accelerator Laboratory, 2575 Sand Hill Road, Menlo Park, California 94025, USA}
\affiliation{Geballe Laboratory for Advanced Materials, Departments of Physics and Applied Physics, Stanford University, Stanford, California 94305, USA}

\maketitle

\tableofcontents

\section{Introduction}

Quantum many-body and relativistic effects are at the heart of modern condensed matter physics: new organizing principles emerge from the collective behavior of a large number of constituents with coupled degrees-of-freedom. As we discover new materials embodying these principles, we are led to new formulations of physics that expand our understanding of matter, providing a pathway to technological paradigms beyond materials like silicon -- where the independent electron approximation works so well -- and raising the prospect of harnessing quantum many-body phenomena for applications. For these reasons, the many-body problem in materials is an extremely rich scientific area of both fundamental and practical pursuit.

\begin{figure}
\centering
\includegraphics[width=\columnwidth]{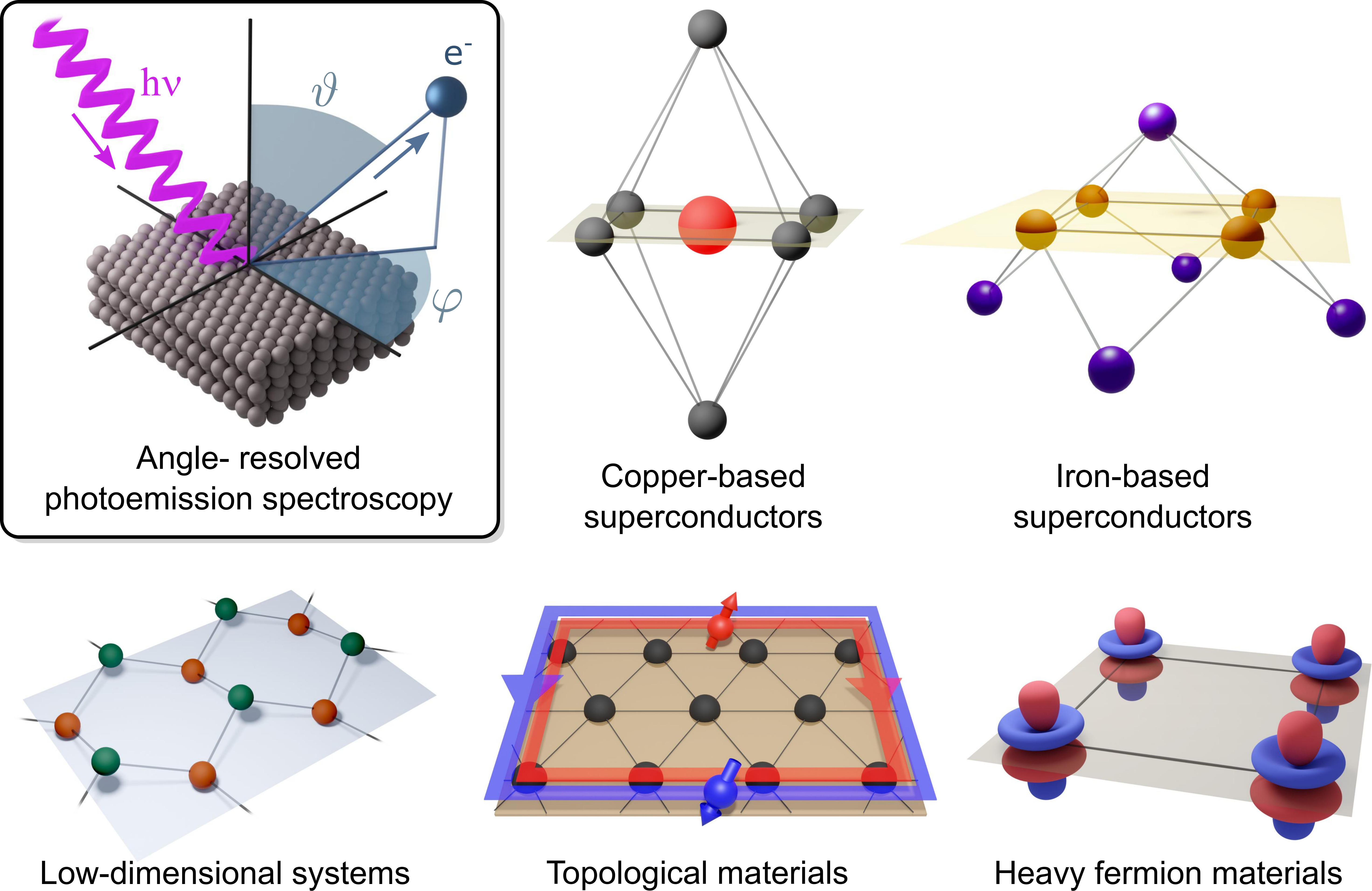}
\caption[]{Top-left: Schematic of angle-resolved photoemission spectroscopy (ARPES). Photons of energy $h\nu$ are used to photoemit electrons into vacuum, where their kinetic energies and emission angles are resolved. Other panels: The scope of quantum materials studied by ARPES, in correspondence to sections in this review (heavy fermion materials are representative of the section on ``other materials''.)
\label{Fig_intro_QMs}}
\end{figure}

 In general a solid state system can be modelled by a Hamiltonian $H$ with its associated eigenvalues and eigenstates. In the simple case where the behavior is dictated by the electron kinetic energy and crystal potential $H_{0}$, the system is well described by the quantum theory of electronic band structure, where the electron wave functions $\psi_\vec{k}(\vec{r})$ are Bloch states with eigenvalues $\epsilon_{\vec{k}}$ representing the electronic band dispersion with respect to wave vector $\vec{k}$~\cite{ashcroft1976solid}. Despite the remarkable success of this theory, as evidenced by the semiconductor revolution it spawned in the middle of the last century \cite{Sze_2006_physics}, for most many-body problems in condensed matter physics such a description is insufficient and/or unsatisfactory. This is because for strongly interacting electron systems, Hamiltonian terms such as $H_{\textrm{e-e}}$ and $H_{\textrm{e-ph}}$ couple electrons to other electrons or to phonons, respectively, and thereby invalidate an independent-electron description \cite{Pines_1966_the,Mahan_2000_many}. This often leads to novel phases featuring surprising phenomena such as high-temperature superconductivity in the cuprates \cite{Bednorz_1988_perovskite} and iron pnictides \cite{kamihara2006iron,kamihara2008iron}. In other cases, interactions such as spin-orbit coupling $H_{\textrm{SOC}}$ encode the electron wave functions with topological properties which are not evident solely from the energy-momentum dispersion $\epsilon_{\vec{k}}$, and require analysis of the geometric phase of $\psi_\vec{k}(\vec{r})$ for a complete understanding \cite{Qi_2011_topological,Hasan_2010_colloquium,Haldane_2017_nobel}. These properties are unusually robust with respect to perturbations, as exemplified by quantized edge conduction in materials exhibiting quantum Hall effects  \cite{Thouless_1982_quantized,Konig_2007_quantum,Chang_2013_experimental,Fei_2017_edge}. The union of these material families have come to be known as \emph{quantum materials.}  Figure~\ref{Fig_intro_QMs} presents an overview of the quantum material families within the scope of this review. 

\begin{figure}
\centering
\includegraphics[width=1\columnwidth]{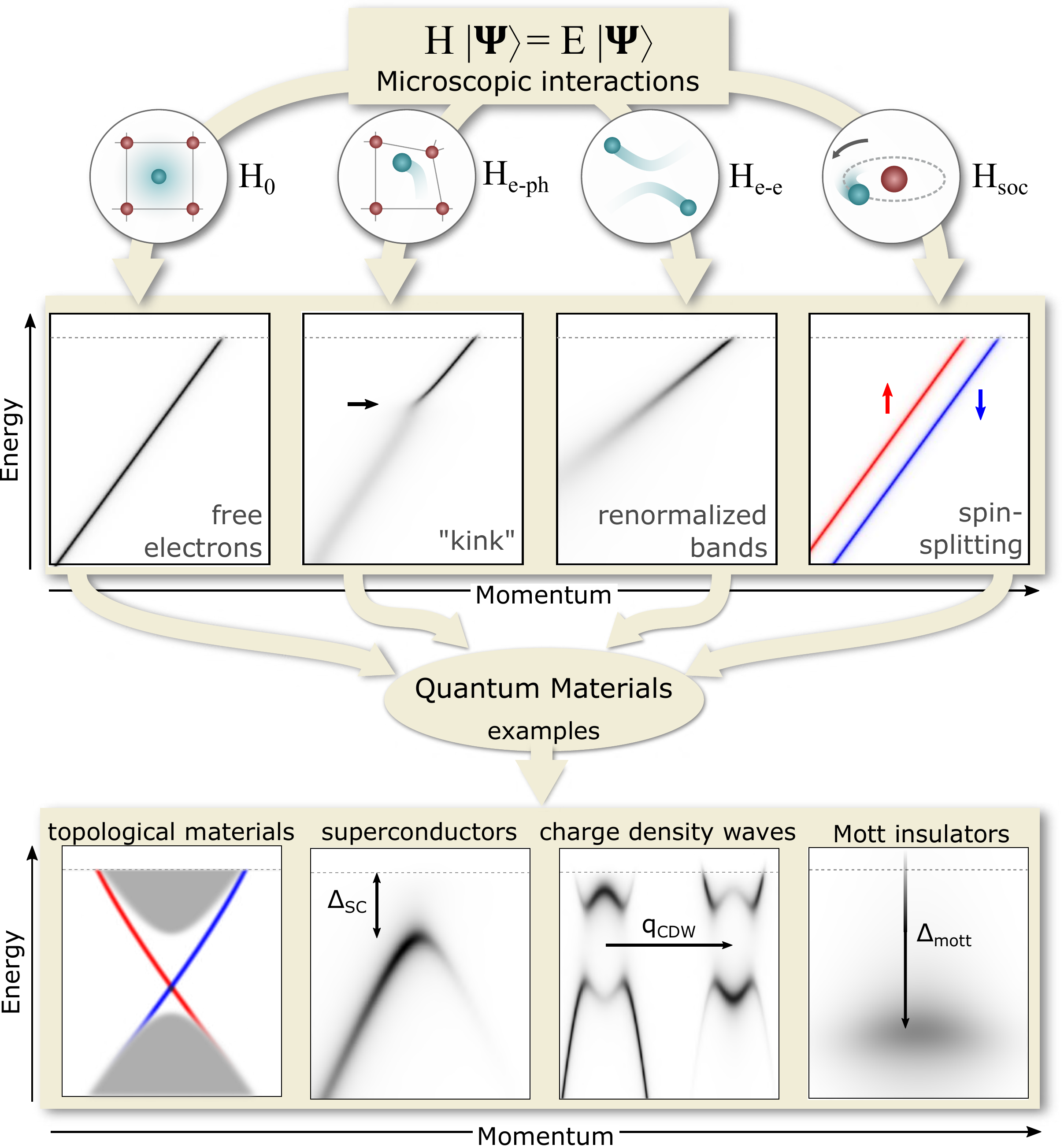}
\caption[]{Schematic depiction of microscopic interactions and their distinct signatures in an ARPES spectrum. Starting with a crystal potential $H_0$, the ARPES spectrum reflects the band structure of independent electrons. Electron-phonon coupling $H_{\textrm{e-ph}}$ introduces a ``kink'' in the band dispersion, with an increased peak width below the kink energy. Electron-electron coupling $H_{\textrm{e-e}}$ leads to an energy-dependent peak width and a renormalized band velocity. Finally, spin-orbit coupling $H_{\textrm{SOC}}$ can split spin or orbital degeneracy. Together these interactions give rise to the physics of quantum materials, including  topological materials (with spin-polarized surface states), superconductors (with superconducting gap $\Delta_{\textrm{SC}}$), charge density waves (with folding wave vector $q_{\textrm{CDW}}$), and Mott insulators (with Mott gap $\Delta_{\textrm{Mott}}$).
\label{Fig_intro_spectralFunction}}
\end{figure}

Fermionic quasiparticles and bosonic elementary excitations are instrumental for describing the rich physics of quantum materials. For fermionic quasiparticles, one of the most descriptive quantities is the \emph{single-particle spectral function}, which is experimentally accessible with momentum and spatial resolutions, respectively, using angle-resolved photoemission spectroscopy (ARPES)~\cite{Smith_1975_angular,Himpsel_1978_experimental,kampf1990spectral,Smith_1991_the,Hufner_2003_photoelectron,damascelli2003angle,Plummer_2007_angle} and scanning tunneling spectroscopy (STS) \cite{binnig1982surface,tersoff1983theory,fischer2007scanning}. Bosonic excitations derived from charge, spin, or lattice degrees of freedom are typically probed by electron-in-electron-out (electron diffraction, electron energy loss spectroscopy)~\cite{egerton2011electron,vig2017measurement}, photon-in-photon-out (x-ray diffraction, inelastic x-ray scattering, optical spectroscopy)~\cite{ament2011resonant,Baron2016high,Basov_2011_electrodynamics} and neutron scattering methods~\cite{lovesey1984theory,shirane2002neutron,furrer2009neutron} through access to the density-density correlation function.  

As shall be elaborated below, ARPES is based on the photoelectric effect, in which light is used to liberate electrons from a material such that their pre-emission energy and momentum distributions can be determined (see Fig.~\ref{Fig_intro_QMs} top-left). Since these are the very same electrons participating in the many-body physics governed by $H$, a wealth of information can be gleaned from these energy-momentum maps, as elaborated in Fig.~\ref{Fig_intro_spectralFunction}. In the case of weakly interacting electrons, the ARPES intensity simply follows the electronic band structure, with the energy-momentum dependence reflecting the band dispersion $\epsilon_{\vec{k}}$. From this information alone is it is possible to extract fundamental properties such as the electron velocities and Fermi surface geometry \cite{Himpsel_1980_experimental}.  The more profound impact of ARPES is due to its ability to detect, quantify and disentangle the various microscopic contributions to $H$ and their combined impact, including those invalidating the independent-electron description. As shown schematically in the middle row of  Fig.~\ref{Fig_intro_spectralFunction}, electron-phonon interactions \cite{Balasubramanian_1998_large,Hengsberger_1999_electron,Valla_1999_many,lanzara2001evidence}, electron-electron interactions \cite{Pines_1966_the},  and spin-orbit coupling \cite{lashell1996spin} each have distinct signatures in the ARPES spectra, visible as renormalized and/or split dispersions with respect to the non-interacting bands.  The amalgamation of these interactions results in the unique physics of quantum materials. The bottom row of Fig.~\ref{Fig_intro_spectralFunction} shows schematic ARPES spectra representative of many quantum materials, including topological states of matter, superconductors, charge-density wave systems and Mott insulators.

\begin{figure}
\centering
\includegraphics[width=1\columnwidth]{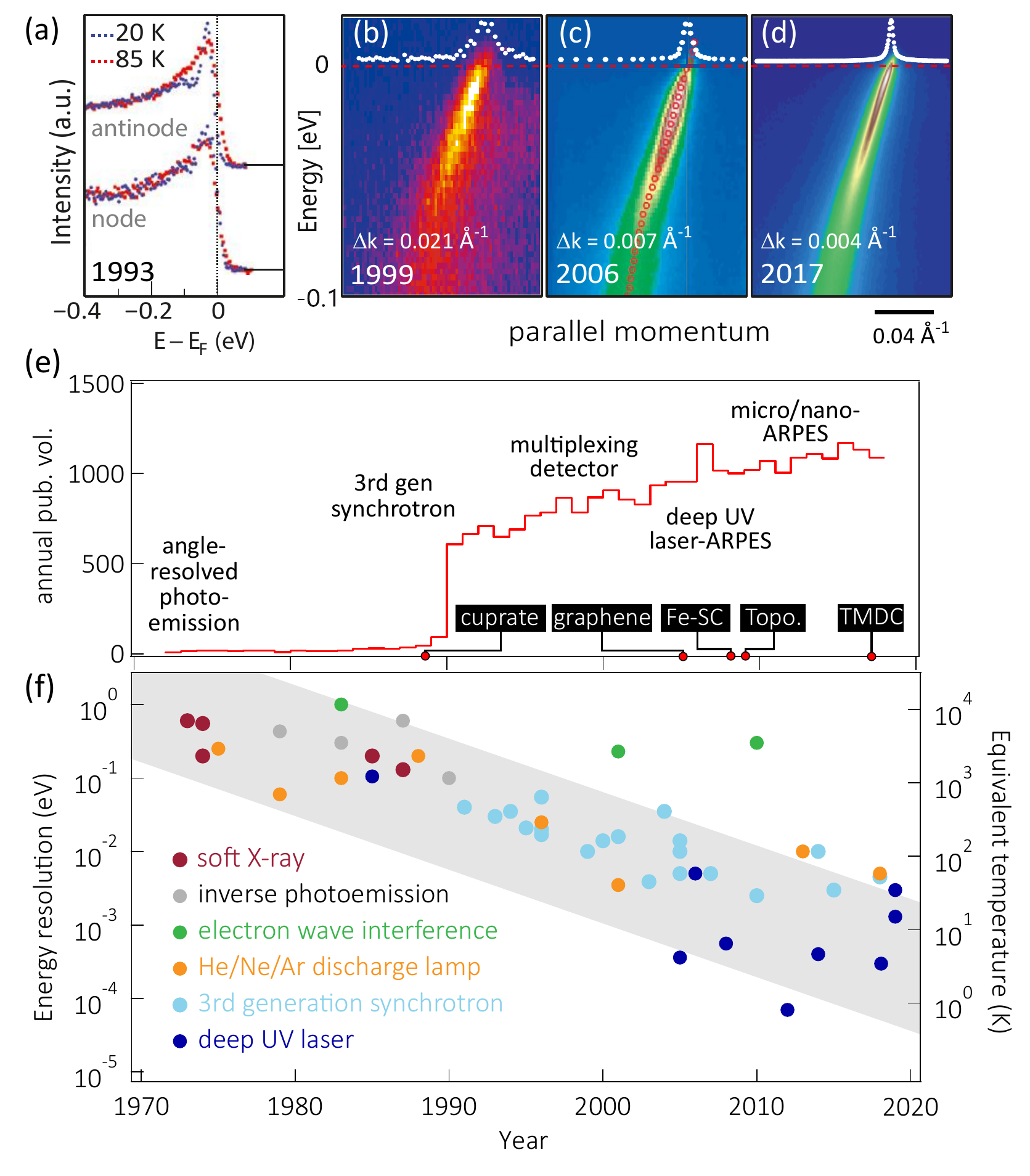}
\caption[]{State-of-the-art in ARPES over time. (a)-(d) Improvement in energy and momentum resolution, as exhibited by the spectral function along diagonal Cu-Cu direction in cuprates. Data in (d) additionally benefited from an improvement in spatial resolution to overcome surface mosaicity. Adapted from ~\cite{shen1993anomalously,Valla_1999_evidence,Koralek_2006_laser,Iwasawa_2017_development}. (e) Approximate number of papers per year utilizing ARPES. Note the jump in the early 1990's due to the advent of 3rd generation synchrotrons and beginning of cuprate research, followed by a steady rise with the introduction of new materials and techniques. Source: Web of Knowledge. (f) Energy resolution sampled from representative publications, showing an overall near-exponential improvement enabled in large part by new light sources. Different light sources are color-coded differently. For a full reference list of (f), see supplement.
\label{fig_intro_resolution}}
\end{figure}

As an experimental technique, ARPES owes much of its utility to the fact that the photoelectric cross-section is at least five to six orders of magnitude higher than other processes such as inelastic light and neutron scattering \cite{Yeh_1985_atomic,Thompson_2001_x}. This remarkable quantum efficiency has served as the foundation upon which the technique could develop into one of the mainstream tools-of-choice for condensed matter physics. The pace of scientific progress has been further accelerated by overcoming technical challenges throughout the past half century. The evolution of the state-of-the-art over this period is illustrated by Fig.~\ref{fig_intro_resolution}. The top row shows representative ARPES spectra of a cuprate superconductor; it is readily seen that the resolution and hence information content of the spectra has increased dramatically. In particular, as the energy resolution has pushed to the scale of liquid Helium temperature, efforts have also been expanded to improve more diverse metrics, such as spin detection efficiency, and spatial and temporal resolutions (see Section~\ref{tech_spinARPES},~\ref{sec_tech_environment},~\ref{sec_tech_trARPES}). Many of these technical developments were catalyzed by the cuprate problem \cite{damascelli2003angle} and have kept apace with the discovery of new families of quantum materials, with the impact evidenced by the steadily increasing number of publications (Fig.~\ref{fig_intro_resolution}(e)). These developments are summarized in Fig.~\ref{fig_intro_resolution}(f), where the crucial role of the light source is highlighted, evidencing the reciprocal stimulation of scientific and technique developments. The routine achievement of meV energy resolution in the past decade is notable as it makes ARPES extremely well-matched to low-temperature phenomena such as superconductivity. As we shall discuss in detail, the energy resolution is but one of many factors contributing to the impact of modern ARPES experiments.  Another important aspect of quantum materials is their dimensionality: systems which are confined to lower dimensions may exhibit substantially altered interactions with respect to their higher-dimensional counterparts, and are now realizable due to advanced synthesis techniques~\cite{lundqvist_1989_physics,saito_2017_highly}. Other developments include tunable sample environments, versatile light sources, and novel spectrometers with efficient multichannel detection capable of resolving multiple quantum numbers of the photoemitted electrons.

The goal of this review is to highlight the exciting advances and future opportunities in quantum materials research unlocked by these ARPES developments, with an emphasis on activities since earlier reviews \cite{Plummer_2007_angle,Himpsel_2009_photoemission, damascelli2003angle,Lynch_2005_photoemission}. Due to the enormity of the field, we have decided to err on the side of breadth rather than depth of coverage. We hope this strategy allows the reader to appreciate the impact of ARPES in the context of quantum materials research, identify the scope and capability of ARPES as a technique, while providing enough references such that the interested reader can easily locate the resources to study any individual topic in greater detail. We also survey the most salient concepts in the theoretical formalism of photoemission, to familiarize the reader with the necessary background for interpreting non-trivial spectroscopic features in complex material systems.  

The review is structured as follows: In Section~\ref{sec_ARPES} we begin with an elementary discussion on the principles of ARPES, followed by Section~\ref{sec_methods} which describes state-of-the-art experimental methods. We then focus on four families of quantum materials (see Fig.~\ref{Fig_intro_QMs}): cuprate superconductors (Section~\ref{sec_cuprates}), iron-based superconductors (Section~\ref{sec_feSC}), low-dimensional materials (Section~\ref{sec_lowD}), and topological materials (Section~\ref{sec_topo}). Section~\ref{sec_otherMat} provides a brief review of ARPES studies on other quantum material families with interesting transport properties, correlation effects, and/or topological properties.  We conclude with a brief discussion and outlook (Section~\ref{sec_outlook}). 

\section{Angle-resolved photoemission spectroscopy}\label{sec_ARPES}

\subsection{General description} \label{sec_ARPES_principles}

\begin{figure}
\centering
\includegraphics[width=1\columnwidth]{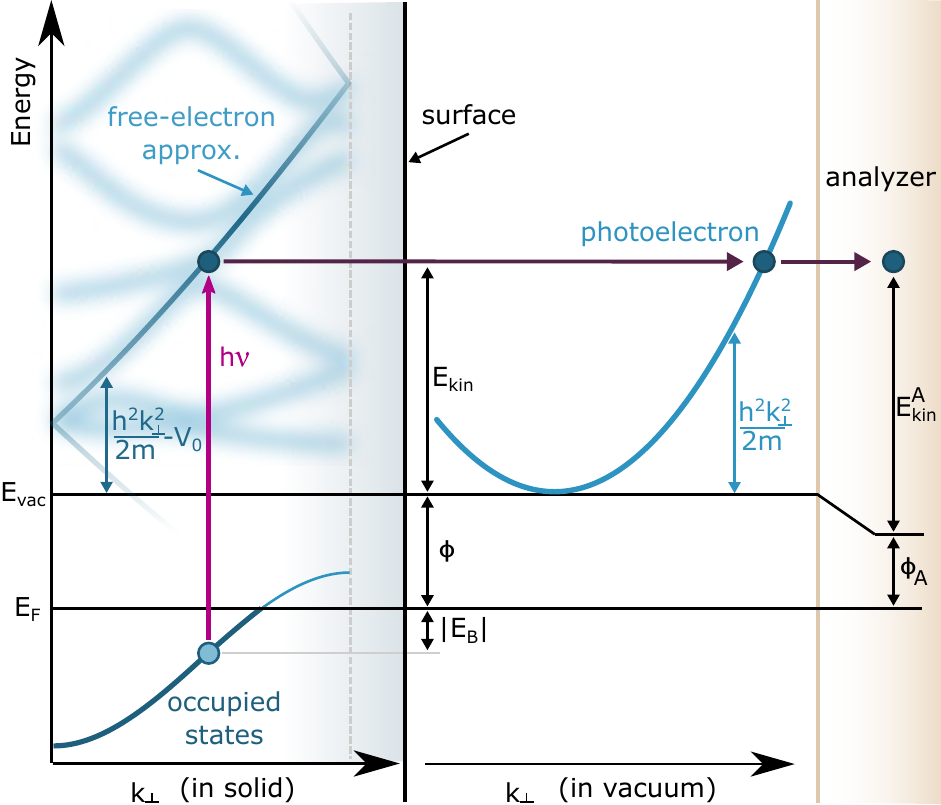}
\caption[]{Kinematics of the photoemission process in the three-step model. An electron is excited from an initial state with binding energy $\vert E_B \vert$ below $E_{\textrm{F}}$ to a final state above $E_{\textrm{vac}}$, with $\vec{k}_{\perp}$ conserved. The final state is often approximated by a free-electron dispersion offset by an inner potential $V_0$, shown here in a reduced zone scheme. After transmission through the surface barrier, the photoelectron has kinetic energy $E_{\textrm{kin}} = h\nu - \phi - \lvert E_B \rvert$, where $\phi$ is the sample work function.  $\vec{k}_{\perp}$ is not conserved during transmission through the sample surface, while   $\vec{k}_{\vert\vert}$ is conserved throughout the entire photoemission process. Note that the detected kinetic energy $E_{\textrm{kin}}^A$ is referenced to the vacuum level in the analyzer, which is determined by the analyzer work function $\phi_A$. 
\label{fig_intro_energetics}}
\end{figure}

ARPES is based on the photoelectric effect, in which a photon impinges on a material and is absorbed by an electron, which then escapes from the material \cite{Cardona_1978_photoemission,Hufner_2003_photoelectron}. The utility of ARPES as a spectroscopic tool derives from the fact that one can exploit  the kinematics of the photoemission process to deduce the binding energy $E_B$ and crystal momentum $\hbar \vec{k}$ of the electron before it was emitted from the material. A generic ARPES measurement consists of a single-crystalline sample irradiated by monochromatic light of energy $h\nu$, resulting in photoemission of electrons in all possible directions. A fraction of these electrons are collected by a photoemission spectrometer (Section \ref{Sec_tech_spectr}) which records the kinetic energy $E_{\textrm{kin}}$ and emission angles $(\vartheta, \varphi)$ of each detected electron. Here $\vartheta$ is the polar angle with respect to the surface normal, and $\varphi$ is the azimuthal angle typically defined with respect to the experimental geometry or crystal axis (see Fig.~\ref{Fig_intro_QMs}). Note that $E_{\textrm{kin}}$ is defined with respect to the sample's vacuum level $E_{\textrm{vac}}$. Based on energy and momentum conservation, one can then derive the following relationships between the pre- and post-emission electronic states:

\begin{equation}\label{Eq_E_conv}
E_{\textrm{kin}} = h\nu - \phi -  E_B 
\end{equation}
\begin{equation}\label{Eq_k_conv}
\hbar \vec{k}_{\vert \vert} = \sqrt{ 2 m E_{\textrm{kin}} } \cdot \sin(\vartheta)
\end{equation}

\noindent where $\phi$ is the sample surface work function and $\hbar \vec{k}_{\vert \vert}$ is the crystal momentum of the electron parallel to the surface in the extended zone scheme. $E_\textrm{kin}$ is the photoelectron kinetic energy, and $E_B$ is the binding energy of the electron prior to emission.\footnote{In writing Eq.~\ref{Eq_k_conv}, we have neglected the momentum of the photon since it is negligible in the ultraviolet range, though it must be considered in the soft and hard x-ray regimes \cite{Fadley_2005_x}}\footnote{Ultimately, the detected kinetic energy is determined by the work function of the analyzer $\phi_A$ rather than that of the sample, as shown in Fig.~\ref{fig_intro_energetics}.} Due to the discrete in-plane periodicity of the crystal structure, $ \vec{k}_{\vert \vert}$ is conserved throughout the photoemission process (modulo an in-plane reciprocal lattice vector $\vec{G}_{\vert\vert})$. The orthogonal component $\vec{k}_{\perp}$ is not conserved during transmission through the surface but can be deduced under certain assumptions (Section~\ref{sec_kz_det}). 

The energetics of the photoemission process are depicted in Fig.~\ref{fig_intro_energetics}.  Rather than directly report $E_{\textrm{kin}}$, which is dependent on $h\nu$, ARPES data is typically plotted with respect to $E - E_\textrm{F} = -E_B$, where $E_\textrm{F}$ is the sample's Fermi level  (see Section~\ref{sec_analysis} for experimental details), to emphasize initial state properties..

Formally, the photoemission process can be described by the transition probability $w_{fi}$ of an $N$-electron initial state $\ket{\Psi_i^N}$  to an excited final state $\ket{\Psi_f^N}$, which can be approximated by Fermi's golden rule:

\begin{equation}\label{eq_intro_fermigolden}
w_{fi} = \dfrac{2\pi}{\hbar} \left| \left\langle \Psi_f^N \vert H_\textrm{int} \vert \Psi_i^N \right\rangle  \right|^2 \delta(E^N_f - E^N_i - h\nu)
\end{equation} 

\noindent where $E^N_i$ and $E^N_f$ are the initial and final state energies of the $N$-electron system. $H_\textrm{int}$ is a perturbative Hamiltonian describing the electron-photon interaction:

\begin{equation}
\begin{split}
H_\textrm{int} &= \frac{1}{2m}(\vec{p} + \frac{e}{c}\vec{A})^2 - e \Phi - \frac{\vec{p}^2}{2m}\\
&\approx \dfrac{e}{2mc}(\vec{A}\cdot \vec{p} + \vec{p}\cdot\vec{A})\\
&\approx \dfrac{e}{mc}\vec{A}\cdot \vec{p}
\end{split}
\end{equation}

\noindent where $\vec{p}$ is the electron momentum operator, and $\vec{A}$ and $\Phi$ are the electromagnetic vector and scalar potentials. In the second line, we enforce the Weyl gauge in which the scalar potential $\Phi = 0$. The first approximation step also disregards two-photon processes $\vec{A}^2$. The second approximation step holds when $\vec{A}$ is constant over atomic dimensions such that $[\vec{A},\vec{p}]\sim\nabla \cdot \vec{A} = 0$ (see Section~\ref{sec_intro_resolution} for a comment on the limitations of this so-called \emph{dipole approximation}).\footnote{Alternatively, one may adopt the Coulomb gauge to directly enforce $\nabla \cdot \vec{A} = 0$. In this case, the scalar potential $\Phi \neq 0$. Only when there is no free charge $\rho$ can the Coulomb and the Weyl gauges be simultaneously satisfied. This condition is also known as the \textit{radiation gauge}, which can be seen via: \begin{equation}
     \frac{\rho}{\varepsilon_0} = \nabla \cdot \vec{E} = -\nabla \Phi - \partial_t(\nabla \cdot \vec{A}) \xRightarrow[\text{gauge}]{\text{radiation}} 0
 \end{equation}} Note that the golden rule formalism is only valid for weak perturbations; for sufficiently intense peak fields such as those achieved in ultrafast pulses, the perturbation expansion must include higher order terms to describe nonlinear effects such as multiphoton absorption  \cite{Lambropoulos_1974_theory}. 

Eq.~\ref{eq_intro_fermigolden} can be rigorously described by the \emph{one-step model}, in which photon absorption, electron excitation, and electron detection are treated as a single coherent process~\cite{Maham_1970_theory,Feibelman_1974_photoemission,Minar_2011_calculation}. Here, the final state of the photoelectron is a time-reversed LEED state, where the wave function rapidly decays into the bulk, and matches a free-electron plane wave form outside the surface~\cite{hopkinson1980calculation,karkare2017one}. Pragmatically it is often more convenient to use the \emph{three-step model}, which phenomenologically divides the photoemission process into three steps \cite{berglund1964photoemission}: (1) The photon drives a direct optical transition for an electron in the bulk of the material. This step contains the information on the intrinsic electronic structure of the material. (2) The electron propagates to the surface. This process is described in terms of an effective mean free path $\lambda_{\textrm{MFP}}$ imposed by both elastic and inelastic scattering processes. (3) The electron is transmitted through the surface barrier, with the electron ultimately occupying a free-electron plane wave state in the vacuum extending to the detector. The three-step model has the advantage of being more tractable since the different steps in the photoemission process are somewhat decoupled; however, in many cases the full one-step formalism may be required to fully explain spectral intensities and matrix element effects~\cite{lindroos1982comparison,Minar_2011_calculation} (see Section~\ref{sec_intro_matrixelem}).

To develop intuition for the photoemission process, we first begin with the simple case of non-interacting electrons. The more general treatment of many-body systems, which is required to describe correlated states, is covered in Section~\ref{many_body_formalism}. As a consequence of the non-interacting condition, the $N$-electron initial and final states can both be trivially factorized: 

\begin{equation}\label{final_decomp}
\ket{\Psi_f^N} = \mathcal{A} \ket{\phi_f^{\vec{k}}} \otimes \ket{\Psi_f^{N-1} }
\end{equation}
\begin{equation}\label{initial_decomp}
\ket{\Psi_i^N} = \mathcal{A} \ket{\phi_i^{\vec{k}}} \otimes \ket{\Psi_i^{N-1}}
\end{equation}

\noindent where $\mathcal{A}$ is an antisymmetry operator enforcing the Pauli principle. $\ket{\phi_i^{\vec{k}}}$ and $\ket{\phi_f^{\vec{k}}}$ are the wave functions of the photoelectron before and after absorbing a photon, which both have the same wave vector $\vec{k}$ due to momentum conservation. We denote their energies as $\epsilon_\vec{k}$ and $\epsilon_f$. $\ket{\Psi_i^{N-1}}$ and $\ket{\Psi_f^{N-1}}$ are the initial and final state wave functions of the remaining $(N-1)$-electron system. The non-interacting limit allows for a dramatic simplification, since the $(N-1)$-electron system is unaffected by the removal of one electron, and thus $\ket{\Psi_i^{N-1}}=\ket{\Psi_f^{N-1}}$. 

We can now calculate the total photocurrent $I = \sum_{i,f} w_{fi}$ by plugging these approximations into Eq.~\ref{eq_intro_fermigolden}. To cast this into an intuitive form, we assume that at most a single transition $(i\rightarrow f)$ occurs at each $\vec{k}$. Then we have:

\begin{equation}\label{eq_intro_photointensity_nonint}
I_{i\rightarrow f}(\vec{k},\epsilon_f) \propto  \lvert M_{f,i}^{\vec{k}} \rvert ^2 \delta(\epsilon_f - \epsilon_\vec{k} - h\nu) 
\end{equation}

\begin{equation}
    M_{f,i}^{\vec{k}} \equiv \langle \phi_f^{\vec{k}} \left| H_{\textrm{int}} \right| \phi_i^\vec{k} \rangle
\end{equation}

\noindent $M_{f,i}$ is the \emph{one-electron dipole matrix element}, described in Section~\ref{sec_intro_matrixelem}. These equations are the central result of this section: the ARPES spectrum of a non-interacting system is a sharp peak which traces the electronic band dispersion $\epsilon_{\vec{k}}$, with its intensity modulated by the dipole matrix element.  This result establishes the capability of ARPES to be used as a band-mapping technique. We stress that this simple picture will be modified in the presence of interactions, as described in Section~\ref{many_body_formalism}. Before we delve into this formalism, we first explore the consequences of steps (2) and (3) of the three-step model, as well as the significance of the matrix elements and photoelectron spin.

\subsection{Final state and $\vec{k}_{\perp}$ determination} \label{sec_kz_det}

Although all components of $\vec{k}$ are conserved during photon absorption (first step of the three-step model), only the surface-parallel component $\vec{k}_{\vert\vert}$ is conserved when the electron transmits through the surface (third step). However, it is possible to recover the orthogonal component $\vec{k}_{\perp\text{solid}}$ if it is assumed that the final-state dispersion of the photoelectron within the crystal can be parametrized by a free electron dispersion offset by a  potential $V_0$: $\epsilon_f = \hbar^2 \vec{k}_\text{solid}^2 / 2m - V_0$  (see Fig.~\ref{fig_intro_energetics})~\cite{pendry1969application,Himpsel_1978_experimental,Chiang_1979_angle,Chiang_1980_angle,Bartynski_1986_angle}. $V_0$, also known as the \textit{inner potential}, was initially introduced as the expectation value of the pseudo-potential experienced by the electrons on the Fermi surface $\bra{\vec{k}_\text{F}}V_\text{ps}\ket{\vec{k}_\text{F}}$~\cite{pendry1969application}, though it is now typically treated as a phenomenological parameter. Assuming the photoelectron suffers no inelastic collisions at the surface, its final state energy in the solid can be equated with the kinetic energy in vacuum $E_\textrm{kin} = \hbar^2 \vec{k}^2_\textrm{vac}/2m$. We see that the inner potential accounts for the discontinuity in $\vec{k}_\perp$ at the surface: $ \hbar^2 \vec{k}_{\perp\text{vac}}^2 / 2m = \hbar^2 \vec{k}_{\perp\text{solid}}^2 / 2m - V_0 $. This leads to the relation:

\begin{equation}
\label{kperp}
 \hbar \vec{k}_{\perp\text{solid}} =  \sqrt{ 2m\left(  E_{\textrm{kin}} \cos^2(\vartheta) + V_0  \right )   }
\end{equation} 

Since $E_{\textrm{kin}}$ varies with $h\nu$, this equation establishes a strategy for using a tunable light source to determine the $\vec{k}_{\perp\text{solid}}$ dispersion. $V_0$ is \emph{a priori} unknown, but can be determined experimentally by combining Eq.~\ref{Eq_E_conv} with the known momentum periodicity in $\vec{k}_{\perp\text{solid}}$: 
\begin{equation}
\label{innerV0}
 E_B(\vec{k}_{\vert\vert},\vec{k}_{\perp\text{solid}}) = E_B(\vec{k}_{\vert\vert},\vec{k}_{\perp\text{solid}} + n\vec{G}_{\perp})
\end{equation} 
\noindent where $\vec{G}_{\perp}$ is the out-of-plane reciprocal lattice vector, and $n$ is any integer. Note that two-dimensional electronic states do not disperse with $\vec{k}_{\perp}$; therefore, $h\nu$-dependent measurements are routinely employed to distinguish surface- from bulk- derived states.

Finally, we note that the free electron final state approximation does not hold in general, especially with low $h\nu$ light sources. The structure of the final states can indeed modulate the intensity of the ARPES spectrum, a fact which must be kept in mind with the increased use of low-$h\nu$ sources in recent years \cite{Miller_2015_resolving,Xiong_2017_three}.

\subsection{Surface-sensitivity, resolution, and $\vec{k}_{\vert\vert}$-range}\label{sec_intro_resolution}

\begin{figure}
\centering
\includegraphics[width=\columnwidth]{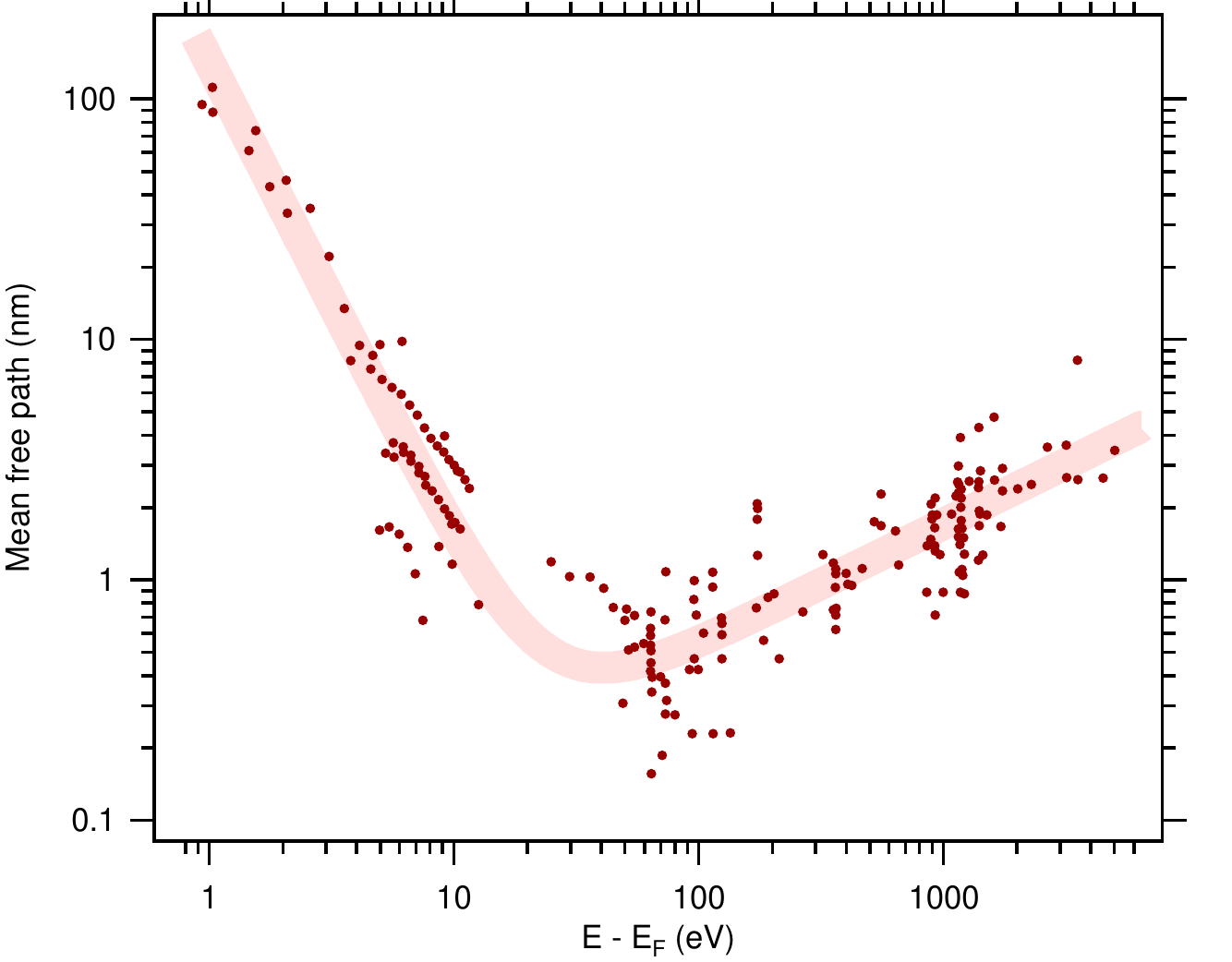}
\caption[]{The ``universal curve'' of inelastic mean free paths for electrons in solids, plotted as a function of electron kinetic energy inside the solid. The data points are compiled from measurements from hundreds of elements. The shaded curve is a best fit: $\lambda_{\textrm{MFP}} = 143/E^2 + 0.054 \sqrt{E}$ in the units of the plot. Adapted from \cite{Seah_1979_quantitative}.
\label{Fig_intro_MFP}}
\end{figure}

It is important to note that ARPES probes the spectral function in the near-surface region of the sample. The photoelectron signal is attenuated from the surface by the inelastic mean free path of electrons in the solid $\lambda_{\textrm{MFP}}$, which is a strong function of kinetic energy but weakly material dependent, with a minimum $< 1$~nm at  $20 \sim 100$~eV, as shown in Fig.~\ref{Fig_intro_MFP} \cite{Seah_1979_quantitative}.  This implies that ARPES performed with ultraviolet light sources is highly surface-sensitive, with the majority of the signal originating from the top few atomic layers. Therefore, the sample surface must be atomically flat and clean to obtain information relevant to the bulk physics of the material, as described in Section~\ref{sec_tech_environment}. 

Other factors impacting the ARPES signal include resolution effects, which can be both intrinsic and extrinsic. An important intrinsic contribution to the $\vec{k}_{\perp}$-resolution is given by the lifetime of the final state, which is finite due to the scattering processes associated with $ \lambda_{\textrm{MFP}}$ (see Section~\ref{many_body_formalism} for a more rigorous description of lifetime effects in photoemission). This can be expressed as a position-momentum uncertainty relation: $\Delta \vec{k}_{\perp} \approx \hbar / \lambda_{\textrm{MFP}} $. For typical $h \nu$ this implies $\vec{k}_{\perp}$-broadening up to $\sim 0.1$\AA$^{-1}$, which can be a significant fraction of the Brillouin zone for layered materials  \cite{Strocov_2003_intrinsic,Feibelman_1974_photoemission}. Another intrinsic factor, often overlooked, is the inapplicability of the dipole approximation $\nabla \cdot \vec{A} = 0$ made in standard treatments of photoemission. In fact, there is an abrupt change in dielectric function at the surface leading to $\vec{A}$ strongly varying on an atomic scale, which can impart momentum to the photoelectron and thereby act as another source of uncertainty in $\vec{k}_{\perp}$ \cite{Levinson_1979_effects, Miller_1996_interference}.

Extrinsic factors impacting resolution include sample and surface quality, which can broaden the momentum resolution due to elastic scattering and by introducing angular uncertainty.  Another important factor is the experimental energy resolution, which  includes the bandwidth of the light source as well as the resolution of the photoelectron spectrometer (Section~\ref{Sec_tech_spectr}): $ \left( \Delta E_{\textrm{tot}} \right)^2 = \left( h \Delta \nu \right)^2 + \left( \Delta E_{\textrm{spec}} \right)^2$. The momentum resolution has negligible contributions from $h \Delta \nu$ and is largely determined by the angular resolution of the spectrometer $\Delta \vartheta$:
\begin{equation}
\hbar \Delta \vec{k}_{\vert \vert} = \sqrt{ 2 m E_{\textrm{kin}} } \cdot \cos(\vartheta) \cdot \Delta \vartheta 
\label{eq_kres}
\end{equation}

Another factor impacting experimental resolution is the \emph{space-charge effect}, which occurs when the density of photoelectrons is high enough such that the Coulomb repulsion between electrons in vacuum cannot be ignored \cite{Zhou_2005_space,Hellmann_2009_vacuum}.  Space-charging leads to an energetic shift and broadening of the ARPES spectra. It is a more severe problem with pulsed light sources, especially those with lower repetition rates, due to the fact that the electrons are more likely to be emitted within the same interval in time.  

Finally,  we note that $h\nu$ determines the range of accessible states in both energy and momentum. The deepest $E_B$ that can be probed is: $E_B^{\textrm{max}} = h\nu - \phi$ while the largest $\vec{k}_{\vert\vert}$ is given by: $\hbar \vec{k}_{\vert\vert}^{\textrm{max}} = \sqrt{ 2 m \left( h\nu - \phi\right) } $. Electrons with $\vec{k}_{\vert\vert} > \vec{k}_{\vert\vert}^{\textrm{max}}$ undergo total internal reflection at the interface and are not photoemitted. 

\subsection{Matrix element effects}\label{sec_intro_matrixelem}

The dipole matrix element $M_{f,i}^{\vec{k}}$ was introduced above as a factor modulating the photoemission intensity within the three-step model and under the dipole and non-interacting electron approximations. ~\footnote{In cases where these approximations break down, one expects additional intensity modulations compared to what is discussed below~\cite{lindroos1982comparison}.} The dipole-transition matrix element has profound consequences for the ability of ARPES to extract microscopic information about the wavefunction of the initial state $\ket{\phi_i^{\vec{k}}}$. For pedagogical discussions, see \cite{Moser_2017_an,karkare2017one} and references therein. Although evaluating the matrix element in general can be complicated due to detailed measurement geometry and orbital hybridization~\cite{day2019computational}, in many cases symmetry or conservation laws provide clear predictions. For example, suppose the electron is photoemitted on a mirror symmetry plane of the sample; this implies that $\ket{\phi_f^{\vec{k}}}$ must be even with respect to the mirror plane  for there to be a finite photoemission intensity (since odd-parity states are zero on a mirror plane). Using $+/-$ to denote even/odd parity, this leads to four possible combinations:

\[ M_{f,i}^{\vec{k}} = \langle \phi_f^{\vec{k}} \left| H_{\textrm{int}} \right| \phi_i^\vec{k} \rangle \propto
  \begin{cases*}
  \matrixel{+}{+}{+} \neq 0 \\
  \matrixel{+}{-}{-}  \neq 0 \\
   \matrixel{+}{+}{-} = 0  \\
   \matrixel{+}{-}{+}  = 0\\
  \end{cases*}\]
  
Thus one can use the parity of  $H_{\textrm{int}} \propto \vec{A} \cdot \vec{p} $ to determine the parity of $ \ket{\phi_i^\vec{k}}$. To see how this works, we can first invoke the dipole approximation to ignore spatial variations of $\vec{A}$ and thus write $\vec{A} \approx A_0 \vec{\hat{e}}$, where $\vec{\hat{e}}$ is the unit polarization vector .  With the commutator relationship $\hbar \vec{p} = -i \left[ \vec{x}, H\right]$ this gives $\lvert M_{f,i}^{\vec{k}} \rvert ^2 \propto \left| \langle  \phi_f^{\vec{k}} \vert \vec{\hat{e}} \cdot  \vec{x}  \vert   \phi_i^{\vec{k}}  \rangle \right|^2$, where $\vec{x}$ is the position operator. Any component of $\vec{\hat{e}}$ orthogonal to the mirror plane has odd parity, while the component in the plane has even parity. Thus, the parity of $\ket{\phi_i^\vec{k}}$ can be deduced by measuring the photoemission intensity for various polarization geometries. Moreover, this concept can be generalized to determine the orbital character of bands throughout the Brillouin zone (see also Section~\ref{FeSC_multiorbital})~\cite{day2019computational,zhang2012symmetry,yi2019nematic,matt2018direct,king2014quasiparticle}. The use of linearly-polarized light to discern orbital character is often referred to as \emph{linear dichroism}.

Another common application of matrix elements is to analyze the contrast in the ARPES spectrum generated between left- and right-handed circularly-polarized light, known as \emph{circular dichroism in the angular distribution of photoelectrons} \cite{Schneider_1995_spin}. This technique can be sensitive to the time-reversal symmetry of electronic states, with recent applications to the spin- and orbital- angular momenta of states in topological insulators \cite{Wang_2011_observation, Park_2012_orbital}, as well as the chirality and Berry phase of Dirac electrons in graphitic materials \cite{Wang_2011_observation, Park_2012_orbital}. However, we caution that straightforward interpretation can be hindered by final-state effects as well as the contribution of $\nabla \cdot \vec{A} \neq 0$ terms, which can lead to geometry- and $h\nu$-dependence of the circular dichroism signal \cite{Jung_2011_warping,Gierz_2012_graphene,Mirhosseini_2012_spin,Scholz_2013_reversal,SanchezBarriga_2014_photoemission,Xu_2015_photoemission}. Furthermore, circular dichroism is generically expected if the experiment is not carefully aligned to the sample's mirror plane. For example, circular dichroism in cuprates \cite{Kaminski_2002_spontaneous} has been largely understood in the context of reflection-symmetry-breaking superstructure, without invoking any phenomena which break time-reversal symmetry \cite{Arpiainen_2009_circular,Borisenko_2004_circular}.

\subsection{Photoelectron spin}\label{intro_electron_spin}
 
 When time-reversal symmetry ($\epsilon_{\vec{k},\uparrow} = \epsilon_{-\vec{k},\downarrow}$) and inversion symmetry ($\epsilon_{ \vec{k},\uparrow} = \epsilon_{-\vec{k},\uparrow}$) are both present, all electronic states are spin-degenerate: $\epsilon_{\vec{k},\uparrow} = \epsilon_{ \vec{k},\downarrow}$. This means spin-polarized photoelectrons may be expected in magnetic materials and in materials with strong spin-orbit coupling and broken inversion symmetry. Due to the increasingly mainstream role of spin-resolved ARPES in quantum materials research (Section~\ref{tech_spinARPES}), it is worth a brief discussion on other factors contributing to the spin of a photoelectron. Here we mention two important considerations: matrix elements and the depth-dependence of photoemission. For more comprehensive reviews we refer to \cite{Kirschner_1985_polarized, Kessler_1985_polarized,Osterwalder_2006_spin,Heinzmann_2012_spin}.

Though the matrix element is often regarded as a higher-order concern for ARPES, it is indispensable when  determining the spin of a photoelectron \cite{Kessler_1985_polarized}.  Light-polarization dependence of the photoelectron spin is ubiquitous in spin-orbit coupled systems due to the fact that the initial state is a linear superposition of different spin states: $\ket{\phi} = \sum_{\alpha} c_{\alpha,\uparrow} \ket{\alpha,\uparrow} + c_{\alpha,\downarrow} \ket{\alpha,\downarrow}$, where $\alpha$ refers to the orbital part of the wavefunction. The light polarization, together with the spatial symmetry of the orbitals, determines which components of the wavefunction are photoemitted (Section~\ref{sec_intro_matrixelem}). This can result in spin-polarized photoelectrons even from unpolarized states, as is well-known for circularly polarized light in GaAs \cite{Pierce_1976_photoemission}. More recently, light polarization has been shown to control the direction of spin-polarization of photoelectrons from topological insulator surface states (Section~\ref{sec_disc_3d_TI}) \cite{Jozwiak_2013_photoelectron, Xie_2014_orbital, Zhu_2014_photoelectron, SanchezBarriga_2014_photoemission}. Furthermore, spin-polarized electrons can be photoemitted even from unpolarized initial states with unpolarized light due to spin-orbit interactions in the final state \cite{Kirschner_1985_polarized,Heinzmann_2012_spin}. Such effects are typically $h\nu$-dependent \cite{Irmer_1995_photon}; see also \cite{Jozwiak_2011_widespread} and references therein. For these reasons, one is cautioned not to take spin-polarized electrons as unambiguous evidence of novel physics, such as topological surface states.

Another important factor concerns the depth with respect to the surface from which the electrons originate. First, the finite mean free path implies that the photoemission signal is weighted strongly towards atomic layers closest to the surface. This has been invoked to explain measurements of ``hidden spin-polarization,'' in which the polarization originates from local inversion symmetry breaking within a unit cell despite the entirety of the unit cell being inversion symmetric \cite{Zhang_2014_hidden, Riley_2014_diirect,Gotlieb_2018_revealing}. At the same time, the photoelectrons originating from different depths can quantum-mechanically interfere, leading to spin polarizations which depend sensitively on geometry and $h\nu$ \cite{Zhu_2013_layer}. 

Though complex, these effects endow spin-resolved ARPES with unique capabilities in unravelling the spatial and orbital structure of the initial state wavefunction. However, they do imply that systematic measurements (as a function of $\vec{k}_{\vert\vert}$, light-polarization, and $h\nu$) with well-defined geometries are required to draw physically meaningful conclusions. Comparison with fully relativistic one-step photoemission calculations can be particularly helpful for interpreting the data  \cite{Braun_1996_the,Minar_2011_calculation,Mirhosseini_2012_spin,Scholz_2013_reversal}.

\subsection{Photoemission from a many-body system}\label{many_body_formalism}

The quantum theory of photoemission has undergone more than a century of evolution since Einstein's theory of photoelectric effect. However, it was not until the 1960s when widely recognized general theories of photoemission for many-body systems began to emerge~\cite{berglund1964photoemission,adawi1964theory}. In general, there are two categorical approaches: multiple scattering~\cite{Maham_1970_theory,bardyszewski1985new} and quadratic-response~\cite{schaich1971model,caroli1973inelastic} theories. The multiple scattering formalism treats photoemission as an inelastic scattering process of electron wave packets, which eventually leave the sample into vacuum according to a partial differential cross-section. In comparison, quadratic-response formalism evaluates the time evolution of the photocurrent operator under the perturbation of an external electromagnetic field on the second order $\sim\vec{A}^2$. These two approaches were later shown to be equivalent in the treatment of photoemission from a many-body system~\cite{almbladh1985theory}. For simplicity, we utilize the same golden rule formula for photoemission used in Section~\ref{sec_ARPES_principles}, which can be derived from the quadratic-response formalism~\cite{hermeking1975derivation}.

For an interacting system, the many-body final and initial states can not be trivially factorized as in Eq.~\ref{final_decomp} and \ref{initial_decomp}. Nevertheless, for a more tractable formalism, we can cautiously adopt these forms as approximations under certain conditions, elaborated below (see also ~\cite{damascelli2003angle}). 

First, we can approximate the removal of the electron as an instantaneous process, known as the \emph{sudden approximation}. In this limit the photoelectron has very high final state energy and no time to interact with the $(N-1)$-electron system, thereby justifying the factorization in Eq.~\ref{final_decomp}. Note that this does not imply that the $(N-1)$-electron system is unaffected by the photoelectron removal. On the contrary, the sudden creation of a hole can be associated with bosonic excitations such as phonons, plasmons, and electron-hole pairs which lead to satellite peaks on the low-energy side of the main photoemission peak  \cite{Aberg_1967_theory,Brisk_1975_shake,Citrin_1977_many}. These are known as \emph{intrinsic losses}, to be distinguished from \emph{extrinsic losses} which the photoelectron may suffer during or even after its transit out of the material~\cite{joynt1999pseudogaps,Hedin_2002_sudden,Rameau_2011_properties}.~\footnote{The electron energy loss function -- measured by EELS -- has been used to quantitatively estimate inelastically scattered electron contributions to the photoemission signal~\cite{norman1999photoelectron,nucker1989plasmons}.} The absence of plasmon satellite peaks is generally taken as evidence for violation of the sudden approximation because in the opposite \emph{adiabatic limit}, the electron is removed slowly enough such that the $(N-1)$-system remains in a ground state and no intrinsic losses can occur  \cite{Gadzuk_1975_excitation,Stohr_1983_transition}.

A transition between adiabatic and sudden regimes is expected with increasing $h\nu$.  It has been argued that this threshold strongly depends on the nature of the interactions, with the crossover at $E_{\textrm{kin}} \sim 1$~keV for long-wavelength plasmons but as low as $10$~eV for strongly correlated localized systems \cite{Lee_1999_transition}. In practice, strongly photon-energy dependent spectra are abundant in many material systems, so it can be difficult to differentiate effects related to the sudden approximation from final-state, matrix element effects, or other loss processes.
Empirically it has been found that ARPES spectra in cuprates measured with $h\nu$ down to 6~eV exhibit similar nodal dispersions to those measured at higher $h\nu$ \cite{Koralek_2006_laser}.  As a result of observations like this, it has become commonplace for most ARPES measurements to be analyzed within the sudden approximation, though we caution that a rigorous justification at low $h\nu$ has not been established to-date.  

Turning now to the initial state, we may utilize the factorization in Eq.~\ref{initial_decomp} by treating interactions in a mean-field approximation (as in Hartree-Fock theory). This approximation does not treat correlation effects self-consistently, and a more rigorous treatment requires the Green's function formalism, to be introduced shortly. 

We now arrive at the most significant deviation from the independent electron-picture: the $(N-1)$-electron system can no longer be regarded as unchanged due to electron removal:  $\ket{\Psi_i^{N-1}} \neq \ket{\Psi_f^{N-1}}$. Instead, under the sudden approximation, the ($N-1$)-electron final state can be left in any number of excited states with eigenfunctions $\ket{\Psi_m^{N-1}}$ and energies $E_m^{N-1}$. The total transition probability is then a sum over excited states:  

\begin{equation}\label{eq_intro_photointensity}
\sum\limits_{f,i} \lvert M_{f,i}^{\vec{k}} \rvert ^2 \sum\limits_{m} \lvert \langle \Psi_m^{N-1} \vert \Psi_i^{N-1} \rangle \rvert ^2 \delta(\epsilon_f + E^{N-1}_m - E_i^N - h\nu)
\end{equation}

\noindent $\lvert \langle \Psi_m^{N-1} \vert \Psi_i^{N-1} \rangle \rvert ^2 $ is the probability that the removal of an electron from state $i$ will leave the $(N-1)$-electron system in the excited eigenstate $m$. For strongly correlated materials, $\ket{\Psi_i^{N-1}}$ will overlap with many eigenstates, leading to rich spectra including satellites and broadened spectral peaks.  These deviations from the non-interacting picture establish the basis for ARPES to investigate  many-body effects in strongly correlated electron systems.

A powerful approach for understanding the structure of Eq.~\ref{eq_intro_photointensity} is provided by the Green's function formalism \cite{Mahan_2000_many}. We shall not treat this formalism in detail, and simply state the most relevant results for establishing a connection to the photoemission intensity. We begin by writing out the form of the \emph{single-electron removal spectral function}  $A^-(\vec{k},\omega)$: 

\begin{equation}\label{Eq_define_Akw}
A^{-}(\vec{k},\omega) = \sum_m \lvert \bra{ \Psi^{N-1}_m  } c_{\vec{k}}  \ket{ \Psi^N_i } \rvert ^2 \delta(\omega-E_m^{N-1} + E^N_i)
\end{equation}

\noindent Physically, $A^{-}(\vec{k},\omega)$ is the probability of removing an electron with energy $\omega$ and wavevector $\vec{k}$ from the interacting $N$-electron system.  $A^-(\vec{k},\omega)$ is related to the full spectral function $A(\vec{k},\omega)$ by: $A^-(\vec{k},\omega) = A(\vec{k},\omega) f(\omega)$, where $f(\omega)$ is the Fermi-Dirac distribution. Note that the annihilation operator $c_\vec{k}$ is applied to the initial state, thus alleviating the earlier restriction to Slater-determinant initial states. By comparing Eqs.~\ref{eq_intro_photointensity} and \ref{Eq_define_Akw}, we are motivated to write the photoemission intensity as:  

\begin{equation}
I(\vec{k},\omega) = I_0(\vec{k},h\nu,\vec{A}) f(\omega) A(\vec{k},\omega)
\end{equation}

 This is the central result of this section: under the sudden approximation (and in the absence of extrinsic losses), the photoemission signal is proportional to the single-particle spectral function. The factor of $f(\omega)$ accounts for the fact that photoemission can only occur from occupied electronic states (an important condition we did not explicitly enforce in the golden rule formulation), which enters through a thermal ensemble average when evaluating the right hand side of Eq.~\ref{Eq_define_Akw}. In practice, this limits application of ARPES to states below the Fermi level. The prefactor $I_0(\vec{k},h\nu,\vec{A})$ accounts for intensity modulations related to matrix element effects, with the $(h\nu,\vec{A})$-dependence explicitly included to highlight the dependence on experimental conditions.

Note that the spectral function obeys an important sum rule (neglecting spin degeneracy): $\int A(\vec{k},\omega) d\omega = 1$ \cite{Hufner_2003_photoelectron}, which is enforced by electron number conservation. This can be related to important physical properties:

\begin{equation}
    n_\vec{k} =  \int f(\omega) A(\vec{k},\omega)  d\omega
\end{equation}

\begin{equation}
    K =  \sum\limits_{\vec{k}}\int 2\epsilon_{\vec{k}} f(\omega) A(\vec{k},\omega)  d\omega
\end{equation}

\begin{equation}
    U = \sum\limits_{\vec{k}}\int (\omega - \epsilon_{\vec{k}}) f(\omega) A(\vec{k},\omega)  d\omega
\end{equation}   

\noindent where $n_\vec{k}$ is the momentum distribution function \cite{randeria1995momentum}, and $K$ and $U$ are the thermal expectation values of the electronic kinetic and potential energies, respectively \cite{norman2000condensation}. In principle these quantities can be computed from the ARPES data, though matrix elements and lack of knowledge of the bare-band dispersion  $\epsilon_{\vec{k}}$ limit the quantitative accuracy. 

In the Green's function formalism, the spectral function is related to the retarded Green's function $G(\vec{k},\omega)$ by: $A(\vec{k},\omega) = -(1/\pi) \textrm{Im} G(\vec{k},\omega)$ \cite{Mahan_2000_many}. Interactions are taken into account via the \emph{proper self-energy} $\Sigma(\vec{k},\omega) =  \Sigma'(\vec{k},\omega) + i\Sigma''(\vec{k},\omega)$, in terms of which the spectral function is given by:

\begin{equation}
A(\vec{k},\omega) = -\dfrac{1}{\pi}  \dfrac{\Sigma''(\vec{k},\omega)}{ \left[ \omega - \epsilon_{\vec{k}} - \Sigma'(\vec{k},\omega)  \right]^2 + \left[ \Sigma''(\vec{k},\omega) \right]^2}
\label{eq_spectral_lor}\end{equation}

It can be seen that $\Sigma'(\vec{k},\omega)$ offsets the electron band energy $\epsilon_{\vec{k}}$ while $\Sigma''(\vec{k},\omega)$ broadens the spectral peak. Physically, the imaginary part of the self-energy represents the single-particle scattering rate, which dictates the lifetime and therefore the energy width of each state~\footnote{It is important to differentiate the single-particle scattering rate, which measures single-particle excitation lifetime, from the transport scattering rate defined in the Boltzmann equation~\cite{ashcroft1976solid} and the de-population lifetime in pump-probe experiments ~\cite{Yang_2015_inequivalence,Kemper_2018_general} .}. Note that for a non-interacting system, $\Sigma(\vec{k},\omega) = 0$ and $A(\vec{k},\omega) = \delta(\omega - \epsilon_{\vec{k}})$, consistent with Eq.~\ref{eq_intro_photointensity_nonint}. For weakly interacting electrons $\Sigma(\vec{k},\omega)$ can be expanded to first order about $\epsilon_{\vec{k}}$, leading to:

\begin{equation}
A(\vec{k},\omega) = Z_{\vec{k}}   \dfrac{\Gamma_{\vec{k}} / \pi}{ \left( \omega - \epsilon_{\vec{k}}   \right)^2 + \Gamma_{\vec{k}}^2} + A_{\textrm{inc}}(\vec{k},\omega)
\label{eq_A_FL}
\end{equation}

\noindent where $ Z_{\vec{k}} = \left(1 - \partial \Sigma' / \partial \omega \right)^{-1}$, $\epsilon_{\vec{k}} = Z_{\vec{k}} \left( \epsilon_{\vec{k}}+\Sigma'  \right)$ and $\Gamma_{\vec{k}} =  Z_{\vec{k}} \lvert \Sigma'' \rvert$, and the self-energy and derivatives are evaluated at $\omega = \epsilon_{\vec{k}}$. This description is valid near the Fermi surface with the conditions $\lvert \Sigma'' \rvert \ll \epsilon_{\vec{k}}$ for small $\omega$ and $\lvert \vec{k} - \vec{k}_{\textrm{F}} \rvert$. Consistent with the predictions of Fermi liquid theory, the concept of a \emph{quasiparticle} survives, as represented by the first term of Eq.~\ref{eq_A_FL} though with a reduced spectral weight $Z_{\vec{k}}$ (also called the coherence factor). $A_{\textrm{inc}}$  is known as the incoherent part of the spectral function. It represents the error introduced by the first-order approximation of  $\Sigma(\vec{k},\omega)$ and must be included to satisfy the sum rule.

The above formalism has to be slightly modified when applied to superconductors, where charge carriers are annihilated in pairs. In this case, the pairing interaction imprints on both the normal self-energy $\Sigma(\mathbf{k},\omega)$ as well as the \emph{anomalous self energy} $\phi(\mathbf{k},\omega)$ \cite{gorkov1958energy,nambu1960quasi}. In this case, ARPES probes the imaginary part of the retarded diagonal component of Nambu Gor'kov Green's function, which indirectly reflects the superconducting pairing via the energy gap. In the case of pairing mediated by exchange of low-energy bosons, the anomalous self energy is related to the superconducting gap $\Delta(\vec{k},\omega)$ by: $\Delta(\vec{k},\omega) = \phi(\vec{k},\omega) / Z_\vec{k}$. The interested reader is referred to \cite{marsiglio2008electron} for a pedagogical discussion on the superconducting spectral function within Migdal-Eliashberg framework. Spectral gap fitting in the superconducting state will be discussed in detail in Sec.~\ref{sec_analysis}.

\section{Experimental methods}\label{sec_methods}

\subsection{Data analysis techniques and conventions}\label{sec_analysis}

\begin{figure*}
\centering
\includegraphics[width=2\columnwidth]{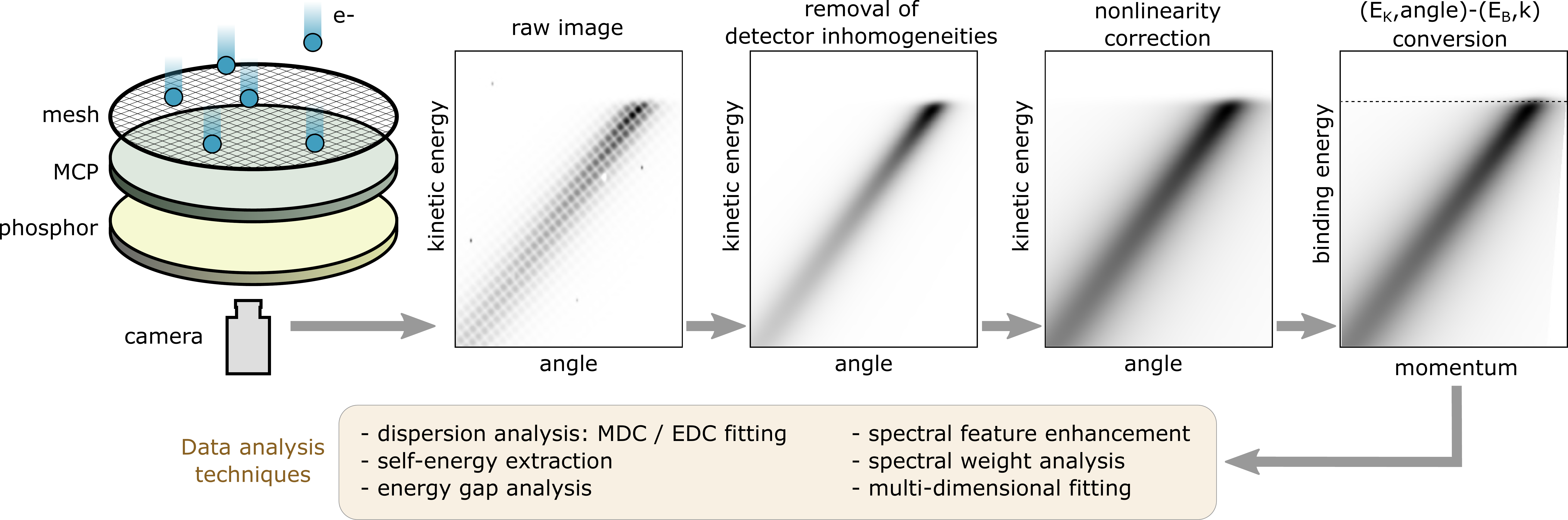}
\caption[]{A working flowchart from detector image to energy-momentum spectrum for modern ARPES with multiplexing detectors. Electron events are amplified in a multichannel plate (MCP) detector and imaged on a phosphor screen using a camera. In many implementations, a wire mesh is used to establish a uniform electric field, but leaves an imprint on the raw data which must be removed. Various methods exist for removing the grid pattern during acquisition (so-called ``dithering'' or ``swept'' modes) or during post-processing. In this flowchart, the effect of the mesh and detector inhomogeneity are exaggerated for clarity (see text for more details of the subsequent analysis steps).  
\label{fig_intro_flowchart}}
\end{figure*}

\begin{figure}
\centering
\includegraphics[width=1\columnwidth]{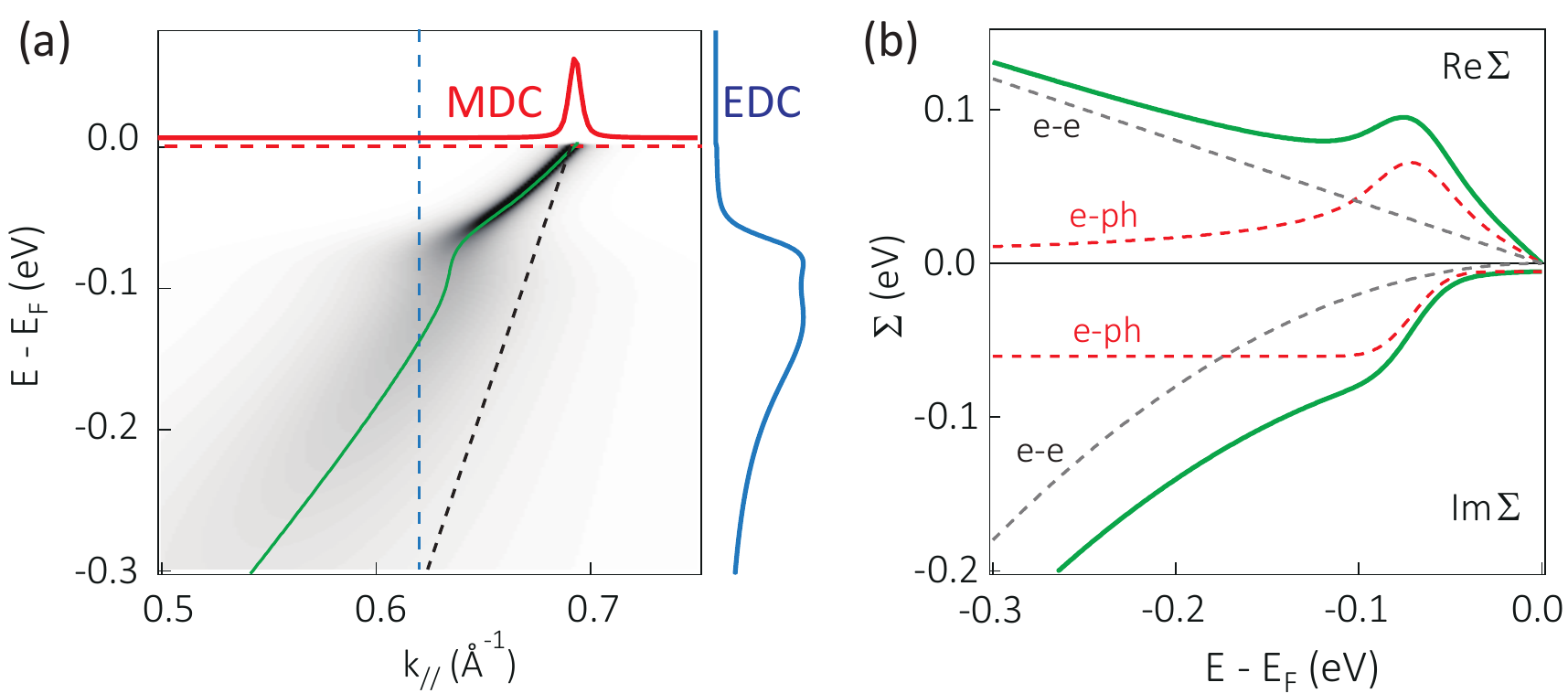}
\caption[]{Dispersion extraction and self energy analysis in ARPES. (a) A \emph{cut} showing the photoemission intensity as a function of energy and momentum. The momentum distribution curve (MDC) and energy distribution curve (EDC) at the red and blue dashed lines are shown above and to the right, respectively. The green line is an MDC fitted dispersion. The black dashed line is the non-interacting bare-band dispersion used in the simulation. (b) By MDC (shown here) or EDC fitting analysis, it is possible to extract $\Sigma'$ and $\Sigma''$ (green lines on the top and bottom panels respectively). From a modeling of their energy-dependence, they may be further decomposed into contributions such as those from electron-electron (e-e) and electron-phonon (e-ph) scattering (dashed lines). 
\label{fig_intro_analysis}}
\end{figure}

Since the advent of multiplexing detectors, ARPES data acquisition and analysis have seen an increasing inclusion of processing techniques. Figure~\ref{fig_intro_flowchart}  lays out the typical workflow of basic data reduction steps from raw camera image to the electron energy-momentum spectra, as well as a partial collection of subsequent model-based analyses. The raw camera images often undergo different image processing procedures (either at the detector level or via post-processing) to remove pixel-to-pixel efficiency variations~\cite{strocov2014soft}. Then detector non-linearity has to be corrected so the recorded intensity is proportional to the true electron counts~\cite{reber2014effects,he2016invited}.  For detailed analysis such as Fermi-function division and spectral weight analysis, discrete structures in the energy spectrum of the light source also have to be deconvolved iteratively. 

The measured $E_\textrm{kin}$ must be converted to the intrinsic energy $E-E_\textrm{F} = -E_B$, entailing a rigid shift of the energy axis of the measured spectrum. In principle the magnitude of this shift is given by $\phi_A - h\nu$, where $\phi_A$ is the analyzer work function (see Fig.~\ref{fig_intro_energetics}). However, empirically it is more accurate to determine the $E_\textrm{F}$ reference by fitting a Fermi-Dirac distribution to the spectrum of a polycrystalline metal which is electrically connected to the sample (thus ensuring that they share the same $E_\textrm{F}$). To account for lensing effects which may occur in the analyzer, this correction must often be performed separately on each angle channel of the detector. Finally, knowing the kinetic energy and emission angles $(E_\textrm{kin},\vartheta_y, \vartheta_y)$, energy-momentum conservation laws (Eqs.~\ref{Eq_E_conv}-\ref{Eq_k_conv}) may be applied to compute the parallel momentum of electrons $\vec{k}_{\vert\vert}$~\cite{ishida2018functions,iwasawa2018accurate}.

The photoemission intensity is most commonly displayed as an \emph{energy-momentum cut} as shown in Fig.~\ref{fig_intro_analysis}(a). Much spectral information can then be extracted depending on the data quality and scope of models (Fig.~\ref{fig_intro_flowchart}). One-dimensional plots of the intensity versus $\vec{k}_{\vert\vert}$ or versus energy are known as momentum distribution curves (MDCs) or energy distribution curves (EDCs), respectively \cite{Maham_1970_theory,schaich1971model,Valla_1999_evidence}. The band dispersion can be approximately obtained by fitting to the peak of the MDCs or EDCs, though in general these procedures will not yield identical results due to the energy and momentum dependence of both the electron self energy and dipole transition matrix element \cite{Norman_2001_momentum}. For example, Fig.~\ref{fig_intro_analysis}(a) highlights a complex EDC lineshape (blue line) due to the energy-dependent electron-phonon self energy. Empirically, MDC fitting is more reliable for steeply dispersing bands, and EDC fitting is preferred for flatter band dispersions. To extract interaction effects, a common approach is to map the fitting results onto Eq.~\ref{eq_spectral_lor} by approximating $\Sigma(\omega)$ to be $k_{\vert\vert}$-independent (only for the momentum along the cut) and reasonably guessing the background lineshape~\cite{valla1999many}. $\Sigma''(\omega)$ can be identified with the product of the energy-dependent peak momentum-width $\Delta k_{\vert\vert}$ with the bare band velocity $\partial \epsilon_{\vec{k}} / \partial \vec{k}$. Extraction of the self-energy can be nuanced since the \emph{bare-band} dispersion $\epsilon_{\vec{k}}$ (black dashed line in Fig.~\ref{fig_intro_analysis}(a)) is in general unknown and  must be empirically estimated from the data \cite{Kordyuk_2005_bare}. After $\Sigma'(\omega)$ and $\Sigma''(\omega)$ are extracted, modeling can be applied to decompose them into contributions from different interaction mechanisms, such as electron-electron and electron-phonon scattering (Fig.~\ref{fig_intro_analysis}(b)); or to even achieve the reconstruction of the anomalous self energy in the superconducting state~\cite{bok2016quantitative,bok2010momentum}, and the Eliashberg function in electron-phonon coupled systems~\cite{shi2004direct,zhou2005multiple,iwasawa2013true}. This modeling can be constrained by invoking the Kramers-Kronig relationship between $\Sigma'(\omega)$ and $\Sigma''(\omega)$ \cite{Norman_1999_extraction}, although this relation relies heavily on assumptions made for the unoccupied side of the dispersion.

Often times particle-hole symmetry $\Sigma(\omega)=\Sigma(-\omega)$ may be conveniently assumed (albeit not necessarily justified), based on which more constrained 2-dimensional spectral function fitting can be executed over the entire energy-momentum cut~\cite{meevasana2008extracting,Li_2018_coherent}. Assuming an energy independent dipole transition matrix element, its momentum dependence may also be extracted from the same 2-dimensional fitting~\cite{meevasana2008extracting}. There exists more complex quantitative data analysis methods, including the tomographic density of states (TDoS) method for impurity scattering removal~\cite{reber2012origin}, marginal Fermi liquid and power law liquid self energy fitting~\cite{reber2015power,leong2017power}, spectral weight moment analysis for energetic evaluation~\cite{kondo2009competition,kondo2011disentangling,hashimoto2015direct}, critical scaling of the spectral function~\cite{wang2009quantum}, and phenomenological self-energy-based superconducting gap fitting~\cite{norman1998phenomenology,franz1998phase,kordyuk2003measuring}. We refer the readers to the respective works for more details.

Given the frequent occurrence in literature, of particular importance is the spectral energy gap determination. One commonly used low-energy phenomenological model to extract the superconducting spectral gap was proposed by~\cite{norman1998phenomenology}, in which the energy-dependence of quasiparticle scattering rate is neglected in most cases. Practically, for energy gaps comparable to the energy resolution, or in spectra which intrinsically lack coherent quasiparticles, model fitting should be treated with extreme caution, as the noise, resolution and scattering effects can easily create an illusion of gap closure~\cite{vishik2018photoemission,he2020xxx}. For situations with clear coherent peaks on the gap edge, one should exercise caution when applying techniques that implicitly enforce particle-hole symmetry (such as symmetrization with respect to the Fermi level). Historically, the shift of the Fermi edge mid-point has been taken as a measure of the energy gap to overcome the lack of superconducting quasiparticle and resolution. This method is vulnerable to momentum misalignment with respect to $\mathbf{k}_F$ and strongly energy-dependent low-energy spectral intensity, therefore should be treated as a qualitative indicator of gaps. In some cases, partially momentum integrated spectra are used to extract superconducting gap via the Dynes fit~\cite{dynes1978direct,reber2012origin}, although the requirement for an energy- and momentum-independent self energy can be challenging to satisfy.

It can be difficult to clearly see the band dispersion in a raw data set if there is a strong background or broad peaks. To qualitatively enhance spectral features for quick visual inspection, a common approach is to take a second derivative of the data (either along the energy or momentum axis) to suppress the slowly-varying background and enhance the peak sharpness. Other contrast-enhancing algorithms include maximum curvature  \cite{Zhang_2011_a}, minimum gradient \cite{He_2017_visualizing}, and more recently super-resolution neural network methods~\cite{peng2020super}. In general high signal-to-noise is required to employ these high-pass filtering algorithms, and they are not guaranteed to faithfully retain the original quantitative spectral information.

\begin{figure*}
\centering
\includegraphics[width=2\columnwidth]{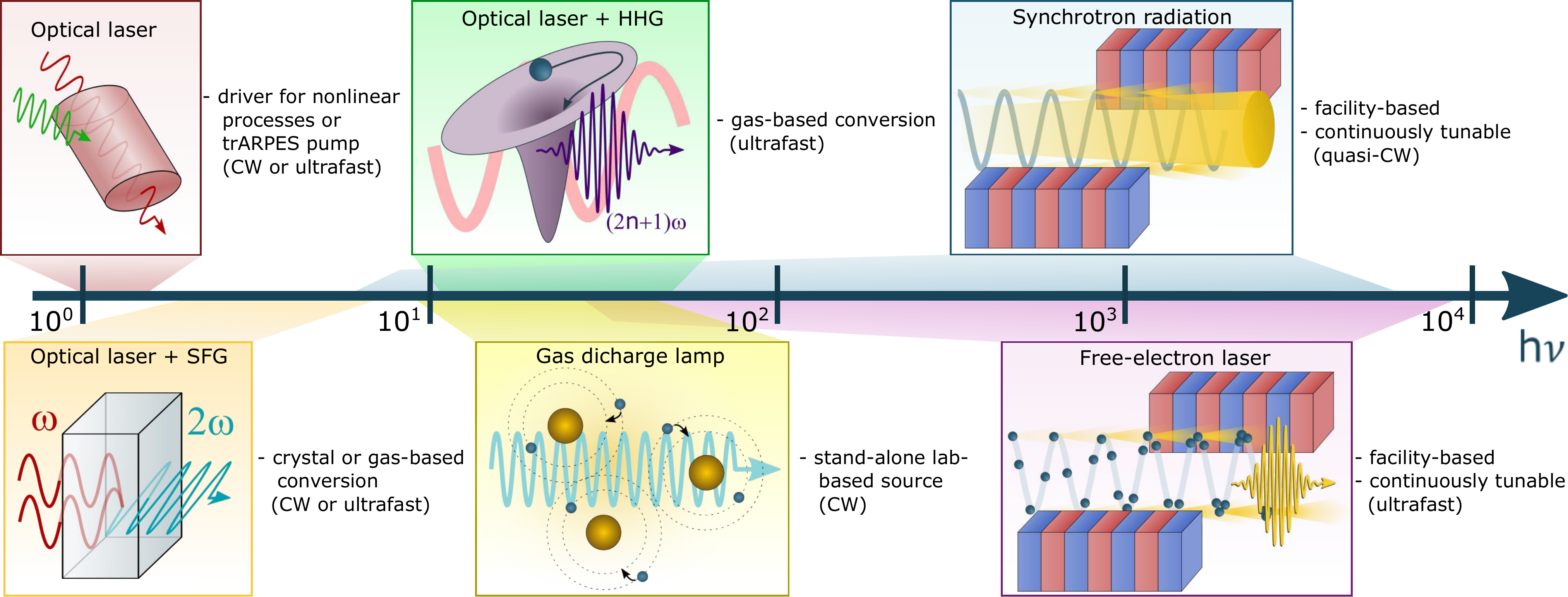}
\caption[Light sources for ARPES]{Light sources for ARPES and their main characteristics and typical photon energy ranges. 
\label{Fig_tech_lightsource}}
\end{figure*}

\subsection{Light sources} \label{sec_tech_lightSources}

The light source is a key factor establishing the capabilities of an ARPES experimental setup. The most relevant specifications are the $h\nu$-range, the bandwidth $h \Delta \nu$, the photon flux, the beam spot size, the polarization control capability, the pulse duration, and the repetition rate \cite{Attwood_2016_x}. The last two are particularly important for time-resolved applications (Section~\ref{sec_tech_trARPES}), the usage of time-of-flight detectors (Section~\ref{Sec_tech_spectr}), and considerations on the space-charging effect (Section~\ref{sec_intro_resolution}). Figure~\ref{Fig_tech_lightsource} arranges major light sources used for ARPES according to their characteristic photon energies. 

$h\nu$ determines important experimental considerations such as the photoelectron escape depth (Fig.~\ref{Fig_intro_MFP}), as well as the range and resolution of both $\vec{k}_{\perp}$ (Section~\ref{sec_kz_det}) and $\vec{k}_{\vert\vert}$ (Section~\ref{sec_intro_resolution}).
In practice, higher $h\nu$ sources are often chosen for new material characterizations due to their large energy-momentum coverage. Laser light sources typically offer lower $h\nu$ but have relatively high flux and can provide either narrow or broad bandwidth. Lower $h\nu$ also implies better momentum resolution at a given detector angular resolution (Eq.~\ref{eq_kres}). Narrow-band lasers are frequently used where high-resolution, energy stability, and statistics are paramount, whereas broadband lasers are employed for time-resolved ARPES (trARPES). Continuously $h\nu$ tunable sources such as synchrotrons are highly desired for $\vec{k}_{\perp}$ mapping in 3D materials and the identification of 2D surface states. Gas discharge lamps provide lab-based options for high energy resolution measurements, albeit with the drawbacks of relatively large beam size, tight sample geometry, and difficult polarization control. ARPES at free-electron lasers is currently a niche technique, but due to its broadly tunable $h\nu$ and ultrafast pulses, may play an important role for trARPES in the future.

We now describe each of these sources in greater detail.

\subsubsection{Synchrotron radiation}

ARPES beamlines exist at synchrotron facilities worldwide. The photon radiation is produced by passing accelerated electrons in a storage ring through a periodic magnetic structure known as an undulator. Subsequent monochromator optics are used to achieve bandwidths down to the meV scale. The greatest advantage of sychrotron radiation is that $h\nu$ is continuously and easily tunable by adjusting the undulator and monochromator, and facilities are available spanning from vacuum ultra-voilet (VUV) to hard x-ray wavelengths. The overwhelming majority of ARPES work is performed in the VUV range due to its much higher cross-section \cite{Yeh_1985_atomic,Thompson_2001_x} and superior resolution for a given resolving power. One advantage of the hard x-ray regime ($>2$~keV) is that $\lambda_{\textrm{MFP}}$ can exceed 10~nm, thus achieving relatively high bulk sensitivity \cite{Gray_2011_probing}. Furthermore, it has been shown that depth-resolution can be achieved with hard x-rays in multilayer samples by creating a standing wave which is scanned vertically through the structure (see Sec.~\ref{ARPES_variants}) \cite{Fadley_2013_hard}.

One disadvantage of undulators is that they can generate unwanted harmonics of the desired $h\nu$, though this effect can be suppressed with quasiperiodic undulators \cite{Sasaki_1995_conceptual, Panaccione_2009_advanced}.  The long-term thermal stability of x-ray optics is an important source of error on a sub-10~meV scale over the course of a typical multi-shift experiment. Modern undulators also offer full linear and circular polarization control \cite{Xi_2013_a,Hand_2016_investigation}. The repetition rate is set by the bunch spacing in the storage ring, and is typically of order 100's of MHz with pulse durations $\sim 10 \sim 100$~ps. Time-of-flight experiments may also be accommodated under reduced-bunch modes, in which the repetition rate and total flux are reduced to a few percent of normal operations. 

While most beamlines operate with a spot size of 10$\sim$100~$\mu$m, there is an increasing movement towards  micro- and nano- ARPES measurements to avoid averaging over inhomogeneous samples \cite{Mino_2018_materials}. Instrumentational approaches include  use of capillaries, Schwarzschild  optics, or Fresnel zone plates to achieve spots down to 120~nm \cite{cattelan2018perspective, Yao_2018_quasicrystalline, Kastl_2019_effects, iwasawa2019buried}. The utility of these techniques is demonstrated in Fig.~\ref{Fig_tech_misc}(a), where an exfoliated WSe$_2$ flake consisting of micron-scale regions of different thicknesses is spatially mapped using ARPES \cite{wilson2017determination}. Concurrently,  synchrotron facilities worldwide are heavily investing in improving beam coherence, enabling decades higher brightness approaching diffraction-limited measurement conditions \cite{Eriksson_2014_diffraction, Maesaka_2015_comparison}. Real-space imaging (PEEM) based approaches have also emerged recently to facilitate spatially resolved ARPES measurements, which is detailed in Section~\ref{Sec_tech_spectr}.

\subsubsection{Laser sources}\label{sec_tech_laser}

UV laser sources offer extraordinary photon flux, energy stability, and excellent energy-momentum resolutions. In a laser, a gain medium is either electrically or optically pumped in an optical cavity and emits in the infrared to visible range. Different schemes can be employed for frequency conversion to ultraviolet to make the source suitable for photoemission: for example, multiple stages of second-harmonic generation (SHG) in nonlinear optical crystals can achieve $h\nu$ up to 6~eV \cite{Koralek_2007_experimental} or 7~eV \cite{Liu_2008_development,Kiss_2008_a}. Higher $h\nu$ up to 9.3~eV \cite{Cilento_2016_advancing} or 11~eV \cite{Berntsen_2011_an, he2016invited} can be achieved by sum-frequency generation (SFG) in a noble gas. Beyond that, $h\nu$ up to $\sim 100$~eV can be achieved by high harmonic generation (HHG) in gas (see Section~\ref{sec_tech_trARPES}), which generates multiple odd harmonics of the driving frequency and therefore requires monochromatization for achieving energy resolution.

The bandwidth is initially set by the laser source, and is related to the pulse duration through the Heisenberg uncertainty limit (Section \ref{sec_tech_trARPES}). Quasi-continuous wave lasers are preferred for achieving sub-meV bandwidth for high energy resolution \cite{Liu_2008_development,Kiss_2008_a}, while $>18$~meV bandwidth is required for sub-100~fs pulses for trARPES. Depending on the mechanism, frequency conversion can either reduce or increase the bandwidth; a notable example of the latter is HHG which has been used to generate 11~fs pulses for trARPES \cite{Rohde_2016_time}. The repetition rate is typically 1~kHz $\sim$ 100~MHz depending on the nature of the laser source and subsequent amplification stages. By placing focusing optics near the sample position, spot sizes $<5$~$\mu$m have been achieved \cite{Iwasawa_2017_development, Cucci_2019_microfocus}. For UV wavelengths, full polarization control is possible using polarizers and waveplates. We refer to \cite{Zhou_2018_new} for a recent review of laser-based ARPES.

\subsubsection{Free-electron lasers}\label{sec_tech_FEL}

Free-electron lasers (FELs) share many properties with both synchrotron sources and optical lasers. In an FEL the gain comes from the synchrotron radiation of free electrons in an undulator; unlike a conventional synchrotron source, the radiation feeds back onto the electron trajectories such that the emitted radiation becomes coherent, allowing for high-intensity femtosecond-scale pulses with $h\nu\approx$100~eV$\sim$10~keV that goes beyond typical table-top HHG wavelengths \cite{Bonifacio_1984_collective,Huang_2007_review}. With coverage of the soft- and hard- x-ray regimes, FELs are in principle extraordinarily flexible sources for trARPES, but the high photon energy and pulsed timing structure have made space-charge effects challenging to overcome \cite{Pietzsch_2008_towards,Hellmann_2012_time,Oloff_2014_time}. Gains can be made by using higher efficiency analyzers \cite{Kutnyakhov_2019_time}, with dramatic improvement promised by the next generation of FEL sources achieving repetition rates up to $\sim1$~MHz \cite{Oloff_2016_femtosecond,Rossbach_2019_10}. The scarcity of available FEL beamtime is another practical challenge for utilizing these sources.  

\begin{figure*}
\centering
\includegraphics[width=2.0\columnwidth]{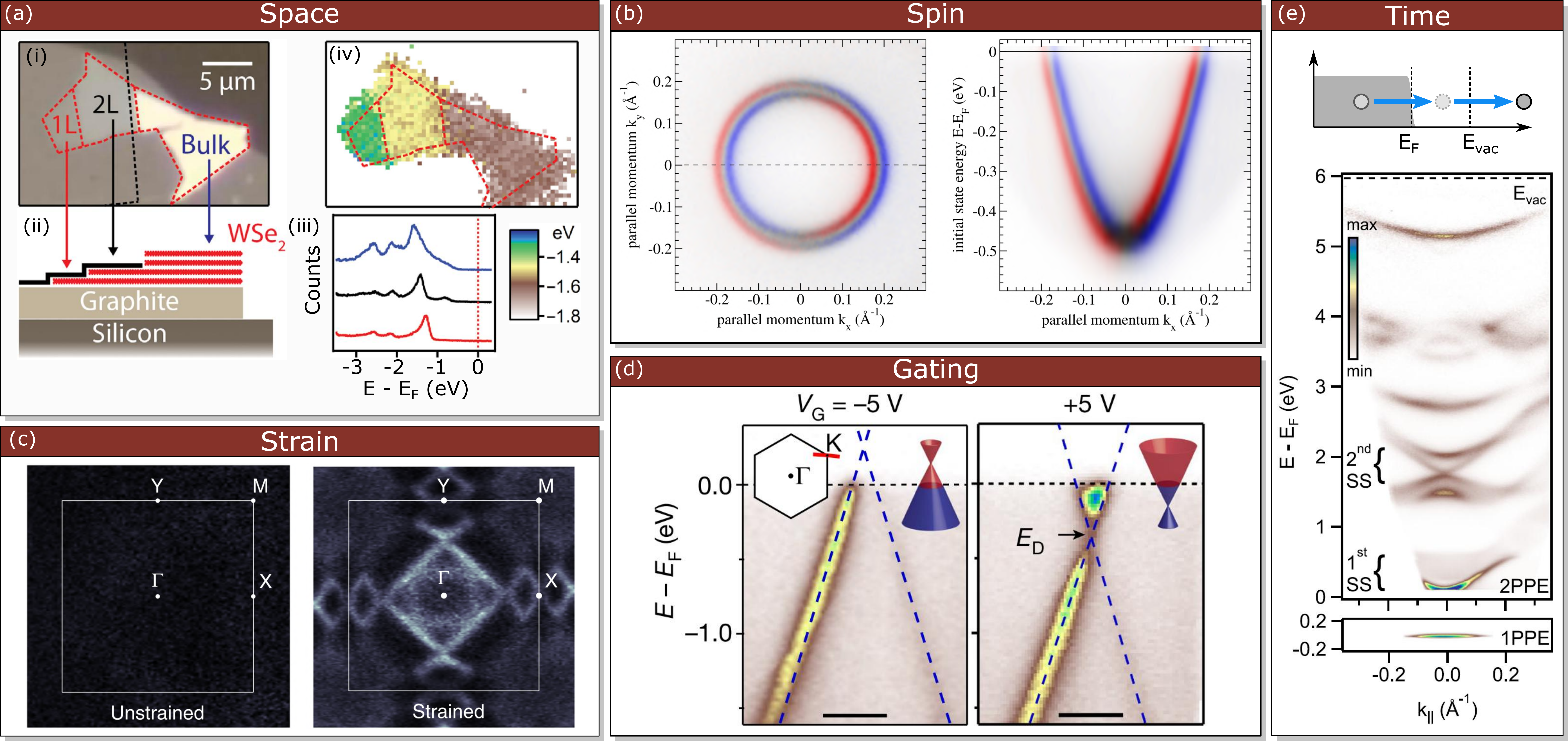}
\caption[State-of-the-art ARPES techniques]{Specialized ARPES techniques. (a) Spatially-resolved ARPES with a 0.6~$\mu$m spot size. (i) The optical image and (ii) schematic cross-section of an exfoliated WSe$_2$ flake with monolayer, bilayer, and bulk regions capped with monolayer graphene. (iii) EDCs from the three spatial regions and (iv) spatial map of the peak energy, demonstrating the ability of $\mu$-ARPES to map heterostructured samples. Adapted from \cite{wilson2017determination}. (b) Spin-resolved ARPES spectra from the Au(111) surface state, demonstrating the efficiency of a state-of-the-art 2D imaging-type spin detector. From \cite{Tusche_2015_spin}. (c) Fermi surface maps for strained and unstrained Ca$_{2-x}$Pr$_x$RuO$_4$ demonstrating an insulator-to-metal transition via non-thermal tuning knobs. From \cite{ricco2018situ}. (d) ARPES cuts at the K-point of graphene using \emph{in-situ} electrostatic gating to shift the chemical potential. Adapted from \cite{Nguyen_2019_visualizing}. (e) Measurement of a second topological surface state above $E_{\textrm{F}}$ in Bi$_2$Se$_3$, demonstrating the ability of trARPES to measure unoccupied states. Adapted from \cite{Sobota_2013_direct}.
\label{Fig_tech_misc}}
\end{figure*}

\subsubsection{Gas discharge lamps}

Gas discharge lamps are commercially available and routinely employed in laboratories. For high intensity sources, microwave radiation is used to excite a gas in a condition known as electron cyclotron resonance (ECR). Although the emission is intrinsically narrow ($\leq1$~meV), a monochromator is used to choose between different discharge lines, and the emission is then focused onto the sample. Most commonly the He I$\alpha$ emission at 21.22~eV is used, supplemented by He II $\alpha$ (40.81~eV), Ne I$\alpha$ (16.85~eV) and Ar I (11.62~eV) emissions. The radiation is intrinsically continuous-wave and unpolarized, though partial polarization control may be realized via a specially designed grating system. Typical spot sizes are $\sim$1~mm, though sizes down to $\sim$200~$\mu$m are achievable with short focal-length capillaries and apertures. 

\subsection{Photoelectron spectrometers}\label{Sec_tech_spectr}

\begin{figure}
\centering
\includegraphics[width=0.9\columnwidth]{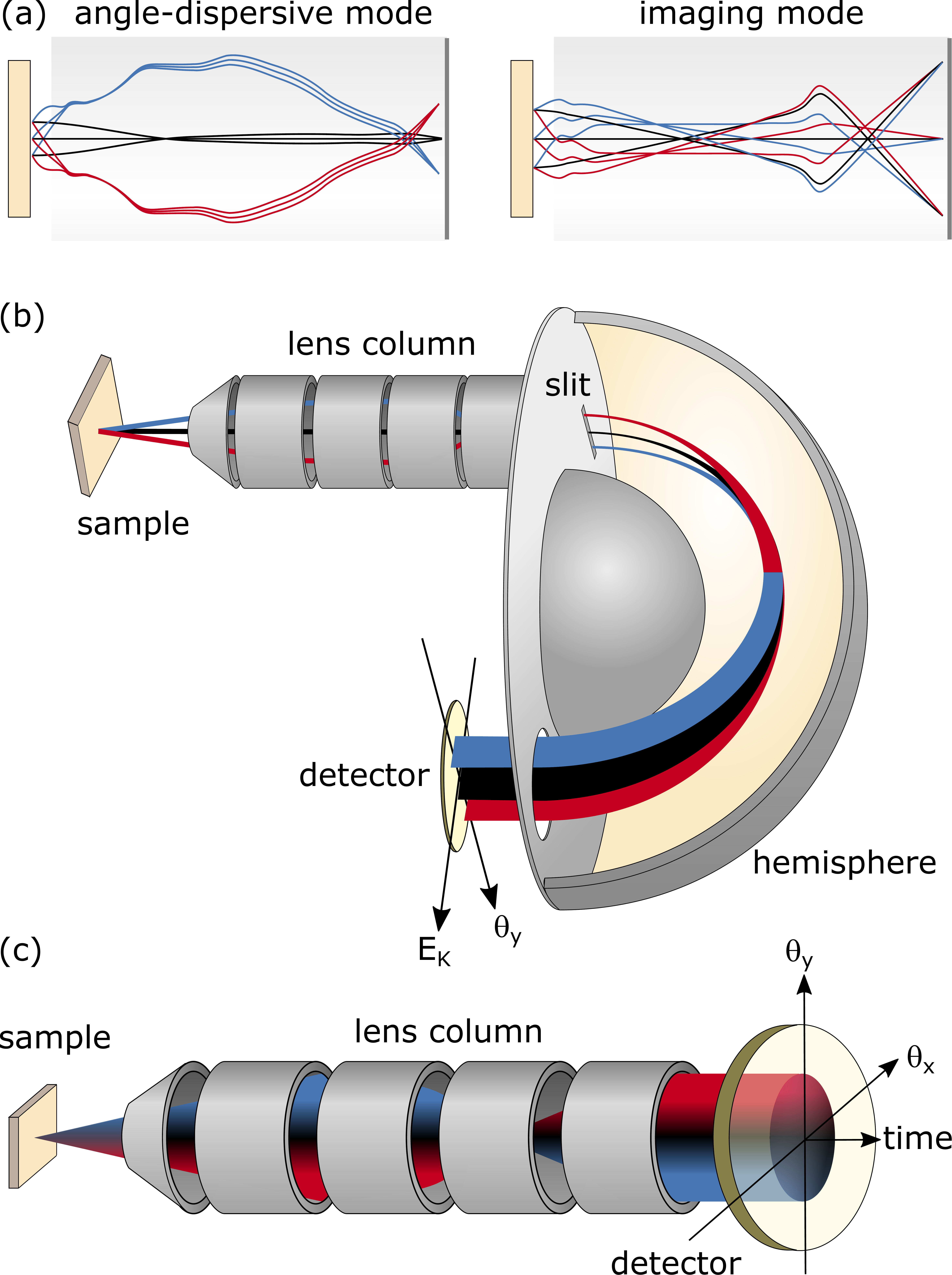}
\caption[Analyzers for ARPES]{Photoelectron spectrometers for ARPES. (a) Simplified depiction of lens modes for a modern analyzer. In angle-dispersive mode, electrons of the same emission angle (denoted by color) arrive at the same spatial position on the detector plane. In imaging mode, the real-space position of the electrons is mapped onto the detector, independent of emission angle. See \cite{Wannberg_2009_electron} for quantitative calculations.  (b) The hemispherical analyzer. For ARPES, the lens column is operated in angle-dispersive mode to map the angular distribution of electrons onto the entrance slit of a hemispherical deflector. A two dimensional detector records the distribution of electrons as a function of emission angle $\vartheta_y$ and kinetic energy $E_{\textrm{kin}}$. (c) The angle-resolved time-of-flight detector. The lens column images the angular distribution of electrons onto a two-dimensional detector with time-resolution, which records the emission angles $(\vartheta_x,\vartheta_y)$ and photoelectron flight times. 
\label{Fig_tech_analyzer}}
\end{figure}

A photoelectron spectrometer uses electrostatic elements to manipulate the trajectory and energy of electrons and impinge them onto a detector. Modern spectrometers feature lensing elements that can be operated to record either the angular or spatial distribution of electrons, as shown in Fig.~\ref{Fig_tech_analyzer}(a), where the angular mode is used for ARPES measurements. The detector records the energetic and angular distributions of the photoelectrons which can be traced back to the electron single-particle spectral function prior to emission. The detector typically consists of a multichannel plate (MCP), which amplifies the signal by converting a single electron into a cloud of $\sim10^6$ electrons while maintaining the spatial distribution of the incident electrons. Most commonly, the MCP output is impinged onto a phosphor from which the resulting luminescence can be read into a computer using a CCD camera (see Fig.~\ref{fig_intro_flowchart}(a)). This scheme allows multiple events to be recorded in parallel \cite{Gelius_1990_a}, but has the disadvantage that the countrate response can exhibit nonlinearities \cite{reber2014effects}. Another approach is the use of a delay-line detector, which individually analyzes each event from the MCP output and thus features a linear response. Another advantage is that these detectors can provide timing information for use in time-of-flight applications. Historically this approach was limited by the ``dead time'' between recording consecutive events, though modern detectors mitigate this problem with multi-hit capabilities \cite{Jagutzki_2002_multiple}.  

In the remainder of this section, we survey the most commonly used photoelectron spectrometers,   as well as auxiliary techniques such as spin polarimetry. 

\subsubsection{Hemispherical analyzers}

The hemispherical analyzer has been the workhorse of the ARPES community for the past two decades \cite{Fellner_1974_New,Wannberg_2009_electron}, as it is highly versatile (compatible with both pulsed and continuous radiation, at energies from eV  to keV) and offers high angle and energy resolutions with moderate throughput. A schematic of a generic hemispherical  analyzer is shown in Fig.~\ref{Fig_tech_analyzer}(b).  It consists of an input lens column, followed by a hemispherical deflector and finally a two-dimensional electron detector. The lens column images the angular distribution of the electrons onto a slit at the entrance of the hemispherical deflector. The deflector consists of two concentric hemispherical electrodes with different electrostatic potentials, resulting in a radial electric field that causes the electrons to undergo elliptical orbits. Thus, electrons with different kinetic energies are dispersed along the radial dimension onto the detector. At the same time, the electron position orthogonal to this axis is determined by its emission angle within the window accepted by the slit. The detector therefore records the two-dimensional photocurrent distribution with respect to $(E_{\textrm{kin}}, \vartheta_y)$ \cite{Martensson_1994_a}. Energy resolutions of order $1$~meV are routinely obtained, though under pristine conditions sub-100~$\mu$eV resolution has been reported \cite{okazaki2012octet,Shimojima_2015_low}.  Typical acceptance angles are $\pm 15^{\circ}$ with resolutions down to $\sim0.1^{\circ}$. For mapping the angular distribution orthogonal to the slit, conventionally the sample is rotated with respect to an axis parallel to the slit. However, state-of-the-art spectrometers now incorporate deflection electrodes within their lens columns, making it possible to electrostatically raster the electron beam within the accepted solid angle and thereby map a portion of the two-dimensional emission cone without any mechanical rotation~\cite{ishida2018functions}.

\subsubsection{Time-of-flight spectrometers}

Time-of-flight (TOF) spectrometers are based on the principle that an electron's kinetic energy can be determined by measuring the duration between its time of photoemission and its time of incidence on the detector. This means the light source must have short pulses and the detector itself must have suitable temporal resolution. Another crucial consideration is the repetition rate: if the period between photoemission events is too short, the slowest photoelectrons in the current cycle will be overtaken by the fastest photoelectrons from the subsequent cycle, leading to ambiguity in the interpretation of the signal. Based on these considerations, suitable pulse durations and repetition rates are $\lesssim100$~ps and $\lesssim1$~MHz \cite{Ovsyannikov_2013_principles}, making TOF spectroscopy suitable for many laser sources, including those configured for pump-probe measurements, but unsuitable for synchrotrons unless the storage ring is operated in a reduced-bunch mode. TOF spectrometers can achieve very low background noise levels because they are intrinsically gated to the pulsed timing structure of the light source. Another attractive feature of TOF spectroscopy is that both axes of the area detector can be used to image the angular distribution of electrons, as shown in Fig.~\ref{Fig_tech_analyzer}(c) \cite{Wannberg_2009_electron, Ovsyannikov_2013_principles}. This allows for higher overall throughput than a hemispherical analyzer, and the collected data constitutes a three-dimensional cube with respect to  $(E_{\textrm{kin}}, \vartheta_x, \vartheta_y)$ containing both two angular directions (typically within $\pm 15^{\circ}$) . 

\subsubsection{Momentum microscopes}

In recent years there has been a growth of ARPES instrumentation based on photoemission electron microscopy (PEEM) focusing optics, also known as momentum microscopes. A large extraction voltage is applied to the front lens element, allowing the spectrometer to collect the complete $\pm90^{\circ}$ cone of photoemitted electrons.  Energy resolution can be obtained using a retarding field (as in typical PEEM) \cite{Kotsugi_2003_microspectroscopic},  TOF analysis \cite{Schonhense_2001_time}, or dispersive energy filtering in a double hemispherical analyzer configuration, which increases the energy resolution and reduces the spherical aberrations introduced by a single hemisphere \cite{Kromker_2008_development}. Notably, the use of PEEM optics allows the spectrometer to be operated in either spatial-imaging or momentum-imaging modes \cite{Barrett_2013_laboratory}. One of the most attractive advantages of momentum microscopes is the high collection efficiency, especially when operated with a TOF detector, since there are no slits or apertures to reduce the throughput, so that virtually all photoemission events are recorded. This efficiency is particularly beneficial for spin polarimetry \cite{Tusche_2015_spin}. At the time of this writing, one challenge of momentum microscopy is that it is currently implemented in a limited number of groups, and requires some specialized expertise to operate. Moreover, it has yet to be demonstrated that it can routinely achieve sub-10~meV resolution. 

\subsubsection{Spin polarimetry}\label{tech_spinARPES}

\begin{figure}
\centering
\includegraphics[width=0.9\columnwidth]{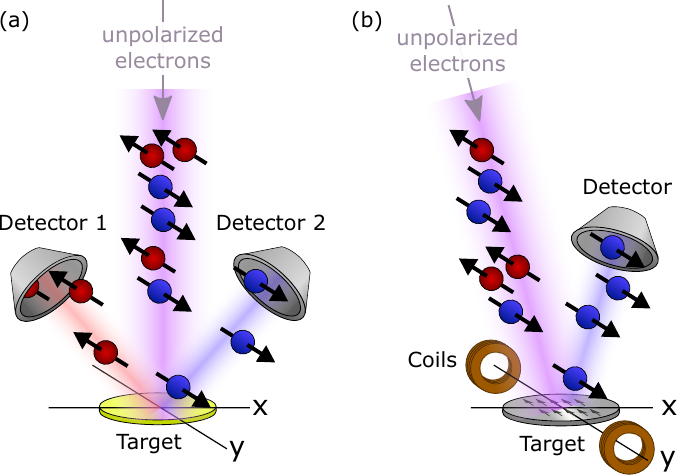}
\caption[Spin detectors]{Common spin polarimetry techniques. (a) Mott scattering employs the spin-orbit interaction between the incoming electrons and heavy nuclei in the scattering target. In the geometry shown here, detectors placed at $\pm x$ will detect contrast attributed to the $y$-component of the electron spin. A pair of detectors can be placed at $\pm y$ to simultaneously detect the $x$-component (not shown). (b) VLEED scattering employs the exchange interaction between the incoming electrons and a ferromagnetic scattering target. The measurement is sensitive to the spin component parallel to the magnetization of the target. Contrast is obtained by flipping the magnetization using Helmholtz coils. 
\label{Fig_tech_spin}}
\end{figure}

In addition to resolving the energy and momentum of the photoelectrons, it is often desirable to measure their spin polarizations (see Section~\ref{intro_electron_spin}). However, the widespread application of spin-resolved ARPES has been hampered by the low efficiency of spin polarimeters. Conventional spin-filtering is performed using a Mott detector, in which high energy ($>10$~keV) electrons are scattered off a heavy-element target such as Au or Th (see Fig.~\ref{Fig_tech_spin}(a)). The spin-orbit interaction results in a spin-dependent spatial asymmetry of the reflected electrons, which is measured by a pair of detectors and used to deduce the spin polarization of the incoming electrons \cite{Gay_1992_mott}.  Alternatively, detectors based on very low energy electron diffraction (VLEED) from a ferromagnetic material have also been developed. VLEED detectors exploit the spin-dependent reflection probability due to the exchange-split band structure of the ferromagnetic scattering target (see Fig.~\ref{Fig_tech_spin}(b)). The spin polarization is deduced by repeating the measurement with opposite magnetization directions of the target \cite{Hillebrecht_2002_high}. 

Spin polarimetry yields no direct information on the electron energy and momentum, so these techniques must be coupled with one of the spectroscopic techniques described above. The most common approach is to install the polarimeter after the exit plane of a hemispherical analyzer, with an aperture in-between which admits only a single energy and momentum channel \cite{Dil_2009_spin}.  Recently multichannel approaches have been explored to dramatically boost the collection efficiency. This has been accomplished by coupling VLEED scattering with TOF analysis (to collect a 1D spectrum vs $E_\textrm{kin}$) \cite{Jozwiak_2010_a}. More recent approaches collect a 2D scattered image $(E_\textrm{kin},\vartheta_y)$ with spin-resolution achieved via   VLEED \cite{Ji_2016_multichannel} or spin-polarized LEED scattering processes, as demonstrated in Fig.~\ref{Fig_tech_misc}(b) \cite{Tusche_2015_spin}. Combination of 2D scattering with TOF analysis enables full 3D $(E_\textrm{kin},\vartheta_x,\vartheta_y)$ spin polarimetry \cite{Elmers_2016_spin}. We refer to recent reports for more in-depth discussions of these emerging technologies \cite{Schonhense_2015_space, Suga_2015_photoelectron, Okuda_2017_recent}.

\subsection{Sample synthesis and measurement environment}\label{sec_tech_environment}

The sample is installed into a manipulator which allows for positioning and orienting the sample with respect to the photoelectron spectrometer. Besides the necessary rotational degrees of freedom, two other crucial components of the manipulator are the cryostat which cools the sample by circulating liquid helium, and the thermal radiation shielding that prevents heating from the environment. While conventional systems routinely achieve $\sim10$~K, sample temperatures below 1~K have been demonstrated using helium-3 \cite{borisenko2010superconductivity}. ARPES measurements must be performed in an ultrahigh vacuum chamber with mu-metal shielding to screen earth's magnetic field in order to reach to $\leq100$~nT at the measurement position. Under such a field, a photoelectron with $\sim2$~eV kinectic energy will be deflected $\sim$0.02$^{\circ}$ from sample to the analyzer ($\sim$1 pixel angular shift on common multiplexing detectors), exemplifying the relevance of this requirement especially for laser-based photoemission.

Due to the surface-sensitivity at ultraviolet wavelengths (Section~\ref{sec_intro_resolution}), ARPES also requires atomically clean sample surfaces to obtain information relevant to the bulk. Preparation procedures entail surface treatment such as sputtering and annealing, or when natural cleavage planes exist, the crystal can be mechanically fractured \emph{in-situ} to expose a fresh surface. After preparation, the surface must be maintained at pressures of order  $1\times10^{-11}$~torr$\sim 1\times10^{-11}$~mbar. At these pressures, residual gas molecules will form a monolayer on the fresh surface over a timescale of approximately 1~day \cite{Hofmann_2013_auger}. Another approach is to synthesize the materials \emph{in-situ} by molecular beam epitaxy. This high level of control over material synthesis has been instrumental for studies of monolayer superconductivity (Section~\ref{sec_sc_filmFeSe}), 2D materials (Section~\ref{sec_lowD}), and topological films (Section~\ref{sec_topo_confinement}).  An additional useful aspect of MBE is that mismatched lattice constants can introduce biaxial strain.  Systematic studies have shown that this can induce changes in Fermi surface topology in nickelates \cite{Yoo_2015_latent} and ruthenates \cite{burganov2016strain}, and tune the overlap of electron and hole pockets in iron-based superconductors \cite{Phan_2017_effects}. 

Although tuning knobs like magnetic field, pressure, and gate-voltage are standard in measurements such as transport, their applicability to ARPES is limited by the fact that photoelectrons must propagate through an obstruction-free, field-free vacuum. Nevertheless, in recent years there have been exciting developments in tuning the sample environment for ARPES measurements. Tensile or compressive strain can also be applied to single crystals via uniaxial mechanical deformation. Single crystals of iron-based superconductors have been detwinned by in-plane clamping, allowing ARPES measurement of a single domain of the otherwise micron-level twinned nematic state \cite{kim2011electronic, yi2011symmetry}. More controlled experiments have been performed using piezoelectric stacks together with a strain gauge, allowing the applied strain to be tuned and quantified \emph{in-situ} \cite{Pfau_2019_momentum}. A separate approach entails bending the crystal, which induced orbital splitting in SrTiO$_3$ \cite{Chang_2013_uniaxial},  and drove an insulator-to-metal transition in Ca$_{2-x}$Pr$_x$RuO$_4$, shown in Fig.~\ref{Fig_tech_misc}(c)  \cite{ricco2018situ}. Finally, strain can be applied by exploiting the differential thermal contraction between dissimilar materials in the sample holder, which was used to drive a Lifshitz transition in Sr$_2$RuO$_4$ \cite{sunko2019direct}.

One other frontier for ARPES is measurement under electrical perturbation. This can be performed in DC non-equilibrium conditions, where an electrical current is passed through the sample during measurement: a notable example is  the current-induced destruction of coherence in a cuprate superconductor, which required not only screening of stray electrical fields, but also consideration of the associated magnetic fields \cite{Kaminski_2016_destroying}. Conversely, measurements can be performed in equilibrium but electrostatically gated conditions. A common approach in this direction is to transfer electrons by \emph{in-situ} deposition of alkali atoms such as potassium, which can be used to tune the near-surface doping \cite{hossain2008situ}. Recently, electrostatic gating with a graphite back gate has been implemented in an ARPES setup, and was used to directly image the carrier density-dependent electronic structure in graphene, shown in Fig.~\ref{Fig_tech_misc}(d), as well as bandgap renormalization in transition metal dichalcogenides \cite{Nguyen_2019_visualizing}. Electrostatic bias can also be applied to increase the effective acceptance-cone angle of the analyzer. Inspired by the high-voltage electron extraction scheme used in PEEM, a biased sample holder with grounded metal mesh cover has been demonstrated to increase the detection range up to $\pm$70$^{\circ}$ solid angle~\cite{yamane2019acceptance}.

\subsection{Time-resolved ARPES}\label{sec_tech_trARPES}

\begin{figure*}
\centering
\includegraphics[width=2\columnwidth]{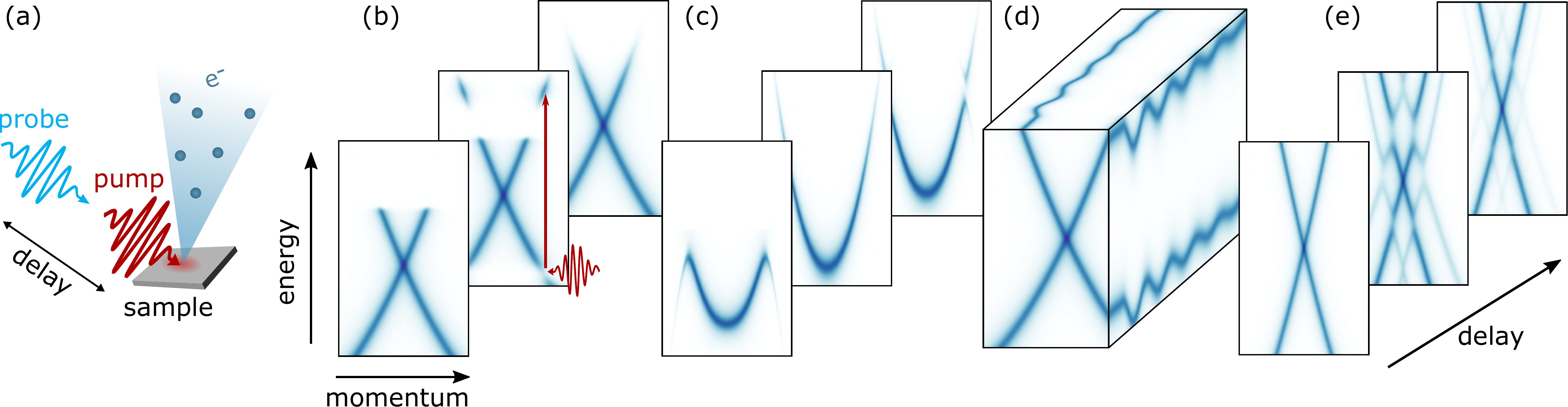}
\caption[trARPES overview]{(a) Schematic of a trARPES experiment: a pump pulse optically excites the sample, and a time-delayed probe pulse generates photoelectrons which are collected by a photoelectron analyzer. (b) -(d) Cartoons of the main classes of excitation phenomena, including (b) direct optical transitions and subsequent population relaxation dynamics; (c) destruction of an ordered phase followed by its recovery; (d) excitation of a coherent phonon mode associated with oscillatory binding energies; and (e) dressing of the electrons by a periodic field, forming Floquet-Bloch states.
\label{Fig_trARPES_overview}}
\end{figure*}

Time-resolved ARPES (trARPES) has undergone rapid development to now play a central role in characterizing the non-equilibrium properties of quantum materials \cite{Bovensiepen_2012_elementary,Smallwood_2016_ultrafast,Zhou_2018_new}. trARPES is performed in a pump-probe configuration using an ultrafast laser system (see also Section~\ref{sec_tech_laser}): the pump pulse excites a sample out of equilibrium, and the probe pulse generates a photoemission signal from the transient system, as shown in Fig.~\ref{Fig_trARPES_overview}(a). The ARPES spectrum is recorded as a function of  pump-probe time delay.

Selection of an appropriate pump pulse is critical because it can generate different excitation conditions; here we introduce four processes relevant to most experiments covered in this review. (1) As shown in Fig.~\ref{Fig_trARPES_overview}(b), absorption of pump photons results in optical dipole transitions from occupied to unoccupied electronic states, with subsequent relaxation governed by intrinsic scattering processes \cite{Fauster_1995_two,Haight_1995_electron,Petek_1997_femtosecond,Weinelt_2002_time,Echenique_2004_decay}. This process forms the basis for \emph{two-photon photoemission spectroscopy}, and allows ARPES to be extended to measure states above $E_{\textrm{F}}$, as shown shown in  Fig.~\ref{Fig_tech_misc}(e) \cite{Sobota_2013_direct} and their population dynamics. (2) When the material is in an ordered state (Fig.~\ref{Fig_trARPES_overview}(c)), the optical excitation can perturb or even destroy the order parameter. trARPES has been extensively employed to study the dynamics of destruction and reformation of states such as charge-density wave and superconducting states  \cite{Hellmann_2010_ultrafast,Rohwer_2011_collapse,Schmitt_2008_transient,Smallwood_2012_tracking}. (3) In other scenarios, the excitation can launch a coherent mode such as a phonon via Raman processes which then manifest as an oscillatory response of the electronic structure (Fig.~\ref{Fig_trARPES_overview}(d)) \cite{Garrett_1996_coherent,Perfetti_2006_time}, providing a novel approach for studying electron-phonon coupling in the time domain \cite{Papalazarou_2012_coherent,Avigo_2013_coherent,Yang_2019_mode}. (4) Finally, the electrons can be ``dressed'' by the periodic structure of the pump's electromagnetic field and form Floquet-Bloch states Fig.~\ref{Fig_trARPES_overview}(e) \cite{Wang_2013_observation}. Note that these four excitation mechanisms are not mutually exclusive, and the pump photon energy $h\nu_{\textrm{pump}}$ is an important knob for tuning their relative contributions. 

Conventionally the fundamental output of a Ti:Sapphire laser at 1.5~eV is used. This $h\nu_{\textrm{pump}}$ primarily interacts with the sample by driving dipole transitions, and exceeds the energy scale of most low-energy collective excitations such as phonons by orders of magnitude \cite{Basov_2011_electrodynamics}. To match $h\nu_{\textrm{pump}}$ to a particular energy scale, nonlinear optical schemes such as optical parametric amplification and difference frequency generation exist for frequency down-conversion, though their widespread application to trARPES is still relatively unexplored due to nontrivial demands on laser technology  \cite{Wang_2013_observation,Gierz_2015_phonon,Mahmood_2016_selective,Reimann_2018_subcycle}.  In general, higher-order nonlinear processes require higher input pulse energies ($\gtrsim100$~$\mu$J), which until recently were only available from laser systems operating at $1\sim10$~kHz repetition rates. Unfortunately, sources operating in this relatively-low frequency regime are undesirable when considering space-charge effects (see Sec.~\ref{sec_intro_resolution}). In recent years, Yb-based lasers have become competitive and may be poised to supplant Ti:Sapphire as a preferred light source for trARPES systems due to their ability to generate $>100$~$\mu$J pulses at repetition rates approaching 1~MHz \cite{Boschini_2014_an,Lorek_2014_high,Reutzel_2019_coherent,Liu_2019_time}. 

The most common approach for probing is to generate the fourth harmonic of the fundamental wavelength of a Ti:Sapphire laser system, resulting in photons with energy $h\nu_{\textrm{probe}}\sim6$~eV. Several groups have implemented HHG in noble gases to generate $h\nu_{\textrm{probe}}$ up to 50~eV \cite{Haight_1994_tunable, Zhou_2018_new}. This approach is usually implemented at lower repetition rates ($\sim$1~kHz) due to limitations in laser technology as mentioned above, though more recent efforts taking advanage of developments in laser technology have pushed the rates significantly higher \cite{Buss_2018_a, Nicholson_2018_beyond,Sie_2019_time,Puppin_2019_time}, even beyond 10~MHz \cite{Mills_2017_time, Corder_2018_ultrafast, Chiang_2015_efficient}.  This is highly desirable for mitigating space-charge, though high repetition rates are deleterious if they provide insufficient time for the sample to fully relax to equilibrium between consecutive pulses, leading to an elevated steady-state sample temperature. The time-averaged laser power and sample thermal conductivity are important factors for quantifying these heating effects \cite{Bechtel_1975_heating}. Finally, we note that soft and hard x-ray FELs will bring trARPES to previously unexplored regimes (Section~\ref{sec_tech_FEL}).

Another consideration is the trade-off between spectral and temporal resolution due to their Fourier reciprocity. The spectral resolution is limited by the bandwidth of the probe pulse, while the time  resolution is determined by the temporal cross-correlation of the pump and probe pulses. For a single pulse, the energy bandwidth and pulse duration  $\Delta E$ and $\Delta t$ (both expressed as a full-width-at-half-maximum) are related by $\Delta E \Delta t \geq 4 \hbar \ln 2 \approx$ 1825~meV$\cdot$fs, where the equality only holds for a transform-limited Gaussian pulse.  Group velocity dispersion management is required to achieve a transform-limited pulse \cite{Trebino_2000_the}. Although schemes have been proposed for bypassing this limit \cite{Randi_2017_bypassing}, it remains an effective constraint which must be considered when designing an experiment.

\subsection{Other variants of ARPES} \label{ARPES_variants}

The majority of contemporary ARPES studies on quantum materials utilize the working principles and assumptions described up to this point. Meanwhile, there have also been parallel technical developments based on variants or extensions of these principles, which often reveal additional material properties otherwise inaccessible to conventional ARPES.

\textbf{Probing surface structures and reconstructions.} Photoelectrons can be scattered and interfere with each other as they escape from the material surface, giving rise to the photoelectron diffraction effect~\cite{liebsch1974theory,bachrach2012synchrotron}. This process goes beyond the three-step model and requires a full multiple scattering treatment. It can be exploited to extract highly surface-sensitive information such as the geometry of surface reconstruction and adsorbate-substrate arrangement~\cite{kevan1978normal,woodruff1978diffraction}, via direct x-ray photoelectron diffraction~\cite{osterwalder2000full} or angle-resolved photoemission extended fine structure (ARPEFS)~\cite{barton1983direct}. In the latter case, the incident photon energy is continuously varied to sweep the outgoing photoelectron wavelength, during which the photocurrent intensity oscillates at a specific emission angle as the diffraction condition is periodically satisfied. Surface structure can then be reconstructed through the analysis of the periodicity and the emission angle~\cite{barton1985surface}.

\textbf{Probing below the surface with depth control.} X-ray standing wave photoemission is an extension of hard x-ray photoemission, which has been used to gain more bulk sensitivity~\cite{dallera2004looking,Gray_2011_probing,sing2009profiling}. Of particular importance is the ability of x-rays to form standing waves inside a crystalline lattice (for hard x-ray)~\cite{batterman1964effect}, or artificially grown layered structures (for soft x-ray)~\cite{yang2000depth,kim2001modified}. Notably here, the dipole approximation breaks down, and strong modulation of the x-ray photon field over an inter-atomic length scale becomes the strength of this technique~\cite{gray2014future}. Photoemission under such conditions not only can provide layer/depth selectivity, but also can be element sensitive when the corresponding resonant x-ray energy is chosen~\cite{gray2013momentum}. Currently this technique mainly focuses on the investigations of core-level electronic structures and magnetism in thin-film hetero-structures (Section~\ref{sec_lowD_TMO}).

\textbf{Probing unoccupied single-particle states.} Inverse photoemission spectroscopy (IPES), also historically known as bremsstrahlung isochromat spectroscopy (BIS), is another technique (other than trARPES) which permits access to the electronic structure of unoccupied electronic states \cite{nijboer1946intensity,dose1977vuv,Himpsel_1990_inverse,Dose_1985_momentum,sawatzky1984magnitude}. Instead of probing the electron-removal spectral function $A^-(\mathbf{k},\omega)$, it measures the electron-addition spectral function $A^+(\mathbf{k},\omega)$, wherein a sample is irradiated with electrons, and the resulting photon emission is recorded. It is also possible to directly inject spin-polarized electrons to probe magnetic and spin-orbit related processes~\cite{unguris1982spin}. The main bottleneck to its widespread application is the low cross section \footnote{The cross section ratio between regular photoemission and inverse photoemission is $\frac{\sigma(\text{IPES})}{\sigma(\text{PES})}\sim\frac{\lambda_e^2}{\lambda_\text{ph}^2}$, where $\lambda$ refers to the electron and photon wavelengths. At an electron energy of 10~eV (1000~eV), this ratio is $\sim10^{-5}$ ($10^{-3}$). The primary reason is due to low photon density of states which limits photon creation phase space in IPES~\cite{smith1988inverse}.}, and consequently a compromise in achievable energy resolution (often $\geq$~200~meV) \cite{Johnson_1985_calculated}. Another drawback in IPES is sample radiation damage due to the large dose of incident electrons ($\sim10^{15}$electron/sec), which may be partly mitigated with near-UV energy electron sources~\cite{pillo1997electronic,yoshida2012near}. Some of its modern applications in 2D materials are discussed in Section~\ref{sec_lowD}.

\textbf{Probing two-particle correlations.} Double-photoemission (DPE) is a process where one photon ejects two electrons simultaneously (also denoted as ($\gamma$,2$e$)). This process is forbidden in an ideal free electron gas, and only becomes possible in the presence of electron-electron interactions~\cite{berakdar1998emission,fominykh2002spectroscopy}. Such a trait makes valence electron DPE a direct probe of electronic correlations in solid state systems~\cite{gazier1970pairs,gollisch2006emission,schumann2009sensing}. Notably, the exchange-correlation hole -- a space of low electron density surrounding a given electron due to Coulomb repulsion -- can be directly measured once two ejected electrons can be traced to one single event~\cite{schumann2006mapping,schumann2007correlation}. Spin polarimetry can also be applied to further analyze the spin correlation between two photoelectrons~\cite{morozov2002spin,samarin2004spin}. Recently, a different type of two-photon-in-two-electron-out coincidence ARPES has been proposed to directly probe the two-particle correlation functions in solids~\cite{su2020coincidence}. One exciting future avenue to apply these two-particle probes is the direct detection of superconducting correlations in electron Cooper pairs~\cite{kouzakov2003photoinduced}. However, the main limitations for current DPE based approaches are the inherently low cross section ($\leq10^{-3}$ of that in single photoemission), and the requirement of low photon flux to reject single-photon photoemission background. As a result, the energy resolution is usually set at $\sim$~eV scale to ensure reasonable electron kinetic energy coverage and signal-to-noise ratio. Recent developments in new detector technology and HHG laser light sources~\cite{wallauer2012momentum,voss2019time,chiang2020laser} continue to advance the state-of-the-art of DPE spectroscopy. 

\section{Copper-based superconductors}\label{sec_cuprates}

\subsection{Overview}

Despite recent developments of pressurized hydrogen-rich superconductors~\cite{drozdov2015conventional,drozdov2019superconductivity,somayazulu2019evidence}, the high transition temperature copper-based superconductors (high-$T_c$ cuprates) still hold the $T_c$ record at 135~K under ambient pressure~\cite{bednorz1986possible,wu1987superconductivity,schilling1993superconductivity}. The intense interest in the cuprates not only lies in their superconducting $T_c$, but also because they provide the simplest effective lattice model to investigate strongly correlated many-body phenomena beyond the canonical Landau quasiparticle description~\cite{imada1998metal,damascelli2003angle,keimer2015quantum}.

The concept of long-lived quasiparticles in Fermi liquid theory lays the foundation for many condensed matter theories, including the BCS theory for superconductivity~\cite{landau1956the,bardeen1957theory}. However, its required premise of weakly interacting fermions is often violated in real materials, where electronic correlation cannot be treated perturbatively~\cite{anderson1987resonating,emery1995superconductivity,laughlin1997evidence,lee2006doping,okazaki2004photoemission,fujimori1992evolution,shen2004missing}. This challenge has stimulated developments in numerical solutions of microscopic models~\cite{white1989numerical,georges1996dynamical,foulkes2001quantum,huang2017numerical,zheng2017stripe,jiang2019superconductivity}, cold atom simulations~\cite{brown2019bad} and application of holographic principles to strongly correlated systems~\cite{zaanen2015holographic,Hartnoll_2018_holographic}. As a strongly correlated single-band material that can be tuned from an insulator to a metal via electron or hole doping, cuprates serve as the most prominent example to refine the description of correlated electronic systems. 
\begin{figure*} 
	\includegraphics[width=1.8\columnwidth]{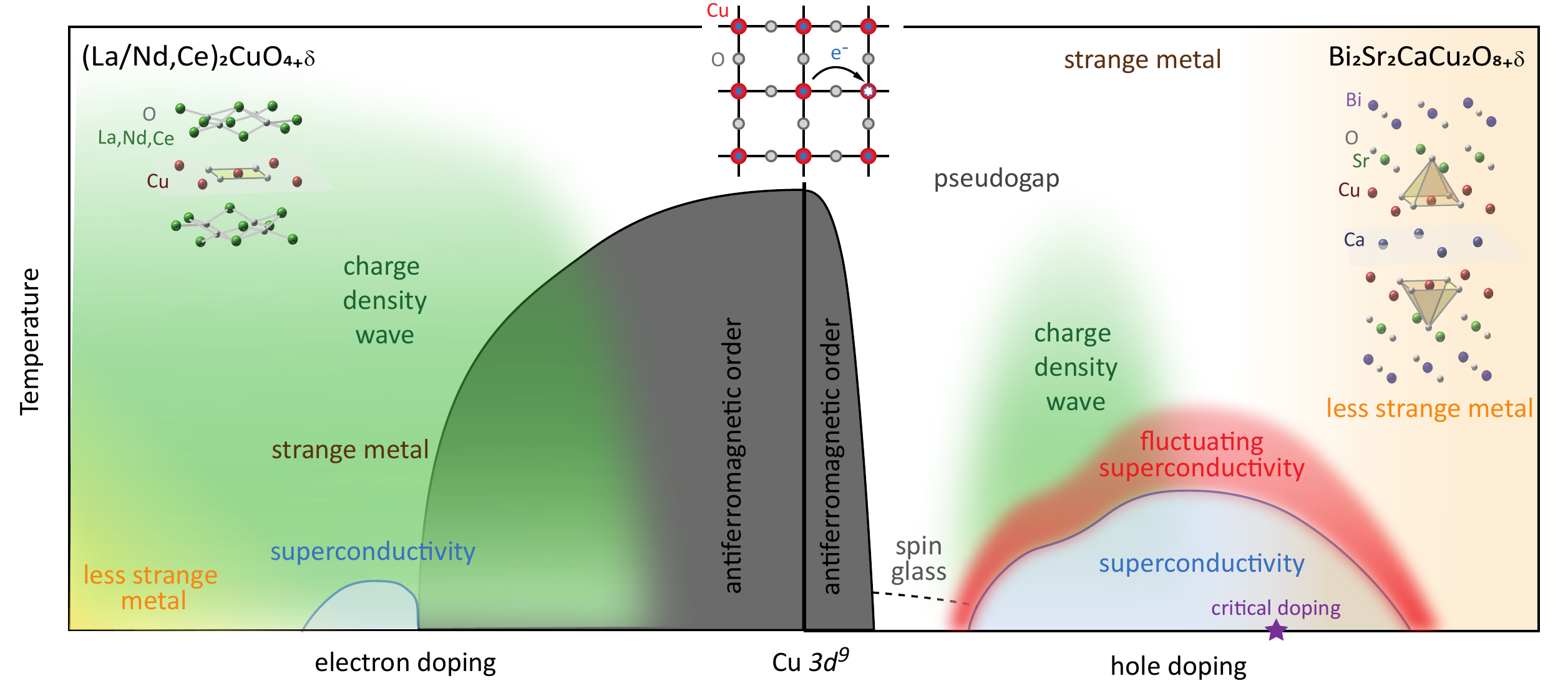}
	\caption{Schematic temperature-doping phase diagram of electron and hole-doped cuprate superconductors. Top left inset: lattice arrangement for one unit-cell of electron-doped cuprate {(La/Nd,Ce)$_2$CuO$_4$} ($T_c^\text{max}\sim$~30~K). Top right inset: lattice arrangement for one half unit-cell of {Bi$_2$Sr$_2$CaCu$_2$O$_8$} ($T_c^\text{max}\sim$~96~K). Middle inset: top view of the CuO$_2^{n-}$ plane in real space. Red circle - copper, grey circle - oxygen, blue circle - electron, white circle - doped hole. Purple star - the critical doping that separates two different metallic regimes.
	}
	\label{fig:fig_cuprate1}
\end{figure*}

The crystal structure of a typical cuprate high-$T_c$ superconductor can be viewed as an alternating stacking of metal-oxygen charge reservoir layers and superconducting CuO$_2^{n-}$ layers (Fig.~\ref{fig:fig_cuprate1} insets), where the latter dominate the low-energy electronic states. In hole-doped systems, each copper atom is usually caged in an oxygen octahedron or pyramid with at least one apical oxygen, forming the ``T-phase'' ~\cite{longo1973structure,eisaki2004effect}. In contrast, the copper atom in electron-doped cuprates generally does not have an apical oxygen, and the charge reservoir layer is heavily buckled, forming the ``T'-phase''~\cite{takagi1989superconductivity,armitage2010progress}. 

The correlation physics of cuprates is best exemplified by the strong dependence of its physical properties on charge doping, as shown in the phase diagram (Fig.~\ref{fig:fig_cuprate1}). Strong Coulomb repulsion on each Cu site makes undoped cuprates correlated insulators with long range antiferromagnetic (AFM) order, despite an odd number of valence electrons for each copper. Both electron and hole doping promote inter-site hopping, thus alleviating the charge localization and giving rise to a plethora of intertwined phases and crossover phenomena in the temperature-doping phase diagram (Fig.~\ref{fig:fig_cuprate1})~\cite{fradkin2015colloquium,kastner1998magnetic,armitage2010progress}. With as little as 2\% hole doping or up to 15\% electron doping, the long range AFM order is destroyed, but substantial short range AFM fluctuations extend to a much broader phase region~\cite{keimer1992magnetic,imai1993spin,drachuck2014comprehensive}. Hole doping induces a series of low temperature phases: a spin glass region characterized by slow spin dynamics~\cite{filipkowski1990observation,julien2003magnetic}, a dome-shaped superconducting region mostly described by $d$-wave gap  symmetry~\cite{shen1993anomalously,hardy1993precision,wollman1993experimental,tsuei1994pairing}, a valence electron charge density wave (CDW, also simply called ``charge order'') and spin stripes~\cite{tranquada1995evidence,abbamonte2005spatially,ghiringhelli2012long,da2016doping}, non-Fermi liquid charge transport (often referred to as a ``strange metal'')~\cite{martin1990normal,batlogg1994normal,harris1995violation,ando2004electronic,hussey2011dichotomy,greene2020strange} and eventually a more coherent metallic region~\cite{ando2000carrier,proust2002heat,vignolle2008quantum}. 

In addition to the ground state properties, correlation effects also manifest at high temperatures. Below a certain critical doping $p_c$, linear resistivity extends beyond the Mott-Ioffe-Regel limit at high temperatures~\cite{gurvitch1987resistivity,martin1990normal,hwang1994scaling,cooper2009anomalous,legros2019universal}, and the energy width of the single-particle spectrum is substantially larger than the electron binding energy and thermal energy at all temperatures above $T_c$~\cite{shen1999novel,tanaka2006distinct,kondo2013formation}. Another hallmark of the normal state is a depletion of the single particle spectral weight near the chemical potential, crossing over at a temperature scale of $T^*$ (broad dash in Fig.~\ref{fig:fig_cuprate1}). This is known as the ``pseudogap'' because it is not universally associated with a symmetry broken phase, and the Fermi surface is not fully gapped. In particular, one should exercise caution when interpreting a wide range of $T^*$'s measured by different techniques (see Section~\ref{sec_PG}). We note that a quantum critical fan-type phase diagram has been often used largely based on transport evidence. More recent transport and single-particle measurements suggest a much less temperature dependent boundary, which is represented here and will be discussed in more detail in Sec.~\ref{sec_emergeQP}. On the electron doped side, the phase diagram is generally similar except for a stronger AFM order and a much less temperature dependent charge ordered region~\cite{damascelli2003angle,armitage2010progress}.

The rest of this section will first cover modern ARPES investigations in order of decreasing electronic energy scales: the normal state, the superconducting state, and the ``zero'' temperature Fermi surface (``Fermiology''). Then, contributions from non-electronic degrees of freedom is discussed in light of electron-boson coupling. The electron- and hole-doped systems will be discussed based on both their unifying phenomenology, and the differentiating electron-hole asymmetry.

\subsection{Normal state}

\subsubsection{Doping evolution of the electronic structure}\label{sec_framework}

\begin{figure} 
	\includegraphics[width=1\columnwidth]{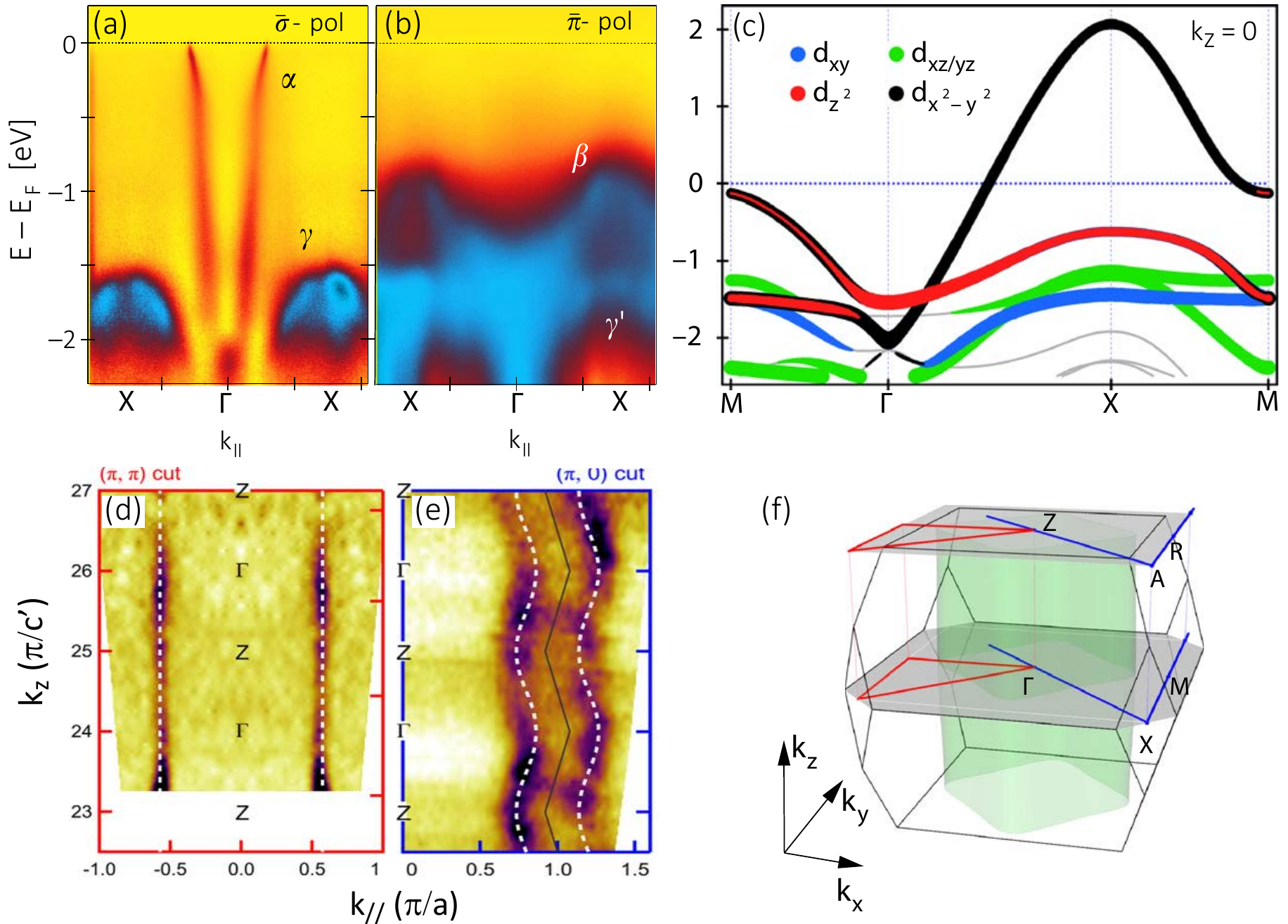}
	\caption{Orbital contents of low energy bands in overdoped La$_{2-x}$Sr$_x$CuO$_4$. (a) $d_{x^2-y^2}$ component selected with in-plane polarized light (b) $d_{3r^2-z^2}$ component selected with out-of-plane polarized light. (c) DFT calculation of $d$-orbital contents in the low energy bands. Adapted from \cite{matt2018direct}. (d) $k_z$ dispersion along nodal in-plane momenta. (e) $k_z$ dispersion along antinodal in-plane momenta. (f) Schematic plot of the 3-dimensional Brillouin zone of LSCO. Adapted from~\cite{horio2018three}.}
	\label{fig:fig_cuprate8}
\end{figure}

 Due to crystal field splitting, the copper $d_{x^2-y^2}$ orbital is the highest partly filled orbital, followed by the $d_{3r^2-z^2}$ orbital (also denoted as $d_{z^2}$, Fig.~\ref{fig:fig_cuprate8}(c))~\cite{pickett1989electronic,yu1987electronically,mattheiss1987electronic,damascelli2003angle}. In one of the more 3-dimensional cuprates La$_{2-x}$Sr$_x$CuO$_4$ (LSCO), polarization-dependent ARPES shows clear dominance of the in-plane $d_{x^2-y^2}$ orbital component near $E_{\textrm{F}}$ (Fig.~\ref{fig:fig_cuprate8}(a)(c)), while the d$_{3r^2-z^2}$ orbital component mostly resides at higher binding energy or near the antinodal momentum (near the ($\pi,0$)-point in the Brillouin zone, also known as the \emph{antinode}) (Fig.~\ref{fig:fig_cuprate8}(b)(c))~\cite{matt2018direct}. Note that a dispersion of 2~eV is observed on the occupied side. Moderate to negligible $k_z$ dispersion in different cuprates is found near the antinode, usually much smaller than the in-plane band width (Fig.~\ref{fig:fig_cuprate8}(d)(e))~\cite{takeuchi2005two,matt2018direct,horio2018three}. Despite its success in describing conventional superconductors, a single-band description of superconductivity is challenged by the vastly different $T_c$'s among different cuprate families, which all share nominally the same CuO$_2^{n-}$ plane. Another potential caveat to the single-band theory lies in the recently discovered heavily hole-doped cuprate superconductors with $T_c$'s exceeding 80~K~\cite{li2018new,gauzzi2016bulk}, where the $d_{3r^2-z^2}$ orbital content contributes more appreciably at $E_{\textrm{F}}$~\cite{maier2018d}. Varying degrees of low-energy $d_{3r^2-z^2}$ orbital content have also been proposed to account for the family-dependence of $T_c$~\cite{sakakibara2012origin}. We focus on the electronic structure associated with the $d_{x^2-y^2}$ orbital for the remainder of this section.

\begin{figure} 
	\includegraphics[width=1\columnwidth]{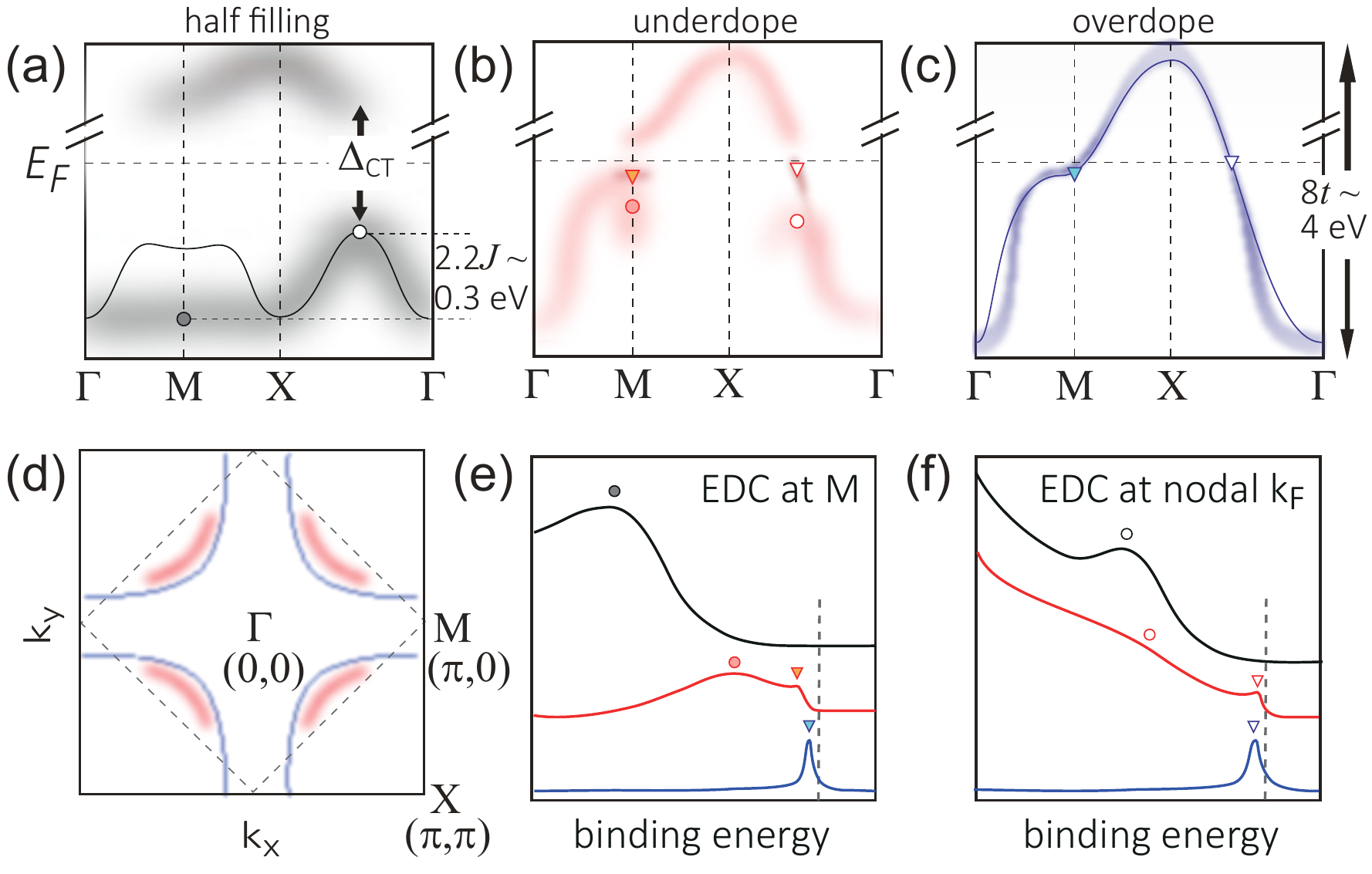}
	\caption{Schematic evolution of the electronic structure in hole-doped cuprate. (a)-(c) Idealized normal state spectral function evolution with hole-doping along high symmetry directions. Solid black line - $t$-$J$ model band calculation at half filling. Solid blue line - tight binding band calculation at heavy hole doping. (d) Fermi surface in underdoped (red) and overdoped (blue) cuprates. The dashed diamond is the AFM zone boundary. (e) ($\pi$,0) and (f) ($\pi$/2,$\pi$/2) or nodal k$_F$ EDC evolution with hole doping. Circles - high energy spectral feature associated with Mott physics. Triangles - low energy spectral features that are, or will evolve into, quasiparticles. Black - half filling. Red - underdoped. Blue - overdoped.}
	\label{fig:fig_cuprate1_sup}
\end{figure}

A free-electron description is inapplicable due to the substantial Coulomb interaction $U$ experienced by the spatially constricted orbitals~\cite{anderson1959new,kanamori1963electron,gutzwiller1963effect,hubbard1963electron,mott1968metal,comanac2008optical,weber2010strength}. Such strong electronic correlation inhibits double charge occupancy and promotes charge localization. In this scenario, the parent compound contains one electron (or equivalently, one hole) per unit cell (``half-filled''). The copper $d_{x^2-y^2}$ orbital heavily hybridizes with the ligand oxygen $p_x$, $p_y$ orbitals (see the CuO$_2$ sublattice in Fig.~\ref{fig:fig_cuprate1} upper inset), and hole carriers doped through oxygenation are postulated to form a singlet on the center copper, known as the Zhang-Rice singlet~\cite{zhang1988effective}. The system gains kinetic energy $t$ when the hole hops between sites, and pays an energy cost $U$ when double occupancy occurs on the same site. Long range AFM order forms on the copper sites~\cite{imada1998metal,lee2006doping} since the electrons gain kinetic energy by virtual inter-site hopping, which is maximized when nearest-neighbor spins are antiparallel to each other. This effective low energy single-band approximation has enabled wide applications of the 2-dimensional single-band Hubbard model to describe the behaviors of doped charge carriers in cuprates~\cite{anderson1987resonating}. To effectively describe the hopping of the singlet, the Hubbard model in the large $U$ limit may be expanded in powers of $t$, leading to the widely used $t$-$J$ model ($J=4t^2/U$) and its extensions~\cite{zhang1988effective,lee2006doping,Spalek_2007_t}. In addition, multi-band models have tested the validity and limitations of this single-band treatment, and found that the charge transfer energy between oxygen and copper dominates the largest low-energy spectral gap~\cite{zaanen1985band,emery1987theory,varma1987charge}. In this single-band Hubbard model description, the role of the Mott gap is played by the charge transfer gap.

One hallmark of the electronic structures of cuprates is their lack of rigidity against carrier doping. The framework of cuprates' normal state electronic structure can be viewed as an evolution from a charge transfer insulator to a metal when doped. Besides the large charge transfer gap $\Delta_{CT}$ at 1$\sim$2~eV scale (sometimes colloquially referred to as the Mott gap for the reason mentioned above), the spectral function measured by ARPES at half-filling can be characterized by (Fig.~\ref{fig:fig_cuprate1_sup}(a))~\cite{wells1995versus}: (1) a dispersive feature with a bandwidth set by $\sim2.2J\sim300$meV; (2) a large energy separation between ($\pi$/2,$\pi$/2) and ($\pi$,0) spectra of about $\sim2.2J$; (3) the feature at ($\pi$/2,$\pi$/2) being very broad in energy, despite being the maximum of the valence band and thus having no electronic decay channels. While (1) aligns well with the predictions of the $t$-$J$ model, (2) can only be described by involving higher order hopping terms or polaron formation, indicating the importance of spin and lattice degrees of freedom~\cite{dagotto1990strongly,nazarenko1995quasiparticle}. Observation (3) is also difficult to explain under model calculations purely based on electronic correlations~\cite{kohno_2012_mott,wang_2015_origin}. Upon doping, sharp quasiparticle dispersion emerges along $\Gamma$-X (Fig.~\ref{fig:fig_cuprate1_sup}(b), see also Section ~\ref{sec_emergeQP},~\ref{sec_LEgap}); the spectral energy difference between ($\pi$/2,$\pi$/2) and ($\pi$,0) rapidly decreases as the ($\pi$,0) spectrum moves towards the Fermi level (Fig.~\ref{fig:fig_cuprate1_sup}(b), see also Section ~\ref{sec_PG}); and the ($\pi$/2,$\pi$/2) spectrum eventually evolves into a well-defined dispersion of 8$t\sim$4~eV energy scale (Fig.~\ref{fig:fig_cuprate1_sup}(c))~\cite{ino2002doping,matt2018direct}. 
Of particular importance, the broad spectra at ($\pi$,0) remain $\sim J$ away from the Fermi level in underdoped systems, which is much deeper than a simple tight-binding dispersion. This signifies the underlying Mott physics on top of which intertwined phases compete and cooperate. This large energy scale -- especially at low doping (circles in Fig.~\ref{fig:fig_cuprate1_sup}(b)(e)) -- is what one would call the  ``high energy pseudogap''~\cite{king1995electronic,shen1995photoemission,marshall1996unconventional}. This gap obliterates the Fermi surface near the zone boundary, giving rise to the ``Fermi arc'' -- an incomplete segment of Fermi surface anchored around ($\pi$/2,$\pi$/2) (Fig.~\ref{fig:fig_cuprate1_sup}(d), red lines). Further investigations in more doped systems reveal a lower energy gap in the normal state near the antinode (solid triangles in (Fig.~\ref{fig:fig_cuprate1_sup}(b)(e)), see also Section~\ref{sec_PG})~\cite{marshall1996unconventional,loeser1996excitation,ding1996spectroscopic}. This low-energy gap, sometimes also referred to as the pseudogap, is likely related to various intertwined orders and their associated fluctuations. These two aspects of ``the'' pseudogap become less distinguishable with doping, and manifest differently in different experiments at different temperatures~\cite{timusk1999pseudogap}. This distinction is often not recognized in the literature, and can be a source of confusion.

\subsubsection{The pseudogap}\label{sec_PG}

The existence of a spectral gap above superconducting $T_c$ -- the ``pseudogap'' -- breaks the conventional wisdom of a coherent metallic normal state that is required to precede a mean-field BCS superconducting transition.
Early experimental indications of the pseudogap are covered in a number of previous reports and reviews~\cite{alloul198989,timusk1999pseudogap,keimer2015quantum,marshall1996unconventional,loeser1996excitation,ding1996spectroscopic,tanaka2006distinct,renner1998pseudogap,campuzano1999electronic,fedorov1999temperature,Lu_2012_angle,hashimoto2014energy}. Here, we specifically discuss this phenomenon in hole-doped cuprates. For electron-doped cuprates, the spectral analogue of the pseudogap is discussed in the context of the antiferromagnetic gap in Sec.~\ref{sec_LEgap} due to much more robust long range AFM order.

\begin{figure} 
	\includegraphics[width=1\columnwidth]{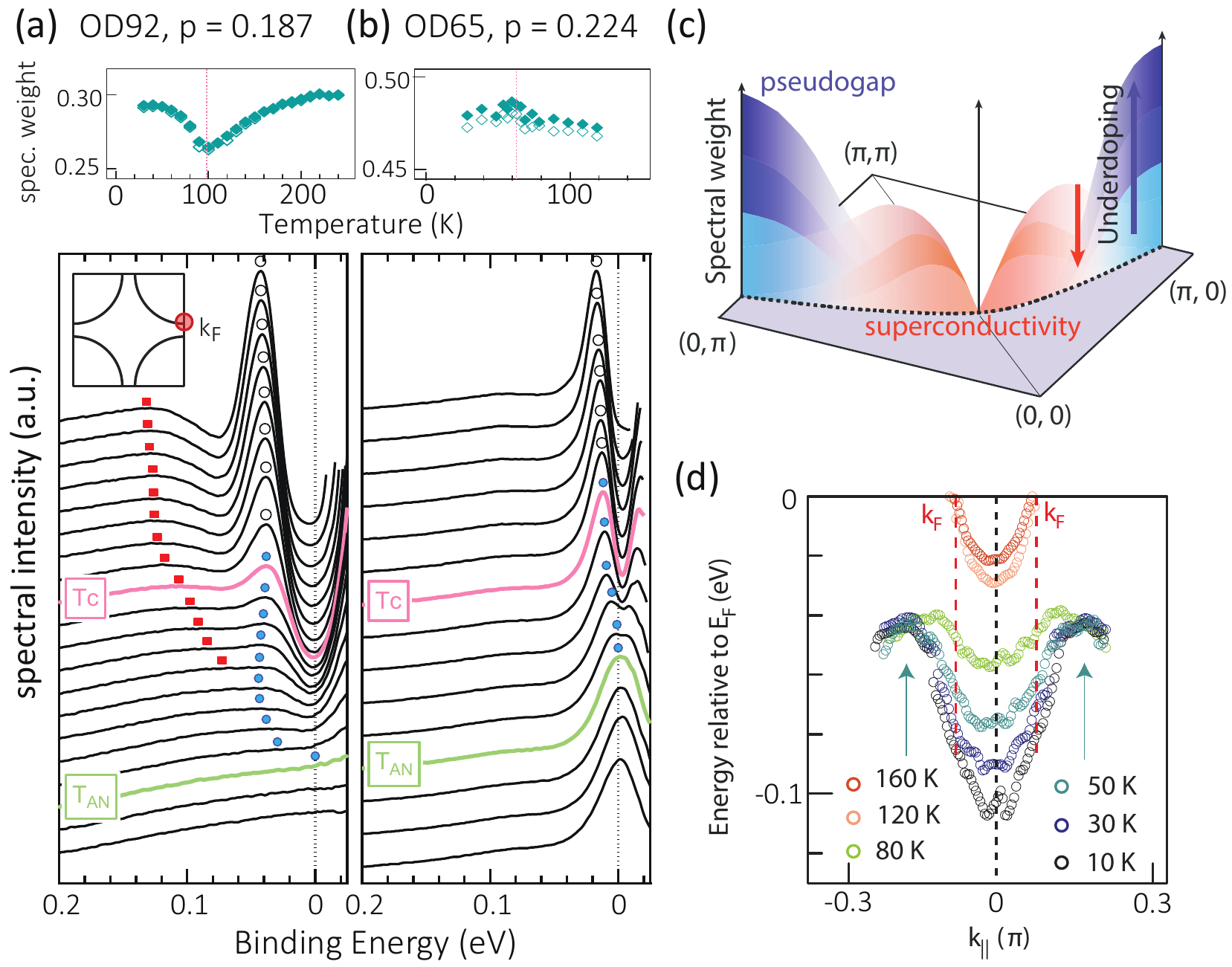}
	\caption{Spectral characters of the pseudogap (red squares) and its relation to superconducting quasiparticle peaks (blue dots). Temperature dependent, Fermi-function divided antinodal spectra for (a) near optimally doped (b) heavily overdoped Bi$_2$Sr$_2$CaCu$_2$O$_8$. $T_{AN}$ is the gap closing temperature at the antinode. Adapted from ~\cite{hashimoto2015direct}. Top panels show the spectral weight evolution within [0,70]~meV binding energy with temperature, normalized by total spectral weight over [0,250]~meV binding energy. (c) Momentum and doping dependent spectral weight competition between superconducting coherent peak (red) and pseudogap (blue) in Bi$_2$Sr$_2$CuO$_6$. Graduated shades correspond to the marked direction of doping. Adapted from ~\cite{kondo2009competition}. (d) Particle-hole asymmetry for the pseudogap in optimally doped Bi$_2$Sr$_2$CuO$_6$. Adapted from ~\cite{hashimoto2010particle}.}
	\label{fig:fig_cuprate5}
\end{figure}

 Pre-formed Cooper pairing is one prominent candidate to explain the pseudogap:  strong superconducting fluctuations above $T_c$ destroy phase coherence but not the pairing~\cite{emery1995importance}. However, progress over the last decade suggests that the pseudogap physics far exceeds this simple picture. First, the fluctuating superconductivity temperature has been shown to be substantially lower than the high energy pseudogap temperature scale $T^*$ in many cuprate systems (see also Section~\ref{SC_fluc})~\cite{kondo2015point,tallon2011fluctuations,bilbro2011temporal}. With high resolution and high statistics photoemission, the pseudogap is identified to contain a both energetically and temperature-wise separated component from superconducting fluctuations (Fig.~\ref{fig:fig_cuprate5})~\cite{tanaka2006distinct,lee2007abrupt,he2009energy,hashimoto2010particle,kondo2009competition,hashimoto2015direct,kondo2015point,chen2019incoherent}.

Further evidence indicating the pseudogap being more than simply fluctuating superconductivity comes from its temperature, momentum and doping dependence. Below $T_c$, the pseudogap manifests as a super-linear deviation from the expected momentum-dependence of a pure $d$-wave superconducting gap form near the antinode (Fig.~\ref{fig:fig_cuprate4}(a)) ~\cite{lee2007abrupt,vishik2012phase,anzai2013relation}. Above the fluctuating superconducting temperature but below $T^*$, a gapless ``Fermi arc'' forms in the nodal region (along the $(0,0)$-$(\pi,\pi)$ direction) that grows with temperature, while the antinodal pseudogapped region gradually shrinks to zero as $T^*$ is approached~\cite{norman1998phenomenology,kanigel2006evolution,vishik2012phase,kaminski2015pairing}. The different gap momentum dependence at different temperatures suggest that the pseudogap is more than a simple extension of the superconducting gap. Additionally, low-energy spectral weight analysis near the antinode shows a pronounced minimum at $T_c$ at under- and optimal doping but not in heavily overdoped side (Fig.~\ref{fig:fig_cuprate5}(a)(b)), indicating competing relation between the pseudogap and superconductivity~\cite{hashimoto2015direct}. Such a distinction is also confirmed by the distinct doping dependence of the low-energy spectral weight near the node and at the antinode (Fig.~\ref{fig:fig_cuprate5}(c))~\cite{tanaka2006distinct,kondo2009competition}.

Moreover, the pseudogap is shown to break particle-hole symmetry as evidenced by the misalignment between the Fermi momentum $\vec{k}_{\textrm{F}}$ in the normal state and the band bending momentum $\vec{k}_{\textrm{G}}$ in the gapped state (Fig.~\ref{fig:fig_cuprate5}(d))~\cite{he2011single,hashimoto2010particle}. Particle-hole symmetry is inherently required for zero-sum-momentum Cooper pairing on the Fermi surface. To account for both the broad linewidth (much larger than the corresponding thermal energy) and particle-hole asymmetry of the pseudogap, proposals based on short range AFM correlation~\cite{rice2011phenomenological} or density wave order~\cite{vershinin2004local,chen2004pair,lee2014amperean,hashimoto2015direct,hamidian2016detection,edkins2019magnetic} have seen limited success largely based on quasiparticle interference and Fermi surface measurements (see Section~\ref{sec_fermiology}). 

 More refined spectral analysis show competition between the pseudogap and superconductivity in both temperature and momentum space (Fig.~\ref{fig:fig_cuprate5}(c)), indicating again the potentially different nature of the high energy pseudogap from fluctuating superconductivity~\cite{kondo2009competition,kondo2011disentangling,hashimoto2015direct}. It should be cautioned that in this strongly correlated region, most of the aforementioned energy gaps lack coherent quasiparticle peaks on the gap edges, hence should \textit{not} be grossly taken as an order parameter of a new phase. For example, non-Fermi liquid self energies such as those in the quantum critical regime have been shown to produce similarly incoherent, gapped single-particle excitation spectra around the chemical potential without breaking any additional symmetry \cite{wu2019special}. These considerations lead to the postulation of pseudogap being a crossover phenomenon.

\subsubsection{Emergence of quasiparticles}\label{sec_emergeQP}

\begin{figure} 
	\includegraphics[width=1\columnwidth]{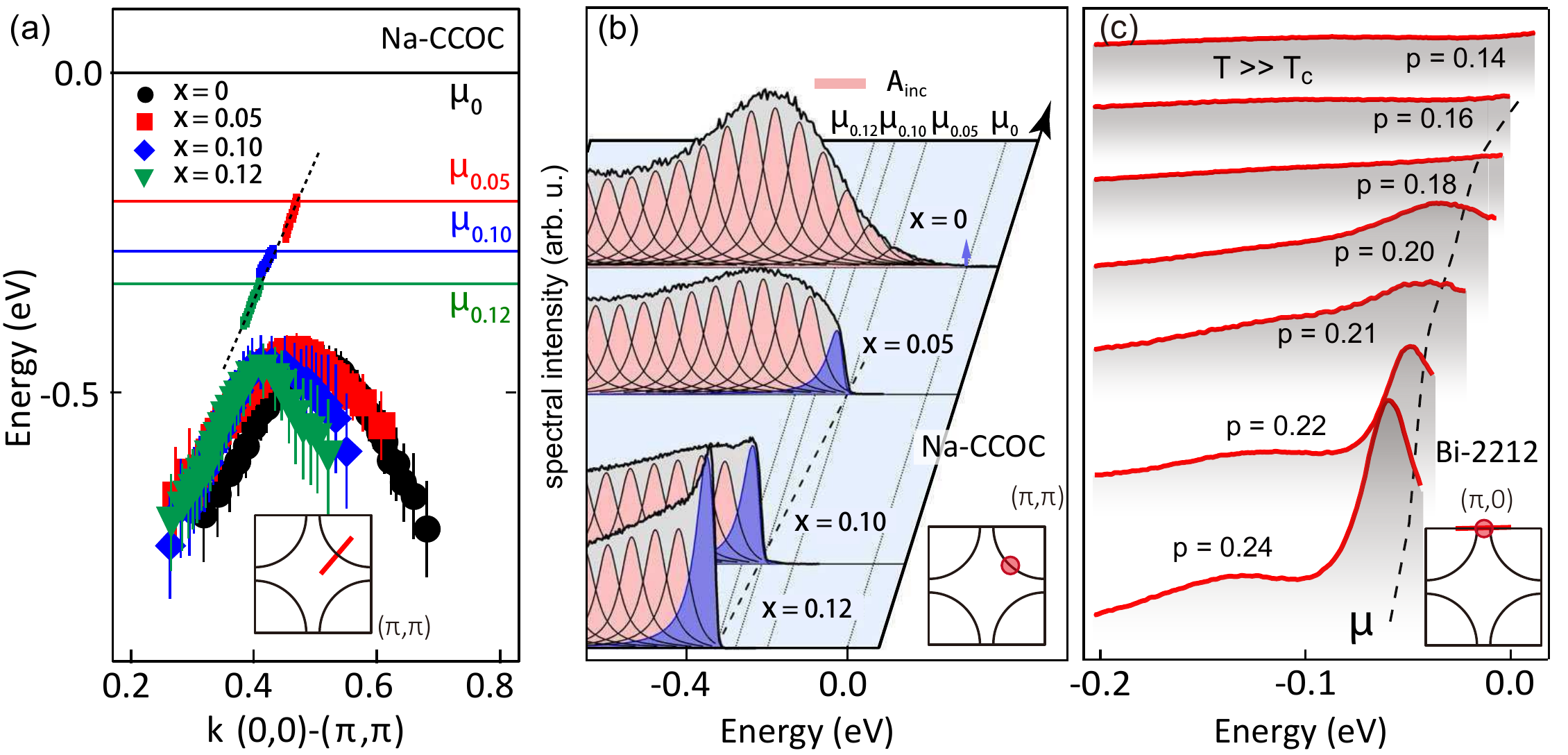}
	\caption{Emergence of quasiparticles in different parts of momentum space with hole doping. (a)(b) Nodal quasiparticles (the straight segment of the dispersions in (a) and purple shade in (b)) are created in Ca$_2$CuO$_2$Cl$_2$ immediately upon Na-doping on Ca sites. Adapted from~\cite{shen2004missing} (c) Doping dependent antinodal spectra in the normal state across the critical doping in Bi$_2$Sr$_2$CaCu$_2$O$_8$ (Bi-2212). Black dashed line is the schematic evolution of the chemical potential in overdoped Bi-2212. Adapted from ~\cite{he2018rapid,van1994electronic}.}
	\label{fig:fig_cuprate2}
\end{figure}

A long standing puzzle on the electronic structure evolution is the apparent lack of chemical potential shift as a function of doping~\cite{allen1990resonant}. This is exemplified by the seemingly doping-independent broad spectra around 0.4-0.5~eV binding energy, which was once considered the as lowest energy ``quasiparticle'' (Fig.~\ref{fig:fig_cuprate2}(a)(b)). Systematic study of the Na-doped Ca$_2$CuO$_2$Cl$_2$ system reveals that such a broad Gaussian spectrum is likely an envelope of polaronic shake-off satellites rather than the quasiparticle itself~\cite{shen2004missing}. This has been proposed to come from correlation-enhanced lattice phonon polaron near half-filling~\cite{mishchenko2004electron,rosch2004electron}. In either case, the spectra along the nodal direction consist of two components as shown in Fig.~\ref{fig:fig_cuprate2}(a)(b) - a weak quasiparticle component and a stronger polaron cascade. The incoherent feature completely dominates the spectra at low doping and shows little doping dependence, yielding an apparent lack of chemical potential shift. The emergence of a quasiparticle component with very low spectral weight changes the picture and provides evidence for a systematic shift of chemical potential with doping as one would expect, shown in Fig.~\ref{fig:fig_cuprate2}(a). This behavior is also seen in other experiments~\cite{hashimoto2008doping,fournier2010loss}. This mechanism naturally explains the valence band top broadening at ($\pi$/2,$\pi$/2) in the half-filled insulating phase (Section~\ref{sec_framework}). It also provides an alternative lattice-supplemented explanation to reconcile the inconsistency between $t$-$J$ model calculation and the exceptionally high binding energy spectrum at ($\pi$,0) in half-filled cuprates. This constitutes the phononic contribution to the high energy pseudogap phenomenon. Upon hole-doping, the coherent spectral weight of the nodal quasiparticle quickly grows while the weight of the polaronic shake-off band continuously shrinks (Fig.~\ref{fig:fig_cuprate2}(b)). This joint manifestation of electron-electron and electron-phonon interactions is generic in cuprates, especially near half-filling. It should be noted that spin polarons are also proposed to contribute to the emergence of nodal quasiparticles near half filling~\cite{martinez1991spin,wang_2015_origin}.

In contrast to the node, the antinodal spectral coherence exhibits a much slower recovery with hole-doping as indicated in {Bi$_2$Sr$_2$CaCu$_2$O$_8$}. The normal state antinodal spectrum remains highly incoherent even at optimal hole doping (Fig.~\ref{fig:fig_cuprate2}(c)), and only abruptly gains coherence past a putative critical doping $p_c\sim$~0.19~\cite{he2018rapid,chen2019incoherent,chatterjee2011electronic}. The strong coupling to the $B_{1g}$ phonon near the antinode is suggested to play a role in this phenomenon~\cite{cuk2004coupling,devereaux2004anisotropic,hashimoto2015direct}. For more discussions of the phonon contribution to the antinodal spectra, see Section ~\ref{sec_EPCAN}.

\begin{figure}
	\includegraphics[width=1\columnwidth]{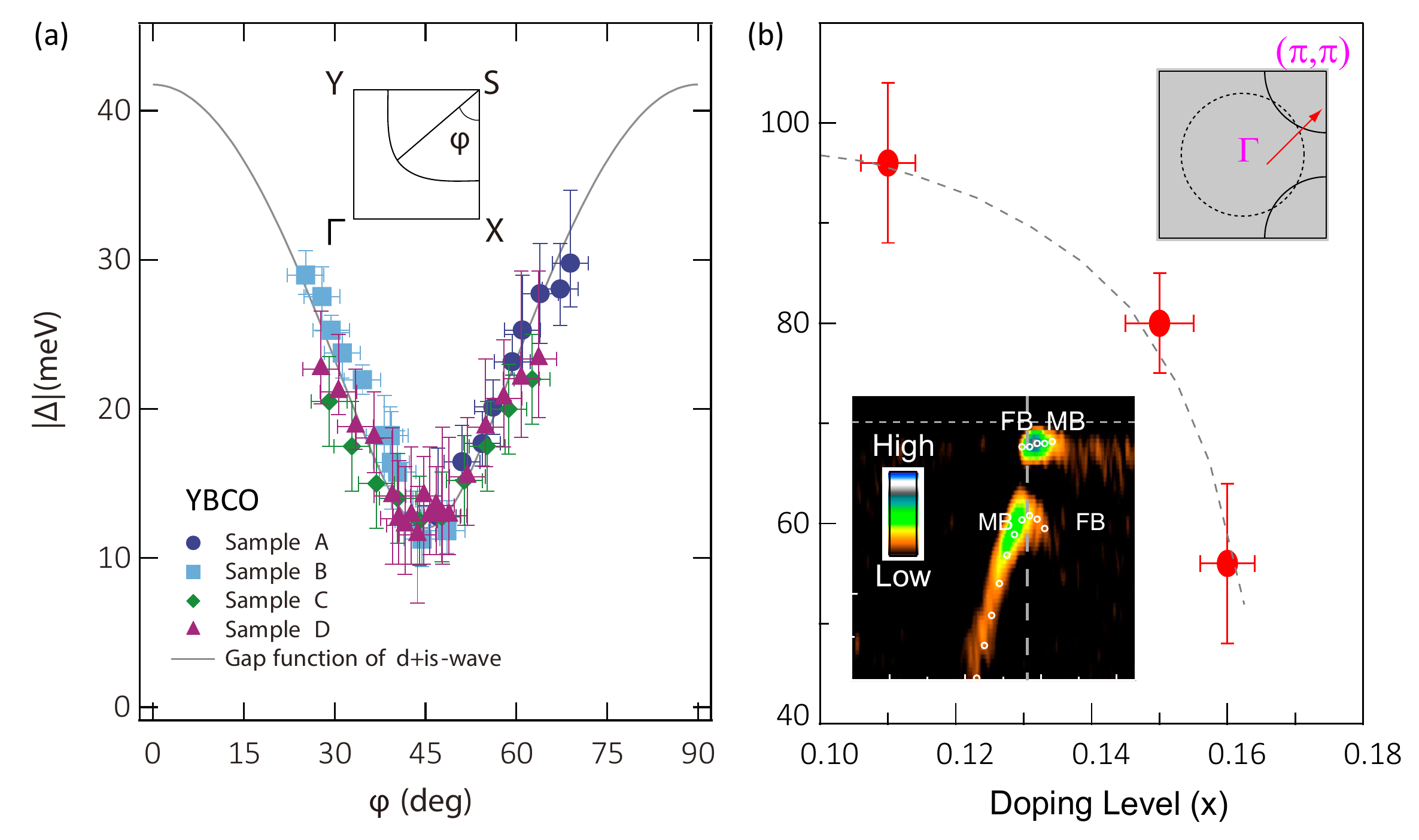}
	\caption{Electronic correlation related energy gaps. (a) Momentum dependence of the gap near the node in underdoped YBCO. Adapted from~\cite{okawa2009superconducting} (b) Energy-momentum spectra (inset) and doping dependence for the AFM gap at the AFM zone boundary in electron-doped cuprate Nd$_{2-x}$Ce$_x$CuO$_4$. Adapted from ~\cite{he2019fermi}.}
	\label{fig:fig_cuprate6}
\end{figure}

\subsubsection{AFM gap and other correlation gaps}\label{sec_LEgap}

Compared to hole doping, AFM order is more robust against electron doping~\cite{motoyama2007spin,armitage2010progress}. In electron-doped Nd$_{2-x}$Ce$_x$CuO$_4$ (NCCO), the highly momentum dependent AFM gap due to ($\pi,\pi$) band folding can be identified at the ``hot spot'' - where the original Fermi surface crosses the AFM zone boundary (cut shown in Fig.~\ref{fig:fig_cuprate6}(b) inset)~\cite{armitage2002doping,he2019fermi}. Surprisingly, this magnetic gap exists even outside the long range AFM ordered phase region, and coexists with superconductivity at least up to 16\% electron doping~\cite{song2012oxygen,song2017electron,horio2016suppression,he2019fermi}. Comparing the doping and momentum dependence of the measured magnetic gap with Hubbard model calculations, the Coulomb interaction strength $U$ is estimated to be as strong as 6$t$ even at a doping where superconducting $T_c$ is maximal (optimal doping)~\cite{he2019fermi}. Topological order has also been proposed as a possibility to explain Fermi surface folding without long range order~\cite{sachdev2018topological}.  

A different form of correlation-induced gap manifests strongly on the hole-doped side. At low hole-doping, the nodal spectrum is gapped by up to $\sim$30~meV from the chemical potential regardless of the presence of superconductivity~\cite{okawa2009superconducting,vishik2012phase,razzoli2013evolution,peng2013disappearance}. This gap is also highly anisotropic, with the gap minimum along the nodal direction (Fig.~\ref{fig:fig_cuprate6}(a)). While proposals for the origin of this gap involve disorder and complex order parameters, the lack of well-defined quasiparticles in this deeply underdoped regime also challenges explanations in terms of a simple quasiparticle gap~\cite{chen2009coulomb,atkinson2012robust,lu2014under}. With further hole doping, this gap at the node gradually disappears before the optimal doping is reached, with the specific values being family dependent.

\subsection{Superconducting properties}

\subsubsection{Momentum dependence}

\begin{figure} 
	\includegraphics[width=1\columnwidth]{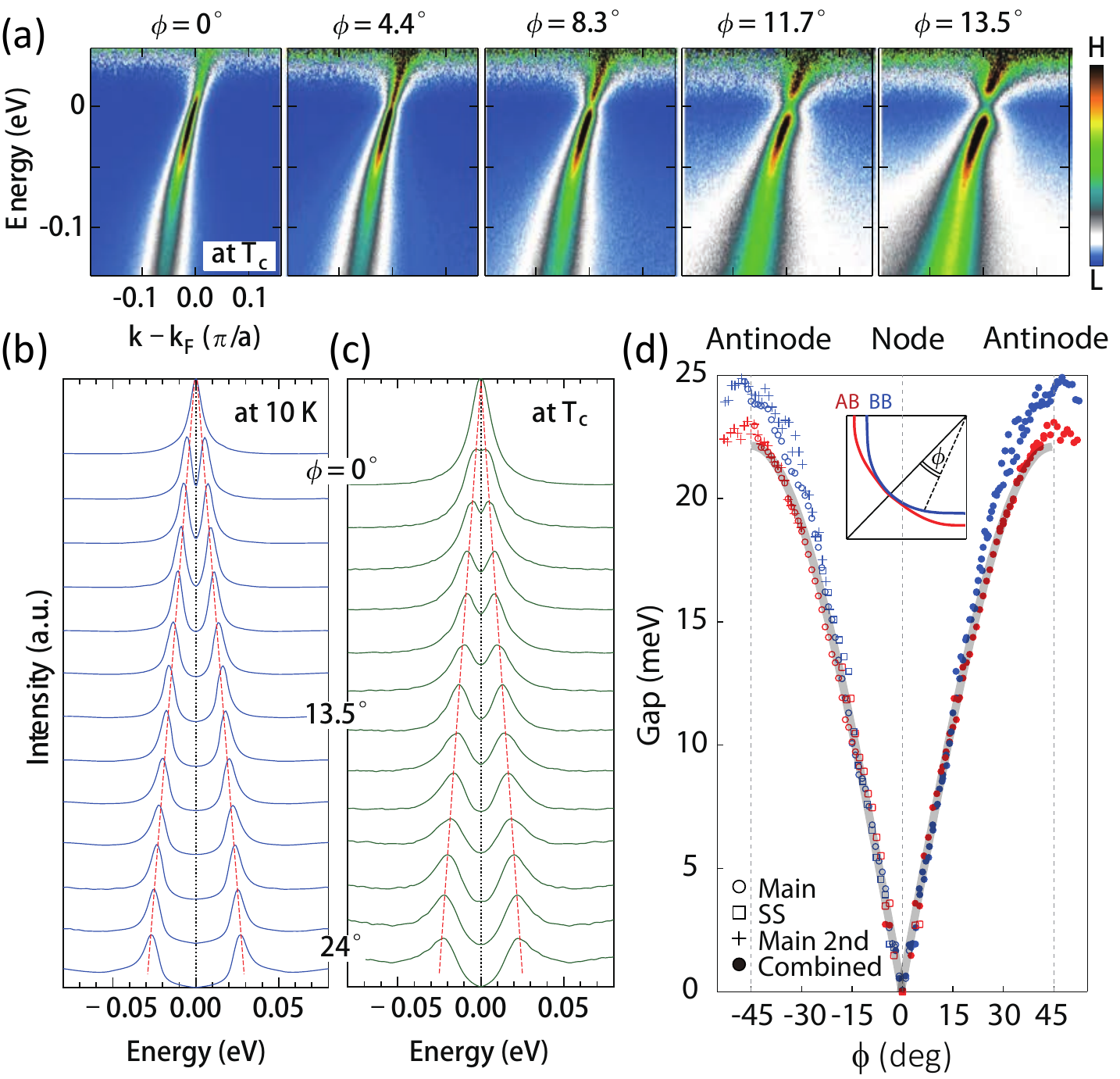}
	\caption{High resolution spectra near the node. (a) Fermi-function divided near nodal cuts in optimally doped Bi$_2$Sr$_2$CaCu$_2$O$_8$ taken at the bulk $T_c$. Note the clearly visible upper Bogoliubov quasiparticle dispersion above $E_\textrm{F}$. (b)(c) Symmetrized energy distribution curves from node to off-node direction at 10~K and the bulk $T_c$ = 92~K. Note the persistent $d$-wave gap at $T_c$ due to superconducting fluctuations. Adapted from ~\cite{kondo2015point}. (d) High resolution measurement examining the $d$-wave form of the energy gaps on both the bonding and antibonding bands in overdoped Bi-2212, measured with VUV laser-ARPES. AB - antibonding band. BB - bonding band. SS - superstructure. Grey line - $d$-wave fit. Adapted from ~\cite{ai2019distinct}.}
	\label{fig:fig_cuprate3}
\end{figure}

A single-particle energy gap is an important marker for various symmetry-broken phases. In a superconductor, the gap represents the energy required to break a Cooper pair. The gap is maximal at the normal state Fermi momentum $\vec{k}_{\textrm{F}}$, where the quasiparticle becomes an  equal admixture of particles and holes, and the nearby spectra satisfy $E(\mathbf{k}_\text{hole}) = -E(\mathbf{k}_\text{electron})$, termed ``\textit{particle-hole symmetry}.'' These composite excitations, known as Bogoliubov quasiparticles, have been  experimentally demonstrated in a momentum-resolved way in Bi$_2$Sr$_2$CaCu$_2$O$_8$~\cite{matsui2003bcs,takahashi2005high,balatsky2009bogoliubov}. Cuprates have a $d$-wave superconducting order parameter which additionally breaks rotational symmetry~\cite{shen1993anomalously,ding1995momentum,tsuei2000pairing,hashimoto2014energy}, giving rise to fully gapped states along the Cu-O bonding (antinodal) direction  and gapless states along the diagonal Cu-Cu (nodal) direction  (Fig.~\ref{fig:fig_cuprate1}, inset). 

\begin{figure} 
	\includegraphics[width=1\columnwidth]{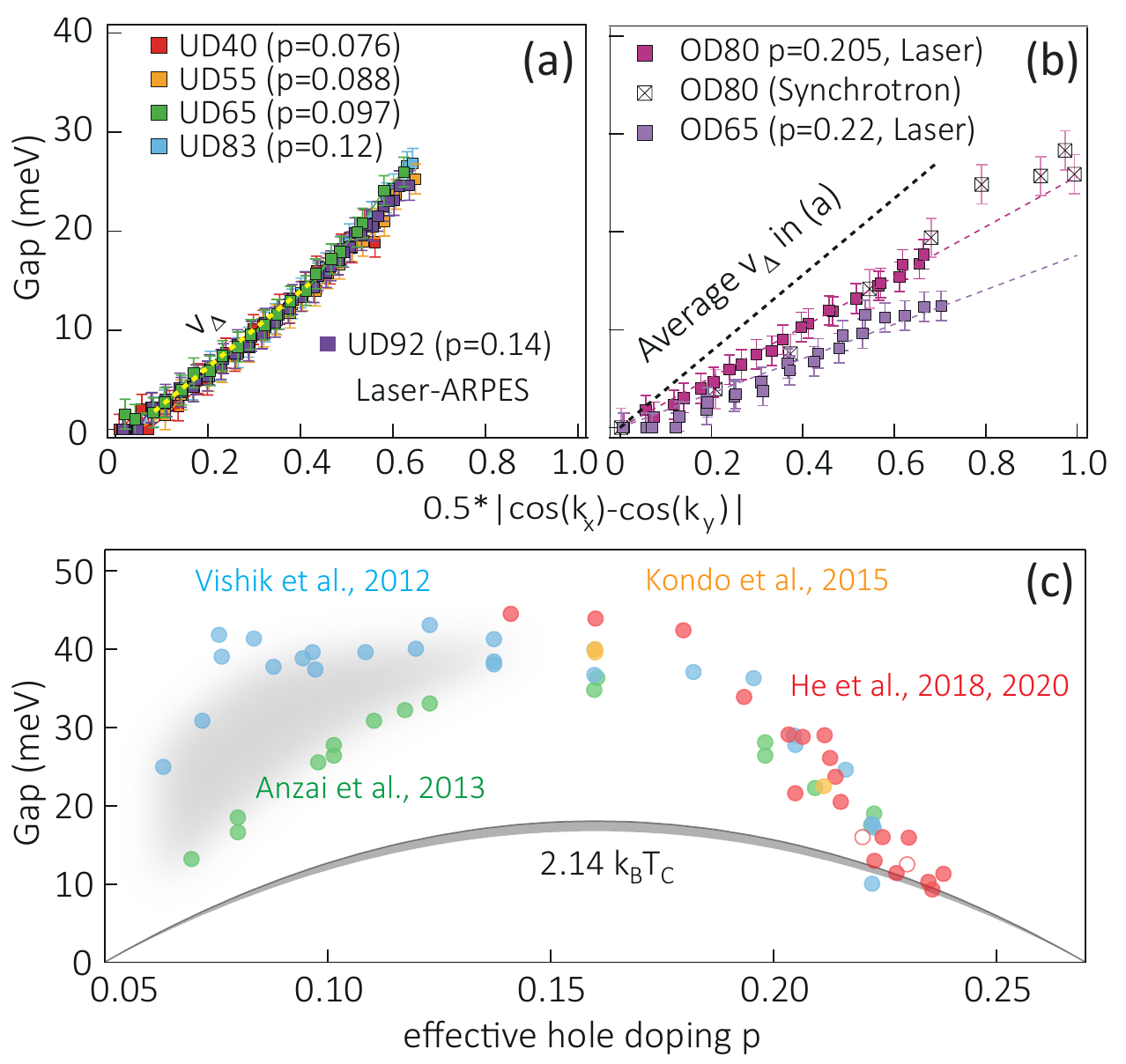}
	\caption{Momentum and doping dependent superconducting gaps in Bi$_2$Sr$_2$CaCu$_2$O$_8$. Near-node gap $\Delta(\vec{k})$ measurements in (a) underdoped and (b) overdoped samples. Assuming negligible pseudogap contribution near the node, the nodal gap velocity $v_\Delta$ may be equated with the superconducting gap at the antinode. Adapted from ~\cite{vishik2012phase}. (c) Doping dependent $d$-wave superconducting gap determined by fitting $v_\Delta$. Superconducting $T_c$ dome (grey crescent) is plotted against hole doping according to the empirical parabolic relation with maximum $T_c$ between 91~K and 98~K~\cite{presland1991general}, and is converted to energy by the weak coupling $d$-wave gap-to-$T_c$ ratio. Data points compiled from \cite{anzai2013relation,vishik2012phase,kondo2015point,he2018rapid,he2020xxx}. The variance in the underdoped region (grey shade) is a combined result of deviation from simple $d$-wave gap form and different fitting momentum ranges for $v_\Delta$.}
	\label{fig:fig_cuprate4}
\end{figure}

Recent generation laser-based ARPES and VUV synchrotron-based ARPES ~\cite{Liu_2008_development,Kiss_2005_photoemission,Berntsen_2011_an,Harter2012a,he2016invited} have mapped the $\vec{k}$-dependent superconducting gap $\Delta(\vec{k})$ in the vicinity of the node  (Fig.~\ref{fig:fig_cuprate3},~\ref{fig:fig_cuprate4})~\cite{kondo2015point,vishik2012phase,vishik2010arpes,sakamoto2016carrier,sakamoto2017superconducting,ai2019distinct}. The measured $\Delta(\vec{k})$ can be fitted to the $d$-wave momentum form $\Delta(\vec{k}) = 0.5 \times v_{\Delta} \lvert \cos{k_x} - \cos{k_y} \rvert$, where $v_\Delta$ is the \emph{gap velocity} (dashed line in Fig.~\ref{fig:fig_cuprate4}(a)(b)). Assuming zero pseudogap contribution near the node, $v_\Delta$ may be identified with the magnitude of the superconducting gap at the antinode. This is considered a plausible way to isolate the superconducting gap component even in the presence of a pseudogap around the antinode. Based on this method, $v_\Delta$ is found to remain at $\sim$40~meV over a wide range of doping (0.07 $< p <$ 0.19) over which $T_c$ changes by a factor of two~\cite{vishik2012phase,sakamoto2017superconducting,zhong2018continuous}. However, it has also been suggested that even the near-nodal region may contain pseudogap contributions especially when underdoped~\cite{anzai2013relation}. The experimental variance in the underdoped regime (grey shade in Fig.~\ref{fig:fig_cuprate3}(c)) reflects the challenge to extract $v_\Delta$ in the absence of a full understanding of the pseudogap's potential impact towards the node. While the difference in the raw data itself is subtle, the extrapolation amplifies variations that depend sensitively on the fitting range and deviations from a simple $d$-wave form.  Regardless of these nuances, the superconducting gap-to-$T_c$ ratio ($2\Delta/k_BT_c$) near optimal doping is found to be $\sim$10, much larger than the weak coupling BCS prediction (Fig.~\ref{fig:fig_cuprate4}(c)). With progressive hole doping past a critical doping $p_c$, this ratio precipitously drops from a large value of $\sim$10 towards the $d$-wave BCS limit, indicating a weaker coupled, more BCS-like superconducting region at sufficiently high hole concentration~\cite{yoshida2011pseudogap,vishik2012phase,he2018rapid,zhong2018continuous}. A similar doping trend is also observed in single-layer cuprate Bi-2201, although the gap-to-$T_c$ ratio is consistently larger than that in Bi-2212 (see also Fig.~\ref{fig:ironSC2}(b))~\cite{sakamoto2016carrier,sakamoto2017superconducting}.

This BCS-like behavior comes together with increased metallicity. ARPES data at heavy hole-doping show quadratic binding energy dependence of the nodal quasiparticle lifetime, which is broadly taken as a trait of Fermi liquid behavior~\cite{chang2013anisotropic,yusof2002quasiparticle,koitzsch2004doping,kaminski2003crossover}. This interpretation is to be cautioned, given that the absolute value of the fitted self energy is still comparable to or larger than the quasiparticle binding energy.

The superconducting gap symmetry in electron-doped cuprates is more challenging to measure, mainly due to the much lower energy scales. Although $d$-wave-like near-node behavior has been reported~\cite{armitage2001superconducting,horio2019d}, the gap maxima is seen near the antiferromagnetic zone boundary instead of the antinode~\cite{matsui2005direct}, as is the case in hole-doped systems.

\subsubsection{Superconducting fluctuations}\label{SC_fluc}

Fluctuating superconductivity above $T_c$ -- in which the pairing amplitude is non-zero but global phase coherence is absent -- has been central to the discussion of the superconductor-insulator transition, the pairing energy scale, and the dimensionality in cuprates. High resolution photoemission measurements on Bi-2212 have revealed substantial fluctuating superconductivity over a temperature range $\Delta T\sim \text{(0.3-0.5)}T_c$ above the superconducting $T_c$ both near the node and at the antinode~\cite{kondo2015point,hashimoto2015direct,chen2019incoherent}. This is in part consistent with transport and tunneling results in the La-214 system~\cite{bovzovic2016dependence,zhou2019electron}. In near-optimally doped Bi$_2$Sr$_2$CaCu$_2$O$_8$ (Fig.~\ref{fig:fig_cuprate5}(a)), a signature of fluctuating superconductivity (blue dots) manifests as a distinct shoulder feature \textit{inside} the pseudogap (red squares) at or above $T_c$ (pink line). A similar feature for fluctuating superconductivity also shows up as a low-energy gap right above $T_c$ in the overdoped metallic region (Fig.~\ref{fig:fig_cuprate5}(b), blue dots)~\cite{hashimoto2015direct,he2018persistent,chen2019incoherent}. Evidence for superconducting fluctuations are also provided by trARPES measurements of the gap dynamics \cite{Smallwood_2012_tracking, Parham_2017_ultrafast,Boschini_2018_collapse}. The relative fluctuation temperature window $\Delta T/T_c$ is barely doping dependent~\cite{chen2019incoherent,tallon2011fluctuations,bilbro2011temporal}, which agrees with the ubiquitous short antinodal superconducting correlation length of only a few lattice constants~\cite{li2018starfish}. Such a short correlation length may also relate to the reduced phase coherence and superfluid density even in overdoped systems that exhibit metallicity. High statistics ARPES shows an exceptionally shallow and flat antinodal band in heavily overdoped cuprates~\cite{chuang2004bilayer,kaminski2006change,he2020xxx}. This departure from a parabolic low-energy band structure, in combination of strong impurity scattering potential, has been associated with a reduced superfluid density~\cite{pitaevskii2016bose,he2020xxx}. 

A major contention regarding the pairing mechanism concerns the pairing energy scale: whether it is due to low-energy pairing bosons on the equivalent energy scale of $T_c$~\cite{scalapino1995case}, or due to high-energy pre-pairing on the scale of $U$~\cite{anderson1987resonating}. In the heavily overdoped cuprates, the superconducting gap opens from a coherent, gapless normal state, at a temperature only modestly higher than the transport $T_c$. This observation suggests weak-coupling pairing mechanism in certain heavily hole-doped cuprates.

\subsection{Fermi surface} \label{sec_fermiology}

  \begin{figure}
  \centering
 	\includegraphics[width=0.7\columnwidth]{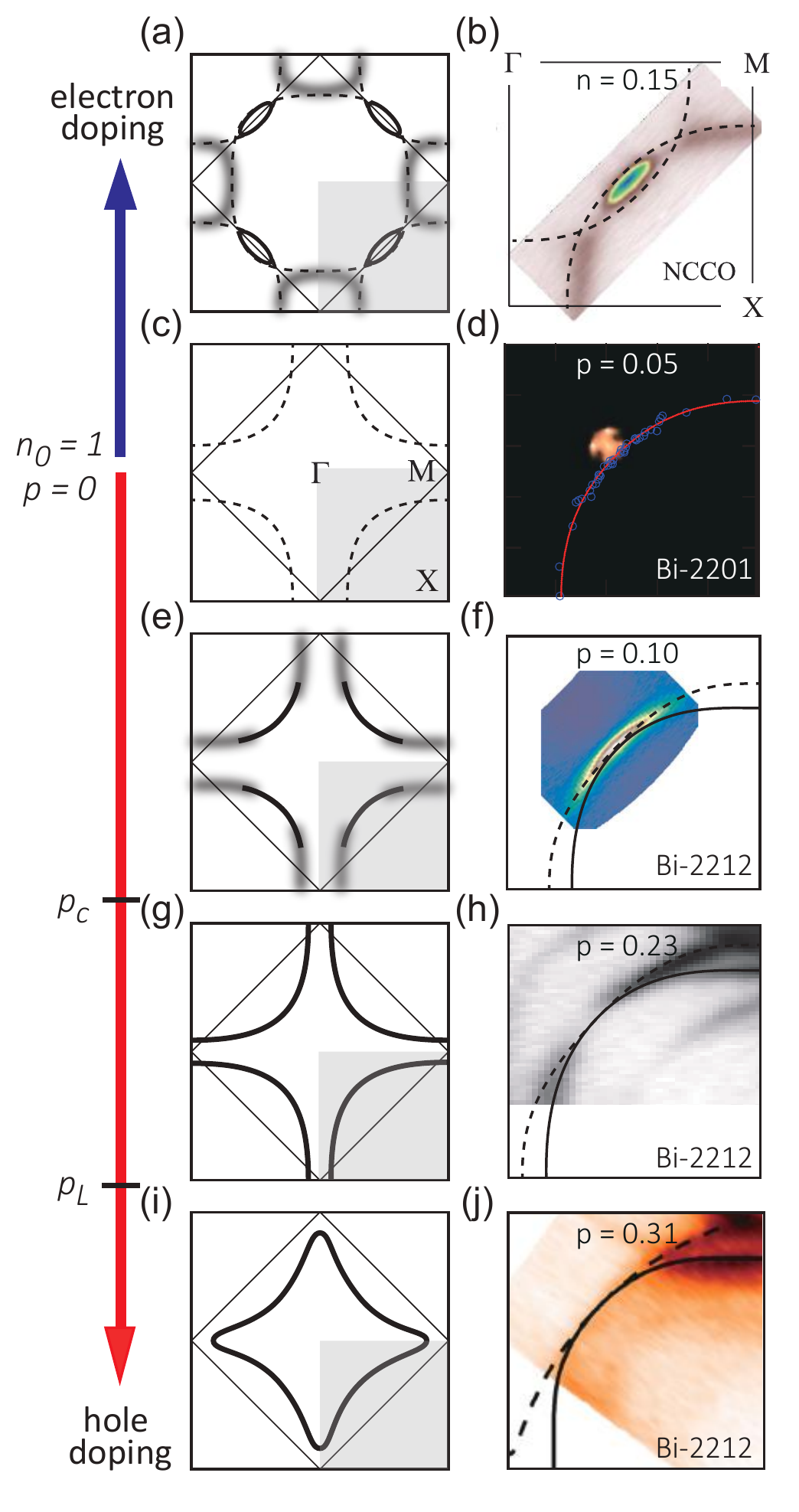}
 	\caption{Cuprate Fermi surface evolution with carrier doping. (a)(b) Schematics and experimental data on the electron doped Fermi surface in (Nd,Ce)$_2$CuO$_4$ with coherent nodal hole pocket (solid black) and less coherent antinodal electron pocket (grey shadow). Dashed lines are Fermi surface sheets before AFM folding. Thin solid diamond is the AFM zone boundary. Adapted from~\cite{song2012oxygen}. (c)(d) Expected tight-binding Fermi surface and the actual Fermi surface close to half-filling in Bi$_2$Sr$_2$CuO$_6$. Adapted from ~\cite{hashimoto2008doping}. (e)(f) Fermi ``arc'' in underdoped Bi$_2$Sr$_2$CaCu$_2$O$_8$ with a coherent node and incoherent antinode. Adapted from ~\cite{reber2012origin}. (g)(h) Above a certain critical doping $p_c$ the entire Fermi surface recovers coherence and connects to a single large hole pocket~\cite{he2018persistent}. The dashed and solid lines represent the antibonding and bonding bands in Bi$_2$Sr$_2$CaCu$_2$O$_8$ respectively. (i)(j) Further hole doping passing the van Hove point at M, undergoing a hole-like to electron-like Fermi surface Lifshitz transition at $p_L$. Adapted from \cite{drozdov2018phase}. All the right panels are zoomed in to the lower-right quadrant of the full BZ on the left.}
 	\label{fig:fig_cuprate7}
  \end{figure}

In a normal metal described by Fermi liquid theory, the Fermi surface is a ground state property that directly influences low-energy transport properties. The area enclosed by the closed Fermi surface measures the total carrier density, and is stable against most perturbative electronic interactions other than an attractive potential~\cite{luttinger1960fermi}. However, without the notion of quasiparticles, as is the case in many underdoped to optimally doped cuprates, one questions whether the Fermi surface remains a well-defined entity. Intriguingly, quantum oscillation measurements in underdoped cuprates provide direct transport evidence for the existence of long-lived quasiparticles at least under a high magnetic field at low temperatures~\cite{sebastian2011quantum}. Determining the existence and the shape of the Fermi surface in cuprates has been central to addressing various correlation effects and a putative quantum critical point (purple star in Fig.~\ref{fig:fig_cuprate1})~\cite{badoux2016change,markiewicz1997survey}.

\subsubsection{Small and large Fermi surface}

A schematic Fermi surface evolution with both electron and hole doping is shown in Fig.~\ref{fig:fig_cuprate7} for a typical cuprate.\footnote{Note that in multilayer systems, there can exist multiple pockets due to interlayer interactions.} We begin by considering the situation at half-filling:  a tight-binding model gives a large Fermi surface centered around $X$ whose area $A_\textrm{FS}$ reflects the carrier density and is therefore exactly half of the full Brillouin zone ($A_\textrm{FS} = 1/2$)  (dashed line in Fig.~\ref{fig:fig_cuprate7}(c)). In reality, as discussed in Sec.~\ref{sec_framework}, strong correlations modify this picture by opening a Mott gap and leading to the complete or partial removal of the Fermi surface (Fig.~\ref{fig:fig_cuprate7}(d)). 

Carrier doping leads to a reemergence of coherent spectral weight, particularly near the node. In electron-doped cuprates one must also consider the robust AFM order, which causes the Fermi surface to be folded by the AFM ordering wave vector ($\pi$,$\pi$) and gapped at the AFM zone boundary (thin-line diamond). These two effects give a small coherent nodal hole pocket and a less coherent antinodal electron pocket coexisting around optimal doping (Fig.~\ref{fig:fig_cuprate7}(a)(b))~\cite{armitage2002doping,armitage2010progress,song2012oxygen,song2017electron,he2019fermi}. Note that the short quasiparticle lifetime near the antinode can make this pocket undetectable by quantum oscillations ~\cite{kartsovnik2011fermi,breznay2016shubnikov,he2019fermi}.

Similarly, on the hole-doped side, coherent quasiparticles appear near the nodal region only inside the AFM zone while the antinodal region remains highly incoherent. However, unlike the electron-doped case, no AFM folding is observed, leading to a construct known as the ``Fermi arc'' (Fig.~\ref{fig:fig_cuprate7}(d)(f)) ~\cite{damascelli2003angle,kanigel2006evolution,hossain2008situ,kondo2015point}. The area of the putative nodal hole pocket -- empirically calculated by assuming that the arc indeed folds across the AFM zone boundary-- is given by the hole doping $p$ ($A_\textrm{FS} \sim p$). Further hole doping extends the coherent quasiparticle towards the Brillouin zone boundary, which eventually connects to form a full hole pocket centered around $X$~\cite{horio2018three,zhong2018continuous,drozdov2018phase,he2018rapid}, and qualitative agreement with the tight-binding model is restored. Note that the total carrier density now includes the electrons which were previously immobilized by correlations, giving $A_\textrm{FS} = 1+p$. The doping-dependent crossover in $A_\textrm{FS}$ is known as the ``$p$ vs $1+p$'' problem~\cite{uchida1991optical,he2014FS,fujita2014FS,badoux2016change}. The critical doping $p_c$ that marks the $p$ to $1+p$ transition -- if it exists -- demarcates an incoherent many-body region from a weakly-coupled quasiparticle region.

Indirect measures of the Fermi surface show mixed results concerning such a critical doping. Scanning tunneling spectroscopy (STS), quantum oscillation, thermodynamic and transport results hint at a universal small-to-large Fermi surface transition around $p_c=12\%$ for Bi-2201, $p_c=19\%$ for Bi-2212, YBCO, LSCO and Tl-2201~\cite{he2014FS,fujita2014FS,badoux2016change}. However, this contradicts existing Hall coefficient measurements at both high and low temperatures~\cite{hwang1994scaling,balakirev2003signature}, possibly due to material dependent Fermi surface anisotropy, dimensionality and topology. Whether the Fermi arc is indeed one side of a small nodal pocket is also contested for hole-doped cuprates. Early photoemission experiments reported evidence for small nodal pockets in hole-underdoped cuprates~\cite{meng2009coexistence,yang2011reconstructed,razzoli2010fermi}, which were later suggested to come from structural distortions and surface reconstructions~\cite{king2011structural,he2011doping,mans2006experimental,nakayama2006shadow,koitzsch2004origin}. More recently, high-resolution ARPES measurements in near optimally doped Bi-2212 showed that the Fermi arc tip does not end on the AFM zone boundary at $T_c$ even with the matrix element effect taken into account~\cite{kanigel2006evolution,hashimoto2011reaffirming,vishik2012phase,kondo2015point}. With increasing hole-doping, an abrupt restoration of quasiparticle coherence directly at the antinode at high temperature has recently been reported at $p_c=19\%$ in Bi-2212~\cite{chen2019incoherent}, which suggests that this crossover does not manifest solely at zero temperature.

The study of the Fermi surface area is also worth noting. Relying on the Luttinger theorem, the carrier doping in a 2D material may be estimated  from the Fermi surface area via ARPES. Strictly speaking, this procedure is only applicable when a well-defined full Fermi surface exists, and becomes problematic when part of the Fermi surface is incoherent or gapped such as in the pseudogap state. In such situations, pragmatically, an ``underlying Fermi surface'' is sometimes used to represent the hypothetical, otherwise ungapped full Fermi surface~\cite{gros2006determining}. Comparing to the nominal doping derived from normal state Hall measurements~\cite{presland1991general,ando2000carrier,balakirev2003signature}, the Fermi surface area measured by ARPES is consistently larger over the entire superconducting doping range~\cite{kondo2004hole,sakamoto2016carrier}. This is further corroborated by recent ARPES results on \textit{in-situ} ozone treated Bi-2212~\cite{drozdov2018phase,zhong2018continuous}. Inferring from the ARPES results in cuprates, the highly anisotropic low-energy electronic structure and strong electron correlation effects affect the simple reciprocal relation between the Hall coefficient and the carrier density.

\subsubsection{van Hove singularity}

While more three-dimensional cuprates like LSCO and YBCO do not have singularities in their electronic density of states, more 2-dimensional cuprates like Bi-2201 and Bi-2212 possess a band structure saddle point at ($\pi$,0), contributing  to a theoretically diverging density of states known as the van Hove singularity (vHs)~\cite{markiewicz1997survey}. Experimentally, this divergence is always broadened either due to correlation effects or $k_z$ dispersion~\cite{gomes2007visualizing}. Sufficient hole doping can lower the Fermi level through the vHs, turning the Fermi surface from a single hole type centered around $X/Y$ to an electron type centered around $\Gamma$ (Fig.~\ref{fig:fig_cuprate7}(i)). This Lifshitz transition at $p=p_L$ has been proposed as another candidate for the pseudogap critical doping $p_c$ in Bi-2212, but photoemission measurements indicate a much larger $p_L$ than $p_c$ in Bi-2212~\cite{drozdov2018phase,he2018rapid}\footnote{Even though the value of $p_L$ varies in ARPES literature in Bi-2212, potentially limited by resolution, it is consistently higher than $p_c$ ~\cite{chuang2004bilayer,kaminski2006change}. Note that the doping here is calculated as an average of both the bonding and antibonding bands' Fermi surface volumes, whereas the doping on the antibonding band itself well exceeds $30\%$.}. Moreover, the logarithmic divergence of density of states at the vHs is considered insufficient to account for the Sommerfeld coefficient divergence at $p_c$~\cite{michon2019thermodynamic,horio2018three}. Photoemission has also shown highly system dependent doping for the Lifshitz transitions: $\sim22\%$ in LSCO~\cite{yoshida2007low,hashimoto2008doping,razzoli2010fermi}, $\textgreater~26\%$ in Tl-2201~\cite{plate2005fermi}, $\sim 35-40\%$ in Bi-2201~\cite{kondo2004hole} and $\textgreater~30\%$ in surface self-doped YBCO~\cite{zabolotnyy2007momentum}. Interestingly in Bi-2201 and Bi-2212, this doping also coincides with the doping beyond which superconductivity vanishes~\cite{kondo2004hole,ding2019disappearance,drozdov2018phase}.

\subsection{Coupling between electrons and collective excitations}\label{sec_cuprate_coupling}

\begin{figure} 
	\includegraphics[width=1\columnwidth]{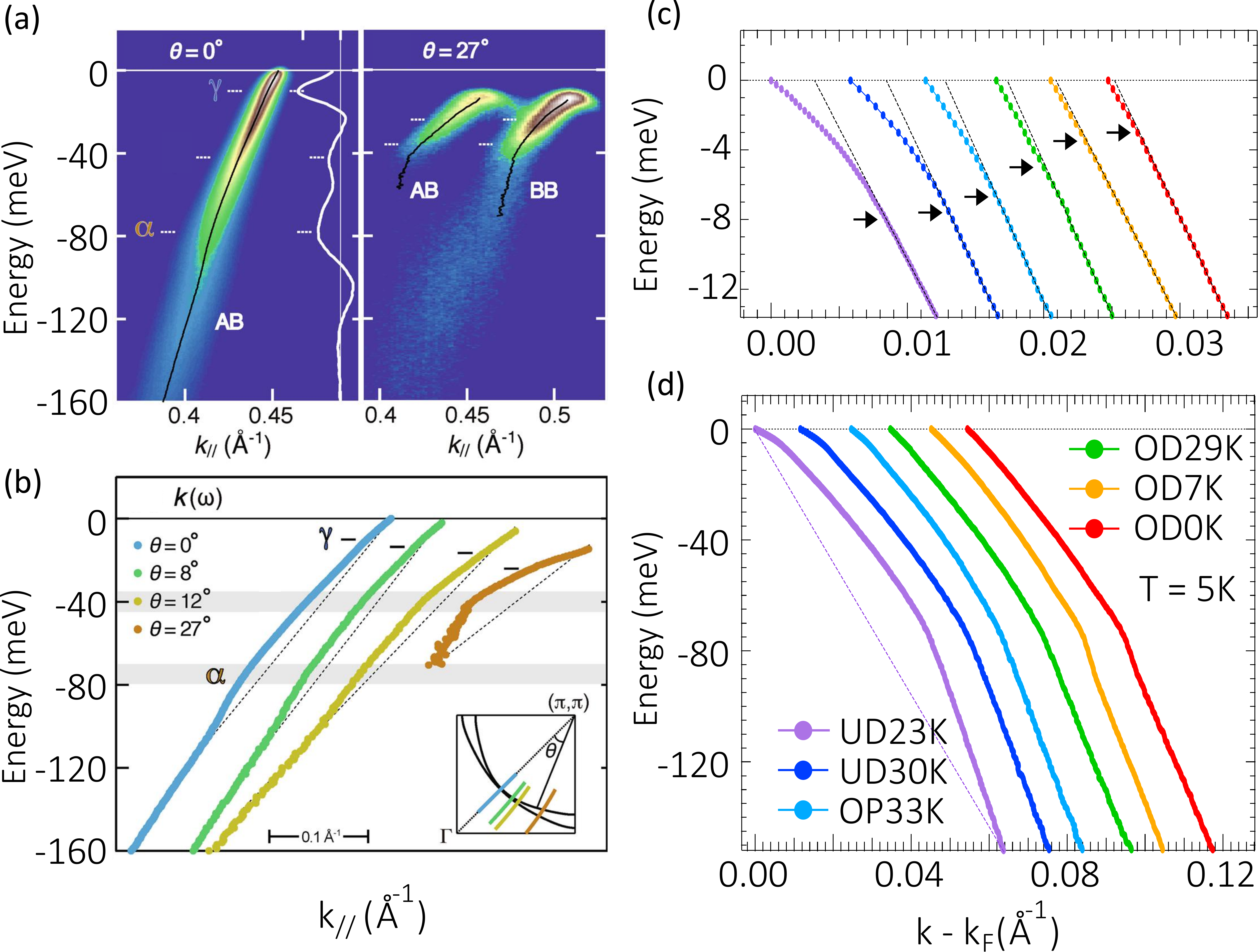}
	\caption{Signatures of bosonic mode coupling on the nodal spectra in cuprates. (a) Mode coupling effects on both the bonding (BB) and antibonding bands (AB) in overdoped Bi$_2$Sr$_2$CaCu$_2$O$_8$ at the node and near-node. (b) Extracted antibonding band dispersions from nodal to near-antinodal region. Adapted from~\cite{anzai2017new}. (c)(d) Doping dependence of the nodal dispersion in Bi$_2$Sr$_2$CuO$_6$, with the low-energy dispersion anomalies enlarged in (c). Adapted from \cite{kondo2013anomalous}.}
	\label{fig:fig_cuprate9}
\end{figure}

In conventional BCS superconductors, bosonic mode coupling provides the quintessential attractive interaction that pairs electrons together~\cite{bardeen1957theory}. The analogous ``pairing glue'' in high-$T_c$ cuprates is still unknown, motivating efforts to spectroscopically characterize electron-boson interactions.  Lattice phonon and spin excitations are two leading candidates to account for strong signatures of highly anisotropic mode-coupling observed in ARPES~\cite{lanzara2001evidence,zhou2003high,cuk2004coupling,lee2006interplay,anzai2010energy,he2013coexistence,plumb2013large,dahm2009strength}.

\subsubsection{Coupling to the node}

With the improved energy resolution of laser-based ARPES, a cascade of dispersion anomalies (``kinks'') have been revealed around the nodal momentum. At 70-80~meV binding energy (feature $\alpha$ in Fig.~\ref{fig:fig_cuprate9}(a)) a major dispersion kink is observed, which is usually interpreted as coupling to the oxygen in-plane breathing modes. It only weakens slightly at high temperatures, and its energy shows little momentum or family dependence~\cite{zhou2003high,borisenko2006kinks,meevasana2006doping,zhang2008identification,lee2008superconductivity,anzai2010energy,ideta2013effect,he2013coexistence,anzai2017new,vishik2014angle}. Laser-ARPES has reported evidence of an isotope effect for this mode up to 3.4~meV ~\cite{iwasawa2008isotopic}, following substantial improvement in resolution and statistics from earlier synchrotron experiments ~\cite{gweon2004an,douglas2007unusual,iwasawa2007re}. This mode is also widely seen in electron-doped cuprates, which lacks an apical oxygen on top of copper~\cite{armitage2003angle,park2008angle}. This further bolsters the in-plane nature of this phonon mode. This mode coupling has also been studied in trARPES by its imprint in the population relaxation dynamics \cite{Graf_2011_nodal,Yang_2015_inequivalence,Rameau_2016_energy} and pump-modulated self-energy \cite{Rameau_2014_photoinduced,Zhang_2014_ultrafast}.

The breathing mode's dominance near the node and weakened coupling off-node (Fig.~\ref{fig:fig_cuprate9}(b)) contrast the coupling behavior of the oxygen buckling B$_{1g}$ mode. The B$_{1g}$ mode was directly identified by Raman spectroscopy and inelastic neutron scattering to show softening across the superconducting $T_c$~\cite{reznik1995q,thomsen1988untwined}. This mode predominantly couples to off-nodal electrons in a way that is highly momentum, temperature and family dependent~\cite{cuk2004coupling,zabolotnyy2007momentum,plumb2013large,anzai2017new}. For example, it cannot effectively couple in single-layer cuprates due to symmetry constraints. In bilayer Bi-2212, where it does couple, the kink shows a strong energy shift with the superconducting gap as the temperature goes from below to above $T_c$~\cite{lee2008superconductivity,johnston2010systematic}. This ``gap-shifting'' behavior reflects the modified phase space for electron-phonon scattering due to the opening of the superconducting gap. Due to the similar energy scales of these optical phonons and the superconducting gap, mode identification from kinks usually requires detailed analysis in the superconducting state~\cite{sandvik2004effect,lee2008superconductivity}.

An even lower energy mode around 10~meV (feature $\gamma$) is observed in Bi-2212 and Bi-2201. Its coupling strength weakens rapidly with hole-doping (Fig.~\ref{fig:fig_cuprate9}(c)(d)), and its momentum-dependent kink energy tracks the $d$-wave momentum dependence of the superconducting gap below $T_c$~\cite{vishik2010arpes,anzai2010energy,kondo2013anomalous,peng2013doping}. Such a momentum-dependent form of gap-shifting is interpreted as coupling to forward scattering channels ($q \approx 0$) either from low-energy phonons or out-of-plane impurities~\cite{johnston2012evidence,hong2014sharp}, which may play a role in enhancing the $d$-wave superconductivity. 

All phonon energies in cuprates have an upper bound at $\sim$100~meV due to oxygen being the lightest composing element. Yet several nodal dispersion anomalies exist at higher binding energies. At 100-150~meV, broad self energy humps are seen in near nodal spectra in Bi-2212~\cite{borisenko2006kinks,zhang2008identification,he2013coexistence}, some interpreted as a final state effect specifically with low energy laser-based photoemission~\cite{Miller_2015_resolving}. At 300-400~meV, a universal nodal spectral ``waterfall'' -- near-vertical, incoherent spectral intensity -- breaks the nodal dispersion, where both dipole-transition matrix element effects and intrinsic strong electronic correlation effects (for example, spinon-holon separation) are proposed to contribute (Fig.~\ref{fig:fig_cuprate1_sup}(c))~\cite{graf2007universal,zhang2008high,rienks2014high,meevasana2008extracting,moritz2010insights}. Because of the steep quasiparticle dispersion, the nodal and near-nodal spectra can not only be used to extract the normal state electron self-energy~\cite{zhou2005multiple,kaminski2005extraction,zhou2007angle,meevasana2008extracting}, but also further invert the anomalous self energy in the superconducting state to directly reveal interactions in the pairing channel~\cite{bok2016quantitative} (see also Section~\ref{sec_analysis}).

\subsubsection{Coupling to the antinode}\label{sec_EPCAN}

\begin{figure} 
	\includegraphics[width=0.8\columnwidth]{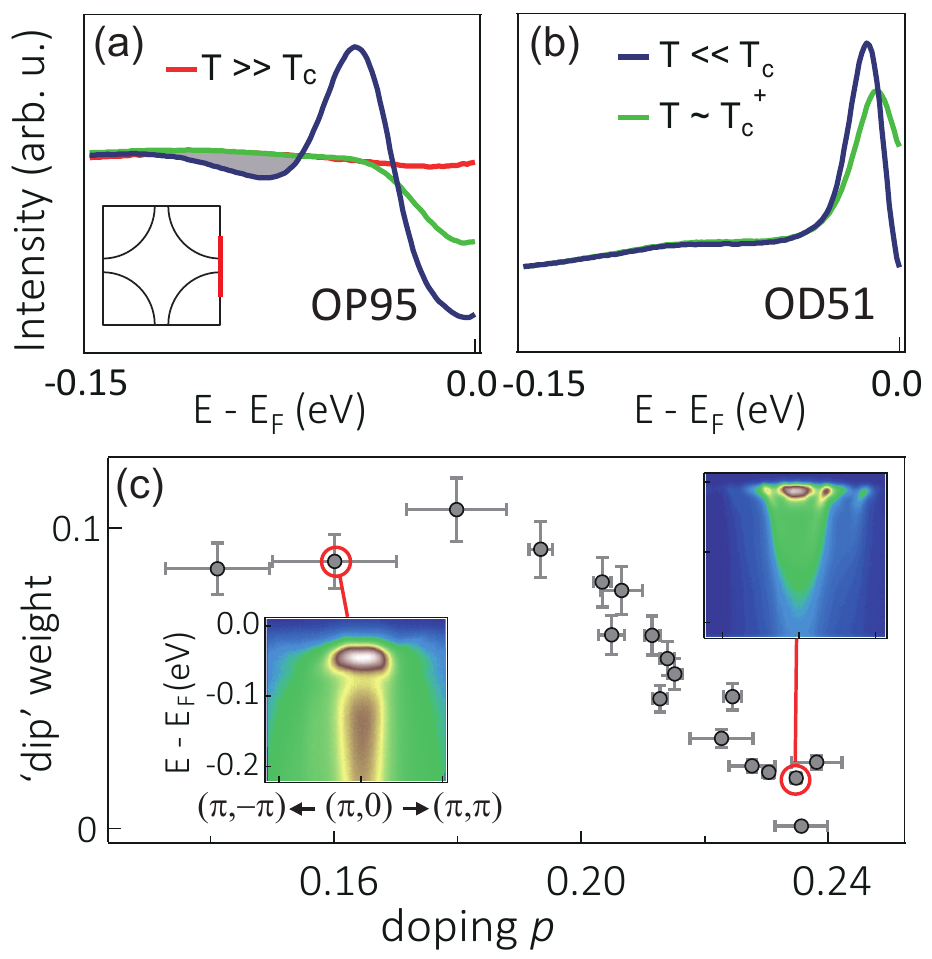}
	\caption{Rapid change of the antinodal mode coupling across the critical doping. Parallel momentum integrated antinodal spectra for (a) optimally doped and (b) heavily overdoped Bi$_2$Sr$_2$CaCu$_2$O$_8$ at different temperatures. The integration window is indicated by the red bar in the inset schematic Fermi surface. (c) Doping dependence of the spectral weight of the antinodal ``dip'' feature, reflecting the mode-coupling strength. The insets show false color plots of the antinodal spectra in the superconducting state for (a)optimally doped and (b) heavily overdoped Bi$_2$Sr$_2$CaCu$_2$O$_8$. Adapted from ~\cite{he2018rapid}.}
	\label{fig:fig_cuprate10}
\end{figure}

Unlike the dispersion anomalies near the node, mode coupling takes the form of spectral weight redistribution near the antinode partly due to the lack of sharp quasiparticle dispersions to directly intercept with. In optimally doped Bi-2212, the antinodal dispersion gradually develops a spectral weight depression around 70~meV below the superconducting pairing temperature (Fig.~\ref{fig:fig_cuprate10}(a)(c)), reminiscent of the density of states in superconducting Pb as a result of strong electron-phonon coupling~\cite{dessau1991anomalous,rowell1965tunneling,cuk2004coupling,devereaux2004anisotropic}. This has been interpreted as either coupling to the oxygen B$_{1g}$ phonon ($\sim$36~meV in Bi-2212) or the spin-resonance mode endemic to $d$-wave superconductors ($\sim$40~meV in Bi-2212), with additional spectral weight contributions from the bilayer splitting in Bi-2212 ~\cite{campuzano1999electronic,borisenko2003anomalous,gromko2003mass,dahm2009strength,devereaux2004anisotropic,he2018rapid}. The strong temperature dependence across $T_c$ of the spectral depression has been used as evidence for coupling to the spin-resonance mode~\cite{abanov1999relation,dahm2009strength}. However, in this scenario, the mode energy must scale with the superconducting gap size. In contrast, a constant energy offset is observed between the superconducting gap and the spectral depression over a wide doping range, suggesting that this mode is instead associated with the oxygen buckling $B_{1g}$ phonon~\cite{he2018rapid}. The substantial spectral weight redistribution from binding energy around $\Delta_{sc}+\Omega_{ph}$ to the quasiparticle peak exemplifies the intricate relation between the mode coupling and the superconducting quasiparticle spectral weight, which is further linked to the superfluid density and phase coherence by an empirical proportionality~\cite{feng2000signature,cho2019strongly}. In contrast, in heavily hole-doped systems, the spectral weight depression at the gap-shifted mode energy is substantially weaker, and most of the superconducting quasiparticle weight comes from within $\Delta_{sc}$ of the chemical potential (Fig.~\ref{fig:fig_cuprate10}(b)(d)).

\subsubsection{Phonons and superconductivity}

The role of electron-phonon coupling in high-$T_c$ superconductivity remains complex yet intriguing. Isotope substitution yields substantial change on the superfluid density but negligible effect on $T_c$~\cite{pringle2000effect,tallon2005isotope}. Phonons are also known to intimately participate in the charge order phenomenon, which directly competes with superconductivity~\cite{blackburn2013inelastic,letacon2014inelastic,chaix2017dispersive,he2018persistent}. These all highlight the challenge for traditional metrics to understand the physics in this complex superconductor, a fact that is not surprising in view of the pseudogap and competing orders, all of which make it difficult to disentangle microscopic contributions using macroscopic quantities. The sign-changing $d$-wave pairing symmetry naturally favors the electronic mechanism with repulsive interaction, rather than the sign conserving breathing phonon that connects the antinodal electrons~\cite{scalapino1995case,sandvik2004effect}. However, small-\textbf{q} coupling phonons like the B$_{1g}$ mode and the $\sim$10~meV low-energy mode are predicted to enhance $d$-wave pairing~\cite{devereaux2004anisotropic,johnston2010systematic,johnston2012evidence,li2016makes}. Indeed, a four-fold change of the superconducting gap size is found to be accompanied by a simultaneous rapid change of both electron-phonon and electron-electron interactions~\cite{he2018rapid}. This is evidenced by the strengthening of mode coupling at the antinode over the same narrow doping window in which the pseudogap vanishes and the superconducting gap-to-$T_c$ ratio normalizes to the BCS limit~\cite{vishik2012phase,he2018rapid,zhong2018continuous}. This indicates that these interactions turn a weaker $d$-wave superconductor to a stronger but more complex superconductor. The involvement of phonons, both in pair-enhancing and pair-breaking in the context of strong electronic correlation, needs to be carefully evaluated further. The current progress from photoemission indicates clear involvement of phonon coupling in the superconductivity, and a potential multi-channel pairing mechanism that resembles that of the monolayer FeSe/SrTiO$_3$ heterostructure (Section ~\ref{sec_sc_filmFeSe}).

\subsection{Outlook}
High-$T_c$ cuprate will continue to receive focused interest especially from ARPES investigation, thanks to the simplicity in its bare electronic structure and the richness in its derivative phases. On the one hand, many classic questions can be addressed with more versatile sample preparation methods and environment control, such as the delineation of intertwined orders and phase boundaries, the family- and layer-dependence of charge distribution and polarizability, quantification of carrier doping beyond the parabolic relation, and interfacial superconductivity tuning. In addition, the pursuit of the superconducting mechanism will see new possibilities for breakthrough in the heavily overdoped region, where the normal state is more coherent. The nature of the superconductor-to-metal quantum phase transition at the far end of the superconducting dome will also receive more attention, in a way akin to the historical development in the superconductor-to-insulator transition at very low doping near the Mott limit. Last but not least, the electron-doped cuprates -- and the role of long range AFM order and apical oxygen -- will receive more focused investigation due to the continued improvement in energy resolution. On the other hand, the search for new emergent states will continue, such as possible spin liquid near the Mott limit and the strange metal phenomena at high temperatures. Cuprate phenomenology also serves as a source of inspiration for out-of-the-box concepts such as correlation driven topological order and AdS-CFT formulation.

\section{Iron-based superconductors}\label{sec_feSC}
\subsection{Overview}\label{sec_FeSC_intro}

\begin{figure}
	\includegraphics[width=1\columnwidth]{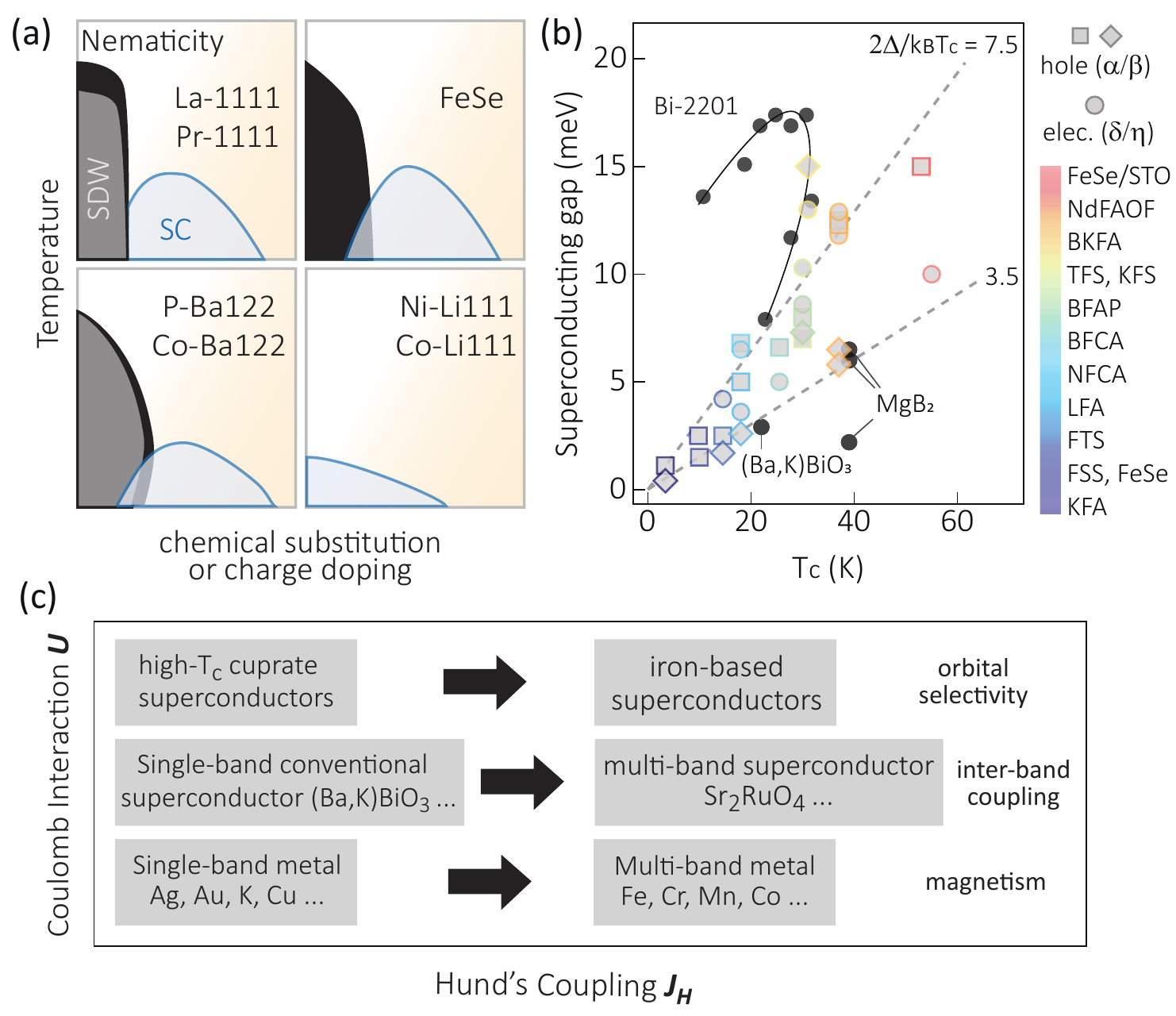}
	\caption{Family dependence and the multi-orbital nature of FeSCs. (a) Schematic temperature-doping phase diagrams for different FeSC families. Black - nematic order, grey - SDW order, blue - superconductivity, orange shade - more metallic transport. (b) Compilation of superconducting gap versus $T_c$ for representative FeSCs (colored markers), single layer cuprate Bi$_2$(Sr,La)$_2$CuO$_{6+\delta}$ (Bi-2201, increasing hole doping clockwise), MgB$_2$ ($\sigma$, $\pi$ and surface bands) and (Ba,K)BiO$_3$. For FeSCs, square and diamond (circle) markers represent superconducting gaps on the hole (electron) pockets. Dashed lines are reference $2\Delta/k_BT_c$ values at 3.5 ($s$-wave limit) and 7.5. Detailed reference list and acronyms are included in Ref[ ]. (c) Coulomb interaction $U$ and Hund's coupling $J_H$ cooperatively generate new phenomena from single-band to multi-band systems.}
	\label{fig:ironSC2}
\end{figure}

The cuprate superconductors were the lone family of unconventional high-$T_c$ superconductors until a new class of high-temperature superconductors was discovered in iron-based pnictide (FePn)~\cite{kamihara2006iron,kamihara2008iron,chen2008superconductivity,rotter2008superconductivity}, and subsequently chalcogenide (FeCh) compounds~\cite{hsu2008superconductivity}. Superconducting $T_c$, defined as the temperature of the Meissner signal onset and zero electrical resistance, reaches above 50~K in bulk single crystals~\cite{ren2008superconductivity,wu2009superconductivity}. Single-particle probes of energy gaps and electrical transport report $T_c$ values in excess of 55~K in thin films~\cite{qing2012interface,he2013phase,ge2015superconductivity}.\footnote{The actual $T_c$ value on this thin film system is still a contested topic, in part due to the difficulty of direct magnetic and transport measurements.} On top of the iron-based superconductors' (FeSC) high $T_c$, the excitement also lies in their rich material systems, wide range of compositional tunability, and the prospect that they can be conceptualized as the multi-band counterpart of the strongly correlated single-band cuprate superconductors. The addition of Hund's coupling $J_H$ to FeSCs of various correlation strengths $U$ unlocks both new experimental paths and theoretical frameworks to capture even richer physics beyond single-band correlated systems (Fig.~\ref{fig:ironSC2}(c)). There exist comprehensive reviews on the materials~\cite{johnston2010puzzle,stewart2011superconductivity,paglione2010high,wang2011electron,chen2014iron,hosono2018recent}, correlation effects including the magnetism and nematicity ~\cite{davis2013concepts,fernandes2014drives,fernandes2016low,si2016high,georges2013strong,haule2009coherence,dai2012magnetism,dai2015antiferromagnetic,dagotto2013colloquium}, superconducting pairing mechanism~\cite{hirschfeld2011gap,chubukov2012pairing,bang2017superconducting,scalapino2012common}, and thin film forms of FeSCs~\cite{pustovit2016metamorphoses,lee2018routes}. This section will focus on the role of ARPES in advancing the understanding of FeSC.

\begin{figure}[!htbp]
	\includegraphics[width=1.0\columnwidth]{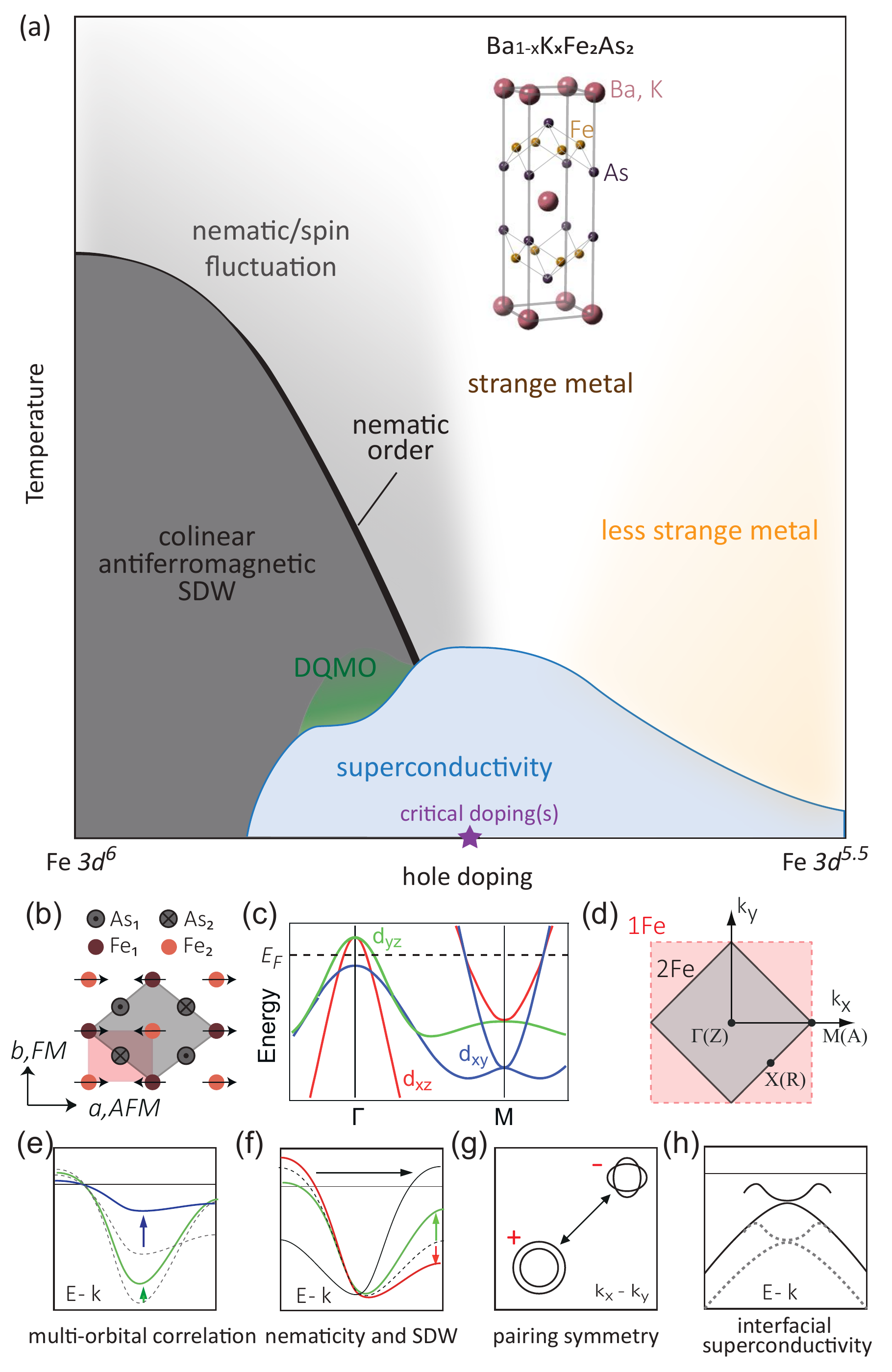}
	\caption{(a) Schematic temperature-doping phase diagram in K-doped Ba-122 FePn system. Adapted from Ref.~\cite{kasahara2010evolution}. Inset: crystal structure of parent compound Ba-122. Adapted from Ref.~\cite{rotter2008superconductivity}. (b) Top view of the Fe-sublattice after both the nematic order and collinear antiferromagnetic order develop. The grey shaded diamond indicates the 2-Fe unit cell, considering the As atoms alternately positioned above and below the Fe plane. The red shaded rectangle is the simplified 1-Fe unit cell that ignores the As atoms. (c) Low energy band structure along Fe-As bond direction in a 3-orbital model of a typical FeSC. (d) Typical Brillouin zone (BZ) of FeSC projected to the FeAs plane. Symmetry labels in parentheses are projected from $k_z$=$\pi/c$. Red - 1Fe BZ. Grey - 2Fe BZ. Schematics for (e) normal state orbital selective band renormalization (f) nematicity and AFM induced band reconstruction (neglecting $d_{xy}$ band) (g) unconventional superconductivity with inter-pocket pairing (h) electron-phonon coupling induced electron shake-off band. Adapted from Ref.~\cite{yi2017role,ye2011phosphor}.}
	\label{fig:ironSC1}
\end{figure}

Fig.~\ref{fig:ironSC1}(a) inset shows the body-centered tetragonal structure of one archetypal iron-pnictide BaFe$_2$As$_2$, where the alkaline earth metal layers and iron-pnictogen layers sandwich each other. Significant charge transfer between these two layers results in Fe $3d^6$ configuration, from which the multi-orbital low-energy electronic states originate. Within each layer, the pnictogen/chalcogen atoms alternately pucker above and below the adjacent Fe plaquettes, creating a tetrahedral crystal field that moderately elevates the $t_{2g}$ orbitals ($d_{xy}$, $d_{yz}$, $d_{xz}$) above the $e_g$ orbitals ($d_{3r^2-z^2}$, $d_{x^2-y^2}$). Fig.~\ref{fig:ironSC1}(c) shows the representative low-energy band dispersions along high symmetry momenta, with the $t_{2g}$ orbital contents most dominant~\cite{lebegue2007electronic,singh2008electronic,lu2008electronic,cvetkovic2009multiband}. Due to the glide-plane symmetry (reflection on the Fe-plane followed by a translation along diagonal Fe-Fe direction) of the pnictogen/chalcogen atoms, each two adjacent Fe atoms are inequivalent, resulting in a 2-Fe Brillouin zone (Fig.~\ref{fig:ironSC1}(b)(d), grey diamond), which by symmetry operations can be ``unfolded'' to a 1-Fe Brillouin zone (Fig.~\ref{fig:ironSC1}(b)(d), red square)~\cite{johnston2010puzzle}. In the latter case, any derived electronic structure has to be folded back into the 2-Fe zone in order to match experimental observations~\cite{kasahara2010evolution}. It should be cautioned that during this virtual folding process, orbital contents can change due to a glide-mirror symmetry on the Fe-plane~\cite{lin2011one,brouet2012impact}.  The low-energy electronic structure in FeSCs generally consists of three hole pockets centered around $\Gamma$, and two electron pockets located at the 2-Fe zone corner~\footnote{The symmetry notation in the Brillouin zones of FeSC can be different, mainly due to the different c-axis stacking between the 122 and 11/111/1111 families. See ~\cite{johnston2010puzzle} for a detailed discussion.}. Such multi-band low-energy electronic structure, together with the interactions therein, sets ground for the extremely rich phases that emerge in FeSC systems.

In FeSCs, proper chemical replacement in practically any atomic site can lead to superconductivity~\cite{chen2009coexistence,nandi2010anomalous,lai2014evolution,jiang2009superconductivity,kasahara2010evolution,parker2010control,liu2010pi,dai2015spin}. Resembling the cuprate superconductors but to a lesser degree~\cite{fukuzumi1996universal}, dopants placed in the charge reservoir layer lead to more robust superconductivity than direct doping in the conduction plane~\cite{ye2014extraordinary}. Meanwhile, the dopants modify the electronic states via changes in not only the carrier concentration, but also the Fe-As bond angle, strength of competing magnetic ground states and the level of disorder~\cite{khasanov2011mossbauer,li2012superconductivity,zhang2014effect}. Fig.~\ref{fig:ironSC1}(a) shows a representative phase diagram in the K-doped BaFe$_2$As$_2$ system. In the stoichiometric parent compound, the system starts off behaving like a bad metal with linear resistivity and magnetic susceptibility in the high temperature tetragonal phase~\cite{rotter2008superconductivity}. Upon cooling, it consecutively undergoes an orthorhombic structural distortion at $T_\textrm{structure}$ ($T_s$) and an antiferromagnetic transition at $T_\textrm{N\'eel}$ ($T_N$) within 1~K of each other~\cite{huang2008neutron}. The ground state is a collinear antiferromagnetic spin density wave (SDW) state with a slightly elongated $a$-axis in which iron moments align antiferromagnetically, and a compressed $b$-axis in which iron moments align ferromagnetically. While the orthorhombicity is smaller than 1\%~\cite{su2009antiferromagnetic,tomic2013plane}, 20\%-100\% in-plane electronic nematicity is observed through resisitivity measurements on detwinned single crystals~\cite{chu2010plane}. The \textit{larger} electrical resistance along the shorter, ferromagnetically ordered $b$-axis also signifies nontrivial low-energy electronic state evolution, emphasizing the need for a direct determination of the electronic structure. Further electron or hole doping will simultaneously suppress $T_s$ and $T_N$, and eventually induce superconductivity~\cite{chen2009coexistence,nandi2010anomalous,lai2014evolution,jiang2009superconductivity,kasahara2010evolution,parker2010control}. Strong nematic fluctuation and strange metal transport behaviors prevail above the superconducting dome, beneath which a critical point is suggested~\cite{fernandes2014drives,si2016high,hussey2018tale}. Further electron doping towards $3d^7$ usually restores metallicity to the normal state, while hole doping towards $3d^5$ steers towards stronger electronic correlation~\cite{yi2017role}. Intriguingly, chemical pressure (via nominal isovalent doping) and hydrostatic pressure on the iron-based parent compounds can also produce similar phase diagrams to those from heterovalent doping ~\cite{jiang2009superconductivity,colombier2009complete,liu2010pi,dong2013phase,klintberg2010chemical}, with an exception in the Fe(Se,S) system. There, the chalcogen height from the iron-plane responds differently to chemical and physical pressure~\cite{matsuura2017maximizing}. This aspect does not have an as comprehensive counterpart in the cuprate phase diagram.

Although intense effort has been invested towards a unified understanding of the magnetism and superconductivity in all FeSC, a wide distribution of family-dependent properties encumbered early efforts. In Pr/Ce/LaFeAsO$_{1-x}$F$_x$ (1111 systems), superconductivity does not coexist with the SDW phase (Fig.~\ref{fig:ironSC2}(a))~\cite{rotundu2009phase,luetkens2009electronic,zhao2008structural}. In (Ba,K)Fe$_2$As$_2$, Ba(Fe,Co)$_2$As$_2$ and BaFe$_2$(As,P)$_2$ (122 systems) and Co/Ni/Cu-doped NaFeAs (Co/Ni/Cu-Na111 system), superconductivity and SDW coexist in the underdoped region. And the structural transition is much more separated from the AFM transition in temperature on the electron doped side than the hole doped side (Fig.~\ref{fig:ironSC2}(a))~\cite{chen2009coexistence,nandi2010anomalous,kasahara2010evolution,parker2010control,wang2013phase}. In the iron-chalcogenide FeSe (11 system), despite strong AFM fluctuations, only the nematic phase exists under ambient pressure (Fig.~\ref{fig:ironSC2}(a))~\cite{wang2016magnetic,matsuura2017maximizing}. Adding to the peculiarity, in Li(Fe,Co)As (Co-Li111 system), both the magnetic phase and the nematic phase are absent, and the superconductivity starts right from the stoichiometric parent compound with $T_c$ = 18~K (Fig.~\ref{fig:ironSC2}(a))~\cite{dai2015spin}. The reported superconducting gap symmetry, gap sizes and fermiology are also widely family-dependent, ranging from nodeless to nodal gap structures on different Fermi surface sheets, where the gap-to-$T_c$ ratios (for the larger gap) roughly fall between the cuprates and more conventional superconductors such as MgB$_2$ and (Ba,K)BiO$_3$ (Fig.~\ref{fig:ironSC2}(b))~\cite{lin2008multiple,ding2008observation,nakayama2009superconducting,de2014selective,medici2015weak,xu2013electronic,zhang2016angle,yoshida2014orbital,drechsler2018mass,okazaki2012octet,borisenko2010superconductivity,chen2010electronic,miao2012isotropic,watson2015emergence,mou2011distinct,zhang2011nodeless,liu2011unconventional,zhang2012nodal,ye2011phosphor,zhang2012nodal}. However, it is believed that a common thread linking them all is the Hamiltonian incorporating moderate Coulomb interaction $U$ and Hund's coupling $J_H$ on the most dominant iron $3d$ orbitals (Fig.~\ref{fig:ironSC2}(c))~\cite{chubukov2014fe,fernandes2016low}. As such, factoring in the vastly different correlation strengths, doping levels, and the associated Fermi surfaces, the hope is to consistently describe the many family-dependent properties of FeSC with a universal, minimal microscopic Hamiltonian.

ARPES played an important role in the dissection of multiple electronic degrees of freedom in FeSCs. This section will first address electronic interaction effects with decreasing energy/temperature: (1) the orbital characters and orbital-selective renormalization on the normal state electronic structure (Fig.~\ref{fig:ironSC1}(e)), (2) the  evolution of the electronic structure in the nematic and SDW states (Fig.~\ref{fig:ironSC1}(f)), (3) the  pairing symmetry, pairing mechanism and other properties of the superconducting state (Fig.~\ref{fig:ironSC1}(g)). Finally, we discuss effects from the lattice degree of freedom, mainly the enhanced electron-phonon interaction due to strong electron correlation, and interfacial superconductivity in thin film iron chalcogenides (Fig.~\ref{fig:ironSC1}(h)). The discussion on possible topological superconductivity in FeChs will be elaborated in Section~\ref{sec_topoSC}. 

\subsection{The normal state}

\begin{figure}
	\includegraphics[width=1\columnwidth]{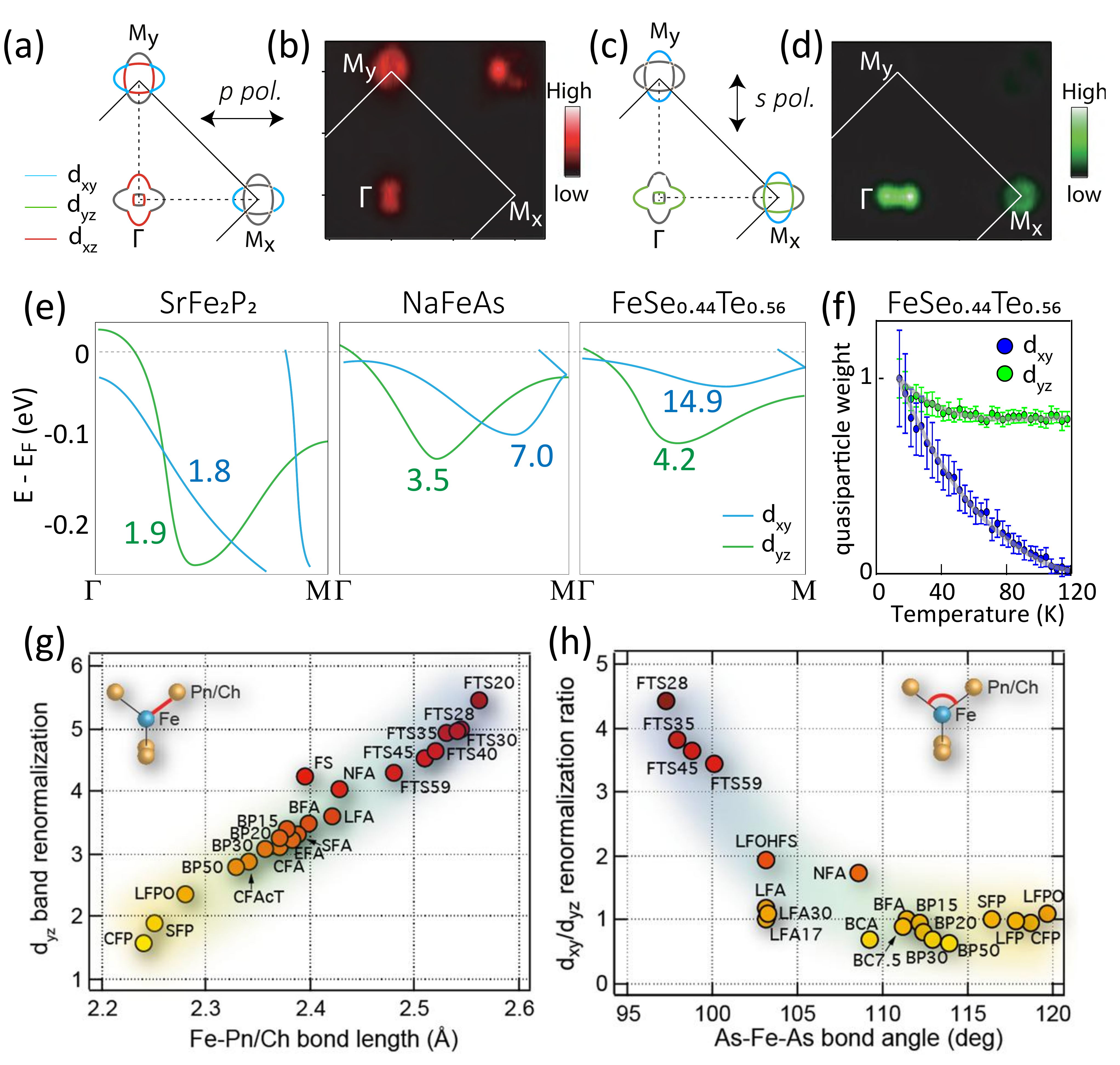}
	\caption{Low energy electronic band structure, Fermi surface, and correlation effects in the normal state of FeSC. Schematics (a)(c) and measured (b)(d) Fermi surfaces in the normal state of NaFeAs. The different experimental geometries are employed to highlight different orbital components based on the dipole transition matrix element effect. Adapted from ~\cite{zhang2012symmetry}. (e) Normal state $d_{yz}$ (green) and $d_{xy}$ (blue) orbital-dominated bands across different families with different correlation strengths. The bands are sketched according to experimental data. The numbers represent the renormalization factors on top of LDA calculated band dispersions~\cite{yi2017role}. (f) Orbital selective quasiparticle decoherence at high temperatures in FeTe$_{0.56}$Se$_{0.44}$. Adapted from~\cite{yi2015observation}. (g) $d_{yz}$ orbital renormalization factor positively correlates with the iron-chalcogen/pnictogen bond lengths. (h) Relative renormalization strength between $d_{xy}$ and $d_{yz}$ orbital-dominated bands as a function of iron-pnictogen bond angle of different FeSC. The acronyms represent different FeSC families, and are fully indexed in ~\cite{yi2017role}.}
	\label{ironSC3}
\end{figure}

\subsubsection{Multi-orbital character} \label{FeSC_multiorbital}
Identification of orbital characters constitutes the crucial first step in the investigation of electronic properties of multi-band metals like the FeSCs. As introduced in Section~\ref{sec_FeSC_intro}, the low energy electronic structure of FeSC most prominently features the $t_{2g}$ \textit{d}-orbitals. In the high temperature tetragonal paramagnetic normal state, the $d_{xz}$ and $d_{yz}$ orbital components near $\Gamma$ rotate into each other under 90$^{\circ}$ rotation, forming two hole-pockets near the Brillouin zone center with alternating orbital contents. The $d_{xy}$-dominant band forms a hole pocket at the 1-Fe BZ corner, which in reality folds into the 2-Fe BZ center~\footnote{In certain systems with strong $d_{xy}$ orbital renormalization or electron doping, some of the zone center hole pockets can be eliminated.}. In the meantime, together with the $d_{xy}$ orbital, the $d_{xz}$ and $d_{yz}$-dominant bands form elliptical electron pockets at the 2-Fe zone corner with alternating orbital components along the Fermi surface. Fig.~\ref{ironSC3} (a)-(d) showcase the orbital content composition on the Fermi pockets in the normal state of NaFeAs~\cite{zhang2012symmetry}. To delineate different orbital components in each band, different combinations of crystal orientation and incident photon polarization are employed to take advantage of the dipole transition matrix element in the photoemission process (Fig.~\ref{ironSC3}(c)(d))~\cite{zhang2012symmetry,yi2011symmetry,lu2008electronic,watson2015emergence,brouet2012impact,wang2012orbital} (Section~\ref{sec_intro_matrixelem}).

The family-dependent normal state electronic structures reveal a close link between the iron-pnictogen/chalcogen atomic arrangements and electronic correlation strengths. While the former can be routinely measured by X-ray diffraction, the latter can be quantified by taking the ratio between DFT calculated and ARPES measured energy bandwidths, namely the band renormalization. Fig.~\ref{ironSC3}(e) shows the schematic measured band dispersions along the $\Gamma-M$ direction in SrFe$_2$P$_2$, NaFeAs and FeSe$_x$Te$_{1-x}$, which reflect an increasing electronic correlations~\cite{yi2017role}. Indeed, reports of strong coupling phenomena such as polaron formation exist in the Fe$_{1.02}$Te system~\cite{liu2013measurement}. The $d_{xz/yz}$ band renormalization is shown to correlate with the iron-pnictogen/chalcogen bond length (Fig.~\ref{ironSC3}(g)), and monotonically increase as Fe filling approaches $3d^{5}$ over a wide range of FeSC families~\cite{yi2017role}.

\subsubsection{Orbital-selective Mottness}\label{sec_osmt}

Due to inter-orbital Hund's coupling, the effect of electronic correlation $U$ on the $d_{xy}$ orbital is relatively independent of that on other $d$ orbitals~\cite{haule2009coherence,georges2013strong}. This enables orbital selective band renormalization and Mott transitions that affect $d_{xy}$ orbital contents more than the rest~\cite{si2016high}. Fig.~\ref{ironSC3}(h) shows that as a function of iron-pnictogen/chalcogen bond angle, the $d_{xy}$-dominant band is more renormalized relative to that of the $d_{yz/xz}$-dominant band especially in the iron-chalcogenides~\cite{yi2017role,brouet2016arpes}. Moreover, temperature dependent studies found an orbital-selective coherent-to-incoherent crossover on the $d_{xy}$ orbital at high temperature, similar to that in other complex oxides (Fig.~\ref{ironSC3}(f))~\cite{yi2013observation,yi2015observation,pu2016temperature,niu2016unifying,miao2016orbital,neupane2009observation}. Such selective multi-orbital correlation has a profound impact on various low temperature phases as discussed in the following sections.

\subsection{Electronic nematicity and magnetic order}

\begin{figure}
	\includegraphics[width=1\columnwidth]{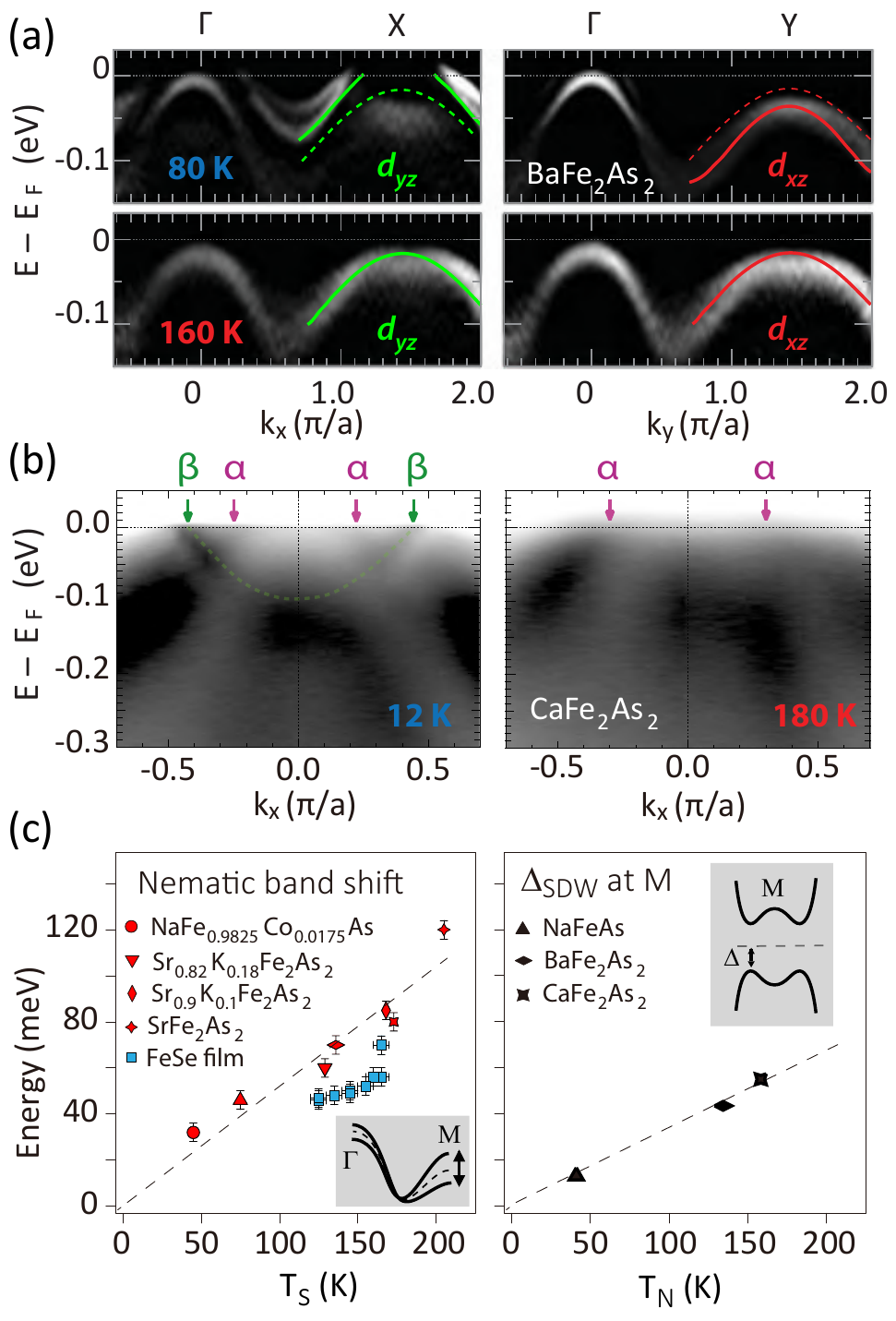}
	\caption{Band reconstructions associated with the nematic order and collinear antiferromagnetic order in underdoped FeSC. (a) Nematic order induced band shift along two high symmetry cuts in BaFe$_2$As$_2$ at 80~K and 160~K ($T_N\sim$138~K). Adapted from~\cite{yi2011symmetry}. (b) Energy momentum cut along $\Gamma$-M in the SDW state. Note the electron pocket folded from the zone corner (green arrows). Adapted from~\cite{kondo2010unexpected}. (c) Nematic band shift energy scale and the SDW gap size at M point plotted against their respective transition temperatures across various FeSC. Adapted from ~\cite{tan2013interface,yi2017role}.}
	\label{ironSC4}
\end{figure}

In underdoped BaFe$_2$As$_2$, cooling from the normal state drives the system into a structural orthorhombic and an electronic nematic phase~\cite{avci2012phase}, even though either or both of these lower symmetry phases may be absent in other families (Fig.~\ref{fig:ironSC2}). The system symmetry is then lowered from $C_4$ to $C_2$, and a subsequent magnetic phase transition brings the system from a paramagnetic to a collinear antiferromagnetically ordered state. Hence, one major early endeavor regarding the electronic nematicity was determining whether it is lattice (structure) driven, spin (magnetic) driven or orbital (charge) driven~\cite{fernandes2014drives}.

Due to the small structural orthorhombicity in this phase region, single crystals naturally form twin domains that are locally $\pi/2$-rotated with respect to each other. With typical domain size on the micron level~\cite{chu2010plane,tanatar2010uniaxial}, early ARPES with $\sim100\mu$m beam spot size inadvertently probed an admixture of electronic structures from both domains, adding to the complexity of an already multi-band system~\cite{liu2009band,yi2009unconventional}. Uniaxial pressure as low as 6~MPa was later applied \textit{in-situ} to detwin the sample~\cite{chu2010plane}, revealing a 30-120~meV band shift between the otherwise degenerate $d_{yz}$ and $d_{xz}$ components near the zone boundary in the nematic phase~\cite{yi2011symmetry,zhang2012symmetry,shimojima2014lifting,watson2019probing,kim2011electronic}. Fig.~\ref{ironSC4}(a) shows this band shift in detwinned Ba-122 along two orthorgonal crystal axes below the orthorhombic transition temperature $T_s$. The nematic energy also scales monotonically with $T_s$ across different FeSC systems (Fig.~\ref{ironSC4}(c), left panel)~\footnote{The original plot interpreted the band shift as the SDW gap, but was later realized to come from the electronic nematicity~\cite{tan2013interface,yi2019nematic}.}. DFT calculation shows that such a large nematic splitting energy scale cannot be accounted for by the $<$1\% orthorhombic lattice distortion~\cite{yi2011symmetry}, disfavoring the lattice-driven scenario for the electronic nematicity. Moreover, piezoresistance measurements indicate diverging electronic nematic susceptibility approaching the structural transition temperature, a further indication of the structural transition being a consequence, rather than the cause, of strong electronic nematicity~\cite{chu2012divergent}.

In the meantime, the resolution of either orbital or spin origin for the electronic nematicity has been considered at best system-dependent. In the orbital-driven scenario, the rising $d_{yz}$ orbital along AFM direction and the sinking $d_{xz}$ orbital along FM direction cause different electron filling, resulting in  the primary order being ferro-orbital order~\cite{lee2009ferro,lv2009orbital,bohmer2015origin}. Such orbital ordering induces magnetic anisotropy, which in turn drives the antiferromagnetic transition at a lower temperature. This scenario is particularly relevant in bulk FeSe, where despite substantial fluctuating magnetic moments, only the nematic order exists, and clear band splitting is observed~\cite{nakayama2014reconstruction,shimojima2014lifting,suzuki2015momentum,zhang2015observation,watson2015emergence,baek2015orbital}. However, simple ferro-orbital ordering implies constant band shift over the entire $\vec{k}$-space due to its localized nature~\cite{kontani2011origin}. This contradicts the highly $\vec{k}$-dependent anisotropic band shift observed in most FeSCs that suggests nematic bond order~\cite{suzuki2015momentum,zhang2016distinctive,Pfau_2019_momentum}. In the meantime, in doped Ba-122 systems, magnetic fluctuation is shown to scale with the orthorhombic fluctuation~\cite{fernandes2013scaling}, and the uniform magnetic susceptibility exhibits in-plane anisotropy~\cite{kasahara2012electronic}, alluding to a spin-driven nematic transition~\cite{avci2014magnetically}. In this case, spin-orbit coupling carries the anisotropy from magnetic fluctuations to break the lattice rotational symmetry at the same or even slightly higher temperature~\cite{fernandes2014drives,xu2008ising}. Indeed at the $\Gamma$-point, a sizeable spin-orbit splitting of otherwise symmetry-protected $d_{xz/yz}$ degeneracy has been observed in many FeSC compounds~\cite{brouet2012impact,suzuki2015momentum,johnson2015spin,borisenko2016direct,day2018influence}. 

While the nematic transition breaks the rotational symmetry and anisotropically shifts energy bands, the SDW transition breaks the translational symmetry, folds $\Gamma$ ($M_y$) and $M_x$ ($\Gamma'$) points into each other, and opens up energy gaps wherever bands cross and symmetry protection is absent (Fig.~\ref{ironSC4}(b))~\cite{kondo2010unexpected,yi2014dynamic}. The detailed band reconstruction schemes are comprehensively summarized in previous reviews by~\cite{yi2017role,zhang2012symmetry,shimojima2010orbital,ran2009nodal}. Here it should be emphasized that the $C_2$ symmetry from the nematic phase and the orbital dependence are evident in the SDW gap anisotropy: it is often the largest on $d_{xy}$ crossings, followed by $d_{yz}$ and $d_{xz}$ segments~\cite{richard2010observation,yi2014dynamic,yi2017role}. The emergence of the antiferromagnetic order can be viewed from both the itinerant perspective (Fermi surface nesting) and the localized perspective (local moment super-exchange)~\cite{fernandes2016low,si2016high,davis2013concepts}. In the former, the family-dependence is rooted in the different Fermi surface topology due to different electron filling and low-energy band structure. Whereas in the latter, the more strongly renormalized bands lose more coherent spectral weight to form localized magnetic moments. Similar to the nematic energy scale, the SDW gap energy also scales with the ordering temperature across different FeSC families (Fig.~\ref{ironSC4}(c), right panel). It should be cautioned that in many FeSC systems, it is non-trivial to cleanly separate the SDW gap contribution from that due to the nematic band shift. This implies that the extracted gap energies in Fig.~\ref{ironSC4}(c) should be viewed only as a general trend.

\subsection{Superconducting properties}
\subsubsection{Pairing symmetry and orbital dependence}
\begin{figure*}
	\includegraphics[width=2\columnwidth]{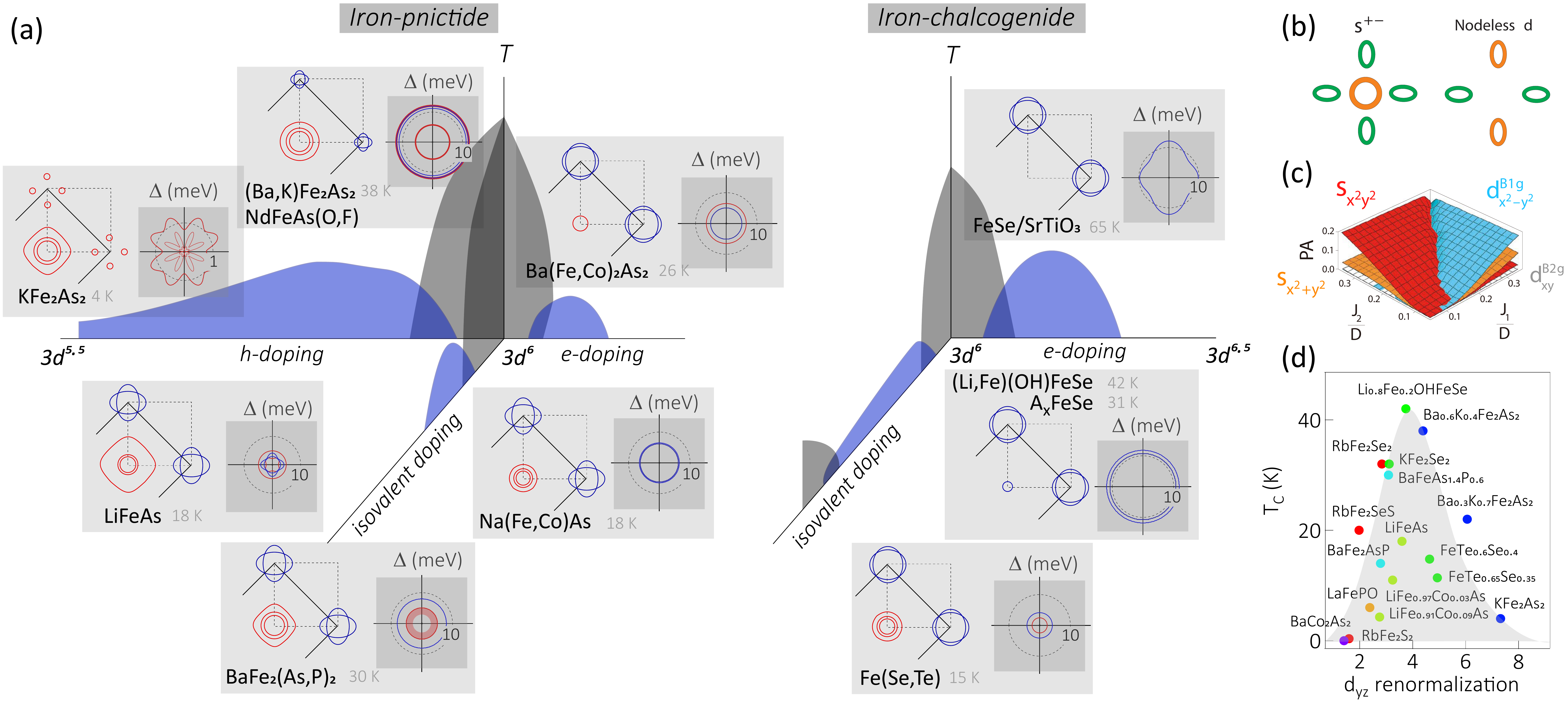}
	\caption{Superconducting gap symmetry and anisotropy across different families of FeSC. (a) Extended temperature-doping phase diagram with corresponding Fermi surfaces and the gap structures in various FeSC. Red circle represents hole pockets, and blue represents electron pockets. Polarplots of the gap amplitude on the hole and electron Fermi surface sheet obey the same color coding respectively. Dashed circles are grid lines for size references. The red shade indicates substantial $k_z$ dependence of the gap when projected onto the \textit{xy}-plane. (b) Proposed inter-pocket pairing mechanisms for $s^{\pm}$- (left) and $d$-wave (right) symmetry. Adapted from ~\cite{hirschfeld2011gap,hirschfeld2016using}. (c) Pairing amplitude computed based on coexisting electron and hole pockets. Different colors represent different symmetries as a function of nearest neighbor and next-nearest neighbor exchange coupling $J_1$ and $J_2$. Adapted from ~\cite{si2016high}. (d) Dependence of superconducting transition temperature $T_c$ on the $d_{yz}$ orbital renormalization strength. Same doping series of materials are represented by markers with the same color.
	Data in (a) are reproduced from ~\cite{ding2008observation,nakayama2009superconducting,okazaki2012octet,miao2012isotropic,watson2015emergence,mou2011distinct,zhang2011nodeless,liu2011unconventional,kondo2008momentum,ye2011phosphor,terashima2009fermi,zhang2012nodal,umezawa2012unconventional,drechsler2018mass,borisenko2010superconductivity,zhang2016superconducting,chen2010electronic,yoshida2014orbital}.}
	\label{ironSC5}
\end{figure*}

The superconducting transitions in bulk FeSC compounds exhibit much more mean-field like behavior in contrast to the strong fluctuations in the cuprates, largely due to the much higher carrier density, smaller superconducting pairing energy, and more three dimensionality~\cite{hardy2010doping}. The superconducting gap opening temperature aligns well with thermodynamic and transport transition temperatures~\cite{chen2008bcs}. STS, ARPES, transport and thermodynamic measurements report multi-gap behavior in the superconducting state of FeSC compounds~\cite{stewart2011superconductivity,kuzmicheva2016andreev}. The superconducting $2\Delta/T_c$ ratio (of the larger gap) varies from family to family, ranging from intermediate to weak coupling BCS limit~\cite{kuzmicheva2014andreev}. A case of BCS-BEC crossover has been made on on the extremely shallow and small $\Gamma$ hole pocket in FeSe$_{1-x}$Te$_x$, where the Fermi energy can be tuned to match the superconducting gap size by doping~\cite{lubashevsky2012shallow,shibauchi2014quantum,rinott2017tuning}. However, little associated thermodynamic evidence has been observed so far.

The pairing symmetry in FeSCs is highly system dependent, partly due to the family-dependent Fermi surface shapes and a vast distribution of (next-)nearest-neighbor exchange interaction strengths (Fig.~\ref{ironSC5})~\cite{davis2013concepts,si2016high,liu2011importance}. Benefiting from both advances in high resolution synchrotron-based and laser-based ARPES, the momentum structure of the superconducting gap amplitude on different Fermi surface sheets can be determined with sub-meV resolution~\cite{okazaki2012octet}. 

The most commonly discussed pairing symmetry is s$^\pm$-wave pairing in systems with hole and electron pockets (Fig.~\ref{ironSC5}(b) left panel). In this scenario, the $\Gamma$ hole pockets and the $M$ electron pockets possess opposite order parameter signs as suggested by the existence of a strong ($\pi,\pi$) spin resonance at $T < T_c$ in neutron scattering~\cite{shamoto2010inelastic,christianson2008unconventional,qiu2009spin}. This is necessitated by stronger Coulomb repulsion for \textit{inter}-pocket channels than \textit{intra}-pocket channels, made possible by the antiferromagnetic spin fluctuation~\cite{wang2013phase,chubukov2014fe,hirschfeld2016using}. ARPES finds nodeless superconducting gaps in Co/K-doped Ba122, NaFeAs~\cite{liu2011unconventional,lin2008multiple,ding2008observation}, undoped and intercalated bulk FeSe~\cite{xu2012evidence,mou2011distinct,wang2011strong} on all Fermi surface sheets (Fig.~\ref{ironSC5}). Accidental or symmetry enforced nodes may still appear in systems with both electron and hole pockets~\cite{wang2011electron}, as suggested in bulk FeSe~\cite{liu2018orbital}, P-doped Ba122 and K122 systems (Fig.~\ref{ironSC5})~\cite{ye2011phosphor,okazaki2012octet}. 

Meanwhile, when the system is heavily electron or hole doped, the intra-pocket repulsion regains dominance either due to increased inter-pocket screening or complete removal of $\Gamma$ hole pockets, and $d$-wave pairing amplitude may be increased (Fig.~\ref{ironSC5}(b) right panel)~\cite{hirschfeld2011gap,lee2018routes}. While nodeless anisotropic gaps are consistently observed in the monolayer system and its bulk counterparts~\cite{zhang2016superconducting,liu2012electronic,zhao2016common,lee2014interfacial,du2018sign,niu2015surface}, both $s$- and $d$-wave symmetry remain viable possibilities as the proposed nodal direction does not always intercept with a Fermi surface~\cite{fan2015plain,du2018sign,ge2019evidence}. However, in heavily electron doped A$_x$FeSe systems, nodeless superconducting gaps have been reported both at the zone center and the zone corner electron pockets~\cite{zhang2011nodeless,mou2011distinct,wang2012observation,xu2012evidence}. Recently, pairing symmetry that breaks time-reversal symmetry has been suggested in heavily hole-doped (Ba,K)Fe$_2$As$_2$ ~\cite{grinenko2018emerging}, although related ARPES study is lacking.

Orbital selective pairing is considered relevant in nematic FeSe systems, where the highly anisotropic $C_2$ gap function on the electron pocket roots in the inequivalence of $d_{xz}$ and $d_{yz}$ orbitals~\cite{liu2018orbital}. In the meantime, the gap anisotropy in underdoped Na(Fe,Co)As has been interpreted either as varying $d_{xy}$ orbital content along the Fermi surface~\cite{zhang2013measurement}, or as a result of momentum-dependent competition with SDW~\cite{ge2013anisotropic}. The role of $d_{xy}$ orbital in FeSe, on the other hand, is suggested to either have extremely weak spectral coherence ~\cite{sprau2017discovery} or move above $E_{\textrm{F}}$ due to nematic splitting and band hybridization at low temperatures or both~\cite{yi2019nematic,huh2019lifted}. Electronic correlation is considered central to the superconductivity in FeSCs, and the superconducting $T_c$ is shown to maximize for systems with intermediate electronic correlation strength across different families and dopings (Fig.~\ref{ironSC5}(d)). However, the impact from the highly family dependent low-energy electronic structures should not be overlooked, because small doping changes can often drive dramatic Fermi surface topology change in FeSCs. Importantly, the superconductivity in FeSCs also shows a wide range of family-dependent isotope effects~\cite{liu2009large,shirage2009inverse,khasanov2010intrinsic}, indicating complex lattice involvement amidst the highly intertwined orbital, magnetic and electronic degrees of freedom. In particular, the role of lattice will be discussed in Section~\ref{sec_FeSCepc}.

\subsubsection{Competition with SDW}

Detailed temperature dependent study of both the superconducting gap and the SDW gap indicates a competition between the two, even though the two gaps are substantially separated in energy~\cite{yi2014dynamic,ge2013anisotropic}. Nuclear magnetic resonance, M\"ossbauer spectroscopy and inelastic neutron scattering experiments find that superconductivity only coexists with SDW when the ordered moment is under 0.3$\mu_B$~\cite{goltz2014microscopic,laplace2012nanoscale,fernandes2010unconventional}, indicating their competing relationship. Other than antiferromagnetic spin fluctuation, nematic fluctuation is also postulated as another candidate to facilitate pairing~\cite{lederer2015enhancement,shibauchi2014quantum}. However, lattice expansivity and high pressure experiments have shown much less direct relation between superconductivity and nematicity in FeSe~\cite{bohmer2013lack,hosoi2016nematic,massat2018collapse}.

\subsection{Coupled lattice and electronic effects}\label{sec_FeSCepc}

On top of the rich electronic effects, the role of the lattice remains intriguing in FeSCs. The superconducting transition temperature $T_c$ shows systematic dependence on the iron-chalcogen bond angle, as reflected in the $T_c$ enhancement from 8~K to 37~K and the corresponding phonon frequency change under modest pressure in FeSe~\cite{mizuguchi2008superconductivity,medvedev2009electronic,margadonna2009pressure,huang2010raman}. In addition, $T_c$ depends sensitively on the pnictogen height, in a fashion similar to its dependence on the band renormalization factor in Fig.~\ref{ironSC5}(d)~\cite{yi2017role}. This suggests an intimate relation between the electronic and lattice degrees of freedom that impact $T_c$. Such relation is further re-enforced by family- and doping-dependent iron isotope effects for both superconducting $T_c$ and the SDW in some FeSC systems~\cite{liu2009large,shirage2009inverse,khasanov2010intrinsic}. From an ARPES perspective, close relationship between band renormalization and local lattice structure, as well as polaronic behavior in iron-chalcogenides are discussed in Section ~\ref{sec_osmt}. In this section, the correlation-enhanced electron-phonon interaction is first discussed in bulk FeSe. Then an interface-enhanced superconductivity in monolayer FeSe serves as an example of the richness in these strongly coupled systems, and the possibility of multi-channel origins of their extreme properties.

\subsubsection{Correlation enhanced electron-phonon interaction}

\begin{figure}
\centering
\includegraphics[width=1\columnwidth]{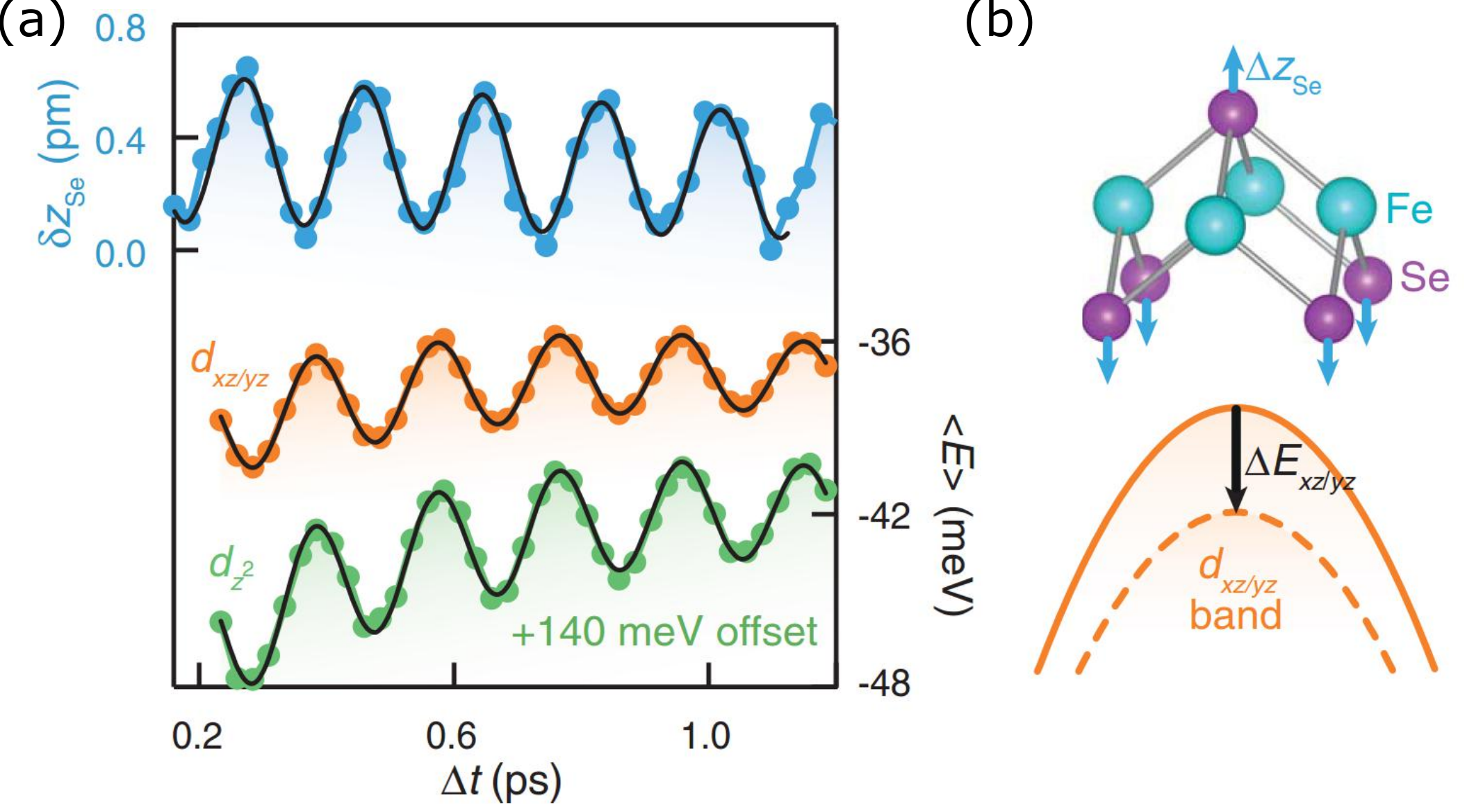}
\caption[Coherent phonon dynamics in FeSe]{Coherent $A_{1g}$ phonon in FeSe. (a) The lattice displacement (top) and band energy shifts (bottom) are resolved by time-resolved x-ray diffraction and trARPES, respectively. (b) Schematic of the lattice (top) and band (bottom) modulation due to the coherently excited mode. From \cite{Gerber_2017_femtosecond}.
\label{Fig_trARPES_FeSe}}
\end{figure}

The role of electron-phonon coupling in multilayer FeSe films was reinforced in a multimodal experiment combining lab-based trARPES with FEL-based time-resolved x-ray diffraction (trXRD). Electron-phonon coupling in a strongly correlated material is very difficult to quantify directly. This time-domain experiment permitted a precision not possible otherwise. In both measurements, an ultrafast laser pulse was used to excite a coherent $A_{1g}$ phonon (see Fig.~\ref{Fig_trARPES_overview}(d)). trXRD tracked the real-space lattice displacement, while trARPES monitored the corresponding shift in the electronic binding energies (see Fig.~\ref{Fig_trARPES_FeSe}). The ratio of these quantities yielded a measurement of the electron-phonon deformation potential far exceeding that predicted by DFT, and highlighed the role of electron correlations for enhancing the electron-phonon coupling strength \cite{Gerber_2017_femtosecond}. This aspect of the electron-phonon interaction was rarely discussed in earlier theoretical assessments of the phonon contribution to superconductivity.

\subsubsection{Interfacial superconductivity in thin-film FeSe}\label{sec_sc_filmFeSe}

\begin{figure} 
	\includegraphics[width=1\columnwidth]{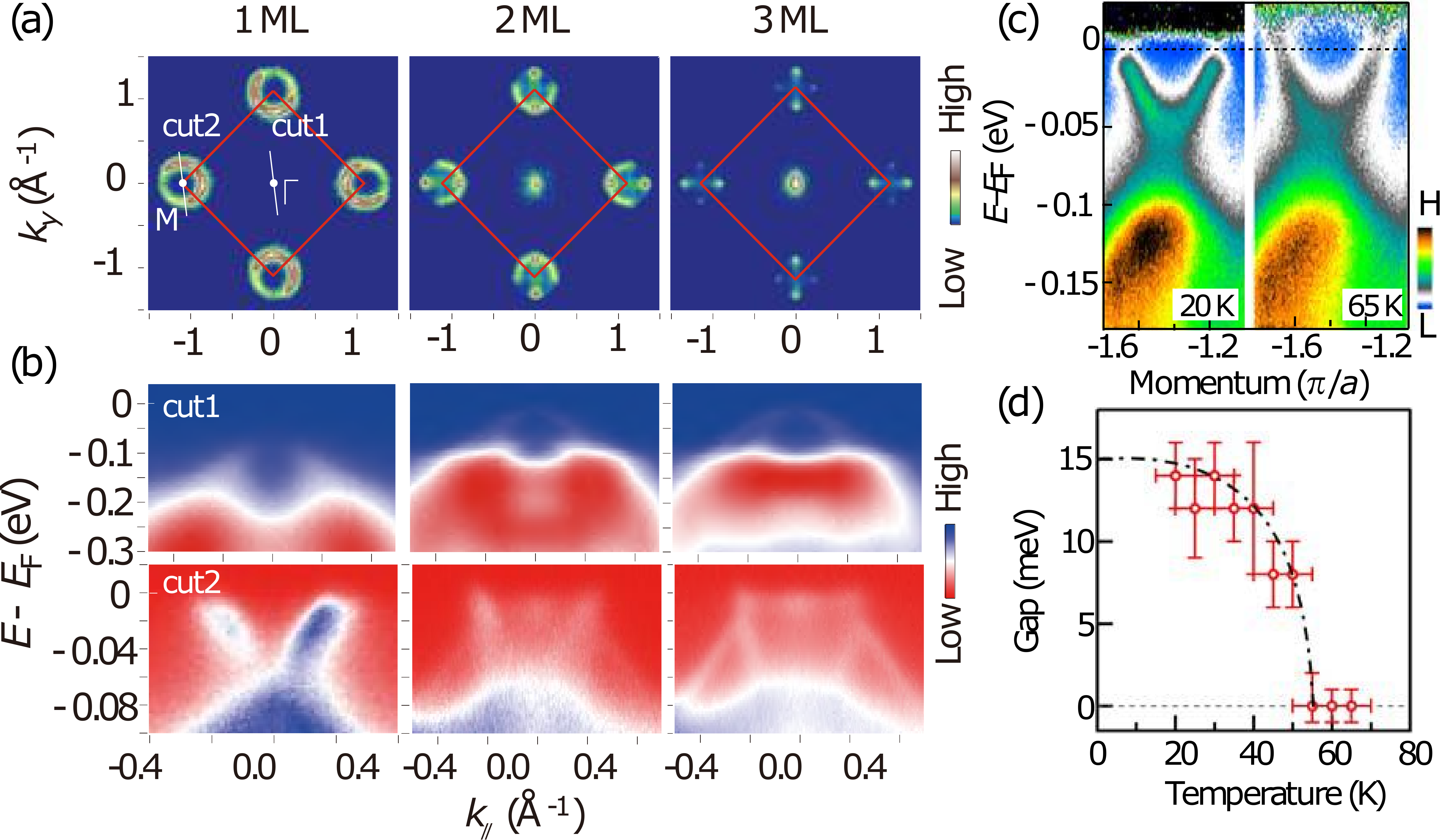}
	\caption{Layer and temperature dependence of the FeSe/SrTiO$_3$ thin film. (a) Fermi surface of monolayer, 2-layer and 3-layer FeSe film at low temperature (left to right). (b) The corresponding energy-momentum cuts along high symmetry direction - cut1 at $\Gamma$ and cut2 at $M$. Adapted from ~\cite{tan2013interface}. (c) The $M$-pocket at 20~K (left) and 65~K (right). (d) Temperature dependence of the superconducting gap on the $M$-pocket. Adapted from ~\cite{liu2012electronic}.}
	\label{ironSC6}
\end{figure}

\begin{figure} 
	\includegraphics[width=1\columnwidth]{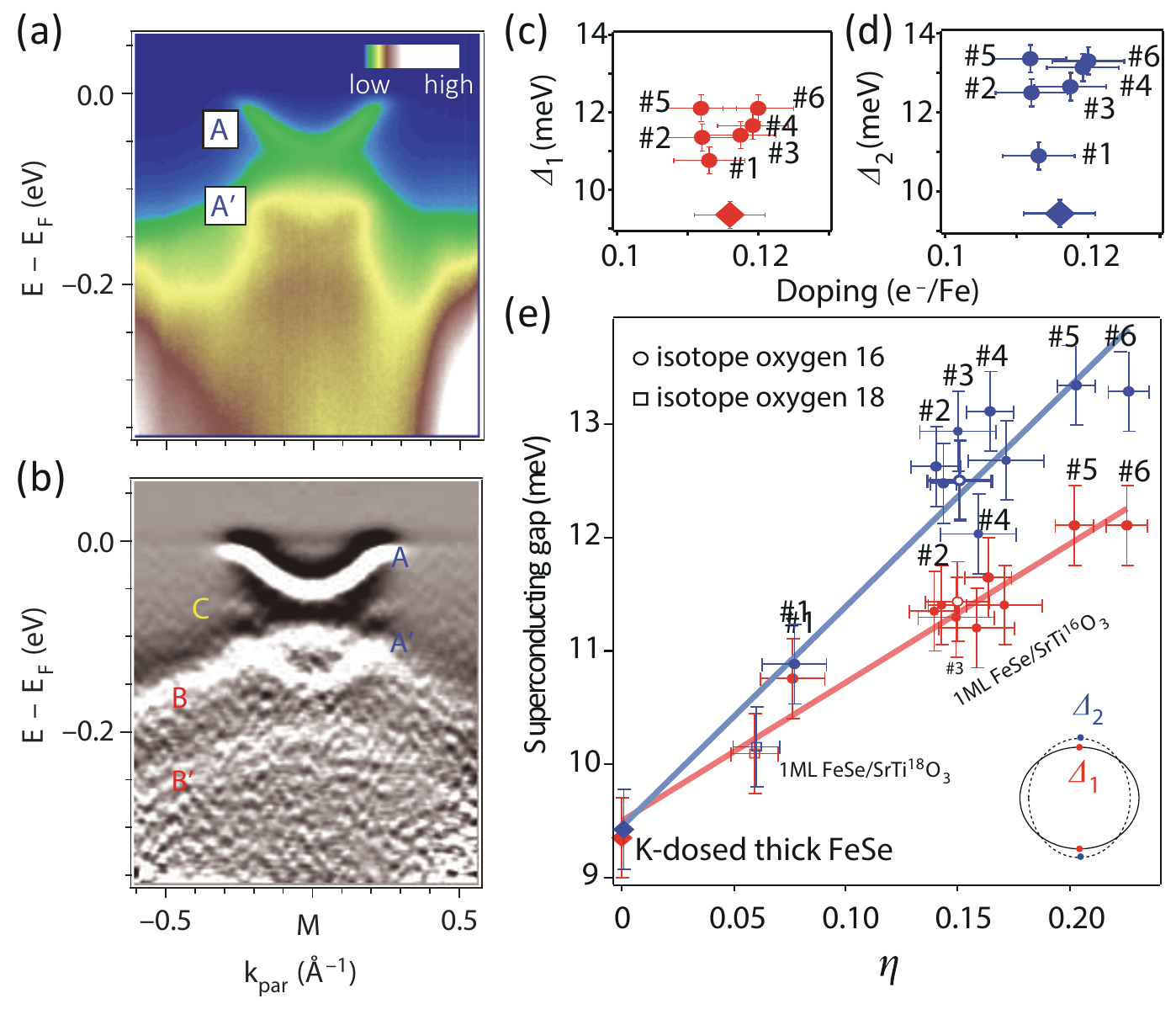}
	\caption{Interfacial electron-phonon coupling in FeSe monolayer film. (a) $\Gamma$-M energy-momentum cut in FeSe/SrTiO$_3$, with the main electron pocket and its shake-off labelled A and A' respectively. (b) Second energy derivative enhanced plot of (a). Adapted from ~\cite{lee2014interfacial}. (c)(d) Superconducting gap sizes on two intercepting M pockets from 7 FeSe/SrTiO$_3$ samples, where 6 contain $^{16}$O (circles) and one is with $^{18}$O (square). (e) Extracted electron-phonon interaction strength $\eta$ positively scales with the superconducting gap size $\Delta$. Adapted from ~\cite{song2019evidence}.}
	\label{ironSC7}
\end{figure}

Monolayer FeSe film grown epitaxially on a SrTiO$_3$ substrate was first measured by STS, showing a single-particle spectral gap as large as $\sim$20~meV which persists above 50~K~\cite{qing2012interface}. Given the maximum $T_c$ of 8~K (37~K under pressure) in bulk FeSe~\cite{hsu2008superconductivity,medvedev2009electronic}, as well as the expectation of stronger phase fluctuations in low dimensional thin films, this initially came as a surprise.

As a result of a large work function difference, substantial charge transfer occurs from the SrTiO$_3$ substrate to the monolayer FeSe film ~\cite{liu2012electronic,he2013phase,tan2013interface,liu2014dichotomy}. ARPES clearly provides such evidence, namely the lack of a nematic order reconstructed Fermi surface (Fig.~\ref{ironSC6}(a), left), the completely occupied $\Gamma$ hole bands, and the absence of an anisotropic nematic band shift that would have broken C4 symmetry (Fig.~\ref{ironSC6}(b), left column)~\cite{liu2012electronic,tan2013interface,liu2014dichotomy}. The absence of $\Gamma$ hole pockets challenges the aforementioned inter-pocket pairing mechanism. Increasing the layer number to two and three readily negates the electron doping effect, and clear consequences from compensated charge carriers and restored nematic order under twinning can be seen (Fig.~\ref{ironSC6}(a)(b), middle and right column). Superconducting gap measurement at the $M$ point shows BCS-type gap closing behavior, with the $2\Delta/T_c$ ratio around the intermediate coupling value of 5 (Fig.~\ref{ironSC6}(c)-(d))~\cite{liu2012electronic,lee2014interfacial}.

Comparing with the optimally electron-doped bulk K$_{0.8}$Fe$_2$Se$_2$ ($T_c\sim$30~K)~\cite{zhang2011nodeless,ying2012observation} and Li$_{0.8}$Fe$_{0.2}$(OH)FeAs ($T_c\sim$41~K)~\cite{zhao2016common}, the putative superconductivity enhancement in FeSe monolayer is proposed to come from either interfacial electron-phonon coupling, or enhanced magnetic exchange $J$ at the interface~\cite{lee2014interfacial,peng2014tuning,cao2014interfacial,huang2017monolayer}.\footnote{It should be cautioned that the transport $T_c$ enhancement in FeSe/SrTiO$_3$ comparing to other intercalated and doped FeSe systems has yet to be firmly established, and the origin of such putative enhancement remains an open question.} Experimentally, a cascade of ``shake-off'' replica electron bands at the $M$ point have been observed, which exist both in the normal and superconducting states (Fig.~\ref{ironSC7}(a)(b))~\cite{lee2014interfacial}. This has been interpreted as evidence for either intrinsic coupling to the $\sim92$~meV SrTiO$_3$ LO4 optical phonon via the Franck-Condon principle in the initial state~\cite{zhang2016role,coh2015large,li2016makes,lee2018routes}, or due to the electron's post-emission interaction with the surface Fuchs-Kliewer phonons.~\cite{li2018electron,jandke2019unconventional}. The energetic separation between the shake-off and primary bands probed by ARPES is shown to be larger than the surface optical phonon energy probed by EELS~\cite{song2019evidence,li2019electronic}. It is also found that a change up to a factor of four in the substrate charge carrier density does not change the replica band behavior~\cite{jia2020xxx}. These observations are consistent with strong electron-phonon coupling effect in the initial states.

FeSe films grown on an orthorhombic rutile-TiO$_2$ substrate are shown to have similar superconducting gap, $T_c$ and shake-off bands to those grown on SrTiO$_3$~\cite{rebec2017coexistence}. This and other substrate-dependent experiments rule out the putative role of nematicity and strain-induced structural distortion in determining the superconducting $T_c$~\cite{rebec2017coexistence,huang2016electronic}. Substrate isotope effect (Fig.~\ref{ironSC7}(c)(d)), as well as a positive correlation between the electron-phonon coupling strength $\eta$ and the superconducting gap size $\Delta$ are observed (Fig.~\ref{ironSC7}(e)), lending support to a multi-channel pairing mechanism~\cite{song2019evidence}. In the meantime, proposals and evidence also exist for a cooperative relationship between 2D-enhanced electron-electron correlation and electron-phonon coupling~\cite{zhao2018direct,he2014electronic,mandal2017correlated}, reminiscing similar correlation-enhanced electron-phonon coupling in bulk FeSe and cuprates.

\subsection{Outlook}
The iron-based superconductors are emerging as an archetypal platform to understand and control multi-orbital correlated physics. As a momentum-resolved single-particle probe, ARPES will continue to drive the understanding of nematicity in the context of orbital-selective physics: Hunds-Mott localization, interplay with magnetism and superconducting pairing. Unification of the itinerant-local perspectives shall be further pursued, with with appreciation of the key role played by the Hund's coupling. With the large number of FeSC families realized via highly systematic and versatile chemical substitution, universal single-particle properties of both the quantum critical phenomena (magnetic and/or nematic) and the correlation effects can be extracted. Interfacial engineering of superconductivity, particularly instigated by the thin-film iron-chalcogenides, will also continue to grow into broader material systems based on the methodology developed and still developing in the FeSCs.

\section{Low dimensional systems}\label{sec_lowD}
\subsection{Overview}
Low dimensional systems have garnered increasing research interest over the past two decades, in part fueled by the discovery of graphene and its half-integer quantum Hall effect at room temperature ~\cite{novoselov2004electric,zhang2005experimental,neto2009electronic}. Due to the spatial confinement and symmetry reduction that are innate to low dimensions, interactions of types and strengths that are uncommon in 3D become possible, giving rise to a wealth of new material properties ranging from interfacial electron gasses to the high temperature quantum spin Hall effect ~\cite{shkolnikov2002valley,ohtomo2004high,wu2018observation}. Operationally, device fabrication and measurements on low dimensional systems -- 2D in particular -- directly benefit from mature technologies such as lithography and physical/chemical vapor deposition perfected through decades of iterations in the semiconductor industry as well as mechanical exfoliation. This also offers a tried-and-tested path from single device physics towards high-density integration and post-Moore's Law electronics~\cite{waldrop2016chips,schaibley2016valleytronics,rhodes2019disorder}. Combining both aspects, unparalleled chemical and physical tunability is brought to low dimensional systems, from which a new era of synthesis-oriented quantum materials research largely stems. 

Due to the substantially reduced material volume comparing to bulk single crystals, traditional thermodynamic (heat capacity, thermal transport) and x-ray/neutron scattering probes face challenges in the study of low dimensional materials.~\footnote{For X-ray and neutron scattering on thin film samples, proper choice of geometry on selected elements with large scattering cross-section can still yield good signal~\cite{need2018quasistatic} even down to the monolayer limit~\cite{fang2017x}.} Meanwhile, electrical transport, optical spectroscopy, and various microscopies remain the major tools to investigate often micron-sized, few-layer-thick samples and devices. Taking advantage of the large interaction cross-section between deep UV light and matter, ARPES has emerged as a powerful technique in quasi-1D and 2D material studies~\cite{mo2017angle,cattelan2018perspective}. Reciprocally, the demand to probe on the length scale of typical low dimensional devices is also spurring the rapid development of ARPES light sources with $\sim$10$\mu$m to $\sim$100~nm sized beam spots (Section~\ref{sec_tech_lightSources}).

This section will emphasize the unique role ARPES has played on (1) graphitic systems  and their correlation effects, (b) spin-orbit coupling, charge order and Mottness in transition metal dichalcogenides, (3) 2D electron gasses and strong coupling effects in complex oxides and their interfaces, and (4)  spin-charge separation in quasi-1D systems.

\textit{In-situ} MBE-ARPES studies on few-layer FeSe and complex oxide films are covered under other sections (see Section ~\ref{sec_sc_filmFeSe}, Section ~\ref{sec_otherMat}), and therefore will not be reiterated here. For a detailed discussion on quantum confinement in topological systems, see Section~\ref{sec_topo_confinement}.

\subsection{Graphene and other single-element monolayers}

\begin{figure} 
	\includegraphics[width=1\columnwidth]{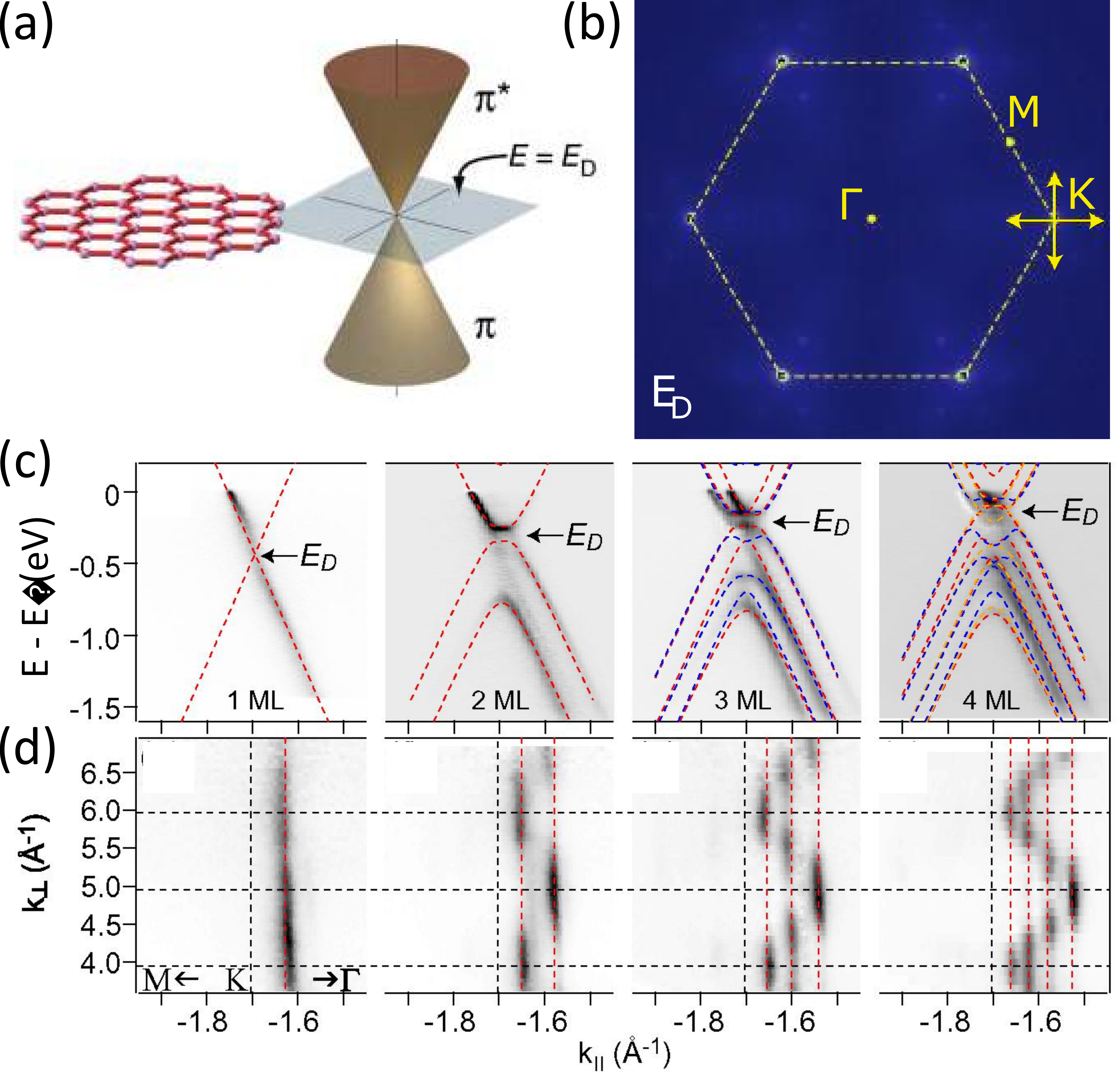}
	\caption{Band structure of graphene. (a) Monolayer graphene and its computed Dirac cone band structure at the Brillouin zone corner. (b) Measured iso-energy contour at the Dirac point energy $E_D$ of monolayer graphene. Adapted from ~\cite{ohta2006controlling}. (c) Band structure near the Dirac point and (d) $k_{||}$-$k_z$ Fermi surface maps for monolayer to 4-layer graphene. Adapted from ~\cite{ohta2007interlayer}.}
	\label{lowDMat5}
\end{figure}

Graphene is a single layer of carbon atoms arranged in a honeycomb lattice via covalent bonding. This lattice structure results in a peculiar low energy electronic structure mainly consisting of carbon $p_z$ electrons ($\pi$ band), with two sets of doubly degenerate Dirac cones (valleys) alternately residing on the six Brillouin zone corners~\cite{neto2009electronic,vafek2014dirac}. Such low energy electronic structure also evinces fundamental concepts such as non-zero Berry's phase in what later became an important component of topological materials research~\cite{novoselov2005two}. Unlike quantum well states in typical semiconductors, low energy charge carriers in graphene exhibit electron-hole degeneracy, vanishing effective mass towards charge neutrality, negligible spin-orbit coupling ($\sim$10$^{-3}$ meV), and approximately follow the Dirac equation of motion~\cite{geim2007rise}. Therefore, in addition to its many intriguing physical properties, graphene is a solid-state platform to interrogate relativistic concepts and phenomena at the thermal energy scale~\cite{stander2009evidence}. Photoemission played a crucial role in determining its electronic structure, dimensional crossover behavior, and various electronic interaction effects. 

Single to few layers of graphene can be synthesized via high temperature vacuum annealing on the (0001) surface of \textit{6H-}SiC~\cite{emtsev2007initial}; or isolated by mechanical exfoliation from bulk graphite single crystals due to the weak inter-plane van der Waals bonding.~\cite{forbeaux1998heteroepitaxial,novoselov2004electric,novoselov2005two}. Synchrotron based ARPES first confirmed the existence of Dirac cone shaped bands in single- and bilayer graphene using the former method (Fig.~\ref{lowDMat5}(a)-(c))~\cite{ohta2006controlling,ohta2007interlayer}. Similar Dirac electrons are also observed in bulk graphite~\cite{zhou2006first}. In particular, adding a second or more layers results in energy splitting of the Dirac cone, rounding out the energy-momentum dispersion from linear to hyperbolic, and giving rise to massive Dirac fermions~\cite{ohta2006controlling}. Surface charge doping via potassium adsorption is demonstrated to continuously modify the band structure near the Dirac point, which is interpreted as a result of broken symmetry between the top and bottom graphene layers~\cite{ohta2006controlling}. Further addition of graphene layers gradually restores the Dirac band's $k_z$ dispersion from zero to $\sim$1~eV via discrete $k_z$ point addition in accordance to the layer numbers (Fig.~\ref{lowDMat5}(c)(d)) ~\cite{zhou2006low,ohta2007interlayer}.

\begin{figure} 
	\includegraphics[width=1\columnwidth]{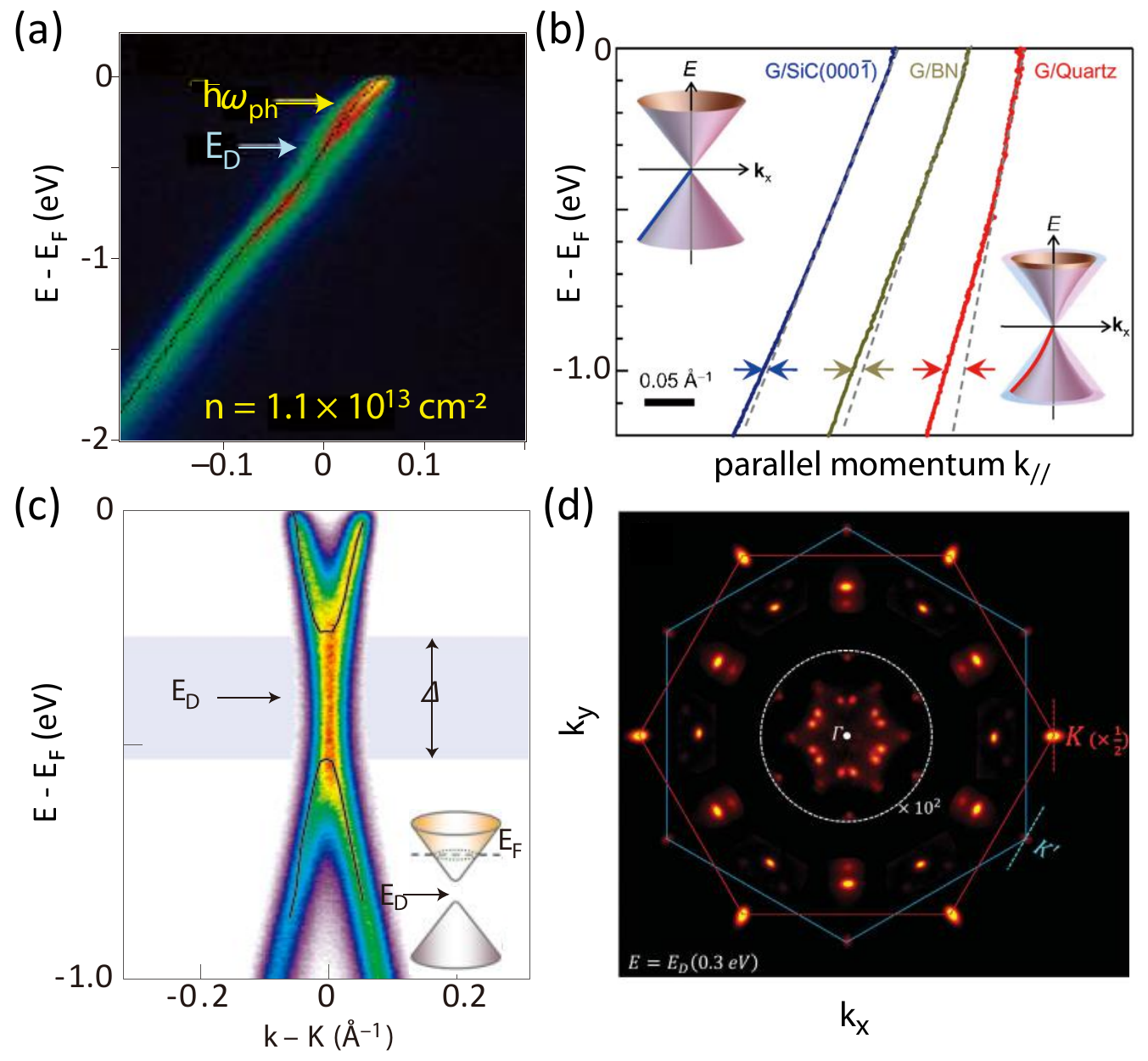}
	\caption{Interaction effect in graphene systems. (a) Low energy dispersion anomalies in doped graphene. Adapted from~\cite{bostwick2007quasiparticle}. (b) Substrate effect causing band renormalization in graphene. Adapted from~\cite{hwang2012fermi}. (c) Substrate induced gap opening on the Dirac point for graphene-6H SiC heterostructure. Adapted from~\cite{zhou2007substrate}. (d) Removal of translational symmetry and the formation of dodecagonal Dirac electron replicas in 30$^{\circ}$ twisted bilayer graphene. Adapted from ~\cite{ahn2018dirac}.}
	\label{lowDMat6}
\end{figure}

In contrast to the early impression of graphene being a purely non-interacting system, a cascade of strong band distortions are observed in doped graphene. Along with thermal, optical and electrical transport measurements, ARPES provides the momentum-resolved single-particle evidence for a hierarchy of quasiparticle dynamics. Figure ~\ref{lowDMat6}(a) shows a series of dispersion anomalies around the K-point at different carrier concentrations~\cite{bostwick2007quasiparticle}. At $\sim$200~meV below $E_{\textrm{F}}$, electron-phonon coupling as strong as $\lambda\sim 0.3$ is observed to disrupt the otherwise linear dispersion. Between 0.5~eV and 1.0~eV below $E_{\textrm{F}}$, signatures of plasmaron formation are manifested through doubling of the Dirac cones~\cite{bostwick2010observation}. Electronic correlation effects are also clearly revealed through inter-layer coupling to different substrates (Fig.~\ref{lowDMat6}(b)(c)). Here, the Dirac cone can be renormalized differently from what one would expect from a mode-coupling induced band renormalization~\cite{hwang2012fermi}, or even gapped out~\cite{zhou2007substrate}. Interactions in graphene have also been studied in trARPES via the photoexcited population dynamics, which can be modelled to evaluate the relative contributions of Auger scattering and impact ionization \cite{Gierz_2013_snapshots,Johannsen_2013_direct}. 

Graphene-based heterostructures, especially those formed with thin film transition metal dichalcogenides and those with an inter-layer twisting angle, are also receiving more investigation with the rising interest in superlattice (``Moir\'e'') engineering on 2D platforms. For example, the superlattice potential between hBN, ruthenium, or iridium and graphene is shown to induce both low energy and high energy band gaps on the Dirac band~\cite{liu2010phonon,wang2016gaps,enderlein2010formation,pletikosic2009dirac}. Twisting two graphene layers relative to each other at small angles is proven  effective in tuning the bandwidth via interlayer band hybridization~\cite{peng2017substrate}. When the angle is exactly 30$^{\circ}$, the rotational symmetry remains while the translational symmetry is removed -- resembling a quasi-crystal (Fig.~\ref{lowDMat6}(d)). Anomalously strong inter-layer potential is shown to scatter Brillouin zone corner Dirac electrons towards the zone center, forming replicas with dodecagonal rotational symmetry~\cite{ahn2018dirac}.

In addition to graphene, borophene (monolayer boron), all of group IV, and group V (with the exception of nitrogen) single-element monolayer systems have been synthesized, covering an extremely diverse set of physical phenomena ranging from the theoretically predicted robust quantum spin Hall insulator~\cite{liu2011low,xu2013large} to tunable band gap semiconductor~\cite{kim2015observation}. Photoemission studies in these systems have mostly focused on eV-scale band structure identification~\cite{feng2016direct,vogt2012silicene,zhu2015epitaxial,mo2017angle,feng2017dirac}, with the major limiting factor being sample stability and availability especially towards the less metallic side. Molten monolayer lead on Cu(111) was used to pioneer the study of the single-particle spectral function in a liquid~\cite{baumberger2004electron}. Surprisingly, the Fermi surface and the Brillouin zone are found to remain in the liquid metal phase, though the radial pair-correlation length becomes exponentially small.

\subsection{Transition metal dichalcogenides}~\label{sec_TMDC}

\begin{figure} 
	\includegraphics[width=1\columnwidth]{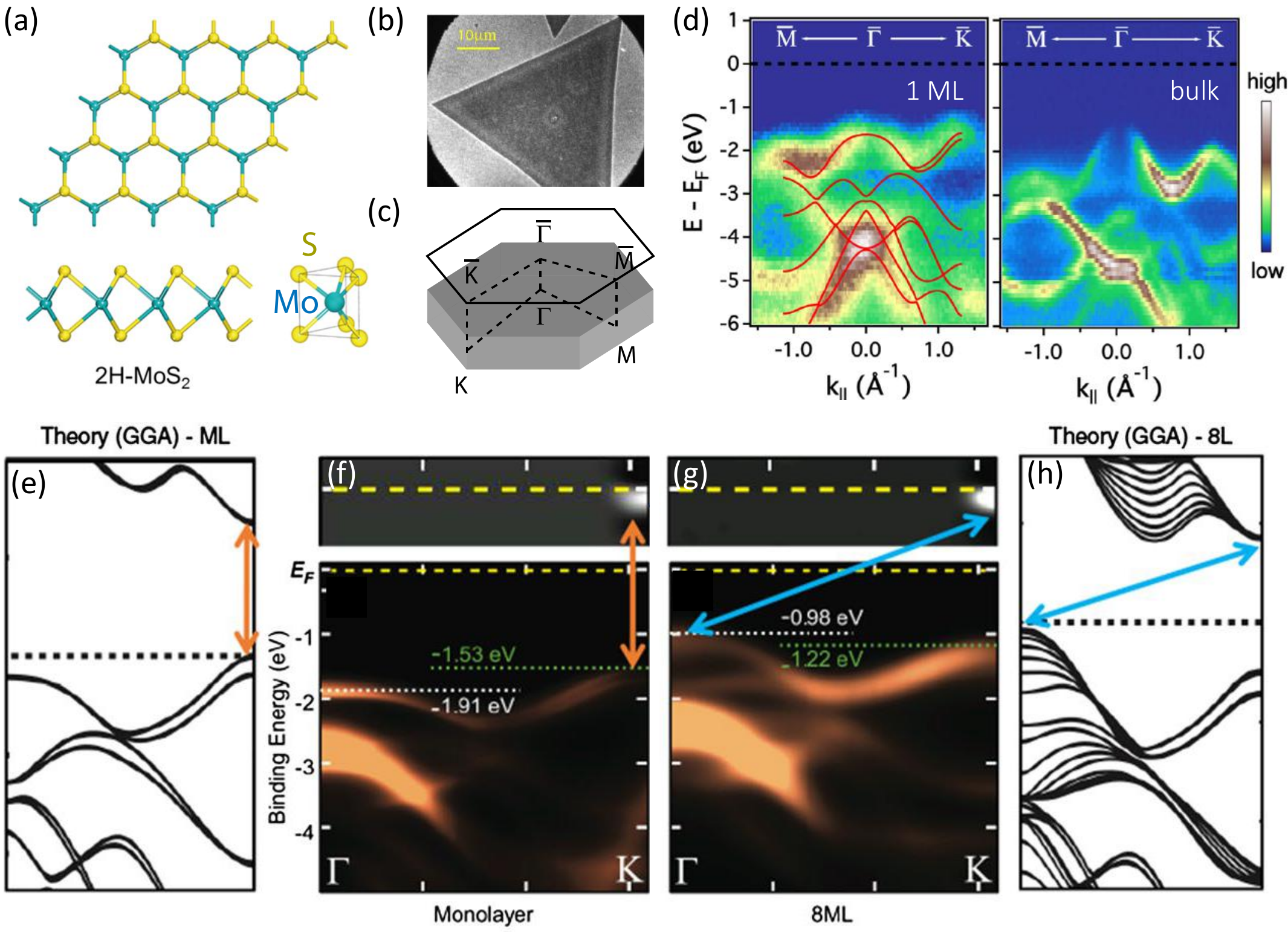}
	\caption{Lattice and electronic structure of layer-dependent 2H-MoX$_2$ (X = S, Se). (a) Top and side view of the lattice structure. (b) PEEM image of a CVD grown flake. (c) Brillouin zone and high symmetry points. (d) Micro-spot ARPES results on monolayer flake and bulk MoS$_2$. Adapted from~\cite{jin2013direct}. (e)-(h) DFT calculation and experimental measurements of monolayer (e-f) and 8 layer (g-h) band structures of MoSe$_2$. Unoccupied states are achieved via surface dosing. Adapted from ~\cite{zhang2014direct}.}
	\label{lowDMat2}
\end{figure}

Many of graphene's intriguing properties come from its low-energy Dirac electrons and weak spin-orbit coupling, which makes it a near zero-gap semiconductor. However, transistors for a logic circuit utilize ``on-and-off''  switching behavior, requiring a non-zero energy gap. Transition metal dichalcogenides (TMDCs) are a family of similarly layered van der Waals materials, which contain semiconductors with tunable eV-scale band gaps and sometimes substantial spin-orbit coupling. Depending on different intra-layer metal-chalcogen bond angle and inter-layer stacking, TMDCs consist of 1T, 1T', 1T'', 2H and 3R phases~\cite{ouyang2015phase,mcdonnell2016atomically}. Mechanical exfoliation of single crystals, chemical vapor deposition and molecular beam epitaxy are all demonstrated to successfully obtain thin films down to monolayer~\cite{bhimanapati2015recent,schafer2016chemical,manzeli20172d,zhou2018library}. In addition to the inherently excellent material tunability at low dimensions, a wide selection of the constituent transition metals and chalcogens further bestows TMDCs a plethora of physical phenomena, including magnetism, charge order, Mottness, topological phases and superconductivity. The simplicity in sample preparation and richness in manipulable physical properties jointly sparked intense research interest in low dimensional TMDCs. ARPES, especially when combined with \textit{in-situ} MBE or a micro-focused beam spot, often provides critical electronic evidence and microscopic guidance for 2D TMD engineering~\cite{mo2017angle}.

Following the serendipitous discovery of up to a factor of 10$^4$ enhancement in luminescence quantum efficiency of monolayer 2H-MoX$_2$ (X = S, Se) compared to its bulk form~\cite{mak2010atomically,splendiani2010emerging}, ARPES observed a clear indirect-to-direct band gap transition going from multi-layer to  monolayer films (Fig.~\ref{lowDMat2})~\cite{jin2013direct,zhang2014direct,yuan2016evolution}. In particular, this was shown to be caused by a rapid rise of the $K$-point valence band top. In the meantime, the lack of inversion symmetry in the monolayer (or odd number of layers) 2H-phase implies spin splitting of the energy bands. Indeed, 140-500~meV spin-orbit splitting of the valence band is observed in monolayer (Mo/W)(S/Se)$_2$~\cite{zhang2014direct,alidoust2014observation}, providing direct electronic evidence for the mechanism of valley-selective optical excitations via circularly polarized light observed via photoluminescence ~\cite{mak2012control,zeng2012valley} as well as trARPES \cite{Bertoni_2016_generation}. Spin-resolved ARPES also reports evidence for spin-polarization on the split valence bands in MoSe$_2$ and WSe$_2$, with a strong dependence on light-polarization and geometry as discussed in Section~\ref{intro_electron_spin}~\cite{mo2016spin}. Spin-orbit coupling also gives rise to topological phases such as the quantum spin Hall state, as discussed in Section~\ref{sec_topo}.

Thin film TMDCs and their associated heterostructures also interact with light strongly, hosting excitons with up to $\sim$30~ns radiative lifetime at room temperature~\cite{liu2015strong,mak2016photonics,mohamed2017long}. Direct determination of the exciton binding energies in monolayer MoS$_2$ and WSe$_2$ on both insulating and conductive substrates is achieved via the combination of optical reflectivity, ARPES, and angle-resolved inverse photoemission (ARIPES) measurements ~\cite{park2018direct}. The exciton binding energy is substantially reduced due to screening on a metallic substrate. Micro-spot ARPES is also utilized to determine inter-layer band alignments and exciton binding energies in MoSe$_2$/WSe$_2$ heterostructures~\cite{wilson2017determination}. 

\begin{figure} 
	\includegraphics[width=1\columnwidth]{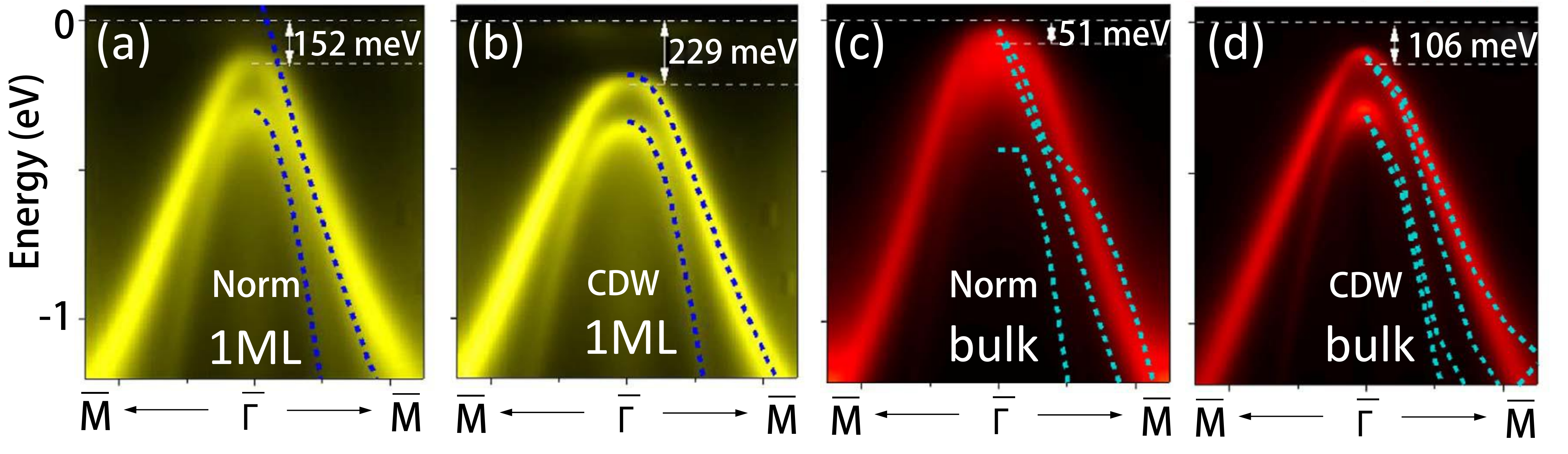}
	\caption{Layer dependent CDW order revealed by band folding in 1T-TiSe$_2$. Single-particle spectra in (a)(c) the normal state and (b)(d) the CDW state of 1T-TiSe$_2$. Yellow - monolayer thin film grown \textit{in-situ}  on bilayer graphene. Red - bulk material. Adapted from ~\cite{chen2015charge}.}
	\label{lowDMat4}
\end{figure}

ARPES also plays an important role in the study of superconductivity and charge order phenomena in Ti/Zr/V/Nb/Ta based TMDCs and their thin films. With the exception of 2H-NbS$_2$, the superconductivity here always occurs in a charge-density-wave (CDW) ordered state. In these systems, electron-phonon coupling usually imprints strongly on the ARPES spectra~\cite{rahn2012gaps}, and the reduction of layer number can modify the system symmetry~\cite{xu2018experimental}, usually leading to a lower superconducting transition temperature and a higher CDW order temperature~\cite{xi2015strongly,ugeda2016characterization,ryu2018persistent,duvjir2018emergence}. Simple Fermi surface nesting scenarios can at best explain a few instances of incommensurate CDW phases such as in VSe$_2$~\cite{borisenko2008pseudogap,shen2008primary,li2018folded}, but is not universally applicable in TMDCs~\cite{johannes2008fermi,Zhu_2015_classification,nakata2018anisotropic}. For instance, ARPES on 1T-TiSe$_2$ shows clear band folding in the CDW state born out of a fully gapped normal state, disfavoring the nesting scenario that would require a Fermi surface to begin with~\cite{chen2015charge}. Combining HSE hybrid functional with \textit{GW} approximation, this CDW formation and associated band gap evolution is ascribed to a band structure origin. Interestingly, the associated ordering temperature and energy gap are considerably larger in the monolayer limit than in the bulk (Fig.~\ref{lowDMat4})~\cite{chen2015charge}. The weaker order in the bulk system is due to dephasing of the CDW between layers. With the indirect band gap comparable to the thermal energy in the normal state, and the abnormally strong intensity of the folded band in the charge ordered state, it has also been proposed that the driving mechanism for this charge order may be excitonic~\cite{cercellier2007evidence}. The band-folding in bulk crystals has also been studied by trARPES, where the timescale for the disappearance of band-folding has been taken as evidence for an excitonic CDW mechanism \cite{Rohwer_2011_collapse,Hellmann_2012_time}. This appears to be supported by simultaneously softened phonon and plasmon modes at the charge ordering wave vector probed with momentum-resolved EELS~\cite{kogar2017signatures}. However, we note that EELS and ARPES measure very different physical quantities regarding exciton condensate. On the other hand, the insulating behavior and CDW order in monolayer 1T-NbSe$_2$ and bulk 1T-TaS/Se$_2$ are attributed to strong electronic correlation (``Mottness''), where the low energy spectra are ubiquitously gapped without any sign of coherent quasiparticles~\cite{lahoud2014emergence,nakata2016monolayer,chen2019visualizing}.

\subsection{2DEG in transition metal oxides}\label{sec_lowD_TMO}

Complex transition metal oxides behave quite differently from the rest of the chalcogenides, mostly because of the exceptionally strong electron negativity of oxygen atoms. As a result, they possess highly ionic bonding (between metal and oxygen ions), much stronger charge transfer (between layers), and higher tendency to form dangling bonds and oxygen vacancies (on the surface and interface). Each trait contributes uniquely to the peculiar properties of a 2D electron gas (2DEG) that appears on their surfaces, and to interfacial coupling on many surfaces and interfaces.

\begin{figure} 
	\includegraphics[width=1\columnwidth]{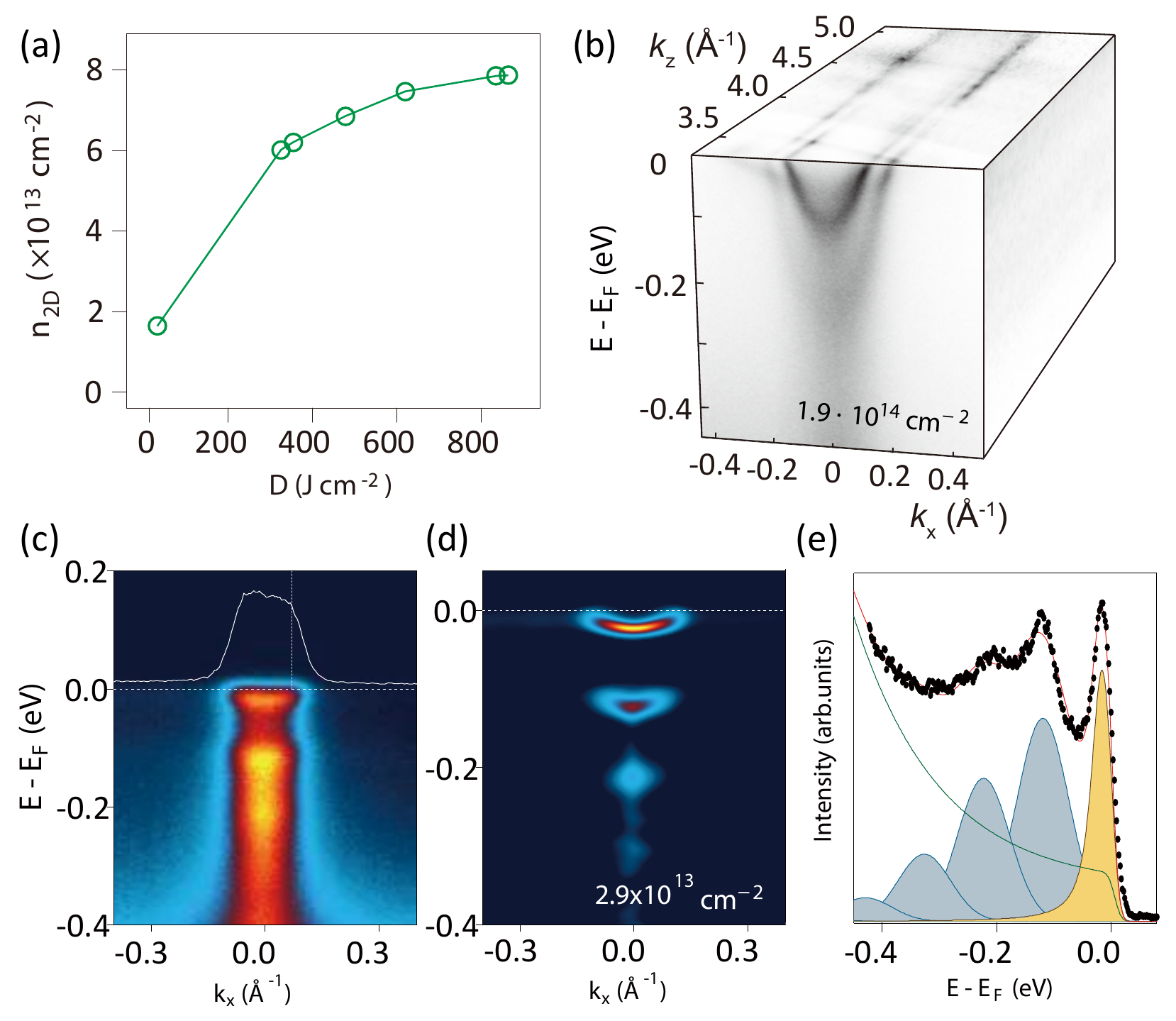}
	\caption{Surface 2D electron gas in SrTiO$_3$. (a) 2DEG carrier density versus UV irradiation dosage on the SrTiO$_3$ (001) surface. Adapted from~\cite{meevasana2011creation}. (b) Electronic structure of the 2DEG on the SrTiO$_3$ (001) surface. Note the high in-plane dispersion and the lack of $k_z$ dispersion. (c) Strong polaronic shake-off on the 2DEG surface state at low carrier concentrations. (d) Maximum curvature plot of panel (c) to highlight the shake-off bands. (e) EDC fitting of the shake-off band consisting of multiple phonon side bands. Adapted from ~\cite{wang2016tailoring}.}
	\label{lowDMat7}
\end{figure}

Following the seminal discovery of a high mobility 2D electron gas at the interface between two insulating perovskites - LaAlO$_3$ and SrTiO$_3$~\cite{ohtomo2004high}, similar 2D conductive states were subsequently discovered and demonstrated by ARPES on SrTiO$_3$ (110), (111), (001) surfaces (Fig.~\ref{lowDMat7}(b))~\cite{king2014quasiparticle,rodel2014orientational,santander2011two,meevasana2011creation,wang2014anisotropic,walker2014control,walker2015carrier}, and on KTaO$_3$ (100) polar surface~\cite{king2012subband,bruno2019band}.~\footnote{Metallic states can also be created on the anatase TiO$_2$ (001) surface. But this state shows strong $k_z$ dispersion, implying a 3D nature~\cite{moser2013tunable}.} The initial photoemission evidence was from the SrTiO$_3$ (001) surface via either bulk crystal cleaving~\cite{santander2011two} or progressive UV irradiation (Fig.~\ref{lowDMat7}(a))~\cite{meevasana2011creation}. Combining surface atomic oxygen treatment with photoemission from oxygen vacancy states, it is revealed that the surface 2DEG comes from UV-induced oxygen vacancies~\cite{walker2014control,walker2015carrier}, and is mostly of Ti-3$d_{xy}$ character~\cite{plumb2014mixed}. The surface states also show clear quantum confinement effects due to surface band bending, with renormalization effects on the $d_{xz}$ and $d_{yz}$ bands~\cite{meevasana2011creation,santander2011two,king2014quasiparticle}. In the wake of surface enhanced superconductivity on monolayer-FeSe/SrTiO$_3$ heterostructures~\cite{lee2014interfacial}, strong electron-phonon coupling induced polaronic shake-off spectra were retrospectively noticed on the SrTiO$_3$ surface state at low carrier concentrations (Fig.~\ref{lowDMat7}(c)-(e))~\cite{chen2015observation,wang2016tailoring}. Such strong coupling is interpreted as a consequence of an exceptionally large Born effective charge associated with the Ti-O bond stretching from the LO4 optical phonon vibration~\cite{lee2014interfacial}.

In an effort to overcome the surface sensitivity and lack of access to buried interfaces, soft X-ray ARPES at Ti $L_3$ and $L_2$ edges ($\sim$460~eV) is employed to penetrate through a 18~\AA~thick LaAlO$_3$ overlayer and probe the interfacial 2DEG on a LaAlO$_3$/SrTiO$_3$ heterostructure ~\cite{cancellieri2016polaronic}. With a 40~meV energy resolution, both the surface states and their polaronic shake-off can be clearly identified. When the soft X-ray photon energy is tuned to match the depth profile of the heterostructure so to form a standing wave, layer-selectivity may be achieved at the antinode of the photon field. This technique was pioneered in probing the buried interface in La$_{0.7}$Sr$_{0.3}$MnO$_3$/SrTiO$_3$ magnetic tunnel junctions to maximize reflectivity at the La 3$d_{5/2}$ absorption edge~\cite{gray2010interface,gray2013momentum}, although the interpretation is still at an early stage.

It should be noted that 2D surface states also frequently exist on semiconductor and metal surfaces~\cite{lashell1996spin,tamai2013spin,jovic2017indium,bianchi2010coexistence}. Due to the inherent broken inversion symmetry on the surface, large spin splitting may be observed if spin-orbit coupling is also strong~\cite{lashell1996spin,tamai2013spin}.

\subsection{Quasi-1D systems}

Further dimensional confinement leads to quasi-1D materials, with even greater electronic instability and the breakdown of the Landau quasiparticle description. Experimentally, these systems are either  bona fide nano-wires, or are effectively 1D because of highly anisotropic valence electronic interactions. The former category requires a carefully self-assembled/aligned nanowire array~\cite{ahn2004mechanism,schafer2008new}, or otherwise nano-spot ARPES at synchrotron facilities with x-ray zone-plate focusing optics~\cite{arango2016quantum} (see also Section~\ref{sec_tech_lightSources}). The latter mostly consists of Luttinger liquid candidate purple bronze Li$_{0.9}$Mo$_6$O$_{17}$~\cite{wang2006new,wang2009quantum,dudy2012photoemission}~\footnote{A close cousin in the molybdenum purple bronze family is K/Na$_{0.9}$Mo$_6$O$_{17}$, which has higher symmetry (C3 rotation) by forming three equivalent chain directions in-plane~\cite{foury1993charge}. This difference causes the system to have a bulk charge order at 115~K, and a separate quasi-2D surface charge order at 220~K~\cite{mou2016discovery}.}, copper spin chain compounds LiCu$_2$O$_2$~\cite{papagno2006electronic}, SrCuO$_2$~\cite{kim1996observation,suga2004high,kim2006distinct}, organic chain compound [Ni(chxn)$_2$Br]Br$_2$~\cite{fujimori2002angle}, and doped vanadium oxides $\beta$-Na$_{1/3}$V$_2$O$_5$~\cite{okazaki2004angle}, V$_6$O$_{13}$~\cite{suga2004high}.

\begin{figure} 
	\includegraphics[width=1\columnwidth]{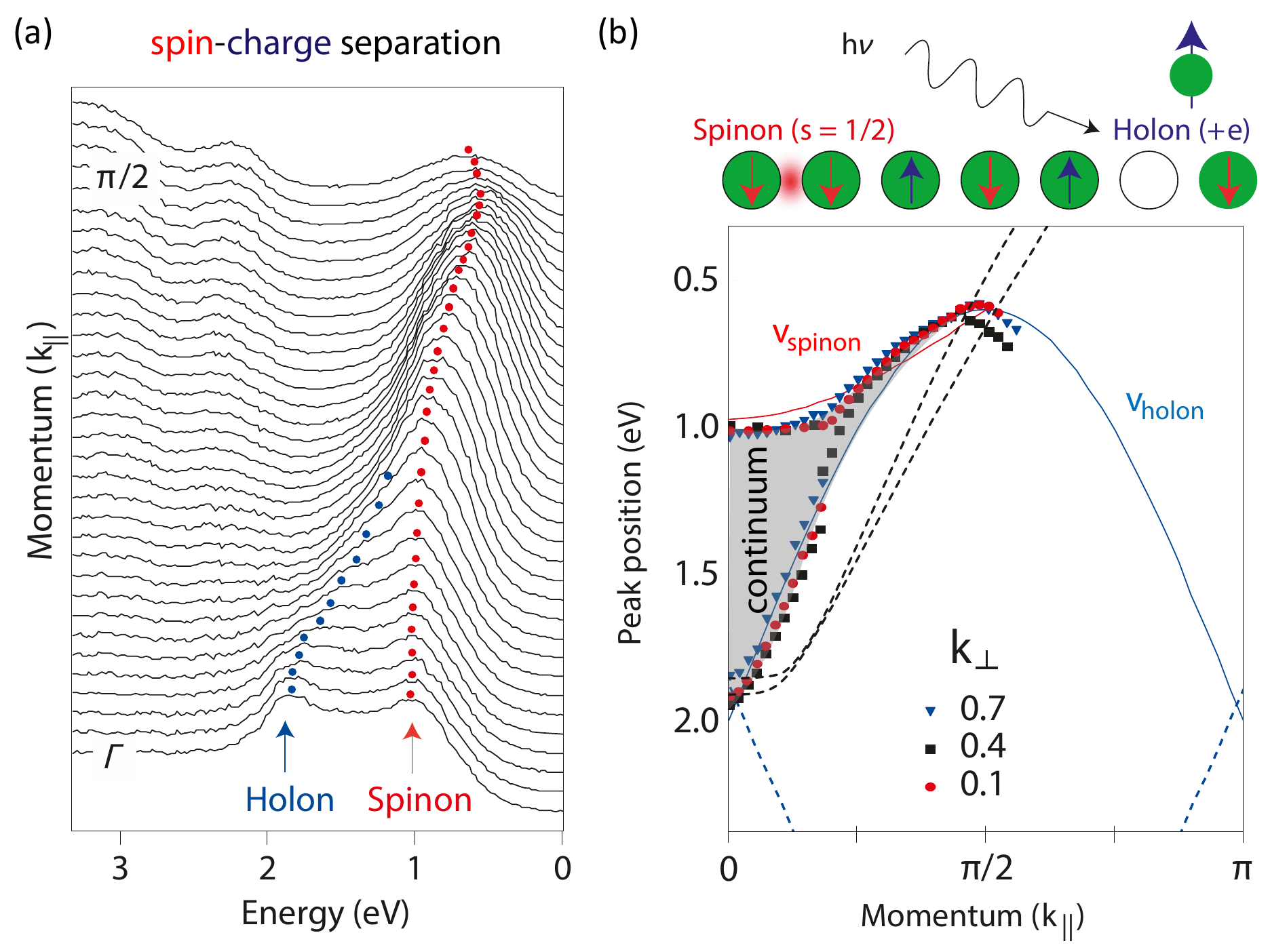}
	\caption{Evidence for spin-charge separation in the 1D chain compound SrCuO$_2$. (a) Raw energy distribution curves for the bifurcated single-particle excitation consisting of spinon (red) and holon (blue) branches. (b) Fitted dispersions collapsed from different perpendicular momenta. The dashed lines are band theory calculations, while the red and blue solid lines are analytical fits to the spinon and holon dispersions. Adapted from~\cite{kim2006distinct}. }
	\label{lowDMat8}
\end{figure}

One main feature of a 1D Luttinger liquid is the fractionalized excitation of a photo-hole into its charge (holon) and spin (spinon) parts -- a phenomenon known as spin-charge separation (Fig.~\ref{lowDMat8}(a))~\cite{nozieres1999theory,giamarchi2003quantum}. In the absence of Landau quasiparticles, the single particle spectral function remains a well-defined quantity to describe the system's response to a single electron removal. Most recent ARPES studies on correlated 1D chain compounds focus on investigating this aspect, often in conjunction with Hubbard model calculations as a function of electron hopping $t$ and Coulomb repulsion $U$~\cite{kim1996observation,fujimori2002angle,suga2004high,papagno2006electronic,kim2006distinct}. In particular, photoemission on SrCuO$_2$ does show evidence for two separate branches of excitation on the single-particle spectrum along the chain direction, bounding a region of excitation continuum in between~\cite{kim2006distinct}. Both branches are non-dispersive perpendicular to the chain direction (Fig.~\ref{lowDMat8}(b)), reaffirming the 1D nature of the electronic structure. The dispersions of the two branches are governed by the charge hopping $t$ (holon), and the spin exchange $J$ (spinon). The fitted value of $J$ = 0.27~eV agrees with respective optical and inelastic neutron scattering results~\cite{kim2006distinct}.

\subsection{Outlook}
Low dimensional materials have seen one of the most prosperous quantum material research scenes in the past decade, with probes tried and tested in graphene research rapidly spilling over to new material systems such as TMDCs and oxide films/interfaces. New device technology enabled new platforms -- such as Moir\'e systems and freestanding oxide films -- will likely continue to spur new diversification in ARPES sample environment and lightsources, such as higher level of integration of micro-ARPES and \textit{in-situ} device manufacturing with electrical characterizations. Low dimensional systems are expected to help shed light on traditionally hard correlation problems thanks to their excellent optical, mechanical, electrical and magnetic tunability. Moreover, they shall expand investigations in dimensionality-specific topics such as those in quantum confinement, exotic topological states of matter, or enhanced/stabilized order comparing to 3D counterparts.

\section{Topological materials}\label{sec_topo}

\subsection{Overview}

\begin{figure}
\centering
\includegraphics[width=1\columnwidth]{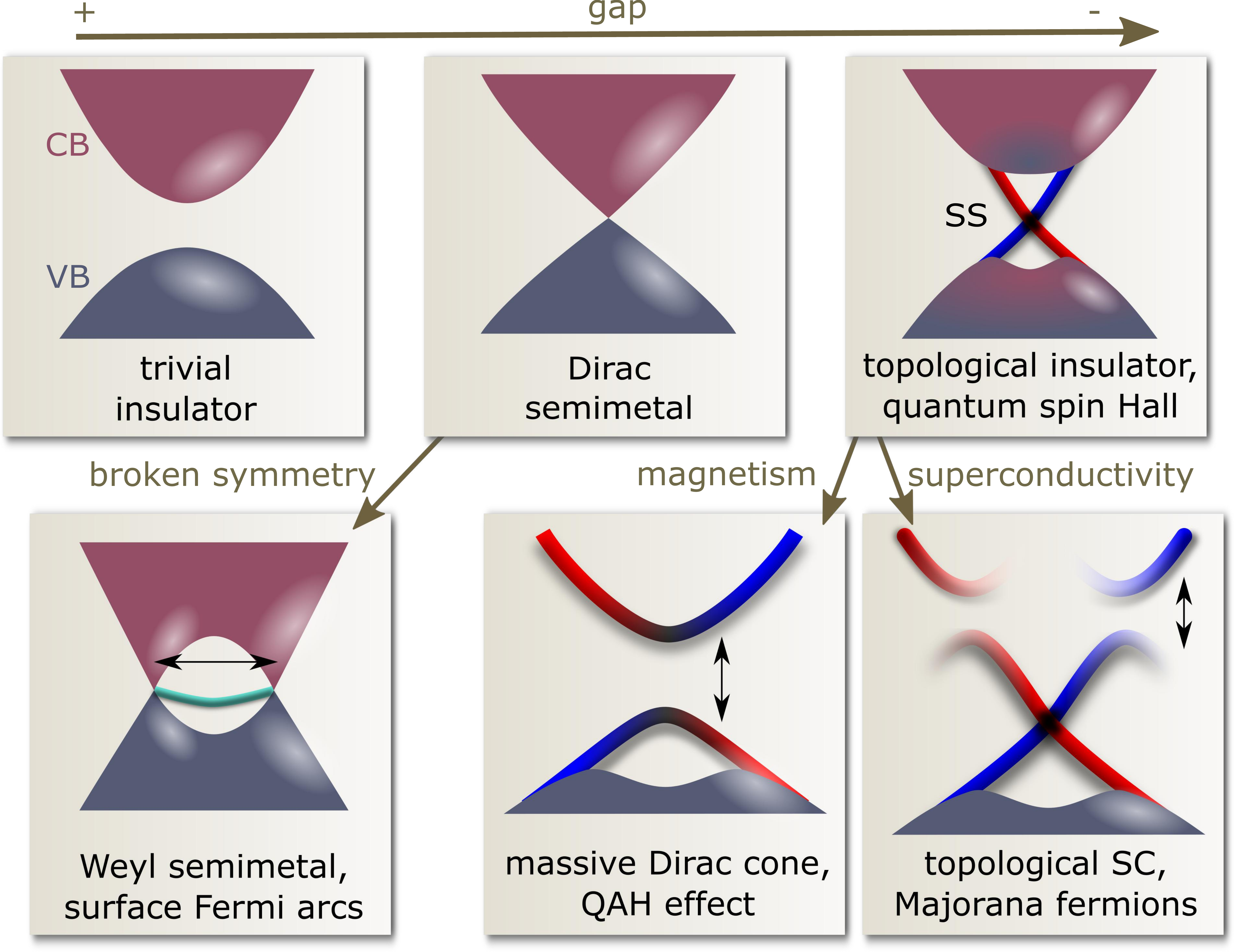}
\caption[Hierarchy of topological materials]{A simplified view of the band structure evolution in topological states of matter studied by ARPES. A trivial insulator has a finite bandgap between conduction (CB) and valence bands (VB). Closure of the gap can produce a Dirac semimetal with linear dispersion. Continued evolution of the gap results in band inversion, which can produce a topological insulator with surface states (SS) bridging the gap.  The Dirac semimetal becomes a Weyl semimetal if either time-reversal or spatial-inversion symmetries are broken, whereas the topological insulator exhibits exotic phenomena when coupled with superconductivity and magnetism.
\label{Fig_Topo_Overview}}
\end{figure}

One of the cornerstones of condensed matter physics is the classification of matter into distinct phases, which is conventionally done by considering spontaneously broken symmetries. In the 1980's a new paradigm emerged in which matter began to be classified according to the notion of topological invariants \cite{Thouless_1982_quantized,Wen_1995_topological}. In mathematics, a topological invariant is a property that is maintained through smooth deformations of an object. Analogously, topological properties of a material are insensitive to smooth deformations of the system's Hamiltonian. (Here ``smooth'' refers to an adiabatic  perturbation which does not close an energy gap). Thus, identifying the topological invariants of a material gives robust predictive power for its physical properties. 

A complete discussion on the topological classification of matter is beyond the scope of this review, and is discussed in detail elsewhere \cite{Hasan_2010_colloquium, Qi_2011_topological,Haldane_2017_nobel}. We instead give a brief conceptual overview, then focus on developments in which ARPES played a pivotal role in discovering or understanding topological phases by identifying nontrivial topological electronic structures \cite{Zhang_2020_angle}. 

\subsection{Quantum Hall states}\label{Section_QuantumHall}

We begin with a brief overview of a family of two-dimensional topologically non-trivial states known as the quantum Hall states, including the quantum Hall insulator (QHI), quantum spin Hall insulator (QSHI), and quantum anomalous Hall insulator (QAHI). Although the experimental study of these states is within the purview of transport rather than ARPES measurements, we begin with this overview because these states embody concepts which are universal to all topological materials. Moreover, the discoveries of these states have been regarded as milestones for our understanding of the topological classification of matter \cite{Haldane_2017_nobel}.
  
The first example of a topological material is the QHI, in which electrons are confined to two dimensions and subjected to a strong out-of-plane magnetic field.  Application of an in-plane longitudinal electric field leads to a transverse voltage drop  given by a non-zero Hall conductance $\sigma_{xy}$. Remarkably, $\sigma_{xy}$ is precisely quantized to integer multiples of $e^2/h$, as has been confirmed for up to one part per billion \cite{vonKlitzing_2005_developments}. This exact quantization, independent of material details, is a reflection of the fact that $\sigma_{xy}$  is a topologically invariant property. The current is conducted along the sample edge due to a general principle known as ``bulk-boundary correspondence,'' which guarantees the existence of gapless conducting states at the interface where a topological invariant changes \cite{Hasan_2010_colloquium}. 

The QSHI, also known as a 2D topological insulator (TI), is analogous to the QHI, except the role of an external magnetic field is replaced by the intrinsic spin-orbit coupling, which acts as an effective magnetic field with equal magnitude but opposite direction for electrons with momenta  $+\vec{k}$ and $-\vec{k}$. Thus, the quantum spin Hall state can be regarded as two copies of the quantum Hall state, with each copy oppositely spin-polarized and counter-propagating along the sample edge. This leads to zero net \emph{charge} Hall conductance, though the \emph{spin} Hall conductance is quantized to multiples of $e/2\pi$ \cite{Kane_2005_quantum, Bernevig_2006_quantum}. The existence of quantized edge channels was experimentally indicated in HgTe quantum wells by measuring the longitudinal four-terminal resistance \cite{Konig_2007_quantum}. Note that time-reversal symmetry is broken for the QHI, but it is preserved for the QSHI. From a band structure point-of-view, the strong spin-orbit coupling causes the valence and conduction bands to invert in energy. Intuitively, one can imagine that if a material with inverted band ordering is brought into contact with one with non-inverted ordering, the gap must close somewhere in the interface between the two materials. It is this gapless interface state which is responsible for edge conduction (even if the second ``material'' is vacuum), again demonstrating the principle of bulk-boundary correspondence.

 The QAHI can be regarded as a QSHI coexistent with magnetic order \cite{Chang_2013_experimental}.  Note that both strong spin-orbit coupling and time-reversal symmetry breaking are important in this case. As in the QHI, the edge conduction leads to $\sigma_{xy}$ quantized to integer multiples of $e^2/h$.  Unlike the QHI, this quantized conduction is due to the intrinsic magnetization of the material, and not imposed by an external magnetic field. 

This discussion exemplifies concepts such as band inversion, the principle of bulk-boundary correspondence, and the roles of spin-orbit coupling and time-reversal symmetry.  As we shall see, the power of ARPES lies in its capability to measure electronic states both in the bulk and at the boundary. In the bulk, ARPES can directly resolve the energetic sequence of bands and whether they exhibit inversion; on the boundary, ARPES can directly resolve electronic states localized at the interface. The influence of spin-orbit coupling is revealed through spin-resolved ARPES measurements, while the role of time-reversal (and other symmetries) is explored by material synthesis and modification via doping/substitution.

\subsection{Topological insulators}

\begin{figure}
\centering
\includegraphics[width=1\columnwidth]{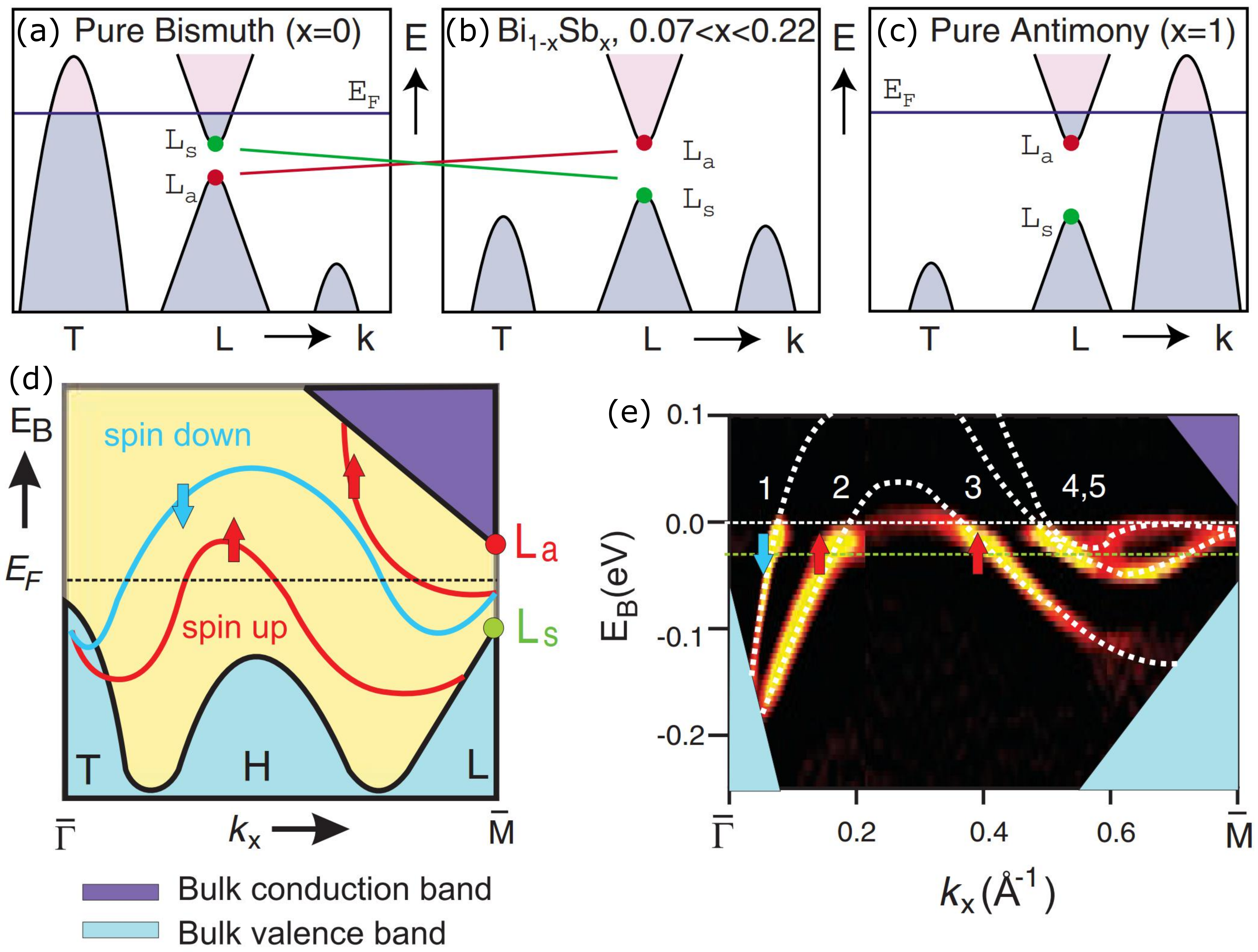}
\caption[The first 3D TI:  Bi$_{1-x}$Sb$_x$]{  Bi$_{1-x}$Sb$_x$ as the first 3D topological insulator. (a)-(c) show the electronic structure for pure Bi, the alloy, and for pure Sb, respectively. The band arrangement in pure Bi is trivial, while the bands at the $L$-point of pure Sb are inverted. Within a critical doping range, the alloy becomes a direct bandgap semiconductor while retaining the topologically non-trivial band inversion. Adapted from \cite{Hasan_2010_colloquium}. (d) Cartoon and (e) ARPES measurement of the surface state dispersion of Bi$_{0.91}$Sb$_{0.09}$, showing five Fermi-level crossings. The arrows denote the spin polarization of the bands, as verified by spin-resolved ARPES measurements. Adapted from \cite{Hsieh_2009_observation}.
\label{Fig_Topo_BiSb}}
\end{figure}

\subsubsection{3D strong TIs} \label{sec_disc_3d_TI}

\begin{figure}
\centering
\includegraphics[width=0.8\columnwidth]{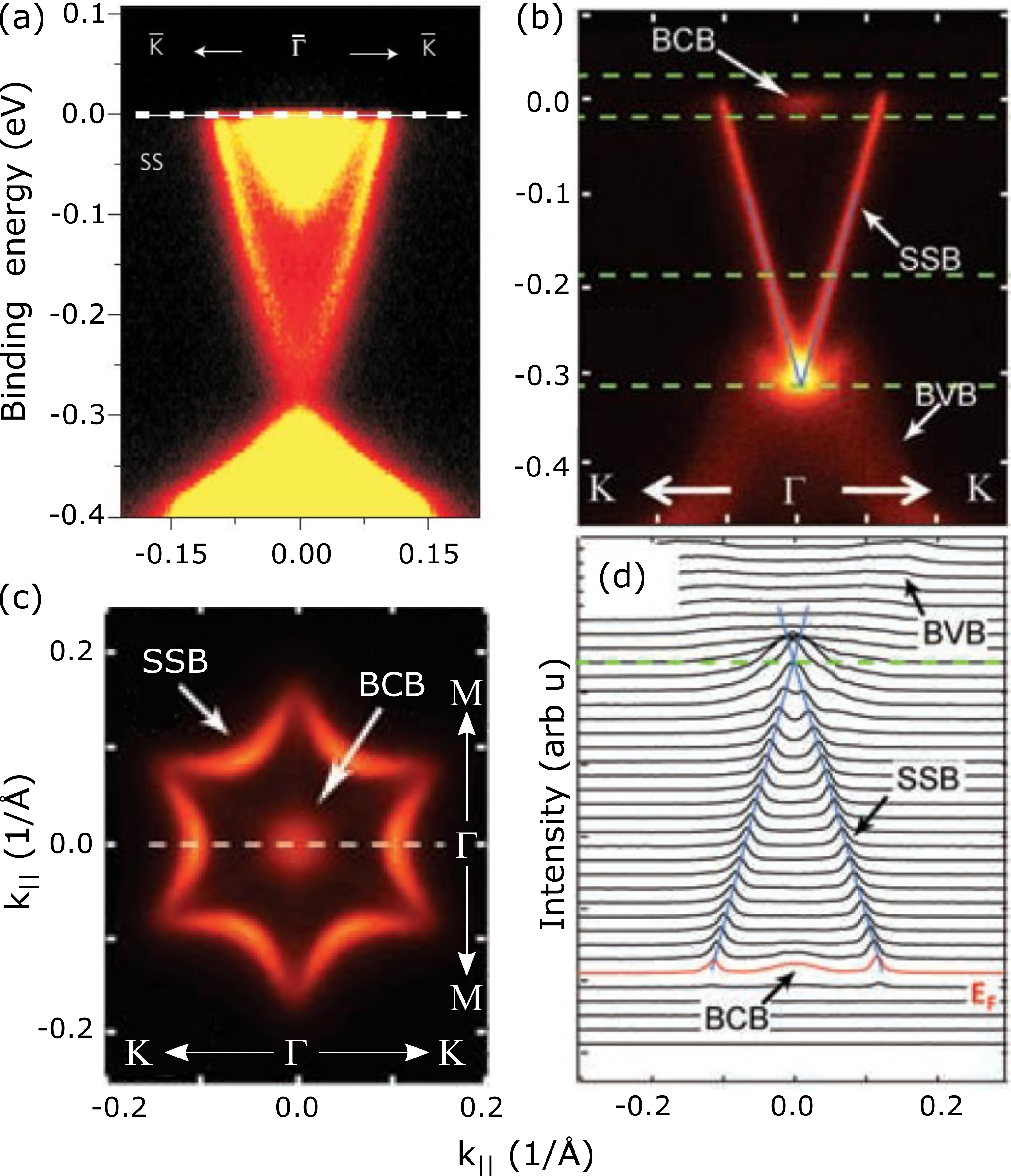}
\caption[3D TIs in Bi$_2$Se$_3$ and Bi$_2$Te$_3$]{ARPES measurements of the 3D topological insulators (a)  Bi$_2$Se$_3$ \cite{Xia_2009_observation} and (b)-(d) Bi$_2$Te$_3$. The topological surface state is observed linearly dispersing across the bulk band gap. (c) The Fermi surface of Bi$_2$Te$_3$, showing the anisotropic dispersion of the surface state. (d) MDCs of the cut shown in (b). Adapted from \cite{Chen_2009_experimental}.
\label{Fig_Topo_BS_BT}}
\end{figure}

\begin{figure*}
\centering
\includegraphics[width=2\columnwidth]{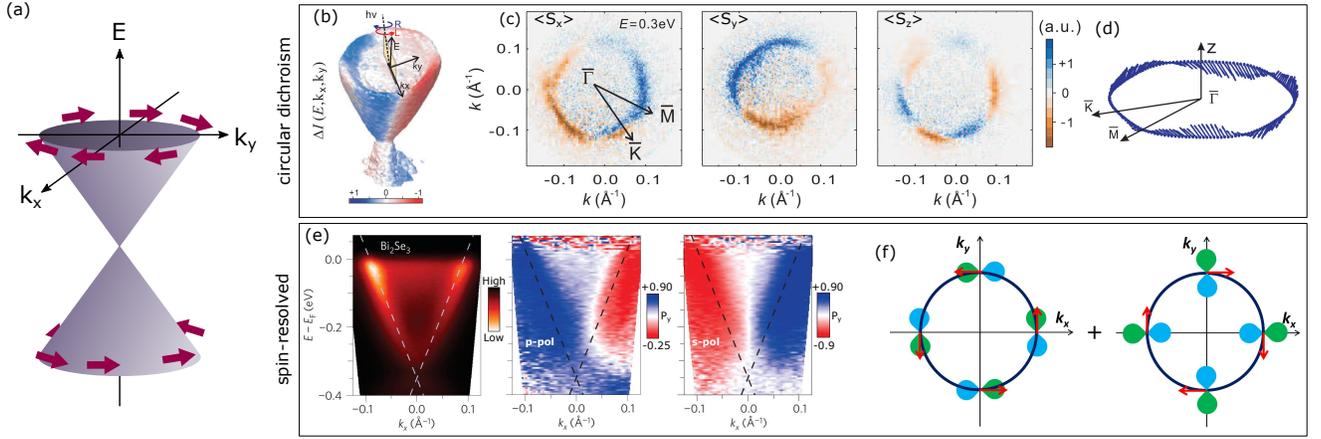}
\caption[Higher order spin effects in 3D Tis]{Deviations from an ideal helical spin texture in 3D TIs. The ideal texture, shown in (a), consists of 100\% polarized in-plane spins tangentially oriented along an isotropic Dirac cone. Top row: Circular dichroism ARPES on Bi$_2$Se$_3$. (b) The difference in photoemission intensity between left-hand and right-hand circularly polarized light. (c) By modeling matrix elements in a spin-orbit coupled system, all three components of the spin-polarization can be calculated from the data.  (d) Summary of the deduced spin-polarization, including an out-of-plane component associated with hexagonal warping. Adapted from \cite{Wang_2011_observation}. Bottom row: Spin-orbital texture of Bi$_2$Se$_3$.  (e) The spin-polarization measured by spin-resolved ARPES reverses sign when the light polarization is rotated. Adapted from \cite{Jozwiak_2013_photoelectron}. (f) This is because the eigenstates are linear combinations of spin (arrows) and orbital (blue/green shapes) components. The experiment measures the spin-polarization  associated with the orbital component photoemitted by the incident light polarization. Adapted from \cite{Zhang_2013_spin}.
\label{Fig_Topo_spinEffects}}
\end{figure*}

The 3D strong TI can be understood as a three-dimensional analog to the QSHI described above \cite{Fu_2007_topological,Moore_2007_topological,Qi_2008_topological}. Conceptually, one begins by considering a trivial insulator with bulk valence and conduction bands of opposite parity separated by an energy gap. In a 3D TI, the spin-orbit interaction causes the bands to become inverted (see Fig.~\ref{Fig_Topo_Overview}). If this inversion occurs at an odd number of points in the Brillouin zone, the material becomes topological (characterized by the so-called $\mathbb{Z}_2$ invariant) and classified as a strong TI. Due to the bulk-boundary correspondence,  a strong TI exhibits gapless surface states which are robust against any perturbation which maintains time-reversal symmetry.  As a consequence, both the 2D surface states of a 3D TI and the 1D edge states of a 2D QSHI are said to be protected by time-reversal symmetry. The distinctive signatures of a 3D TI are encoded in quantized magnetoelectric responses \cite{Qi_2008_topological,Essen_2009_magnetoelectric}  detectable in a high-precision optical measurement \cite{Wu_2016_quantized}. Unlike for the 2D QSHI, the transport signature is subtle \cite{Hasan_2010_colloquium}, making ARPES an indispensable tool for identifying 3D TI materials.

Following theoretical prediction \cite{Fu2_2007_topological}, the first experimentally observed 3D TI was the Bi$_{1-x}$Sb$_x$ alloy \cite{Hsieh_2008_a}.  Both Bi and Sb are semimetals with negative indirect gaps, but finite direct gaps throughout their entire Brillouin zones. In Sb, the valence and conduction bands are inverted at the three equivalent $L$ points, but the absence of a global bandgap precludes it from being classified as a TI (Fig.~\ref{Fig_Topo_BiSb}(a) and (c)). However, there is a small range ($0.07 < x < 0.22$) in which the alloy is a direct bandgap semiconductor at the $L$ points while still retaining the band inversion of Sb, as shown in Fig.~\ref{Fig_Topo_BiSb}(b). ARPES experiments identified surface states, as verified by their lack of $k_z$ dispersion \cite{Hsieh_2008_a}. Confirmation of their topological origin was based on two criteria: (1) the bands cross $E_{\textrm{F}}$ an odd number of times between the two time-reversal invariant momenta $\overline{\Gamma}$ and $\overline{\textrm{M}}$  (Fig.~\ref{Fig_Topo_BiSb}(e)), and (2) the bands are spin-polarized and thus non-degenerate, as confirmed by spin-resolved ARPES \cite{Hsieh_2009_observation}. These observations, summarized in Fig.~\ref{Fig_Topo_BiSb}(d), together indicate that the surface bands cannot be  eliminated by any perturbation that maintains time-reversal symmetry.

The next 3D TIs to be theoretically predicted and experimentally discovered were Bi$_2$Se$_3$ and Bi$_2$Te$_3$ \cite{Zhang_2009_topological,Xia_2009_observation, Chen_2009_experimental}. The advantages of these materials over Bi$_{1-x}$Sb$_x$ are that they are free of alloying disorder and exhibit an exceptionally clean electronic structure with a single Dirac cone surface state.  As shown in Figs.~\ref{Fig_Topo_BS_BT} (a) and (b) for Bi$_2$Se$_3$ and Bi$_2$Te$_3$, respectively, the bulk bands are semiconducting with a gap $>200$~meV, while the surface states bridge the gap near the $\Gamma$-point (Fig.~\ref{Fig_Topo_BS_BT}(d)) and do not disperse with $k_z$ \cite{Xia_2009_observation, Chen_2009_experimental}.  As shown in Fig.~\ref{Fig_Topo_BS_BT}(c), the Fermi surface of Bi$_2$Te$_3$ consists of a hexagonally-warped pocket from the surface state and a bulk pocket from the conduction band which can be tuned away from $E_{\textrm{F}}$ by doping \cite{Chen_2009_experimental}.  Spin-resolved ARPES measurements confirmed that the surface state of these materials has the requisite helical spin texture, with spins oriented predominantly in-plane and tangential to the Fermi surface \cite{Hsieh_2009_a}. Bi$_2$Se$_3$ and Bi$_2$Te$_3$ have been the key material platforms to much of the later work on topological materials, including the discovery of the quantum anomalous Hall effect \cite{Chang_2013_experimental}, and therefore appear prominently in this review despite the abundance of newer materials.

\subsubsection{Topological protection and spin-polarization}\label{sec_topo_protect}

While it is difficult to establish unambiguous proof, ARPES does provide compelling evidence for the unusual robustness of the topological surface state.  Trivial surface states are often found to be exquisitely sensitive to disorder and surface adsorbates  \cite{Damascelli_2000_Fermi, Yang_2005_Fermi, Noh_2009_anisotropic}. In contrast, the surface states of Bi$_2$Se$_3$  have been clearly observed in ARPES  even after exposure to atmosphere \cite{Benia_2011_reactive, Chen_2012_robustness}. More aggressive disorder, such as removal of atoms by sputtering, causes the surface state to migrate toward the deeper, unperturbed layers \cite{Queiroz_2016_sputtering}. While these results are intriguing, we caution that these demonstrations of robustness cannot be taken as definitive proof of the topological nature of the bands.  

The robustness of the topological surface state is only one aspect of a notion known as ``topological protection.'' Another consequence is  the fact that nonmagnetic backscattering from  momentum $\vec{k}$ to $-\vec{k}$ on the surface state is suppressed due to the fact that these states have opposite spin orientations \cite{Roushan_2009_topological}. However, scattering to any state other than the one at $-\vec{k}$ is still permitted. Via self-energy analysis, ARPES has detected signatures of interband scattering with bulk states \cite{Park_2010_quasiparticle} as well as intraband electron-phonon scattering, with reported coupling strengths ranging from $\lambda=0.076$ up to $\lambda\sim3$  \cite{Hatch_2011_stability, Pan_2012_measurement,Chen_2013_tunable,Kondo_2013_anomalous}, with the disparate results likely attributed to differing experimental resolution and sensitivities. Inter- and intra- band scattering processes have also been documented in the time-domain by trARPES \cite{Sobota_2012_ultrafast, Wang_2012_measurement}. 

Another important consideration for the scattering properties of the surface state is its deviation from the ideal helical spin-texture shown in Fig.~\ref{Fig_Topo_spinEffects}(a). The dispersion of the surface states of  Bi$_2$Te$_3$ is hexagonally warped at energies away from the Dirac point (see Fig.~\ref{Fig_Topo_BS_BT}(c)), which opens scattering channels associated with  out-of-plane components of the spin polarization \cite{Fu_2009_hexagonal,Alpichshev_2010_STM}. The out-of-plane spin was directly measured by spin-resolved ARPES in Bi$_2$Te$_3$ \cite{Souma_2011_direct}, and subsequently deduced by circular-dichroism ARPES in Bi$_2$Se$_3$ (see Section~\ref{sec_intro_matrixelem} for a discussion on the relationship between spin-polarization and circular dichroism) \cite{Wang_2011_observation}, as shown in the top row of Fig.~\ref{Fig_Topo_spinEffects}. 

An additional deviation from the ideal helical texture is its partial spin polarization ($<$100\%) due to the substantial spin-orbit coupling in these materials  \cite{Yazyev_2010_spin}.  Early spin-resolved ARPES measurements reported polarizations ranging from 25\% \cite{Souma_2011_direct} to 75\% \cite{Pan_2011_electronic}, though subsequent theory work showed that the photoelectron spin-polarization is not equivalent to that of the initial state \cite{Park_2012_spin}. In fact, later experiments showed that the measured spin-polarization could be reversed or even rotated out-of-plane by controlling the excitation photon polarization \cite{Jozwiak_2013_photoelectron, Xie_2014_orbital, Zhu_2014_photoelectron, SanchezBarriga_2014_photoemission}, as shown in Fig.~\ref{Fig_Topo_spinEffects}(e). This is explained by  the effect of spin-orbit coupling: since spin is not a good quantum number, the eigenstates are linear combinations of spin and orbital components, as shown in Fig.~\ref{Fig_Topo_spinEffects}(f) \cite{Zhang_2013_spin,Cao_2013_mapping}.  These experiments highlight the inherent complexity of a spin-resolved ARPES measurement as expounded in Section~\ref{intro_electron_spin}; here, the measurement is only sensitive to the spin-polarization of the photoemitted orbital component, which is controlled by light polarization due to matrix elements. 

\subsubsection{Quantum confinement}\label{sec_topo_confinement}

\begin{figure}
\centering
\includegraphics[width=1\columnwidth]{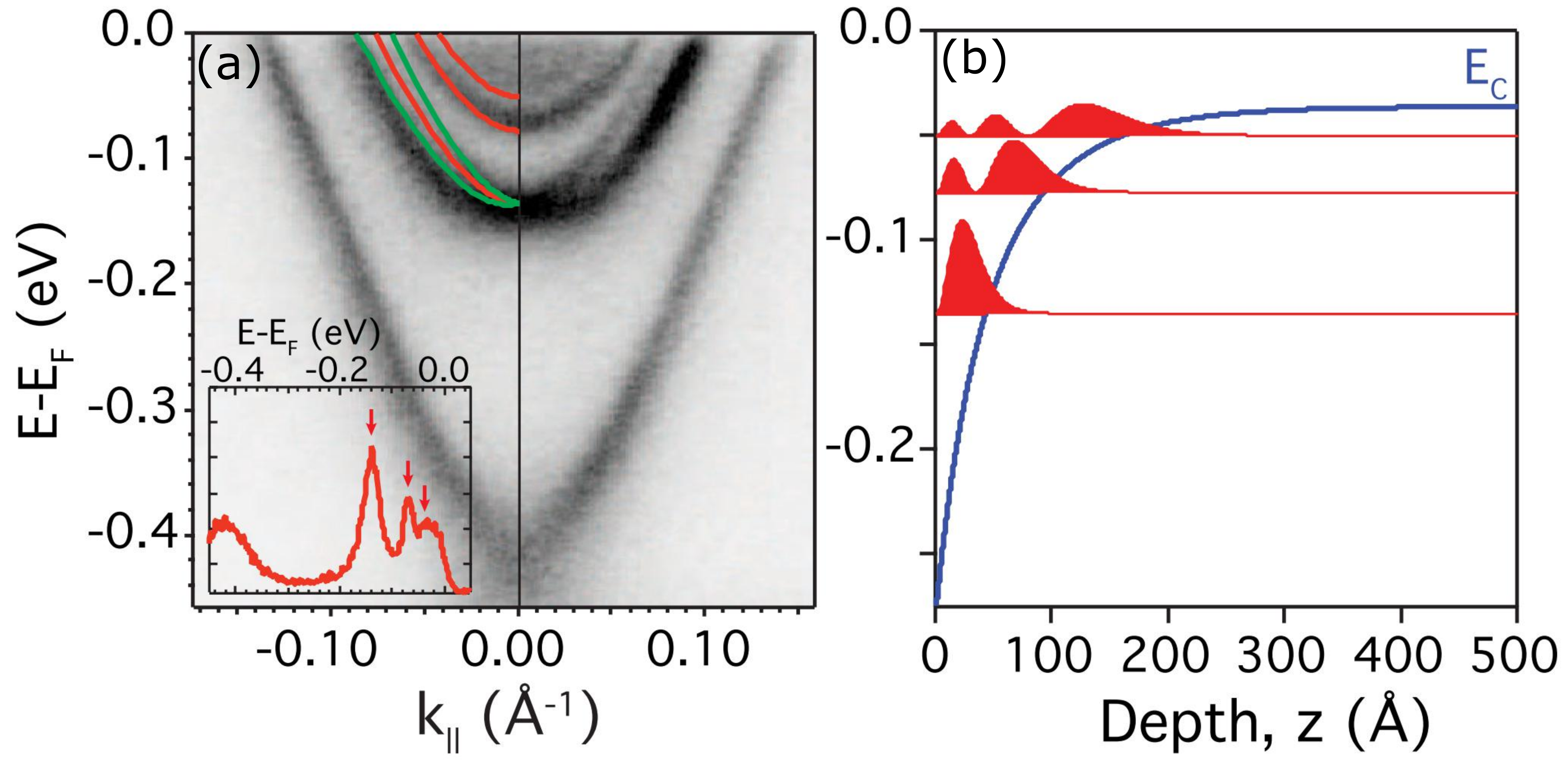}
\caption[Spin-polarized 2DEG in Bi$_2$Se$_3$]{(a) A series of three 2DEG states on a Bi$_2$Se$_3$ surface contaminated with adsorbates. The lowest sub-band exhibits Rashba-splitting due to strong spin-orbit coupling. (b) Model for the formation of 2DEGs. The conduction band energy $E_c$ is subject to a band-bending potential near the surface, leading to quantum confinement of the bulk wave function. Adapted from  \cite{King_2011_large}.
\label{Fig_Topo_Rashba}}
\end{figure}

\begin{figure*}
\centering
\includegraphics[width=2\columnwidth]{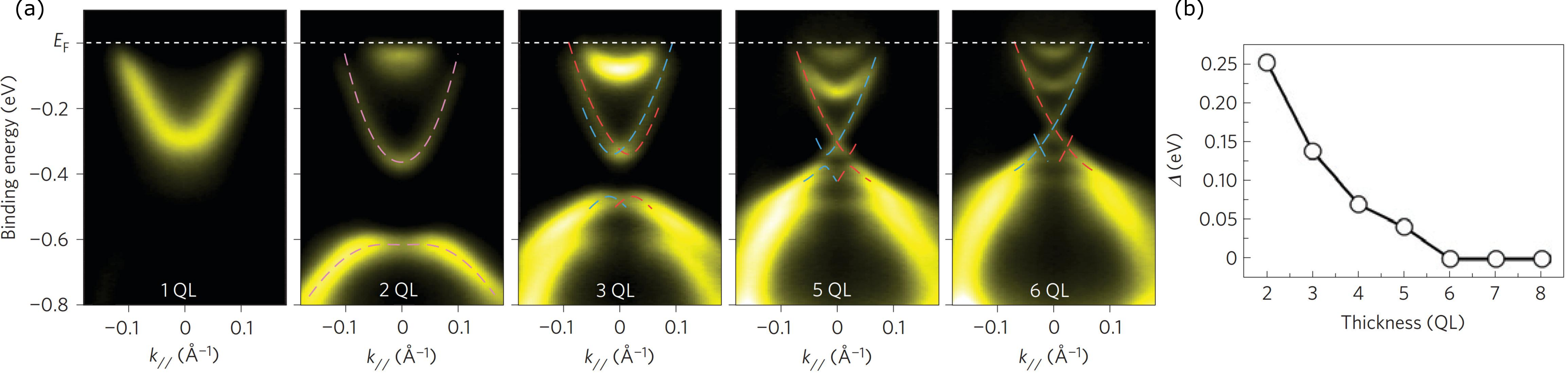}
\caption[3D TI in the 2D limit]{ (a) ARPES spectra of Bi$_2$Se$_3$ thin films synthesized to the thickness shown (QL = Quintuple Layer). There is no Dirac cone observed for sufficiently thin samples. (b) The gap at the Dirac point extracted as a function of film thickness. The gapless topological surface state recovers for  thicknesses $> 6$~QL. Adapted from \cite{Zhang_2010_crossover}.
\label{Fig_topo_thinFilm}}
\end{figure*}

The short mean free path for photoelectrons allows ARPES to probe phenomena which occur exclusively at the surface. For example, the chemical potential probed by ARPES deviates from that measured in bulk-sensitive transport measurements due to a band-bending potential near the surface, as determined by comparison to quantum oscillations \cite{Analytis_2010_bulk}. This potential continues to evolve after cleaving due to residual adsorbates in the UHV environment \cite{bianchi2010coexistence}, and can be accelerated by deposition of impurities \cite{Wray_2010_a,King_2011_large}. In both cases, sufficient band-bending leads to quantum confinement of the bulk wave function near the sample surface, which manifests as a two-dimensional electron gas (2DEG) degenerate with the bulk bands~\cite{chiang2000photoemission}. As shown in Fig.~\ref{Fig_Topo_Rashba}, the 2DEGs form a series of quantum well states which are spatially localized in the band-bending potential well. Moreover, due to the strong spin-orbit coupling in the system, the 2DEGs exhibit Rashba splitting with a predominantly in-plane spin texture, as verified by spin-resolved ARPES measurements \cite{King_2011_large}.  This splitting is significantly larger than that observed in semiconductor heterostructure 2DEGs and Au(111) surface states \cite{King_2011_large}, and comparable to the giant Rashba-type splitting observed for bulk bands in non-centrosymmetric semiconductors \cite{Ishizaka_2011_giant}. In addition to their potential application for spintronics, these 2DEGs need to be considered when interpreting transport measurements in TIs, since they can contribute 2D conduction channels in addition to the topological surface state \cite{Bansal_2012_thickness}. 

Quantum confinement can also be induced by exfoliation or by fabricating thin film samples via layer-by-layer growth using molecular beam epitaxy. In the latter approach, the samples are grown under UHV conditions and typically transferred to an ARPES measurement chamber without subjecting them to atmospheric exposure. It was theoretically predicted that for a sufficiently thin film, a 3D TI will transition to a 2D QSHI in an oscillatory fashion as a function of the film thickness \cite{Liu_2010_oscillatory,Lu_2010_massive}. This is because the surface state wavefunctions on opposite sides of the sample begin to overlap and hybridize, opening up a gap at the Dirac point. ARPES measurements on Bi$_2$Se$_3$ have observed this gapped Dirac point for film thicknesses $<6$ quintuple layers (QLs), as shown in Fig.~\ref{Fig_topo_thinFilm}, while at the same time the bulk band structure is quantized into a series of quantum well states due to the spatial confinement \cite{Sakamoto_2010_spectroscopic, Zhang_2010_crossover}.  The oscillatory behavior was not observed, possibly because the oscillations dominantly manifest in the sign rather than the magnitude of the gap \cite{Liu_2010_oscillatory}.  It was later shown that topologically protected surface transport is  diminished in sufficiently thin films \cite{Taskin_2012_manifestation,Wu_2013_a}. Interestingly, not all topological states are gapped in the thin-film limit: in Sb, the film-substrate interaction breaks the symmetry between the top and bottom surfaces, which inhibits the hybridization that would open a gap \cite{Bian_2012_interfacial}.
 
\subsubsection{Magnetic topological insulators}

\begin{figure}
\centering
\includegraphics[width=0.8\columnwidth]{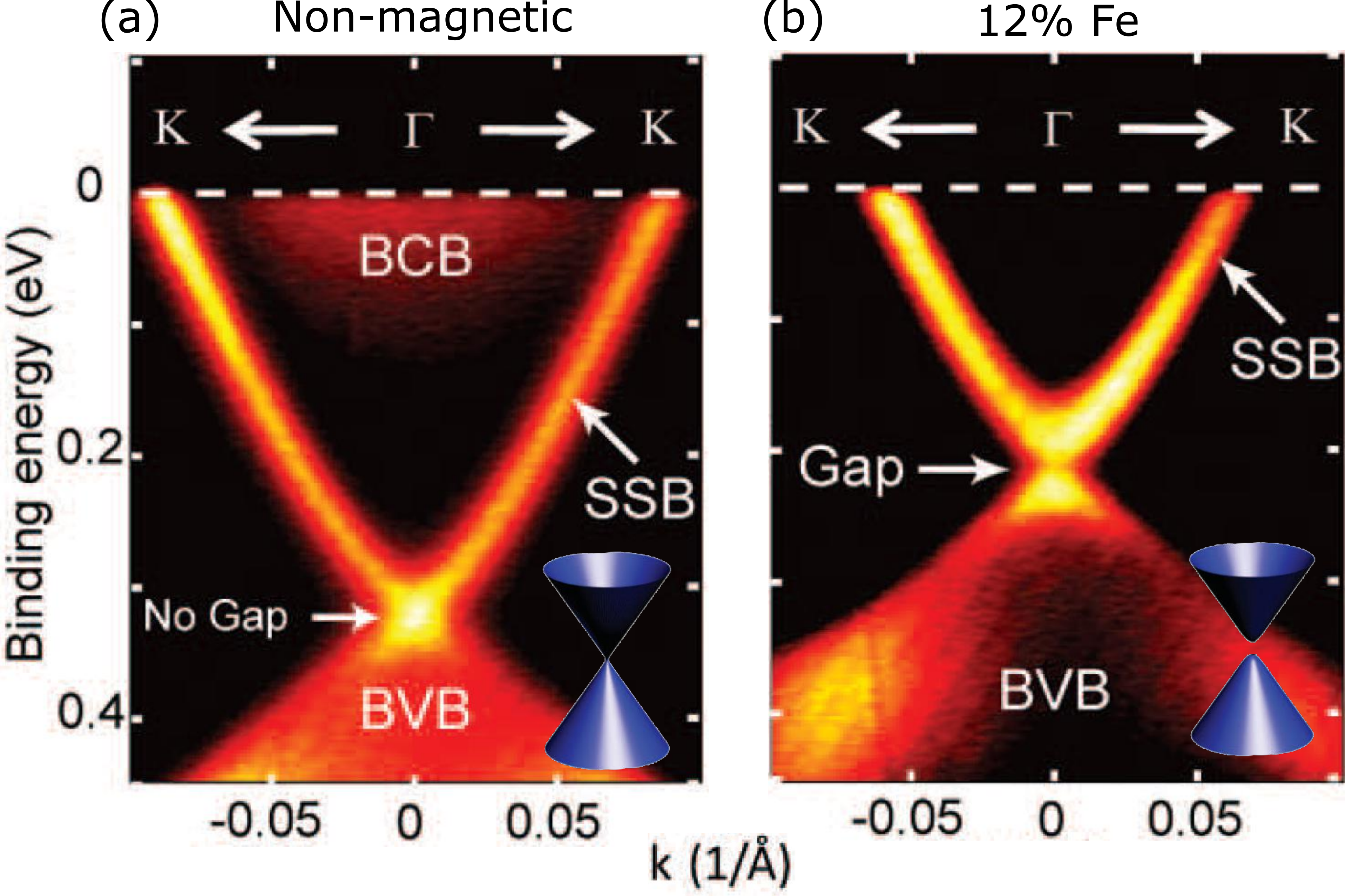}
\caption[Magnetically doped topological insulator]{Magnetically doped topological insulator. (a)  Ungapped Dirac cone of Bi$_2$Se$_3$, and (b) gapped Dirac cone of Fe-doped  Bi$_2$Se$_3$. Adapted from \cite{Chen_2010_massive}.
\label{Fig_Topo_magDope}}
\end{figure}

A 3D TI subject to broken time-reversal symmetry is  associated with nontrivial magnetoelectric effects such as image magnetic monopoles and topological Kerr/Faraday rotations \cite{Qi_2008_topological,Essen_2009_magnetoelectric,Wu_2016_quantized}. When reduced to the 2D limit, magnetic TIs become a platform for studying the quantum anomalous Hall effect \cite{Yu_2010_quantized}, as was experimentally demonstrated in thin films of Cr-doped (Bi,Sb)$_2$Te$_3$ \cite{Chang_2013_experimental} and MnBi$_2$Te$_4$ \cite{Deng_2020_quantum}. 

In ARPES, a signature of broken time-reversal symmetry in a 3D TI is the opening of a gap at the Dirac point, as shown in the lower row of Fig.~\ref{Fig_Topo_Overview} \cite{Liu_2009_magnetic}. Experimentally this has been investigated by both bulk doping \cite{Chen_2010_massive} and surface doping \cite{Wray_2010_a}  of magnetic impurities. As shown in Fig.~\ref{Fig_Topo_magDope}, a spectral weight suppression is clearly observed at the Dirac point of Bi$_2$Se$_3$ when Fe dopants are introduced to the bulk, even in the absence of bulk ferromagnetic order \cite{Chen_2010_massive}.  Subsequent work on Mn-doped Bi$_2$Se$_3$ thin films revealed a gap derived from out-of-plane ferromagnetic order, as demonstrated by closure of the gap above the Curie temperature. Further evidence for the magnetic nature of the gap was provided by spin-resolved ARPES, which revealed an out-of-plane component of the spin-polarization at the $\Gamma$-point. No out-of-plane spin component was observed for systems doped with non-magnetic impurities \cite{Xu_2012_hedgehog}.  

Despite these positive observations, there remains a number of important uncertainties on how Dirac cones are gapped in the presence of magnetism, and even in how gapped Dirac cones should be interpreted in general. First, in certain circumstances it has been demonstrated that Dirac point gapping may be completely unrelated to the existence of magnetism \cite{Bianchi_2011_simultaneous,SanchezBarriga_2016_nonmagnetic}.   At the same time it remains unclear in what conditions magnetism is sufficient to open a gap, since  other groups have reported that deposition of magnetic impurities does not open a gap \cite{Scholz_2012_tolerance,Valla_2012_photoemission, Schlenk_2013_controllable}. A recent development which promises to shed light on this issue was the prediction of MnBi$_2$Te$_4$ as an antiferromagnetic TI \cite{Li_2019_intrinsic,Zhang_2019_topological,Otrokov_2019_unique}, with the first published ARPES results reporting a gap of $\sim100$~meV \cite{Zeugner_2019_chemical,Vidal_2019_surface,Lee_2019_spin}. However, a series of works contradicted this claim with reports of a gapless surface state \cite{Li_2019_Dirac,Hao_2019_gapless,Chen_2019_intrinsic,Chen_2019_topological,Swatek_2020_gapless}.  It seems likely that these discrepant results are attributed to a $h\nu$-dependent photoemission cross-section for the surface states \cite{Hao_2019_gapless,Chen_2019_intrinsic}. If it indeed proves to be the case that the surface state is gapless, it may imply that the antiferromagnetic order is modified, multiple types of magnetic orders coexist, or that the magnetism is disordered near the surface \cite{Hao_2019_gapless}.

\subsubsection{Topological phase transitions}

\begin{figure}
\centering
\includegraphics[width=\columnwidth]{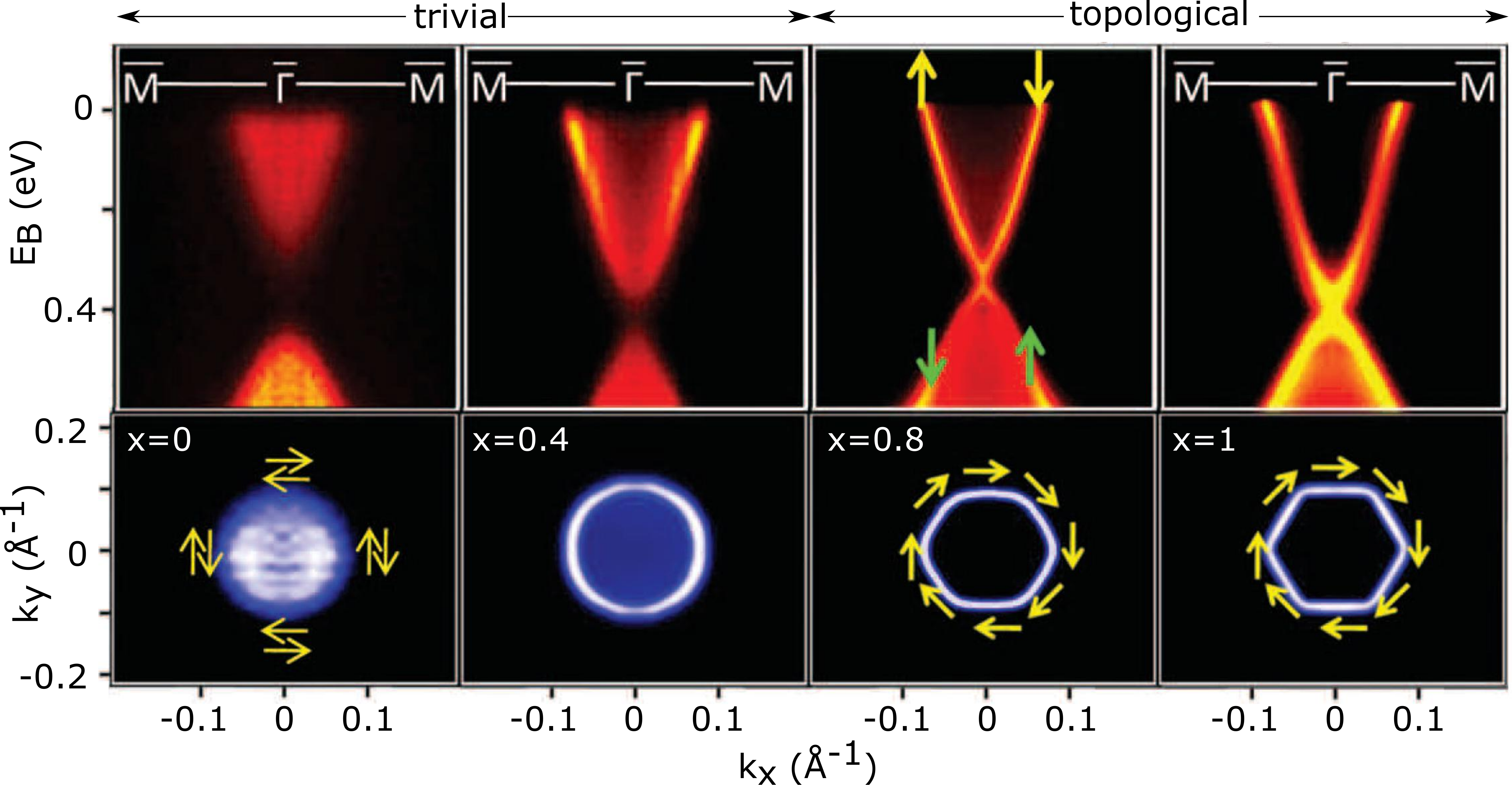}
\caption[Topological phase transition driven by chemical substitution in TlBi(S$_{1-x}$Se$_x$)$_2$]{ Topological phase transition driven by chemical substitution  in TlBi(S$_{1-x}$Se$_x$)$_2$. The doping level $x$ is indicated in the lower panels. For $x\lesssim 0.6$ the material is a trivial semiconductor, while for $x\gtrsim 0.6$ the bandgap becomes inverted and a topological surface states forms.  Adapted from \cite{Xu_2011_topological}.
\label{Fig_topo_PhaseTransition}}
\end{figure}

A topological phase transition can be driven by continuously tuning a material parameter through a range which results in band inversion. Unlike a conventional phase transition, this process does not  involve a broken symmetry; instead, the phase transition is characterized by a change in the topological invariant. A classic example of a topological phase transition is the crossing of a Landau level in a 2DEG under a changing magnetic field, as in the quantum Hall effect \cite{Hasan_2010_colloquium}. In the context of 3D TIs, the  most widely studied system by ARPES is TlBi(S$_{1-x}$Se$_x$)$_2$ with the chemical substitution $x$  varied to tune both the spin-orbit interaction strength and lattice parameter. For $x=0$ the material is a trivial semiconductor, while for $x=1$ it is a 3D TI. As shown in Fig.~\ref{Fig_topo_PhaseTransition}, at the intermediate value $x\sim0.6$ the bandgap closes, inverts, and a spin-polarized topological surface state emerges \cite{Xu_2011_topological}. Interestingly, multiple groups have reported that despite the band inversion the surface state remains gapped up to $x\sim 1$, an observation which is difficult to reconcile with its topological classification.  Potential explanations given by these groups include the roles of spontaneously broken symmetry \cite{Sato_2011_unexpected}, surface termination \cite{Niu_2012_topological}, bulk-surface scattering \cite{Souma_2012_spin}, or surface disorder \cite{Pielmeier_2015_response}, though a comprehensive understanding is still lacking. Another open issue is whether the surface states appear discontinuously with $x$, or evolve smoothly through the phase transition. ARPES measurements near the topological critical point suggest the latter possibility, as a gapped, spin-polarized surface state begins to develop spectral weight even  on the trivial side of the phase transition \cite{Xu_2015_unconventional}. In Bi$_2$Se$_3$, a signature of this trivial surface state remains as a spin-polarized surface-localized state degenerate with the bulk bands deep into the topological phase \cite{Jozwiak_2016_spin}.

\begin{figure}
\centering
\includegraphics[width=1\columnwidth]{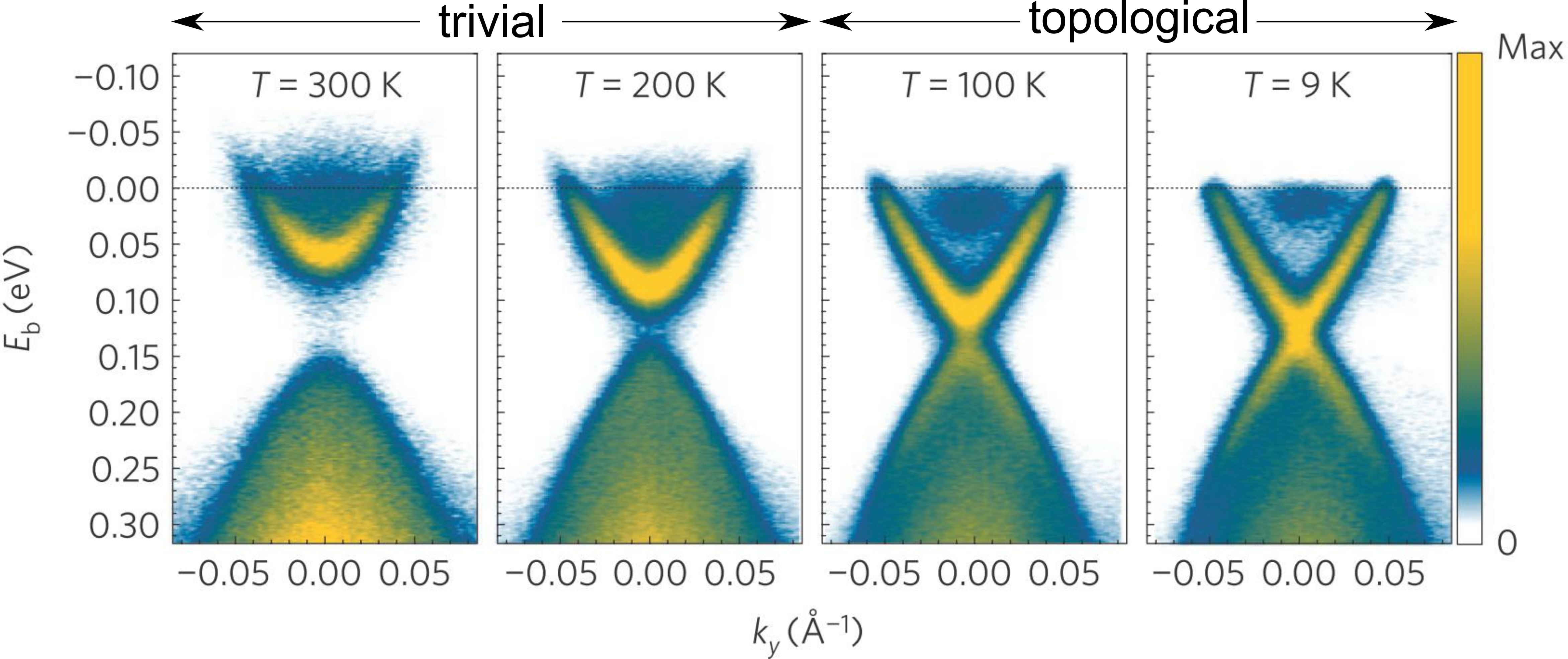}
\caption[Topological phase transition in the crystalline topological insulator Pb$_{1-x}$Sn$_x$Se]{Topological phase transition driven by temperature in the crystalline topological insulator Pb$_{1-x}$Sn$_x$Se ($x=0.23$). At a temperature between 100~K and 200~K the bulk band gap inverts, leading to the formation of topological surface states. Reproduced from \cite{Dziawa_2012_topological}.
\label{Fig_Topo_CTI}}
\end{figure}

Another platform for studying topological phase transitions is provided by the Pb$_{1-x}$Sn$_x$Y (Y=Se,Te)  class of  topological crystalline insulators (TCIs). TCIs represent a distinct topological phase from TIs because they are protected by the point-group symmetry of the crystal structure, in contrast to time-reversal as in the case of TIs \cite{Fu_2011_topological}.  SnTe was predicted to be a TCI protected by mirror symmetry, endowing the high-symmetry surfaces with an even number of Dirac cones, in contrast to the odd number required for $\mathbb{Z}_2$ TIs.  PbTe and PbSe were predicted to be topologically trivial but susceptible to band inversion by application of pressure, strain, or alloying \cite{Hsieh_2012_topological}. Indeed, as shown in Fig.~\ref{Fig_Topo_CTI}, Pb$_{1-x}$Sn$_x$Se ($x=0.23$) undergoes a topological phase transition due to a temperature-dependent inversion of the bulk bands  \cite{Dziawa_2012_topological}. Unlike the earlier example of TlBi(S$_{1-x}$Se$_x$)$_2$, here the bulk bands are clearly resolved on both sides of the transition.  ARPES has also verified TCI phases in SnTe \cite{Tanaka_2012_experimental} and Pb$_{1-x}$Sn$_x$Te \cite{Xu_2012_observation}. 

\subsubsection{The quantum spin Hall effect revisited}

\begin{figure}
\centering
\includegraphics[width=1\columnwidth]{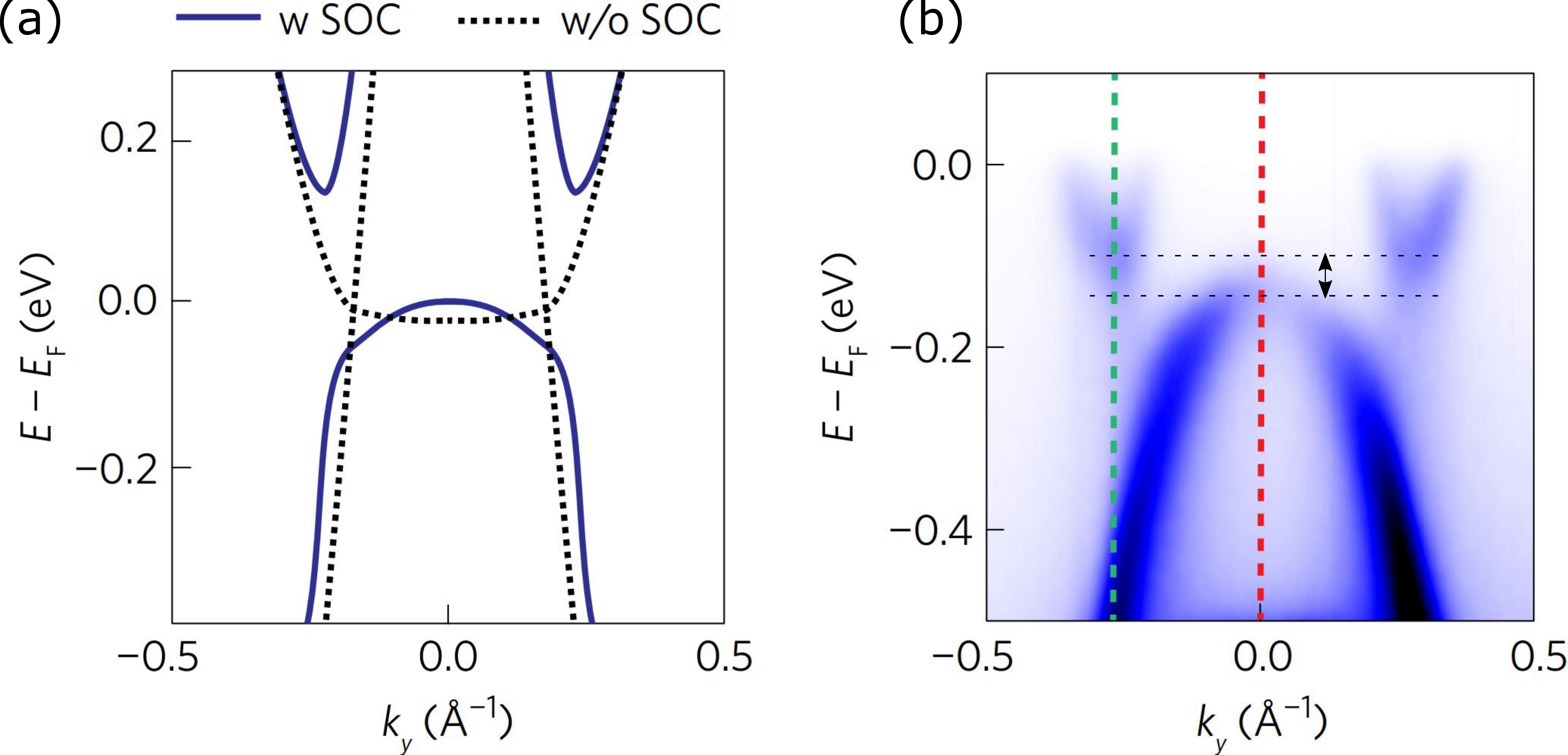}
\caption[Quantum spin hall insulator in monolayer  1T'-WTe$_2$ ]{Monolayer 1T'-WTe$_2$ as a quantum spin Hall insulator. (a) Calculated band structure, showing the band inversion enabled by spin-orbit coupling. (b) ARPES measurement revealing a 45~meV bandgap, as indicated by the horizontal dashed lines. Adapted from \cite{Tang_2017_quantum}.
\label{Fig_topo_WTe2}}
\end{figure}

As stated in Section \ref{Section_QuantumHall}, some of the first 2D topological systems studied included QSHIs such as HgTe quantum wells at milliKelvin temperatures \cite{Konig_2007_quantum}.  Recently there has been a renewed search for QSHIs which are not dependent on a heterostructured design and which exhibit a larger bandgap suitable for application at higher temperature. Such a model system could help reconcile some of the experimentally puzzling aspects which remain for the HgTe quantum wells \cite{Ma_2015_unexpected,Nichele_2016_edge}. One notable development was the prediction of the quantum spin Hall effect in monolayer transition metal dichalcogenides \cite{Qian_2014_quantum}. ARPES work on monolayers of 1T'-WTe$_2$ synthesized by molecular beam epitaxy supported this prediction by measuring an inverted bandgap of 45~meV, as shown in Fig.~\ref{Fig_topo_WTe2} \cite{Tang_2017_quantum}, with simultaneous reports of edge conduction in exfoliated films \cite{Fei_2017_edge}.  The quantum spin Hall effect, including quantized edge conduction, was  subsequently confirmed up to 100~K in transport measurements \cite{wu2018observation}. In later experiments, the gap of monolayer 1T'-WSe$_2$ was found to be up to 130~meV \cite{Ugeda_2018_observation,Chen_2018_large}.  Other systems promising room-temperature applications include ultrathin Na$_3$Bi \cite{Collins_2018_electric}, Bismuthene on SiC \cite{Reis_2017_bismuthene}, and Stanene on Cu \cite{Deng_2018_epitaxial}. These discoveries lie at the intersection of 2D and topological materials research fields and make quantum spin Hall platforms more readily available.

A closely related phase is the 3D ``weak'' TI, which can be regarded as a stack of QSHI layers \cite{Fu2_2007_topological,Fu_2007_topological}. For a bulk crystal one expects the top and bottom surfaces to be insulating, while the side surfaces would exhibit 1D surface states. Experimental verification by ARPES has been hindered by the difficulty in measuring photoelectrons from the side surfaces of cleaved crystals. For candidate materials such as ZrTe$_5$ much of the supporting evidence has been limited to showing that the top surfaces are insulating \cite{Xiong_2017_three}, though such reports have been controversial due to the small gap \cite{Manzoni_2016_evidence}. Possible indications of the side surface states have manifested as one-dimensional features superposing the spectrum, possibly attributed to photoemission from the edges of cracks in the sample surface \cite{Zhang_2017_electronic}. Nano-ARPES with sub-1~$\mu$m spatial resolution has been utilized to separately resolve the signal from the top (001) and side (100) surfaces of $\beta$-Bi$_4$I$_4$, revealing a 1D state associated solely with the side surface \cite{noguchi2019weak}. This is suggestive of a one-dimensional edge state, but as the first measurement of its kind,  follow-up work is required to confirm that this is indeed an incontrovertible hallmark of a weak TI.

\subsubsection{Topological superconductors}\label{sec_topoSC}

The theory of topological superconductors is analogous to that of TIs, with the role of the insulating band gap replaced by the particle-hole symmetric superconducting gap \cite{Schnyder_2009_classification,Qi_2011_topological}. Similarly, the gapless edge modes of TIs are replaced by gapless Majorana states in topological superconductors, with the form of the wavefunction constrained by the particle-hole symmetry of the Bogoliubov-de Gennes Hamiltonian. An interesting situation arises when one of these states is bound to the interface between normal and superconducting regions at a vortex. This state, known as a Majorana zero mode, is an equal admixture of electrons and holes, represents a quasiparticle which is charge-neutral, has exactly zero energy, and is its own antiparticle. It is therefore analogous to the Majorana fermion, which has been hypothesized as a elementary particle in nature but not yet experimentally observed \cite{Qi_2011_topological,Hasan_2010_colloquium}. In contrast, Majorana zero modes are non-fermionic since they obey non-Abelian statistics, which allows them to form the basis for the field of topological quantum computation \cite{Nayak_2008_non}.

The simplest theoretical proposal for a system exhibiting Majorana zero modes involves a superconductor with spinless $p_x + i p_y$ pairing. It was later realized that the non-spin-degenerate bands of a 3D TI surface state, if driven to superconduct, would resemble a  spinless $p_x + i p_y$ superconductor which maintains time-reversal symmetry \cite{Fu_2008_superconducting}. Many efforts focused on inducing superconductivity in  TIs via the proximity effect. For example, scanning tunneling spectroscopy measurements on Bi$_2$Se$_3$ thin films fabricated on superconducting NbSe$_2$ revealed a superconducting gap $>0.5$~meV for film thickness $<3$~QLs, though  complementary ARPES experiments showed that the Dirac cone features a sizable hybridization gap  \cite{Wang_2012_the}. Further experiments  showed that even the hybridization-gapped Dirac cone hosts spin-polarized carriers, and exhibits a superconducting gap of $\sim 0.5$~meV up to $T_c \sim 7~K$, with the superconducting gap size decreasing with film thickness  \cite{Xu_2014_momentum}. In a related development, superconducting gaps were demonstrated in up to 10~QL Bi$_2$Se$_3$ films on polycrystalline Nb substrates \cite{Flototto_2018_superconducting}. Finally, we mention that superconducting gaps up to 15~meV were reported up to 60~K in Bi$_2$Se$_3$ films grown on a cuprate superconductor \cite{Wang_2013_fully}, though follow-up studies brought this observation into question, citing unfavorable conditions due to mismatched Fermi surface topologies, incompatible lattice symmetries, and a short coherence length \cite{Xu_2014_Fermi,Yilmaz_2014_absence}.

Another approach is to identify superconducting materials which intrinsically exhibit topological surface states. An intensely investigated potential platform is the iron-based superconductor FeTe$_{1-x}$Se$_x$ ($x=0.45$), where ARPES revealed possible signatures of a topological surface state crossing the gap between bulk bands near the $\Gamma$-point. These states are spin-polarized, and exhibit an isotropic superconducting gap up to 1.8~meV with $T_c$=14.5~K \cite{Zhang_2018_observation}.  Scanning tunneling microscopy measurements discovered a zero bias peak in vortex cores, which is a necessary but not sufficient condition for establishing a Majorana bound state \cite{Wang_2018_evidence}. Similar ARPES observations have been reported in other materials in the iron-based superconductor family \cite{Liu_2018_robust, Zhang_2019_multiple, Liu_2019_a}. In all of these materials, the identification of the topological surface state is not as unambiguous as in the Bi$_2$Se$_3$-family due to small bulk gaps and nearly-overlapping bulk bands, and there remains considerable controversy in the interpretation of the ARPES data \cite{Borisenko_2019_strongly}. Recently, topological surface states were also suggested in superconducting MgB$_2$ ~\cite{zhou2019observation}, TaSe$_3$~\cite{nie2018topological}  and 2M-WS$_2$~\cite{fang2018coexistence}, which all remain to be  further scrutinized.

\subsubsection{Topological Kondo insulator candidates}\label{sec_topoKI}

In a topological Kondo insulator, the role of a bulk insulating gap is played by the hybridization gap between itinerant carriers and localized $f$-electrons in a heavy Fermion material (see Section~\ref{sec_heavyFermion}) \cite{Dzero_2010_topological}. These materials are noteworthy in that electron correlations play a central role in the formation of the inverted band structure. One prominent yet controversial example is SmB$_6$.  Some measures of the Fermi surface, such as quantum oscillations, provide evidence for two-dimensional states \cite{Li_2014_two}, though additional signals attributed to the bulk indicate that much is not yet understood about quantum oscillations in a Kondo insulator \cite{Tan_2015_unconventional}. Several ARPES works supported the existence of topological states with the observation of surface states within the bulk Kondo gap \cite{Xu_2013_surface, Neupane_2013_surface, Jiang_2013_observation} which are spin-polarized \cite{Xu_2014_direct}. However, doubts have been raised about the interpretation of these features, especially due to the role of surface termination \cite{Zhu_2013_polarity},  bending of the chemical potential in the near-surface region \cite{Frantzeskakis_2013_Kondo}, and coexistence of topologically trivial Rashba-split surface states \cite{Hlawenka_2018_samarium}. Due to these and other open questions \cite{Dzero_2016_topological}, it remains unclear whether SmB$_6$ can be considered a topological Kondo insulator.

\subsubsection{Other TIs}

Several dozen TIs have been experimentally studied in the years following the initial discovery of Bi$_{1-x}$Sb$_x$ in 2008 \cite{Ando_2013_topological}. trARPES has been a useful tool for its ability to resolve topological states even when unoccupied in equilibrium \cite{Niesner_2012_unoccupied,Sobota_2013_direct,Yan_2015_topological,Zhang_2017_topologically}. It is now recognized that topological materials are not nearly as rare as one might expect: large-scale theoretical searches have predicted thousands of topological materials, estimating that up to 30\% of materials in nature are topologically non-trivial, with $\sim12$\% being TIs \cite{Vergniory_2019_a,Zhang_2019_catalogue, Tang_2019_comprehensive}. These studies have published freely accessible, searchable databases, thus bringing end to the era in which topological materials are evaluated on a case-by-case basis.

\subsubsection{Platform for Floquet physics}

\begin{figure}
\centering
\includegraphics[width=1\columnwidth]{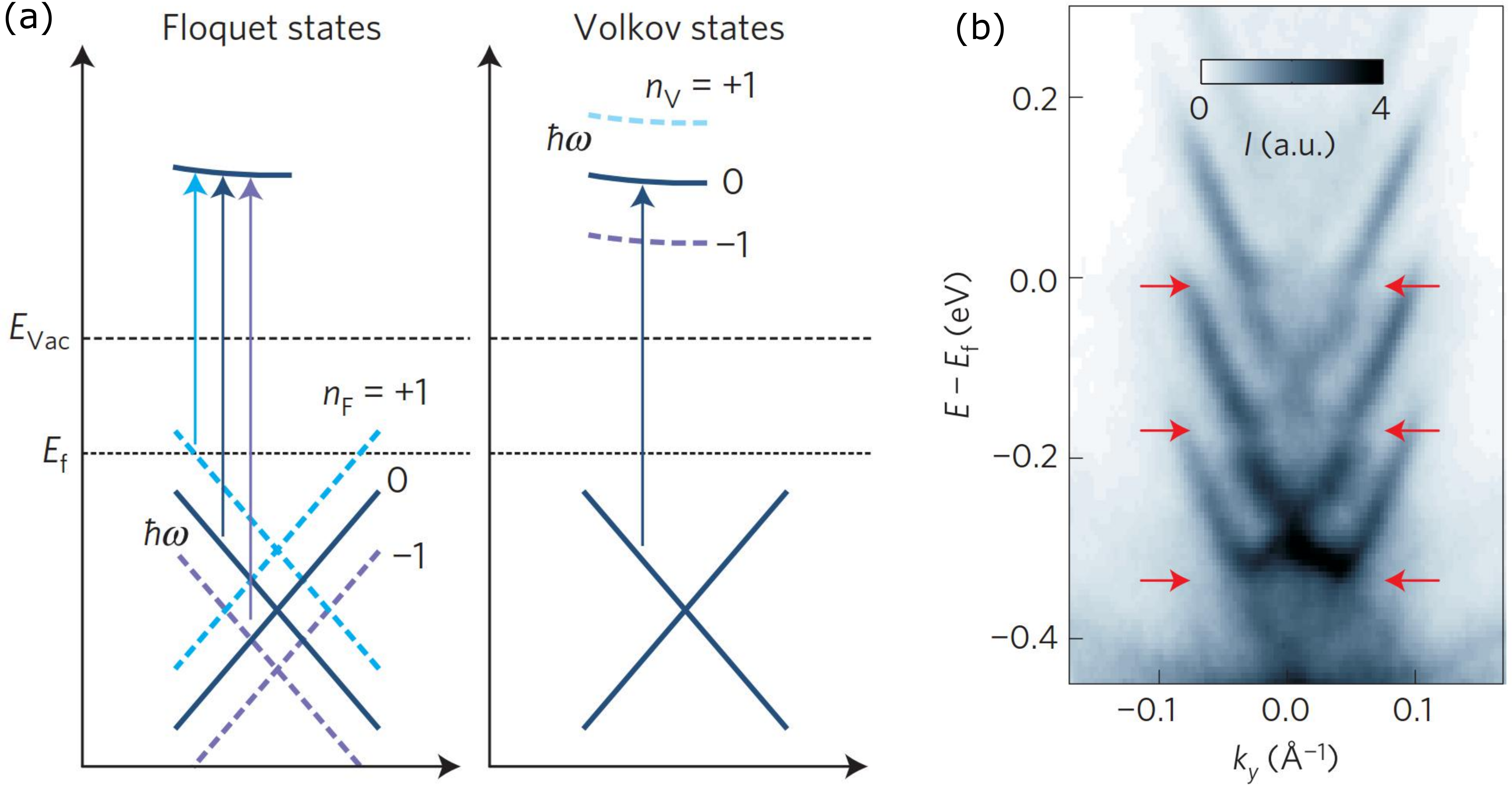}
\caption[Floquet-Bloch states in Bi$_2$Se$_3$]{ (a) Distinction between Floquet states and Volkov states, which are dressing of electronic states in the solid and vacuum, respectively. Both states can exist simultaneously. (b) Experimental measurement of Floquet-Bloch bands on Bi$_2$Se$_3$. Red arrows highlight avoided crossings between neighboring Floquet states. Adapted from  \cite{Mahmood_2016_selective}.
\label{Fig_trARPES_Floquet}}
\end{figure}

TIs have served as a platform for studying non-equilibrium \emph{Floquet-Bloch} states using trARPES (see Fig.~\ref{Fig_trARPES_overview}(e)). These states follow from the Floquet theorem, which shows that a periodic perturbation with period $T$ leads to quasistatic eigenstates that are evenly spaced in energy by $2\pi\hbar/T$ \cite{Shirley_1965_solution}. In a trARPES experiment, the periodic perturbation is applied by the electric field of the pump pulse. 

Floquet-Bloch bands were first demonstrated for the topological surface states of Bi$_2$Se$_3$ excited with a mid-infrared pump. Replica bands were observed, with avoided-crossing gaps between neighboring Floquet states. In addition, it was shown that  dressing of the Dirac cone with circularly-polarized light broke time-reversal symmetry, and thus opened a gap at the Dirac point \cite{Wang_2013_observation}. One subtlety in studying Floquet-Bloch states is that they are difficult to distinguish from laser-assisted photoemission (LAPE), in which the photoelectron emits or absorbs photons into so-called Volkov states. Both effects lead to replica bands spaced by the photon energy; the distinction is that Floquet-Bloch states are dressed in the solid, while Volkov states are dressed in the vacuum. The polarization dependence of the intensities and avoided-crossing gaps allows for discriminating these effects, and even suppressing the Volkov states, see Fig.~\ref{Fig_trARPES_Floquet} \cite{Mahmood_2016_selective}. We note that LAPE at photoexcited surfaces is  quite generic \cite{Miaja_2006_laser}; therefore, observation of replica features alone is insufficient for identifying Floquet-Bloch states, and hybridization between the sidebands must be observed \cite{Mahmood_2016_selective}. The sub-gap photon energy and clean Dirac structure is what made  Bi$_2$Se$_3$ an ideal platform for demonstrating this distinctly non-equilibrium phenomenon, though it should be noted that the topological property itself was not strictly relevant.

\subsection{Topological semimetals}

Since the discovery of TIs, it has been recognized that the topological classification of matter can be extended to semimetals. The first example we shall discuss is the Dirac semimetal, which hosts a point of fourfold degeneracy about which the bands disperse linearly in all three momenta dimensions (see Fig.~\ref{Fig_Topo_Overview}, upper row). If inversion or time-reversal symmetry is broken, the nodal point splits into two doubly degenerate nodes separated in momentum space, creating what is known as a Weyl semimetal (Fig.~\ref{Fig_Topo_Overview}, lower row). We shall discuss the basic concepts underlying the topology of these phases, and highlight the role of ARPES in identifying the phases and their characteristic surface states. For the interested reader we refer to more comprehensive reviews on these topics \cite{Turner_2013_beyond,Armitage_2018_Weyl}. 

\subsubsection{Dirac semimetals}

\begin{figure}
\centering
\includegraphics[width=\columnwidth]{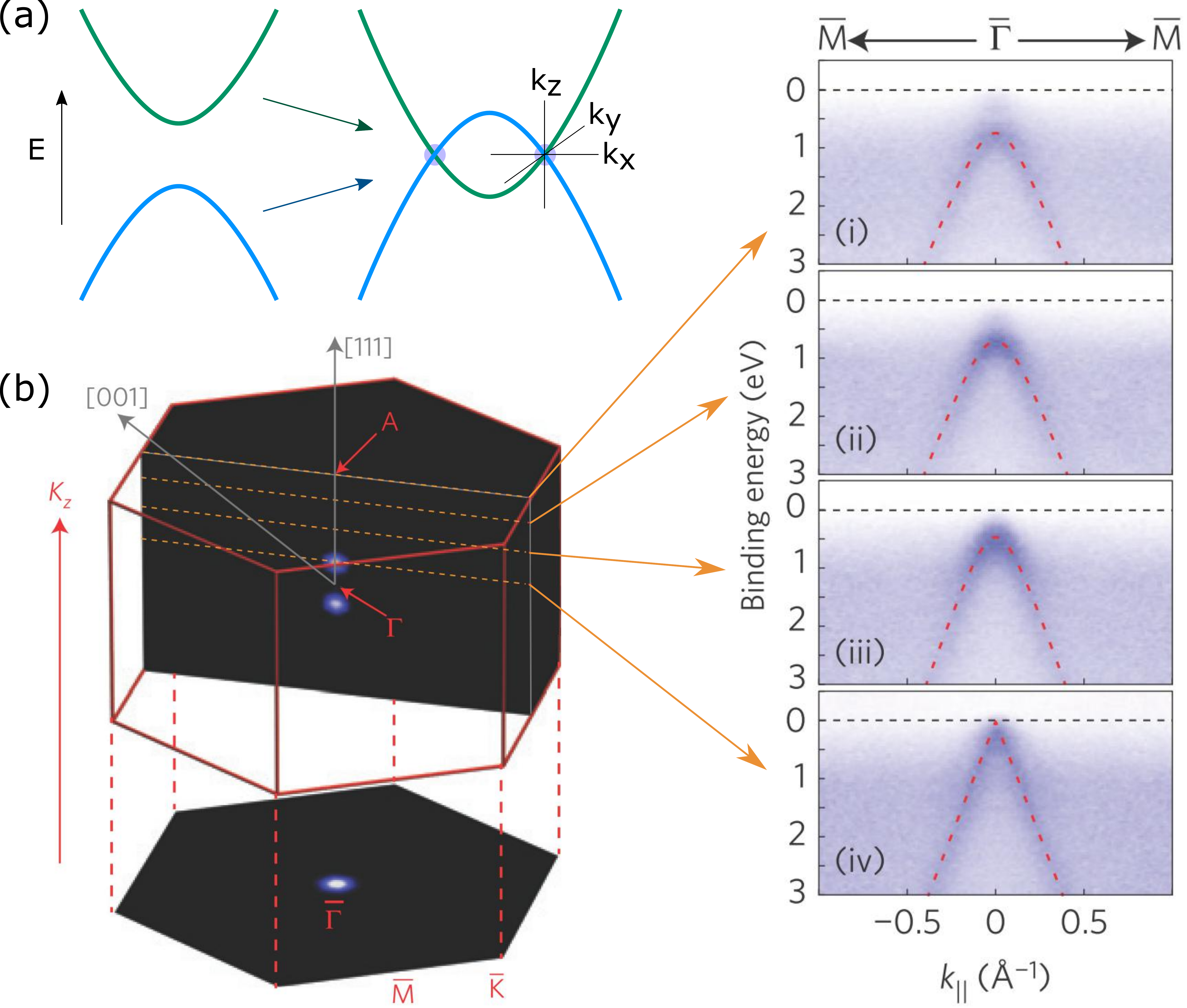}
\caption[Dirac semimetals]{Dirac semimetals. (a) Cartoon of the band-inversion mechanism of Dirac semimetal formation. If two 3D bands of opposite parities are driven to invert, and are symmetry-forbidden from mixing, they will necessarily form a pair of fourfold degenerate Dirac nodes which disperse linearly as a function of $k_x$, $k_y$, and $k_z$. (b) Experimental realization in Cd$_3$As$_2$. The 3D Fermi surface consists of a pair of nodal points. A cut through a node reveals a linear band dispersion (panel (iv)).  Adapted from \cite{Liu_2014_a}.
\label{Fig_topo_DSM}}
\end{figure}

Dirac semimetals are realized at the topological phase transition in 3D TIs, when the bulk bandgap closes and a fourfold degeneracy occurs. However, this degeneracy is accidental since an infinitesimal change of the tuning parameter will re-open the gap. The question arises whether a Dirac semimetal can be realized as a more robust electronic state. Indeed, it can happen when a band inversion occurs between two bands which cannot be mixed due to symmetry, as shown in Fig.~\ref{Fig_topo_DSM}(a). Note that the fourfold degeneracies necessarily appear in pairs, and can be gapped by breaking additional symmetries \cite{Yang_2014_classification,Armitage_2018_Weyl}. 

Dirac semimetals were first observed by ARPES in Na$_3$Bi \cite{Liu_2014_discovery} and Cd$_3$As$_2$ \cite{Neupane_2014_observation, Borisenko_2014_experimental, Liu_2014_a}, with the salient features summarized in Fig.~\ref{Fig_topo_DSM}(b). In both cases, the band crossings are protected by bulk $c$-axis rotational symmetries. By mapping the electronic structure as a function of $k_x$, $k_y$, and $k_z$, the experiments confirm that the Fermi surface consists of a pair of nodes, while the bands disperse linearly along all three momenta directions.  We note that while topological surface states have been observed in Dirac semimetals \cite{Yi_2014_evidence, Xu_2015_observation}, they do not enjoy the same level of protection as in Weyl semimetals due to the fact that the surfaces can break the spatial symmetries that preserve the crossings in the bulk \cite{Potter_2014_quantum,Kargarian_2016_are}.

\subsubsection{Weyl semimetals}

As mentioned above, a Weyl semimetal is created when a Dirac semimetal is subjected to broken inversion and/or time-reversal symmetry. Each Weyl node is associated with an integer-valued topological index known as chirality. Since chirality is conserved, a Weyl node is stable unless annihilated with a node of opposite chirality  \cite{Turner_2013_beyond,Armitage_2018_Weyl}. Thus, Weyl nodes are intrinsically more robust than Dirac nodes, which are chirality-neutral and therefore depend on additional symmetries to protect against gapping. 

Like TIs, Weyl semimetals are associated with topological surface states, but they have the unusual property that their Fermi surfaces form arcs in momentum space. These arcs must connect Weyl points of opposite chirality, and are therefore topologically protected as long as the Weyl points avoid annihilation by remaining separated.  Viewed as a geometrical construct, a Fermi surface must be a closed contour, so the existence of a Fermi arc appears anomalous. The key here is that the surface states are not isolated since they merge into the bulk at the Weyl points. In fact, the Fermi surface contour is globally closed if one considers the arcs on opposite sides of the sample \cite{wan2011topological,Turner_2013_beyond,Armitage_2018_Weyl}. This is analogous to a real-space lattice dislocation that propagates to opposite sides of a crystal. 

\begin{figure}
\centering
\includegraphics[width=0.9\columnwidth]{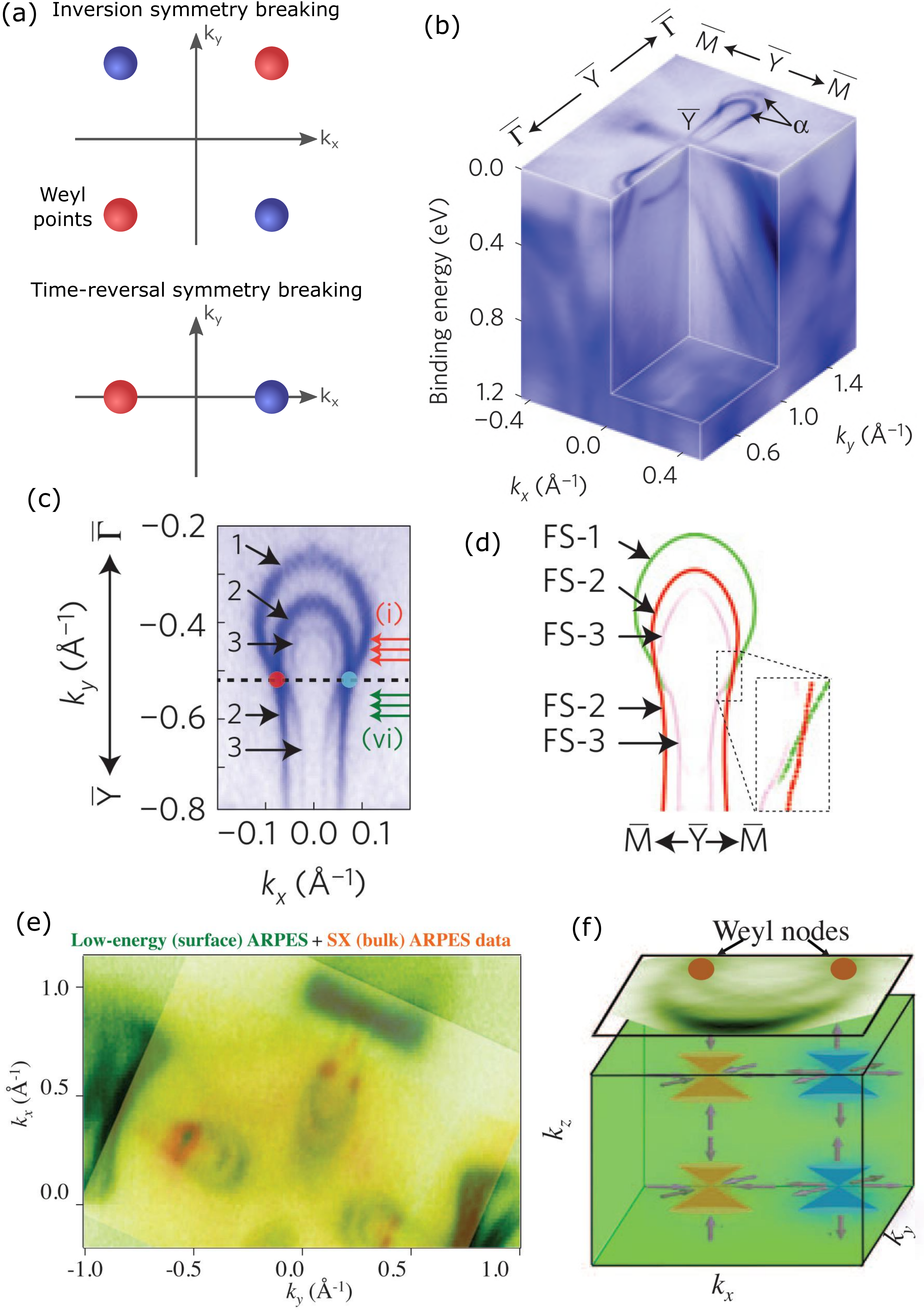}
\caption[Weyl semimetals]{Weyl semimetals. (a) The simplest inversion symmetry breaking Weyl semimetal has four Weyl nodes, and the simplest time-reversal symmetry breaking Weyl semimetal has two Weyl nodes. (b) Band-mapping of the inversion symmetry breaking Weyl semimetal TaAs, with (c) a close-up of the surface Fermi arcs compared to (d) a calculation. Adapted from \cite{Yang_2015_Weyl}. (e) Fermi surface mapping of TaAs combining low-energy, surface-sensitive ARPES (green) with soft x-ray, bulk-sensitive ARPES (orange). The soft x-ray measurements isolate the bulk Weyl nodes, while the low-energy measurements reveal the surface Fermi arcs connecting them. (f) Schematic of this electronic structure, including the bulk Weyl nodes and surface Fermi arcs. Adapted from \cite{Xu_2015_discovery}.
\label{Fig_topo_Weyl}}
\end{figure}

The first Weyl semimetals discovered were associated with broken inversion symmetry. In this case the Weyl nodes exist in multiples of four because time-reversal symmetry maps a node at $\vec{k}$ onto a node at $-\vec{k}$ with the same chirality; thus, another pair must exist to achieve net zero chirality, as shown in Fig.~\ref{Fig_topo_Weyl}(a). The first materials studied by ARPES include TaAs \cite{Lv_2015_experimental, Xu_2015_discovery, Yang_2015_Weyl}, NbAs \cite{Xu_2015_discovery2}, and TaP \cite{Xu_2015_experimental,Liu_2016_evolution}. A band-mapping of the bulk and surface bands is shown in Fig.~\ref{Fig_topo_Weyl}(b), with a zoom-in of the Weyl nodes and surface Fermi arcs in (c)-(d) \cite{Yang_2015_Weyl}. To verify the bulk and surface assignment of these features, Fig.~\ref{Fig_topo_Weyl}(e) shows an overlay of Fermi surfaces from surface-sensitive low-energy ARPES with bulk-sensitive soft x-ray ARPES. The Weyl node structure is clearly associated with the bulk, while the arcs are associated with the surface (Fig.~\ref{Fig_topo_Weyl}(f)) \cite{Xu_2015_discovery}.  There has been some discrepancy in the identification and interpretation of the Fermi arcs between various groups; this may be because only their existence is topologically protected, while the detailed dispersion can be highly sensitive to the surface condition \cite{Sun_2015_topological,Yang_2019_topological}.  

In the case of broken time-reversal symmetry the Weyl nodes are created in pairs of opposite helicity (Fig.~\ref{Fig_topo_Weyl}(a)). The experimental evidence for Weyl semimetals with broken time-reversal symmetry has been more elusive \cite{Kuroda_2017_evidence}, though recently there was compelling evidence from ARPES and STM that Co$_3$Sn$_2$S$_2$ is a ferromagnetic Weyl semimetal with three pairs of Weyl nodes \cite{Liu_2019_magnetic,Morali_2019_fermi}.  Similarly, Co$_2$MnGa was found to be a magnetic Weyl semimetal, exhibiting so-called line-nodes rather than nodal points \cite{Belopolski_2019_discovery}. For all these materials, to-date only measurements in the ferromagnetic phase have been reported.

For the Weyl node semimetals discussed thus far, the Weyl fermions feature closed, nearly-circular constant-energy contours with vanishing density-of-states at the node. If the dispersion around the Weyl node tilts sufficiently, the constant-energy contours become open, and there is a finite density-of-states at the energy of the node. These two cases have been classified as type I and type II. While type II Weyl semimetals have been reported in ARPES measurements \cite{Huang_2016_spectroscopic, Deng_2016_experimental}, subsequent work has shown that unambiguous identification is not straightforward since the distinction between topologically trivially and non-trivially Fermi arcs in the ARPES data can be remarkably subtle \cite{Bruno_2016_observation}. Just as is the case for type I semimetals, measurements on a time-reversal symmetry breaking type II Weyl semimetal have been recently reported \cite{Borisenko_2019_time}. 

Finally, we emphasize that this is not an exhaustive review of topological semimetallic states. Other exotic states, including  drumhead surface states in line-node semimetals \cite{Burkov_2011_topological,Belopolski_2019_discovery} and helicoid surface bands in chiral semimetals \cite{Fang_2016_topological,Schroter_2019_chiral,Sanchez_2019_topological}, among others, continue to be experimentally investigated, with ARPES playing the leading role in characterizing their nontrivial band topology.

\subsection{Outlook}\label{sec_outlook_TIs}

The pace of research on topological materials in just over a decade has been breathtaking, with ARPES playing a central role not only mapping their band dispersions, but also projecting out the spin-orbital components of their wavefunctions. In the near term, there will continue to be a strong effort toward identifying magnetic topological materials with unequivocal Dirac point gapping. The development of small-spot ARPES will pave the way for discovering novel topological physics at edge channels and domain walls. The pursuit of a robust topological superconductor will continue, with ARPES evidencing not only the topological surface states and superconducting gap, but also measuring the dispersion of the elusive Majorana quasiparticles. Finally, investigations will increasingly advance beyond systems described by single-particle theory, with an emphasis on the relationship between topological phenomena and strong correlations \cite{Rachel_2018_interacting}.

\section{Other materials}\label{sec_otherMat}

The sections above have exemplified the roles of electronic/lattice interactions, multiorbital Hund's couplings, dimensionality, and spin-orbit interactions for giving rise to a variety of rich electronic phenomena. However, the material systems described by these ingredients are certainly not limited to the examples discussed so far. To demonstrate the breadth of ARPES's impact in condensed matter physics, we now provide a brief overview of how ARPES has facilitated a microscopic understanding of various other material families. These families include metals/semimetals with remarkable transport properties, superconductors, $f$-electron systems exhibiting Kondo physics, and charge density wave systems. This section is primarily organized by physical phenomenon, though material families which share similar compositions are grouped where appropriate. 

\subsection{Conventional superconductors}
The copper- and iron-based superconductors comprise a significant portion of ARPES studies on superconductivity partly because of their large superconducting energy scales ($\Delta_{sc}\sim$~10-50~meV), matching with the ARPES energy resolution at the time of the materials' discoveries. With the advent of high-resolution ($\sim$70~$\mu$eV) low-temperature (sub-K) small-spot ($\sim$100~nm to 1~$\mu$m) laser- and synchrotron-based ARPES, spectroscopic features in superconductors with T$_c$'s at single digit Kelvins begin to receive more investigations. This includes superconducting Boron-doped diamond ($\Delta_{sc}$ = 0.78~meV, $T_c$ = 6.6~K)~\cite{ishizaka2007observation}, $\beta$-pyrochlore superconductor KOs$_2$O$_6$ ($\Delta_{sc}$ = 1.63~meV, $T_c$ = 9.6~K)~\cite{shimojima2007interplay} and Sn ($\Delta_{sc}$ = 0.52~meV, $T_c$ = 3.7~K)~\cite{okazaki2012octet}. This subsection will focus on a collection of such superconducting systems, with an emphasis on the determination of superconducting gap size and momentum structure.

\subsubsection{MgB$_2$ and graphite intercalation compounds}

MgB$_2$ with $T_c = 39$~K holds the $T_c$ record for any binary compound under ambient pressure~\cite{nagamatsu2001superconductivity}. It has a clear isotope effect that is dominated by the boron atoms~\cite{bud2001boron}, and the electronic structure is characterized by bands associated with highly covalent in-plane $\sigma$-bonds and out-of-plane $\pi$-bonds~\cite{belashchenko2001coexistence}. The superconductivity is postulated to be attributed to the highly anharmonic $E_{2g}$ optical phonon involving mainly boron motion~\cite{hinks2001complex,yildirim2001giant}. These observations indicate that MgB$_2$ is a conventional phonon-mediated multi-band superconductor.  

\begin{figure} 
	\includegraphics[width=1\columnwidth]{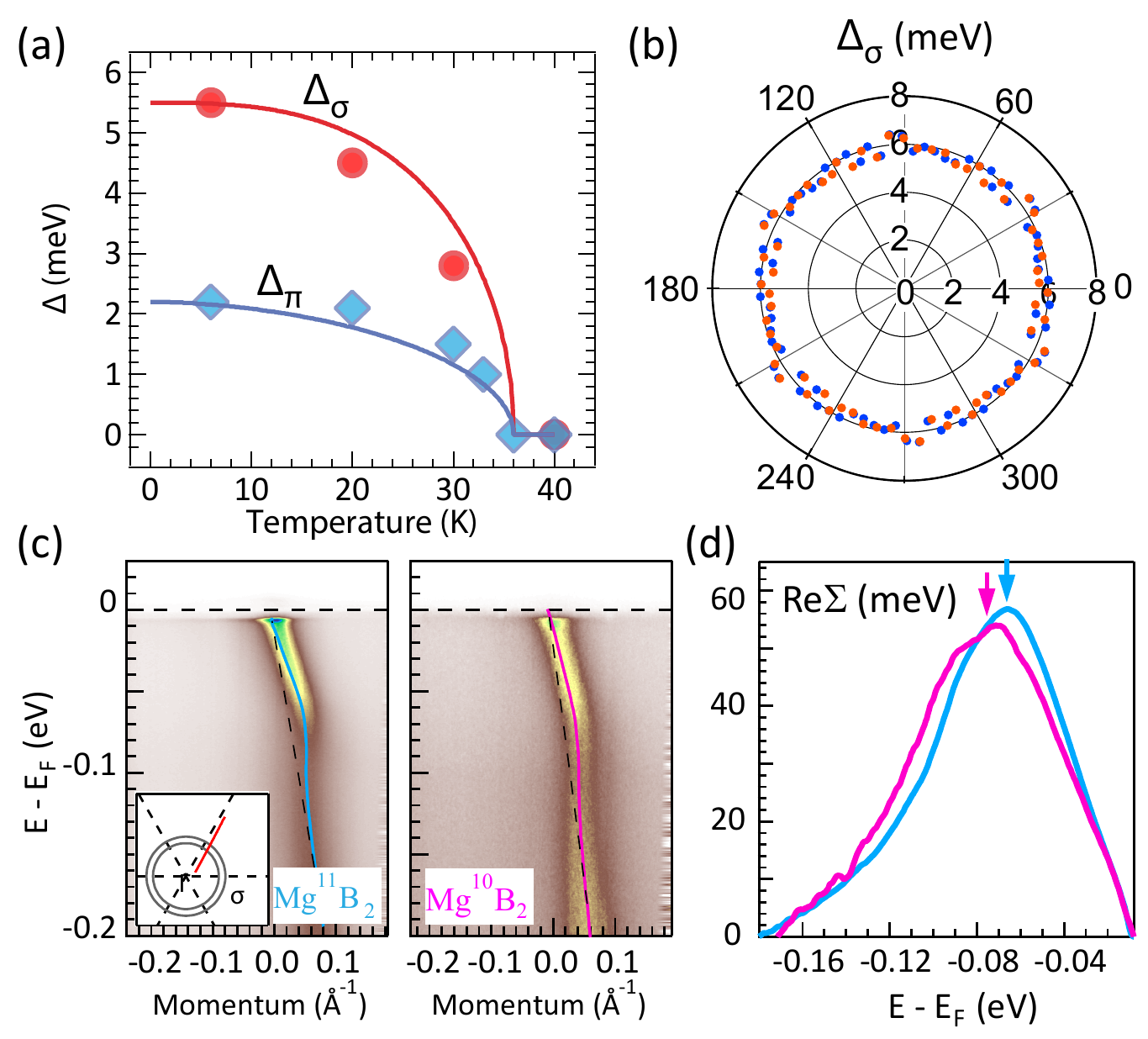}
	\caption{Spectral properties of superconducting MgB$_2$. (a) Temperature dependent energy gaps on the $\sigma$ and $\pi$ Fermi surface sheets. Adapted from ~\cite{tsuda2003definitive}. (b) Angle dependent superconducting gap on two closely positioned $\sigma$ bands around $\Gamma$ (red and blue). Adapted from ~\cite{mou2015momentum}. (c) $\Gamma$-M energy-momentum dispersion anomaly of on B-isotope substituted Mg$^{11}$B$_2$ (cyan) and Mg$^{10}$B$_2$ (magenta). (d) Electron self energy (real part) extracted for two B-isotope MgB$_2$ $\sigma$-bands. Adapted from ~\cite{mou2016isotope}.}
	\label{Fig_others_MgB2}
\end{figure}

Early momentum-integrated superconducting gap measurements reported a wide range of gap values~\cite{rubio2001tunneling,sharoni2001tunneling,takahashi2001high}, which were later realized to contain contributions from both the $\sigma$ and $\pi$ bands~\cite{buzea2001review,sologubenko2002thermal,yelland2002haas}. ARPES first reported direct measurement of both $\sigma$ and $\pi$ bands~\cite{uchiyama2002electronic} and their respective superconducting gaps: 5.5-6~meV on the $\sigma$ band and a nearby surface band, and 1.5-2.2~meV on the $\pi$ band (Fig.~\ref{Fig_others_MgB2}(a))~\cite{souma2003origin,tsuda2003definitive}. Subsequently, interband pairing was considered to play a significant role in determining the coupling strength and $T_c$~\cite{choi2002origin,dolgov2009interband}. With sub-meV energy resolution and the tightly focused beam offered by a VUV laser lightsource, detailed doping, angle, and isotope substitution dependence of the superconducting gaps were subsequently examined to great detail especially near the Brillouin zone center~\cite{tsuda2005carbon,mou2015momentum,mou2016isotope}. The $\sigma$ bands exhibit an isotropic superconducting gap and a strong mode-coupling feature around 66.5~meV (Mg$^{11}$B$_2$) and 70~meV (Mg$^{10}$B$_2$) (Fig.~\ref{Fig_others_MgB2}(b-d)).

A closely related material family includes graphite intercalated compounds such as CaC$_6$ and YbC$_6$~\cite{emery2005superconductivity}. In analogy to MgB$_2$, these are multiband superconductors but with 2D $\pi$ bands derived from stacked graphite sheets, and 3D free-electron-like interlayer bands derived from the $s$-orbitals of the intercalant atoms. The superconductivity was theoretically proposed to be a result of electron-phonon interactions between the $\pi$ and interlayer bands \cite{Boeri_2007_electron,Sanna_2007_anisotropic,Calandra_2005_theoretical}. Early ARPES work reported mode-coupling on the graphite bands \cite{Valla_2009_anisotropic} and evidence for the existence of the interlayer bands \cite{Sugawara_2009_Fermi}, with later work evidencing the interband electron-phonon coupling and superconducting gap on both bands \cite{yang2014superconducting}. More recently, similar mechanisms have been invoked to explain superconductivity in decorated monolayer graphene~\cite{Fedorov_2014_observation,ludbrook2015evidence}.

\subsubsection{Bismuthates}

Bismuthate superconductors, most in the doped forms of Ba$_{1-x}$K$_x$BiO$_3$ and BaPb$_{1-x}$Bi$_x$O$_3$, have the highest $T_c$ (34~K and 13~K respectively) of all oxide superconductors predating the  cuprates~\cite{sleight1993high,cava1988superconductivity}. The stoichiometric parent compound is a Bi$^{3+/5+}$ mixed valence perovskite semiconductor at room temperature, where the differing ionic radii of Bi$^{3+/5+}$ cause the BiO$_6$ octahedra to buckle alternately into a robust charge order~\cite{cox1976crystal}. Ion replacements not only introduce charge carriers, but also cause an orthorhombic to tetragonal structural transition that accompanies the emergence of superconductivity~\cite{sleight2015bismuthates}.

The low energy electronic structure of BaBiO$_3$ is predominately composed of O 2$p$ electrons, where a band gap in excess of 0.4~eV is seen by ARPES on \textit{in-situ} grown thin films~\cite{plumb2016momentum}. Strong $k_z$ dispersion and an isotropic 3-dimensional single Fermi surface are observed in slightly K-overdoped bulk compound ($T_c$ = 22~K), where long range Coulomb interaction is postulated to account for the \textit{expanded} bandwidth and enhanced electron-phonon coupling~\cite{wen2018unveiling}. Superconducting gap measurements also show highly isotropic momentum structure; a $2\Delta/k_BT_c$ ratio at the $s$-wave BCS limit is observed. Strong electron-phonon coupling causes a dispersion anomaly around 50~meV binding energy, with $\lambda\sim$ 1.3~\cite{wen2018unveiling}. Taking into account moderate electronic correlation enhancement, recent $GW$ perturbation theory calculations successfully reproduced such a strong electron-phonon coupling constant, thus ascribing the superconducting mechanism to conventional phonon-mediated $s$-wave BCS type~\cite{li2019electron,wen2018unveiling}.

\subsection{Cobaltates and Rhodates}

Cobaltates and rhodates exhibit many symmetry breaking phases in their temperature-doping phase diagrams, but are most well known for their thermoelectric properties~\cite{foo2004charge}. Their excellent figure of merit is interpreted as the combined effect of a large Seebeck coefficient, high electrical conductivity and low thermal conductivity. While the low thermal conductivity is usually explained by anharmonic lattice rattling~\cite{voneshen2013suppression}, the large Seebeck coefficient used to be understood as a consequence of large spin-orbital entropy~\cite{wang2003spin}. However, direct ARPES measurement of the quasiparticle dispersion instead shows that the Seeback coefficient is attributed to the combined effects of a peculiar flat band top, with  electronic correlation and electron-phonon coupling induced mass renormalization~\cite{chen2017large,kuroki2007pudding}. In particular, by comparing the fully occupied $t_{2g}$ bandwidths between the stronger correlated sodium cobaltate and the weaker correlated potassium rhodates, the electronic correlation is shown to double the Seebeck coefficient from the rhodates to the cobaltates~\cite{chen2017large}. Via similar approaches, the presence of well-defined quasiparticles in Sr$_2$RhO$_4$ also enabled a direct, quantitative derivation of thermodynamic properties from low-energy single-particle spectra~\cite{baumberger2006fermi}.

The doping evolution of the Fermi surface shape and volume in Na$_x$CoO$_2$ also highlights various many-body effects. Single particle hopping as small as 10~meV and strong band renormalization are argued in the Curie-Weiss metallic phase~\cite{hasan2004fermi,yang2004arpes}, although the total bandwidth is oberved to be much larger. At $x$ = 1/3 doping, the hexagonal Fermi surface is shown to exhibit a strong CDW instability at a nesting wave vector that corresponds to the cobalt sublattice ~\cite{qian2006quasiparticle,yang2007angle}. At the actual charge ordered doping $x$ = 1/2, the Fermi surface is argued to better trace the low-temperature-ordered sodium sublattice~\cite{qian2006quasiparticle}.  Measurements of both the non-superconducting and a hydrated superconducting variant indicate a single large Fermi surface originating from cobalt's $a_{1g}$ band, whereas the zone corner $e_g^{'}$ band always remains below, though near, the chemical potential~\cite{hasan2004fermi,shimojima2006angle}. Similar observation has also been made in the weaker correlated potassium rhodates~\cite{chen2017large}. The missing $e_g^{'}$ pocket has since been suggested as an extrinsic surface termination artifact~\cite{pillay2008electronic}, a disorder effect~\cite{singh2006destruction}, and a correlation effect~\cite{bourgeois2009dynamical}; but no consensus has been reached. 

\subsection{Ruthenates}

\begin{figure} 
	\includegraphics[width=1\columnwidth]{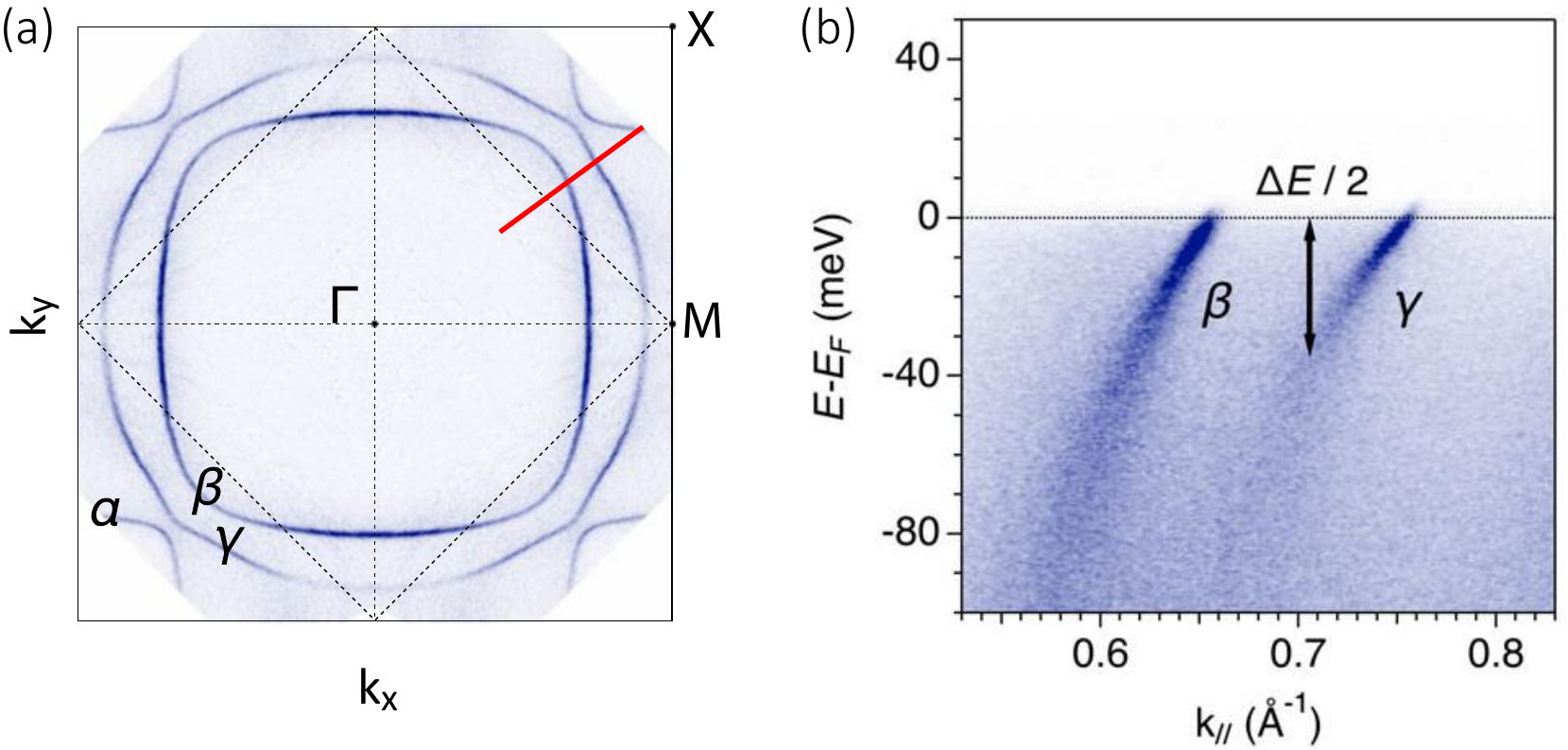}
	\caption{High resolution laser ARPES on Sr$_2$RuO$_4$. (a) Fermi surface on a CO passivated Sr$_2$RuO$_4$. (b) Spectral intensity on the energy-momentum cut along the red line in (a), where the band splitting is used to extract the spin-orbit coupling strength. Reproduced from ~\cite{tamai2019high}} 
	\label{Fig_others_iridate}
\end{figure}

In contrast to $3d$ transition metal compounds which are dominated by electronic correlation effects, the heavier and more orbitally-extended 4$d$/5$d$ transition metal compounds have a moderate Coulomb interaction $U$ at a comparable scale to the Hund's coupling $J_H$ and spin-orbit coupling $\lambda$. As one prominent example, ruthenates have driven the development of new experimental techniques since their discovery~\cite{maeno1994superconductivity}. Sr$_2$RuO$_4$ exhibits putative unconventional superconductivity that is highly tunable by strain~\cite{ishida1998spin,pustogow2019pronounced,hicks2014strong,steppke2017strong}, has one of the cleanest 2-dimensional Fermi liquid normal states up to $\sim$25~K~\cite{mackenzie1996quantum}, a strange high temperature phase that overshoots the Mott-Ioffe-Regel limit~\cite{tyler1998high,cao2004violation}, and strong spin-orbit interaction~\cite{mackenzie2003superconductivity}.

Early photoemission experiments observed a rather complicated Fermi surface in \textit{in-situ} cleaved Sr$_2$RuO$_4$ single crystals. Not fully accounted for within the quantum oscillation results, some of the Fermi pockets were later recognized as surface reconstruction effects~\cite{okuda1998fermi,damascelli2001fermi,ding2001band,shan2012effect}. After deliberate surface passivation, the surface states disappear and three sharpened bulk bands - $\alpha$, $\beta$, $\gamma$ (Fig.~\ref{Fig_others_iridate}(a)(b)) remain~\cite{tamai2019high}. A cascade of low-energy dispersion anomalies are identified between 15-80~meV on both the $\alpha$ and $\beta$ bands~\cite{aiura2004kink,ingle2005quantitative,iwasawa2005orbital,kim2011self}, and a cuprate-like high energy anomaly at 700-800~meV also signals the presence of electronic correlation effects~\cite{iwasawa2012high}. High resolution ARPES measurements enabled by a new generation of deep UV laser light source allow for detailed band structure measurement and full Fermi surface mapping covering almost the entire Brillouin zone~\cite{he2016invited,tamai2019high}. This makes it possible to perform full-momentum extraction of the electronic self energy via a band-orbital basis transformation, leading to the revelation that the anisotropic self energy is mainly a result of momentum-dependent orbital content mixing~\cite{tamai2019high,haverkort2008strong}. This is further supported by spin-resolved ARPES measurements on all three bulk bands, where the apparent spin-orbit coupling strength $\lambda$ is estimated to be $\sim$130$\pm$30~meV ~\cite{veenstra2014spin}. By comparing the energy and momentum splitting of $\beta$ and $\gamma$ bands with DFT and dynamic mean-field theory (DMFT) calculations, $\lambda$ is re-evaluated at 200~meV, accounting for both electronic correlation enhancement and quasiparticle coherence factor renormalization~\cite{tamai2019high,haverkort2008strong,kim2018spin}. It is worth pointing out that due to the momentum-independence of the spin-orbit self energy in an orbital basis, it also offers an ideal material platform to benchmark DMFT.

The physical properties of the ruthenate family are also heavily influenced by a low-lying van Hove point, both in Sr$_2$RhO$_4$~\cite{shen2007evolution} and Sr$_3$Ru$_2$O$_7$~\cite{tamai2008fermi}. Tunneling and ARPES experiments consistently identify high electron densities within $\sim$6~meV of the Fermi level, which is interpreted as heavy Ru 4$d_{xy}$ electrons~\cite{iwaya2007local,lee2009heavy,allan2013formation}. In Sr$_2$RuO$_4$, the $\gamma$ band van Hove singularity near the zone boundary was first predicted then observed to cross the Fermi level when subjected to strain applied via substrate lattice mismatch on thin film samples~\cite{burganov2016strain} or mechanically strained bulk crystals \cite{sunko2019direct}. \textit{In-situ} tuning of compressive strain up to -4.1\% on the closely related (Ca,Pr)$_2$RuO$_4$ single crystal also causes the quasiparticles to appear on the Fermi surface, inducing a insulator-to-metal transition~\cite{ricco2018situ} (see Fig.~\ref{Fig_tech_misc}(c)).

\subsection{Iridates}

The Ruddlesden Popper (RP) series iridate Sr/Ba$_{n+1}$Ir$_n$O$_{3n+1}$ with perovskite structure attracted great research interest due to its variety of exotic magnetic and electronic phases. The single-layer compound ($n=1$) is considered a spin-orbit coupled Mott insulator. The system antiferromagnetically orders at 240~K (with Sr) or 230~K (with Ba) with a total angular momentum $J_\text{eff}$ = 1/2~\cite{kim2008novel,kim2009phase,moser2014electronic,uchida2014correlated,moon2008dimensionality}. As a system that also defies expected metallicity and possesses correlation induced antiferromagnetism, Sr$_2$IrO$_4$ has been widely considered as a second gateway to illuminate the cuprate high-$T_c$ problem~\cite{wang2011twisted,watanabe2013monte}. The bilayer compound ($n=2$) is a semiconductor that forms $c$-axis collinear antiferromagnetic order at 285~K with a weak ordering moment~\cite{cao2002anomalous}. And at n = $\infty$, similar to the situation in ruthenate and manganite RP series, the system develops metallicity and becomes a correlated metal~\cite{moon2008dimensionality,nie2015interplay,cao2018challenge}. While its magnetism has been mainly investigated with resonant x-ray scattering and neutron scattering techniques~\cite{kim2014excitonic,rau2016spin}, photoemission plays an important role in revealing the corresponding evolution in the electronic structures.

\begin{figure} 
	\includegraphics[width=1\columnwidth]{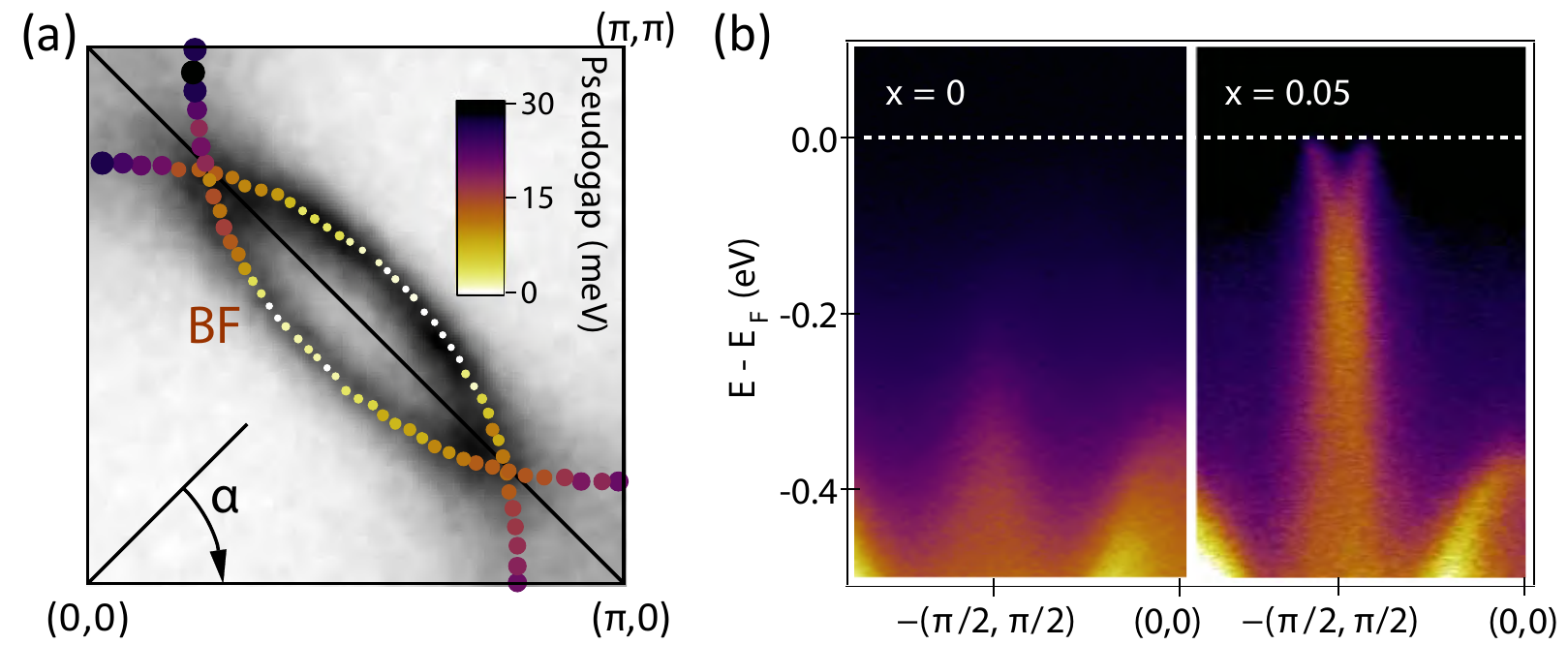}
	\caption{Fermi surface and development of quasiparticles in iridates. (a) Fermi surface overlaid with the anisotropic energy gap (colored dots) in Sr$_2$IrO$_4$. (b) Emergence of quasiparticles with electron doping in Sr$_2$IrO$_4$. Adapted from ~\cite{de2015collapse}.}
	\label{Fig_others_ruthenate}
\end{figure}

Large incoherent spectral gaps are observed in both stoichiometric Sr$_2$IrO$_4$ and Sr$_3$Ir$_2$O$_7$, which may possess topologically nontrivial surface states after \textit{in-situ} sample cleaving~\cite{wojek2012insulator,liu2014spin,brouet2015transfer,de2014coherent}. Doping La on the Sr site or Rh on the Ir site drives a metal-insulator transition~\cite{ge2011lattice,li2013tuning,chen2015influence}, during which clear low energy quasiparticle spectral weight develops~\cite{brouet2015transfer,de2014coherent,de2015collapse,he2015fermi} (Fig.~\ref{Fig_others_ruthenate}(b)). The stabilizing role of spin-orbit coupling in the Mott insulating state was recently disentangled from the doping effect via controlled Ru- and Rh- doping and careful analysis of orbital contents~\cite{zwartsenberg2020spin}.  The Fermi surface develops a band folding that coincides with both the antiferromagnetic ordering wave vector and a structural distortion due to IrO$_6$ octahedron rotation. The latter also exists in non-magnetic isostructural rhodates (Fig.~\ref{Fig_others_ruthenate}(a))~\cite{he2015fermi,de2014coherent,de2015collapse,baumberger2006fermi}. This folded Fermi surface manifests as an ``arc''-like feature that resembles the Fermi arc in cuprates (see Section~\ref{sec_fermiology})~\cite{kim2014fermi,de2014coherent,he2015fermi}. With either bulk or surface carrier doping, the low energy spectra indeed exhibit an anisotropic energy gap that lacks clear quasiparticles, prompting comparison to the pseudogap phenomenon or even superconductivity in the cuprates (Fig.~\ref{Fig_others_ruthenate}(b))~\cite{de2015collapse,battisti2017universality,kim2016observation,yan2015electron}. Measurements above the N\'eel temperature show no sign of spectral gap closing, indicating the Mott insulating nature of the system~\cite{moser2014electronic}. The electronic correlation may also give rise to negative electronic compressibility -- a lowering chemical potential with electron addition, providing a potential microscopic explanation for the tendency towards phase separation in doped iridates~\cite{he2015spectroscopic,chen2015influence}.

Another major iridate family \textit{Ln}$_2$Ir$_2$O$_7$ (Ln = lanthanide series or Bi) is of the pyrochlore structure. With increasing lanthanide radius, the ground state gradually evolves from magnetic insulator (\textit{Ln} = Lu - Nd) to paramagnetic correlated metal (\textit{Ln} = Pr)~\cite{matsuhira2011metal}. ARPES experiments in this family remain challenging, mainly due to the lack of an easy cleaving plane. In the metallic compound Pr$_2$Ir$_2$O$_7$, a cubic and time-reversal symmetry protected Fermi node is observed by ARPES, supporting the system as a correlated topological material~\cite{kondo2015quadratic,wan2011topological}. Further temperature dependent experiments on the all-in-all-out spin ordered compound Nd$_2$Ir$_2$O$_7$ shows a highly 3-dimensional metallic normal state with a similar Fermi node~\cite{nakayama2016slater,guo2016direct}. However, while the single particle gap only opens below the magnetic transition, the quasiparticle also gradually loses spectral coherence approaching zero temperature. This indicates successive transitions from a metal to a Slater insulator, then to a Mott insulator~\cite{nakayama2016slater}.

Quantum spin liquid candidate honeycomb iridates Na$_2$IrO$_3$ and $\alpha$-Li$_2$IrO$_3$ are formed by edge-sharing IrO$_6$ octahedra. This particular geometry contains 90$^{\circ}$ Ir-O-Ir bonds that promote Kitaev interaction, in contrast to the near-linear bond that promotes Heisenberg interaction in the corner-sharing perovskite iridates~\cite{jackeli2009mott}. While most of the focus remains on the magnetic degrees of freedom, photoemission confirms that the size of the Mott gap ($\sim$340~meV) is comparable to the spin-orbit coupling strength ($\sim$500~meV)~\cite{comin2012Na}. With spatially resolved synchrotron ARPES, spurious conductive surface states are observed to occur on Na-terminated regions, postulated to locally kill the Mott gap~\cite{alidoust2016observation,moreschini2017quasiparticles}.

\subsection{Delafossite oxides}

The materials PdCoO$_2$, PtCoO$_2$, and PdCrO$_2$ are anisotropic metals with remarkably high in-plane conductivity. The conduction occurs in the Pd/Pt layers separated by insulating CoO$_2$/CrO$_2$ layers; we refer to \cite{Mackenzie_2017_the} for a comprehensive review.  The bulk electronic structure as measured by ARPES consists of a single 2D band crossing the Fermi level with a hexagonal cross-section which exhibits very weak correlation effects \cite{Kushwaha_2015_nearly}. This unusually clean electronic structure makes these materials an excellent model system for studying electronic interactions using ARPES.

The cleaved surface is polar, leading to the formation of surface states clearly visible in  ARPES. They are not protected as in topological materials, as evidenced by the relative ease with which they can be removed by disorder \cite{Noh_2009_anisotropic, Sobota_2013_electronic}. The  surface states of PtCoO$_2$ exhibit a large Rashba-like spin splitting attributed to the atomic spin-orbit coupling of the relatively lightweight Co, which is unlocked by the unusually large magnitude of inversion symmetry breaking at the CoO$_2$-terminated surface \cite{Sunko_2017_maximal}. Another interesting aspect of the surface states is that they can host ferromagnetic order  \cite{Mazzola_2018_Itinerant}. Independently, the bulk of these materials can also exhibit magnetic order: in PdCrO$_2$, the localized Cr$^{3+}$ ions ($S=3/2$) exhibit a $120^{\circ}$ spin structure, making it possible to study the interaction of itinerant electrons with a localized antiferromagnetic structure \cite{Takatsu_2009_critical}. In ARPES this interaction manifests as a folding of the itinerant Pd bands with respect to the antiferromagnetic zone boundary, which vanishes above the N\'eel temperature $T_N=37.5$~K \cite{Noh_2015_direct}. This folding was proposed to arise from a novel mechanism which convolves the ARPES spectrum of the itinerant layer with the spin susceptibility of the AFM layer \cite{Sunko_2018_probing}.

\subsection{Heavy fermion systems}\label{sec_heavyFermion}

Heavy fermion systems are typically  rare earths or actidines with partially filled 4$f$ or 5$f$ orbitals, in which the charge carriers exhibit an effective mass up to three orders of magnitude times larger than that of a bare electron. The essential physics can be described in terms of a lattice of localized $f$-electron moments interacting with an interpenetrating sea of conduction electrons. Mediated by coherent Kondo scattering, the local moments form a many-body spin singlet with the conduction electrons. This results in the formation of a composite quasiparticle which inherits the mass of the $f$-electrons, and is expected to increase the Fermi surface volume due to incorporation of the $f$ degrees-of-freedom. At the same time, the interplay between RKKY interactions and Kondo screening leads to an antiferromagnetic quantum critical point, often accompanied by unconventional superconductivity \cite{Hewson_1993_the,Si_2010_heavy}. Since the antiferromagnetism and superconductivity manifest at relatively low ($<10$~K) temperature scales, these phases and their associated quantum criticality have been extensively studied by transport and thermodynamics measurements, while ARPES has focused on higher-energy physics such as the degree of $f-d$ hybridization.

\begin{figure}
\centering
\includegraphics[width=\columnwidth]{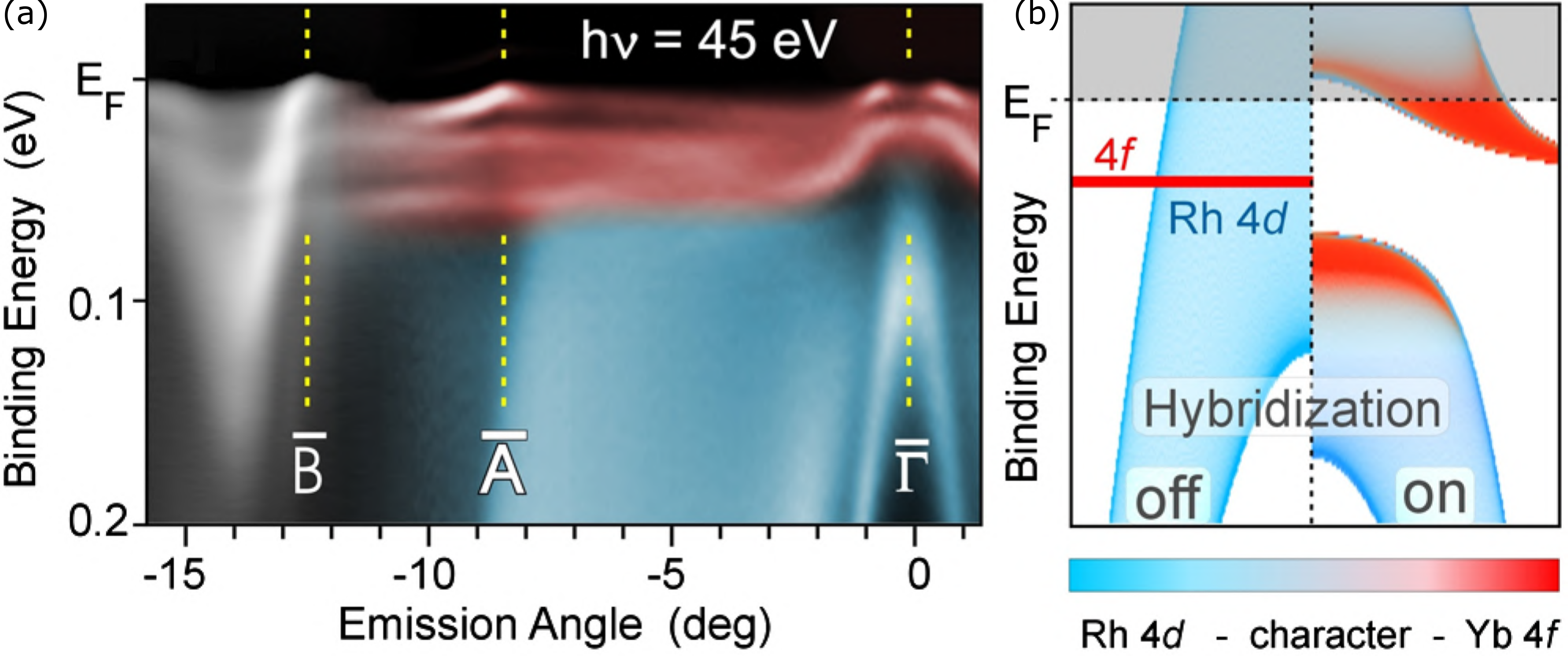}
\caption[Heavy fermion materials]{(a) ARPES spectrum from the heavy fermion system YbRh$_2$Si$_2$, and (b) cartoon of the spectral function as described by the periodic Anderson model. The renormalized $f$-level, or Kondo resonance, exists below $E_{\textrm{F}}$ (shaded red) and hybridizes with the dispersive Rh 4$d$ conduction electrons. From \cite{Danzenbacher_2011_insight}.
\label{Fig_Others_HeavyFermion}}
\end{figure}

Since the $f$-electron cross section is strongly $h\nu$-dependent, with diminishing intensity below $\sim50$~eV \cite{Yeh_1985_atomic}, $h\nu> 100$~eV is routinely employed to exploit the enhanced $f$-electron signal. Moreover, soft x-ray ($>500$~eV) ARPES has been useful for suppressing the contribution of surface states and achieving true bulk sensitivity, albeit at the cost of compromised energy resolution \cite{Yano_2007_three}. Another technical yet important limitation is imposed by safety protocols concerning transuranic compounds, spurring the development of separate dedicated ARPES facilities for these materials \cite{Graham_2013_integrated}.

Much of the ARPES work on $f-d$ hybridization has investigated $4f$ Ce- and Yb-based compounds. For these materials, the spectral function is generally well-described by the periodic Anderson model: the binding energy of the bare $f$-electrons is renormalized by correlations, forming a non-dispersive band near $E_{\textrm{F}}$ (also known as the Kondo resonance) which then hybridizes with the dispersive $d$-electron bands \cite{Denlinger_2001_comparative}. One advantage of Yb- over Ce- compounds for ARPES is that the Kondo resonance is below $E_{\textrm{F}}$ \cite{Fujimori_2016_band}. Fig.~\ref{Fig_Others_HeavyFermion}(a) and (b) show the ARPES spectrum of a prototypical heavy fermion material YbRh$_2$Si$_2$, together with a cartoon of the periodic Anderson model: the 4$f$ Kondo resonance below $E_{\textrm{F}}$ (shaded red) is incorporated into the Fermi surface by hybridizing with the Rh 4$d$ bands (shaded blue) \cite{Danzenbacher_2011_insight}. Here a multitude of flat bands are observed due to crystal-field splitting of the 4$f$ levels \cite{Vyalikh_2010_k}. One of the central questions concerns the temperature scale associated with the $f-d$ hybridization. ARPES measurements on YbRh$_2$Si$_2$ found no significant changes from 1~K to 100~K \cite{Kummer_2015_temperature}, while measurements on  CeCoIn$_5$ suggest that dehybridization occurs above $\sim$200~K \cite{Chen_2017_direct,Jang_2017_evolution}.  Surprisingly, these values vastly exceed the temperature scales for coherent Kondo scattering inferred from resistivity measurements \cite{Trovarelli_2000_YbRh2Si2, Petrovic_2001_heavy}. Further understanding is required to reconcile temperature-dependent thermodynamic and transport properties with the single-particle spectral function measured by ARPES.

$5f$-electrons have been studied in U-based compounds. While some materials such as UPd$_3$, UGe$_2$, and USb$_2$ do seem well-described by the periodic Anderson model \cite{Beaux_2011_electronic}, other materials such as UFeGa$_5$ are better understood in an itinerant $5f$-electron model \cite{Fujimori_2006_itinerant}. One U-compound which has attracted significant attention is URu$_2$Si$_2$ due to the observation of a phase transition in the specific heat at $T_{\textrm{HO}}=17.5$~K \cite{Palstra_1985_superconducting}.  Though apparently of magnetic origin, this has come to be known as the ``hidden order'' phase since magnetic order remains mysteriously unobserved \cite{Durakiewicz_2014_photoemission}. ARPES revealed the emergence of a flat band near $E_{\textrm{F}}$ in the hidden order phase \cite{Syro_2009_Fermi}, and attributed it to a doubling of the unit cell along the $c$-axis which folds the $\Gamma$-point to the Z-point \cite{Yoshida_2010_signature} and leads to dramatic Fermi surface reconstruction \cite{Bareille_2014_momentum}. Other recent results suggest that these flat bands already exist at higher temperatures, but below $T_{\textrm{HO}}$ they rapidly hybridize with the conduction electrons to form sharp spectral features \cite{Chatterjee_2013_formation}. This contrasts sharply with the high-temperature gradual onset of hybridization discussed in $4f$-systems above, and therefore appears to be a distinct signature of the hidden order transition.

\subsection{Extreme magnetoresistance semimetals}\label{sec_XMR}

Since 2014 there has been a surge of research on semimetals which exhibit large magnetoresistance such as WTe$_2$ \cite{Ali_2014_large}, Cd$_3$As$_2$ \cite{Liang_2015_ultrahigh}, LaSb \cite{Tafti_2016_resistivity}, among others in related families. This effect has been termed extreme magnetoresistance (XMR) due to the large magnitude ($>10^6$\%) and non-saturating behavior up to remarkably high magnetic fields. The mechanism for XMR has been under some debate; with many of these materials exhibiting some form of topological order, some works have suggested that the XMR is associated with the lifting of topological protection by the external magnetic field \cite{Shekhar_2015_extremely,Tafti_2016_resistivity,Liang_2015_ultrahigh}. Others have argued for a conventional carrier compensation picture \cite{Pippard_1989_magnetoresistance}, in which the XMR derives from a nearly equal concentration of electrons and holes such as in WTe$_2$ \cite{Pletikosic_2014_electronic} and LaSb \cite{Zeng_2016_compensated}. In topologically-trivial materials such as YSb, the XMR is explained in terms of imbalanced carrier concentrations complemented with substantially different electron and hole mobilities  \cite{He_2016_distinct}. More evidence against an ostensible clean role of topology is provided by a comparative ARPES study of LaX (X=Bi,Sb,As) which showed that these materials belong to different topological classes despite all exhibiting XMR  \cite{Nummy_2018_measurement}. On the other hand, the carrier concentrations were shown to be strongly imbalanced in the topologically-non-trivial LaBi  \cite{Jiang_20118_observations}, further raising questions about whether universal conclusions can be drawn on the relative roles of topology and carrier-compensation in this class of materials.

\subsection{Rare-earth tritellurides}\label{sec_RTe3}

\begin{figure}
\centering
\includegraphics[width=\columnwidth]{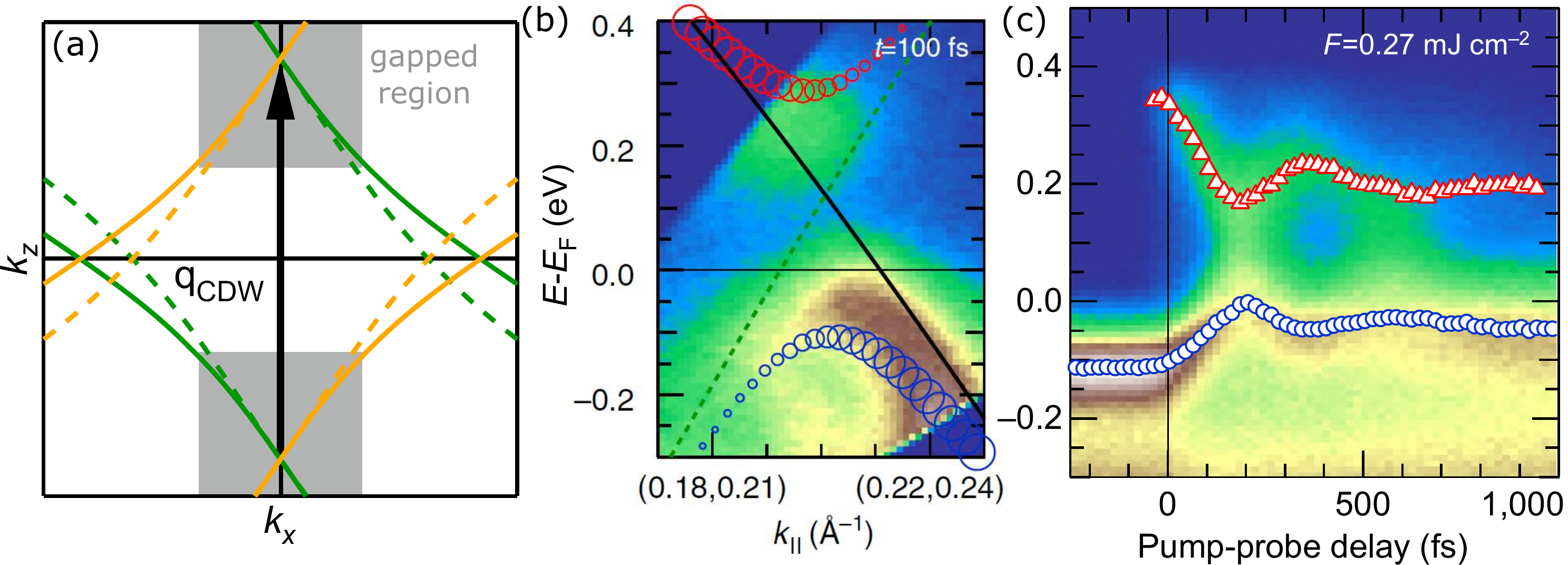}
\caption[Rare-earth tritellurides]{trARPES studies on the CDW in RTe$_3$. (a) Tight-binding Fermi surface for DyTe$_3$, where  orange and green represent the $p_x$ and $p_z$ orbitals, respectively. Dashed lines represent the folded shadow bands. The CDW ordering vector $q_{\textrm{CDW}}$ and gapped region are indicated. (b) Cut through the gapped region after pumping, showing the gap and shadow bands. (c) Due to photoexcitation of carriers, both sides of the gap can be tracked, revealing a partial closure and oscillation of the gap amplitude. Adapted from \cite{Rettig_2015_persistent}.
\label{Fig_others_CDW}}
\end{figure}

The CDW is a prototypical ordering phenomenon in condensed matter which exemplifies the role of electron-lattice interactions. In a canonical Peierls scenario, the electronic energy gain of a lattice distortion overwhelms the elastic energy cost, leading to a divergence of the electronic susceptibility at the wave vector which nests the Fermi surface ($q = 2k_{\textrm{F}}$) and an accompanying inter-unit cell charge modulation \cite{Gruner_1994_density}. However, doubts have been raised whether this concept of ``Fermi surface nesting'' can be applied to real materials with finite temperature, scattering rates, and imperfect nesting geometries; instead, the $q$-dependent electron-phonon coupling for all occupied states must be considered \cite{johannes2008fermi}. We further note that strong electronic correlation effects can also give rise to exotic valence electron CDW or excitonic insulating states, which have a less pronounced influence on the lattice than the aforementioned mechanisms (see Sections~\ref{sec_cuprates} and \ref{sec_TMDC}).

The rare-earth tritellurides (RTe$_3$) have been a model system for ARPES to study CDWs. Structural studies have revealed an incommensurate CDW in a broad range of materials (R=La,Sm,Gd,Tb,Dy,Ho,Er,Tm) with transition temperatures in the range $\sim240\sim420$~K \cite{Ru_2008_effect,DiMasi_1995_chemical}.  The normal-state electronic structure is well-described by a tight-binding model with weakly-hybridized quasi-1D $p_x$ and $p_z$ orbitals (see Fig.~\ref{Fig_others_CDW}(a)). Upon entering the CDW state, ARPES shows that the bands are folded by $q_{\textrm{CDW}}$ leading to gapping of the Fermi surface and the formation of shadow bands \cite{Gweon_1998_direct,Brouet_2004_fermi,Brouet_2008_angle,Moore_2010_fermi}. Although many of these results are discussed in a canonical nesting-driven scenario, it has been pointed out that $q$-dependent electron-phonon coupling can have important contributions to determining $q_{\textrm{CDW}}$ \cite{Eiter_2013_alternative,Maschek_2015_wave}. 

trARPES has been extensively employed to investigate the the dynamics of the order parameter \cite{Schmitt_2008_transient,Rettig_2014_coherent,Leuenberger_2015_classification}. Fig.~\ref{Fig_others_CDW}(b) shows a cut through the gapped region after pumping, where the normally-unoccupied side of the gap (above $E_{\textrm{F}}$) is visible due to a non-equilibrium population of electrons.  As shown in Fig.~\ref{Fig_others_CDW}(c), the gap not only reduces in magnitude but also oscillates, reflecting a coherent modulation of the order parameter known as the \emph{amplitude mode}. Interestingly, it has been suggested that the nesting conditions themselves are dynamically modified, raising the prospect of stabilizing order using ultrafast excitations \cite{Rettig_2015_persistent}.

\subsection{Manganese oxides}

The manganese oxides (manganites) have been the subject of intense investigation because they exhibit colossal magnetoresistance (CMR), in which the conductivity changes by orders of magnitude upon application of a magnetic field \cite{Ramirez_1997_colossal}. Unlike the XMR effect in semimetals, this phenomenon is a manifestation of the competition between many-body interactions including structural, orbital, and spin degrees-of-freedom, resulting in a complex phase diagram which hosts a variety of magnetic phases. Several ARPES studies have focused on both the single layer La$_{1-x}$Sr$_x$MnO$_3$~\cite{lev2015fermi,horiba2016isotropic} and the bilayer manganite La$_{2-2x}$Sr$_{1+2x}$Mn$_2$O$_7$~\cite{Mannella_2005_nodal} in the doping range around $x\sim0.4$, which is a ferromagnetic metal below $T_c \sim 120$~K and a paramagnetic insulator above $T_c$.  Many of the microscopic ingredients are generally agreed upon: the ferromagnetic state is mediated by double-exchange interactions between Mn moments which simultaneously favors electron delocalization.  In the high-temperature paramagnetic state, the electrons undergo self-trapping due to the strong electron-phonon interaction, forming  small polarons with a tendency to become localized at impurities. The fact that a magnetic field can tip the balance between these states is the origin of the colossal magnetoresistance \cite{Millis_1998_lattice}. There is, however, some controversy on the role of polarons.  Early ARPES work on the $x=0.4$ doping reported a pseudogapped Fermi surface with well-defined quasiparticle peaks only in the (0,0)-($\pi$,$\pi$) direction below $T_c$, suggesting a common phenomenology with the Fermi arc state in cuprates \cite{Mannella_2005_nodal}. The incoherent spectral weight was assigned to localized polarons, while the sharp peaks were taken as evidence for a ``polaronic metal'' in which polaron condensation acts in concert with double-exchange interactions to foster metallic conductivity. This hypothesis was supported by a direct correlation between the quasiparticle spectral weight and the dc conductivity  \cite{Mannella_2007_polaron}.  However, the universality of this observation was challenged by other experiments for $x = 0.36 \sim 0.38$ reporting quasiparticle peaks in the ($\pi$,0) direction persisting well above $T_c$ \cite{Sun_2006_quasiparticlelike, Sun_2007_a, deJong_2007_quasiparticles}.  More recently, a combined STM-ARPES study argued that the intrinsic Fermi surface is gapped throughout $k$-space both below and above $T_c$ in the broad doping range $x = 0.3 \sim 0.425$. Quasiparticle peaks were only observed in ARPES for $<5$\% of the cleaved surface, and were assigned to regions with stacking-fault intergrowths as separately observed in STM measurements \cite{Massee_2011_bilayer}.  Due to the reported phase separation, the Fermi surface volume is an important metric for evaluating the purity of the phase being measured. While a unifying picture is still lacking in this class of materials hosting many nearby phases, the spectral signatures of highly polaronic physics appear to be robust.

\section{Conclusion and outlook}\label{sec_outlook}

The past two decades have witnessed an explosive growth of research on quantum materials, with ARPES playing the central role as a direct probe of electronic structure in materials exhibiting strong interactions, multiorbital Hund's coupling, low dimensionality, and strong spin-orbit coupling. In parallel, theoretical calculations from first principles and model Hamiltonians have developed rapidly, greatly adding value to the scientific output of ARPES experiments. In strongly correlated states of matter, ARPES has allowed for detailed investigation of  Fermi surfaces, order parameters, and mode-couplings, in phases ranging from conventional Fermi liquid to spin-ordered, charge-ordered, or superconducting states. In topological materials,  it has directly visualized the bulk electronic structure responsible for  non-trivial topology, as well as the associated boundary states, particularly in cases in which they are inaccessible to transport probes. Advanced synthesis techniques have made it possible to explore these materials in lower dimensions, unlocking interactions and phenomena that are non-existent in their three-dimensional forms. These scientific pursuits have been accelerated by increasingly sophisticated ARPES measurement capabilities with space, spin, and time resolutions, all while pushing to ever lower temperatures, thus resolving all the  quantum numbers of the photoemitted electrons in the relevant regions of the phase diagram. Finally, while the material systems have conventionally been limited to the two-dimensional plane of temperature vs doping, the phase space for ARPES is now a multidimensional landscape featuring non-thermal tuning knobs such as mechanical strain and electrostatic gating. 

ARPES is rapidly evolving from a stand-alone characterization technique toward a platform for discovering and controlling new quantum phenomena. Conventional ARPES as a technique has matured to the stage that complete-system solutions are now commercially available, which will fast-track the popularization of its basic functionalities to the broader materials research community as well as industry \cite{Shallenberger_1998_recent,Cabuil_2007_process}. Given the rapidly growing volume of research, standardization of data acquisition, open-source data analysis routines, and data storage is imminently desired. At the same time, from the instrumentation point-of-view, we foresee a major impact due to continued development of light sources. Fourth generation synchrotrons are being built and coming online, offering much improved brightness and thus spectral and spatial resolution \cite{Maesaka_2015_comparison}. Ultrafast laser technology appears to be on the verge of a breakthrough, with Yb-based fiber sources becoming increasingly turnkey and achieving ever higher pulse energies at higher repetition rates. With continued development, it is not difficult to envision HHG deep into the VUV range in a regime where space-charge is mitigated to a sub-10~meV level, making these tunable sources  attractive for ARPES and trARPES alike. In parallel, the development of high-repetition rate FELs at large-scale user facilities will provide access to the  soft- and hard- x-ray regime, granting true bulk sensitivity \cite{Gray_2011_probing}. The photoexcitation energies in both lab-based and FEL-based trARPES will continue to push toward the mid-IR and THz range \cite{Reimann_2018_subcycle}, allowing for surgical excitation of select collective modes and driving of non-equilibrium phases.  Continued development of quasi-continuous wave lasers extending deeper into the VUV will pave the way for $\mu$eV-level resolution with sufficient $\vec{k}_{\vert\vert}$-space access to cover the entire Brillouin zone of all materials of interest \cite{He_2016_invited}. At the same time, addressing physics on the $\mu$eV-scale such as conventional superconductivity or long timescale fluctuations will require correspondingly low sample temperatures. Sub-1~K temperature has been demonstrated using helium-3 cryostats \cite{borisenko2010superconductivity}, with continued improvement expected with developments in thermal insulation and radiation shielding.

On the photoelectron spectrometer side, it is clear that multiplexing detectors will play an increasingly prominent role, especially in the domain of spin-resolved ARPES, where these efficiency gains will be used to routinely map the spin- and orbital- part of  wavefunctions, with exquisite sensitivity to local symmetries. Meanwhile, in sync with the deepening understanding of photoemission theory, expansion of single photoemission to a multi-particle probe via interference effects and multi-electron emission will spearhead the effort to directly address many-body correlation and entanglement effects~\cite{kouzakov2003photoinduced,huth2014electron,trutzschler2017band}. On the sample side, \emph{in-situ} synthesis and environment tuning will permit access to phases previously thought to be beyond the scope of an ARPES experiment~\cite{shen2017situ,trotochaud2016ambient,cattelan2018perspective,yamane2019acceptance}. Enabled by the in-lens deflector of modern electron spectrometers, the combination of electrified sample environment with nanoscale spatial-resolution will unlock a new era of \emph{in-operando} studies of fabricated devices and exfoliated heterostructures.  The pace of concerted scientific and technique co-development will continue, with the rate accelerating with more rapid iteration between experiment and theory. 

ARPES will continue to be a leading tool to push the frontier of quantum materials research, help set the intellectual agenda by testing new ideas, discovering surprises, and challenging orthodoxies.  There is little doubt that this technique is going to be at the focal point of the necessary debates leading to new paradigms of physics. 

\begin{acknowledgments}
We thank N.P. Armitage, F. Baumberger, R. Birgeneau, S.-D. Chen, X. Chen, Y.-L. Chen, T.-C. Chiang, T.P. Devereaux, A.V. Fedorov, D.L. Feng, M. Hashimoto, J.-F. He, R.-H. He, E.W. Huang, Z. Hussain, R. Kaindl, P.S. Kirchmann, D.-H. Lee, I. Lindau, Z.-K. Liu, D.-H. Lu, E.Y. Ma, N. Mannella, R.M. Martin, C.E. Matt, S.-K. Mo, J. Osterwalder, L. Perfetti, H. Pfau, W.E. Plummer, S. Shin, O. Tjernberg, A. Vishwanath, Y. Wang, S. Wu, S.-Y. Xu, M. Yi, J. Zaanen, Y. Zhang and X.J. Zhou for helpful discussions and/or a careful reading of this manuscript. We thank T.P. Devereaux, A. Fujimori, M. Hashimoto, Z. Hussain, P.S. Kirchmann, W.-S. Lee, D.-H. Lu, S.K. Mo, R.G. Moore, and B. Moritz for long term collaborations. We thank H. Eisaki, I.R. Fisher, M. Greven, A. Kapitulnik, Z.Q. Mao, J. Mitchell, T. Sasakawa, S. Tang, Y. Tokura, S. Uchida, Q.K. Xue, and Y. Zhang for sample collaborations. We thank  R.J. Birgeneau, Y.-L. Chen, M. Kastner, C. Kim, R.B. Laughlin, S.A. Kivelson, D.-H. Lee, N. Nagaosa, X.L. Qi, S. Sachdev, D.J. Scalapino, J. Zaanen, and S.C. Zhang for stimulating discussions and collaborations. The SLAC/Stanford work was supported by the DOE Office of Basic Energy Sciences, Division of Materials Science. The Stanford work was also supported by the Betty and Moore Foundation’s EPiQS Initiative through Grant GBMF4546. Y. He thanks the support from the Miller Institute for Basic Research in Science. Z.-X. Shen would like to thank the hospitality of the Radcliffe Institute of Advanced Study and Physics Department of Harvard University for his sabbatical during which this work was completed.
\end{acknowledgments}

\bibliography{RMP_Ref,bib_arpes_rmp2019_cuprates,bib_arpes_rmp2019_feSC,bib_arpes_rmp2019_lowDmat,bib_arpes_rmp2019_othermat}

\end{document}